\ifx\pdfoutput\undefined
	\documentclass[english,final,dvips]{aipproc}
\else
	\documentclass[english,final,pdftex]{aipproc}
\fi

\layoutstyle{8x11single}

\usepackage{graphicx}

\usepackage{amsmath}
\usepackage{amsfonts}
\usepackage{amssymb}

\usepackage{babel}
\usepackage{dcolumn}

\newcolumntype{e}[1]{D{.}{.}{#1}}

\begin{document}

\title{The Blackholic energy: long and short Gamma-Ray Bursts (New perspectives in physics and astrophysics from the theoretical understanding of Gamma-Ray Bursts, II)\thanks{We recall that ``New perspectives in physics and astrophysics from the theoretical understanding of Gamma-Ray Bursts, I'' was already published in {\it COSMOLOGY AND GRAVITATION: X$^{th}$ Brazilian School of Cosmology and Gravitation; 25$^{th}$ Anniversary (1977-2002)}, M. Novello, S.E. Perez-Bergliaffa (eds.), {\it AIP Conf. Proc.}, {\bf 668}, 16 (2003), see Ref.\cite{Brasile}.}}

\author{Remo Ruffini}{address={ICRA --- International Center for Relativistic Astrophysics}, altaddress={Dipartimento di Fisica, Universit\`a di Roma ``La Sapienza'', Piazzale Aldo Moro 5, I-00185 Roma, Italy.}}

\author{Maria Grazia Bernardini}{address={ICRA --- International Center for Relativistic Astrophysics}, altaddress={Dipartimento di Fisica, Universit\`a di Roma ``La Sapienza'', Piazzale Aldo Moro 5, I-00185 Roma, Italy.}}

\author{Carlo Luciano Bianco}{address={ICRA --- International Center for Relativistic Astrophysics}, altaddress={Dipartimento di Fisica, Universit\`a di Roma ``La Sapienza'', Piazzale Aldo Moro 5, I-00185 Roma, Italy.}}

\author{Pascal Chardonnet}{address={ICRA --- International Center for Relativistic Astrophysics}, altaddress={Universit\'e de Savoie, LAPTH - LAPP, BP 110, F-74941 Annecy-le-Vieux Cedex, France.}}

\author{Federico Fraschetti}{address={ICRA --- International Center for Relativistic Astrophysics}, altaddress={Universit\`a di Trento, Via Sommarive 14, I-38050 Povo (Trento), Italy.}}

\author{Vahe Gurzadyan}{address={ICRA --- International Center for Relativistic Astrophysics}, altaddress={Yerevan Physics Institute, Alikhanian Brothers Street 2, 375036, Yerevan-36, Armenia.}}

\author{Luca Vitagliano}{address={ICRA --- International Center for Relativistic Astrophysics}, altaddress={Dipartimento di Fisica, Universit\`a di Roma ``La Sapienza'', Piazzale Aldo Moro 5, I-00185 Roma, Italy.}}

\author{She-Sheng Xue}{address={ICRA --- International Center for Relativistic Astrophysics}, altaddress={Dipartimento di Fisica, Universit\`a di Roma ``La Sapienza'', Piazzale Aldo Moro 5, I-00185 Roma, Italy.}}

\begin{abstract}
We outline the confluence of three novel theoretical fields in our modeling of Gamma-Ray Bursts (GRBs): 1) the ultrarelativistic regime of a shock front expanding with a Lorentz gamma factor $\sim 300$; 2) the quantum vacuum polarization process leading to an electron-positron plasma originating the shock front; and 3) the general relativistic process of energy extraction from a black hole originating the vacuum polarization process. There are two different classes of GRBs: the long GRBs and the short GRBs. We here address the issue of the long GRBs. The theoretical understanding of the long GRBs has led to the detailed description of their luminosities in fixed energy bands, of their spectral features and made also possible to probe the astrophysical scenario in which they originate. We are specially interested, in this report, to a subclass of long GRBs which appear to be accompanied by a supernova explosion. We are considering two specific examples: GRB980425/SN1998bw and GRB030329/SN2003dh. While these supernovae appear to have a standard energetics of $10^{49}$ ergs, the GRBs are highly variable and can have energetics $10^4$ -- $10^5$ times larger than the ones of the supernovae. Moreover, many long GRBs occurs without the presence of a supernova. It is concluded that in no way a GRB can originate from a supernova. The precise theoretical understanding of the GRB luminosity we present evidence, in both these systems, the existence of an independent component in the X-ray emission, usually interpreted in the current literature as part of the GRB afterglow. This component has been observed by Chandra and XMM to have a strong decay on scale of months. We have named here these two sources respectively URCA-1 and URCA-2, in honor of the work that George Gamow and Mario Shoenberg did in 1939 in this town of Urca identifying the basic mechanism, the Urca processes, leading to the process of gravitational collapse and the formation of a neutron star and a supernova. The further hypothesis is considered to relate this X-ray source to a neutron star, newly born in the Supernova. This hypothesis should be submitted to further theoretical and observational investigation. Some theoretical developments to clarify the astrophysical origin of this new scenario are outlined. We turn then to the theoretical developments in the short GRBs: we first report some progress in the understanding the dynamical phase of collapse, the mass-energy formula and the extraction of blackholic energy which have been motivated by the analysis of the short GRBs. In this context progress has also been accomplished on establishing an absolute lower limit to the irreducible mass of the black hole as well as on some critical considerations about the relations of general relativity and the second law of thermodynamics. We recall how this last issue has been one of the most debated in theoretical physics in the past thirty years due to the work of Bekenstein and Hawking. Following these conceptual progresses we analyze the vacuum polarization process around an overcritical collapsing shell. We evidence the existence of a separatrix and a dyadosphere trapping surface in the dynamics of the electron-positron plasma generated during the process of gravitational collapse. We then analyze, using recent progress in the solution of the Vlasov-Boltzmann-Maxwell system, the oscillation regime in the created electron-positron plasma and their rapid convergence to a thermalized spectrum. We conclude by making precise predictions for the spectra, the energy fluxes and characteristic time-scales of the radiation for short-bursts. If the precise luminosity variation and spectral hardening of the radiation we have predicted will be confirmed by observations of short-bursts, these systems will play a major role as standard candles in cosmology. These considerations will also be relevant for the analysis of the long-bursts when the baryonic matter contribution will be taken into account.
\end{abstract}

\maketitle 

\section{Introduction} 

In the last century the fundamental discoveries of nuclear physics have led to the understanding of the thermonuclear energy source of main sequence stars and explained the basic physical processes underlying the solar luminosity (see e.g. M. Schwarzschild \cite{msch}).

The discovery of pulsars in 1968 (see Hewish et al. \cite{hbpsc68}) led to the first evidence for the existence of neutron stars, first described in terms of theoretical physics by George Gamow as far back as 1936 \cite{gam36}. It became clear that the pulsed luminosity of pulsars, at times $10^2$ -- $10^3$ larger than solar luminosity, was not related to nuclear burning and could be simply explained in term of the loss of rotational energy of a neutron star (Gold \cite{g68,g69}). For the first time it became so clear the possible relevance of strong gravitational fields in the energetics of an astrophysical system.

\begin{figure} 
\centering 
\includegraphics[width=\hsize,clip]{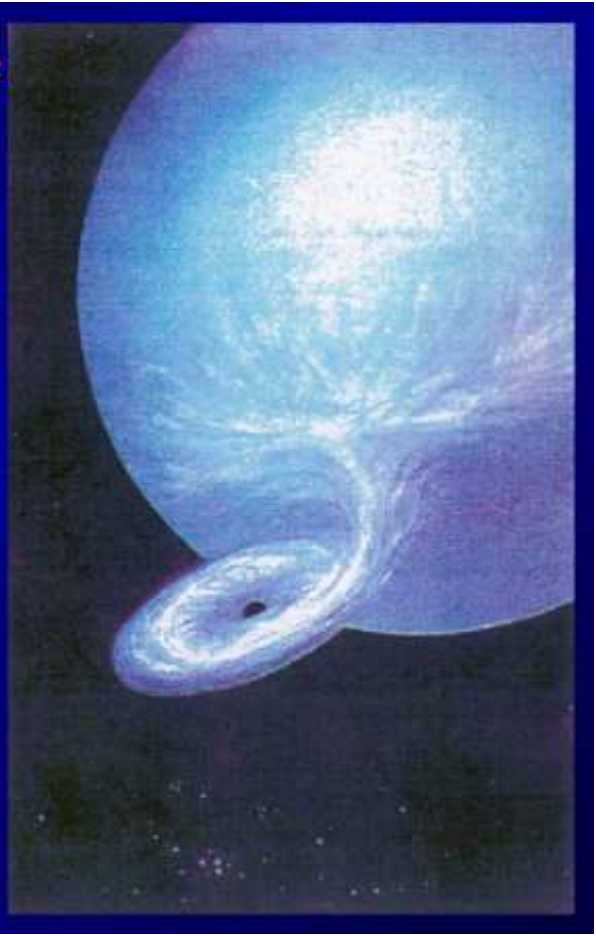} 
\caption{Cygnus X-1 offered the possibility of identifying the first black hole in our galaxy (Leach and Ruffini \cite{lr73}). The luminosity $\Phi$ of $10^4$ solar luminosities points to the accretion process into a neutron star or a black hole as the energy source. The absence of pulsation is naturally explained either by a non-magnetized neutron star or a Kerr-Newmann black hole, which has necessarily to be axially symmetric. What identifies the black hole unambiguously is that the mass of Cygnus X-1, larger than $9M_\odot$, exceeds the absolute upper limit of the neutron star mass, estimated at $3.2M_\odot$ by Rhoades and Ruffini \cite{rr74}.}
\label{CygX1} 
\end{figure}

The birth of X-ray astronomy thanks to Riccardo Giacconi and his group (see e.g. Giacconi and Ruffini \cite{gr78}) led to a still different energy source, originating from the accretion of matter onto a star which has undergone a complete gravitational collapse process: a black hole (see e.g. Ruffini \& Wheeler \cite{rw71}). In this case, the energetics is dominated by the radiation emitted in the accretion process of matter around an already formed black hole. Luminosities up to $10^4$ times the solar luminosity, much larger then the ones of pulsars, could be explained by the release of energy in matter accreting in the deep potential well of a black hole (Leach and Ruffini \cite{lr73}). This allowed to probe for the first time the structure of circular orbits around a black hole computed by Ruffini and Wheeler (see e.g. Landau and Lifshitz \cite{ll2}). This result was well illustrated by the theoretical interpretation of the observations of Cygnus-X1, obtained by the Uhuru satellite and by the optical and radio telescopes on the ground (see Fig. \ref{CygX1}).

The discovery of gamma-ray bursts (GRBs) sign a further decisive progress. The GRBs give the first opportunity to probe and observe a yet different form of energy: the extractable energy of the black hole introduced in 1971 (Christodoulou and Ruffini \cite{cr71}), which we shall refer in the following as the blackholic energy\footnote{This name is the English translation of the Italian words ``energia buconerale'', introduced by Iacopo Ruffini, December 2004, here quoted by his kind permission.}. The blackholic energy, expected to be emitted during the dynamical process of gravitational collapse leading to the formation of the black hole, generates X-ray luminosities $10^{21}$ times larger than the solar luminosity, although lasting for a very short time.

The extreme regimes of GRBs evidence new and unexplored regimes of theoretical physics. It is the aim of this talk to outline the progress achieved in understanding these astrophysical systems and the theoretically predicted regimes for the first time submitted to direct observational verification. 

It is a pleasure to present these results in Brazil. While sitting at the Casino de Urca, George Gamow and Mario Schoenberg in 1939 identified the basic process leading to the formation and cooling of a newly born neutron star (see \ref{gamow}). They called this process essentially related to the emission of neutrinos and antineutrinos the Urca process. It is a welcomed coincidence that, during the preparation of my talk at the tenth Marcel Grossmann Meeting \cite{r03mg10}, examining the data of the recently observed GRB 030329, we have received a confirmation of a scenario we have recently outlined in three papers giving the theoretical paradigms for the understanding of GRBs (Ruffini et al. \cite{lett1,lett2,lett3}).

We have clear evidence, first advanced in the system GRB980425/SN1998bw (Ruffini et al. \cite{cospar02}, Fraschetti et al. \cite{f03mg10}) and now confirmed also in the system GRB030329/SN1003dh, that there are in these systems three different components: 1) the GRB source, generated by the collapse to a black hole, 2) the supernova, generated by the collapse of an evolved star, 3) an additional X-ray source which is not related, unlike what is at times stated in the literature, to the GRB afterglow. In honor of the work done in the town of Urca by George Gamow and Mario Schoenberg, identifying in the neutrino emission of the Urca process the basic mechanism leading to the process of gravitational collapse and the formation of a relativistic compact star, we named these two X-ray sources URCA-1, the one formed in the system GRB980425/SN1998bw, and URCA-2, the one formed in the system GRB030329/SN2003dh. We shall now recall some of the main steps in reaching this understanding out of the GRB phenomenon and explore possible explanation of the origin of these two sources.

We then turn to the analysis of the short GRBs. We first review some progress in the study of the general relativistic collapse of a shell of matter endowed with electromagnetic fields, which has been motivated by the study of the short GRBs. We then deduce from these theoretical developments some consequences for the interpretation of the mass-energy formula of the black hole, as well as some conceptual consequences for the relation between general relativity and thermodynamics. We also apply some recent progress on the solution of the Einstein-Maxwell-Blasov equations to the thermalization process occurring in the electron-positron plasma generated by the vacuum polarization process.

We finally give a very specific theoretical prediction on the burst structure to be expected in short GRBs. We then proceed to the conclusions and some general considerations about this novel astrophysical scenario.

\section{The energetics of gamma-ray bursts} 

It is well known how GRBs were detected and studied for the first time using the {\em Vela} satellites, developed for military research to monitor the non-violation of the Limited Test Ban Treaty signed in 1963 (see e.g. Strong \cite{s75}). It was clear from the early data of these satellites, which were put at $150,000$ miles from the surface of Earth, that the GRBs did not originate either on the Earth nor in the Solar System.

\begin{figure} 
\centering 
\includegraphics[height=\hsize,clip,angle=90]{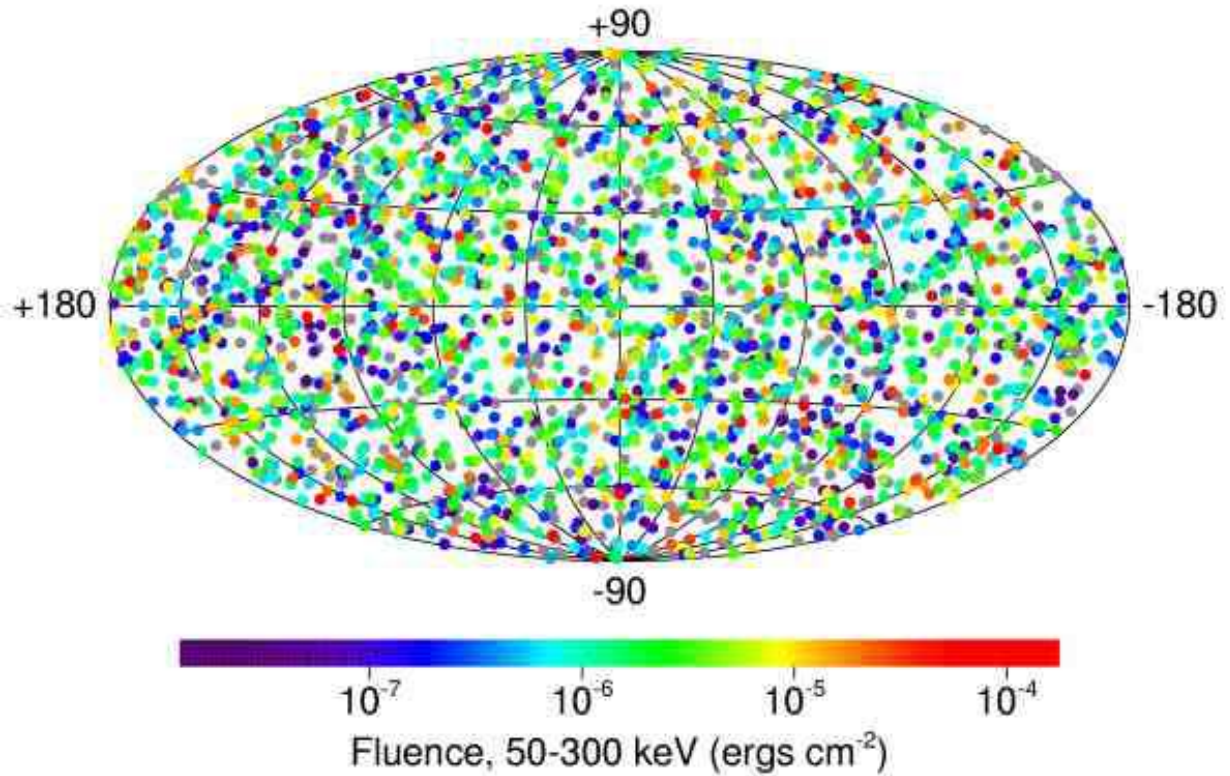} 
\caption{Position in the sky, in galactic coordinates, of 2000 GRB events seen by the CGRO satellite. Their isotropy is evident. Reproduced from BATSE web site by their courtesy.}
\label{batse2k} 
\end{figure} 

\begin{figure} 
\centering 
\includegraphics[width=\hsize,clip]{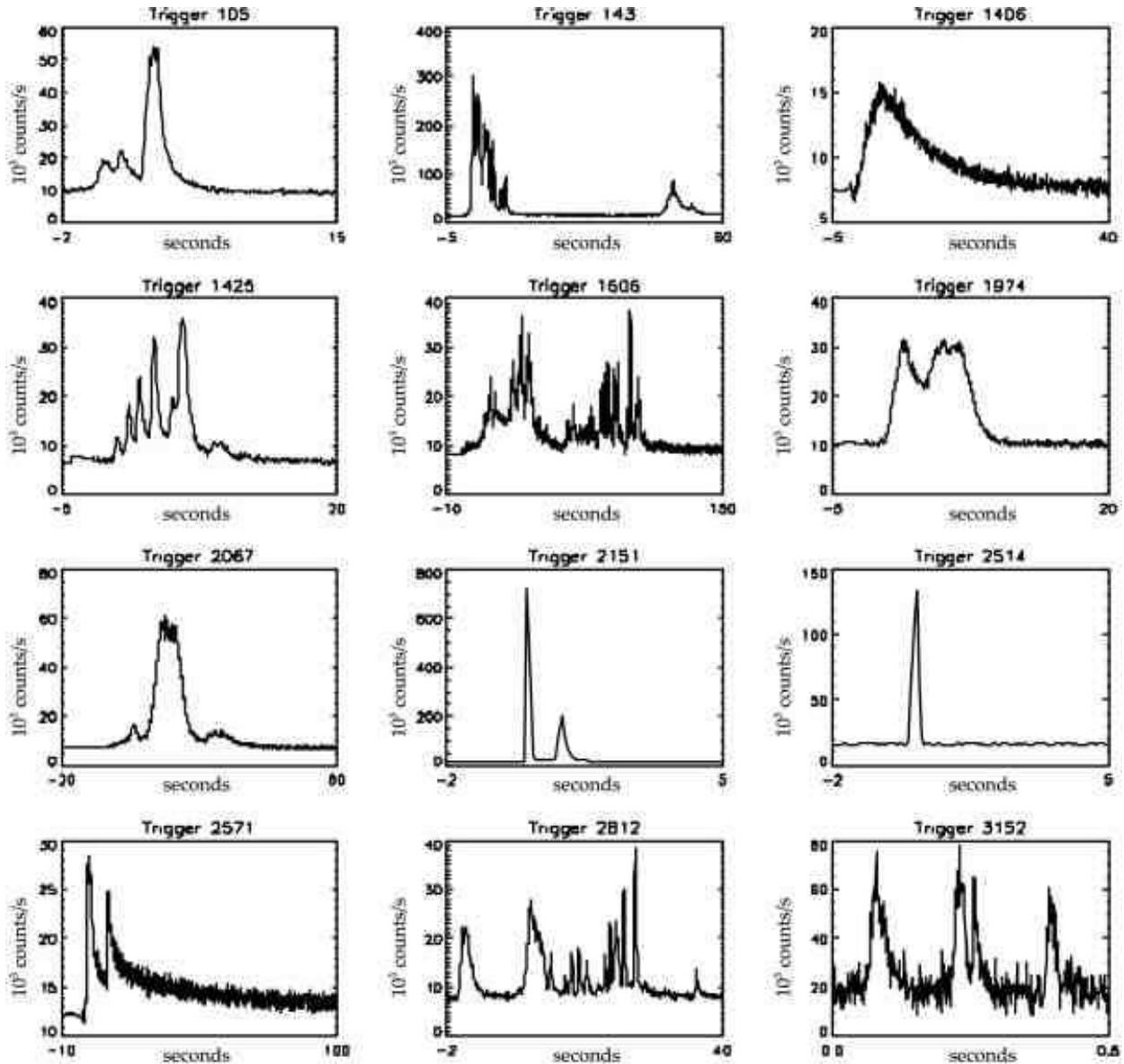} 
\caption{Some GRB light curves observed by the BATSE instrument on board of the CGRO satellite.}
\label{grb_profiles_eng} 
\end{figure}

\begin{figure} 
\centering 
\includegraphics[width=\hsize,clip]{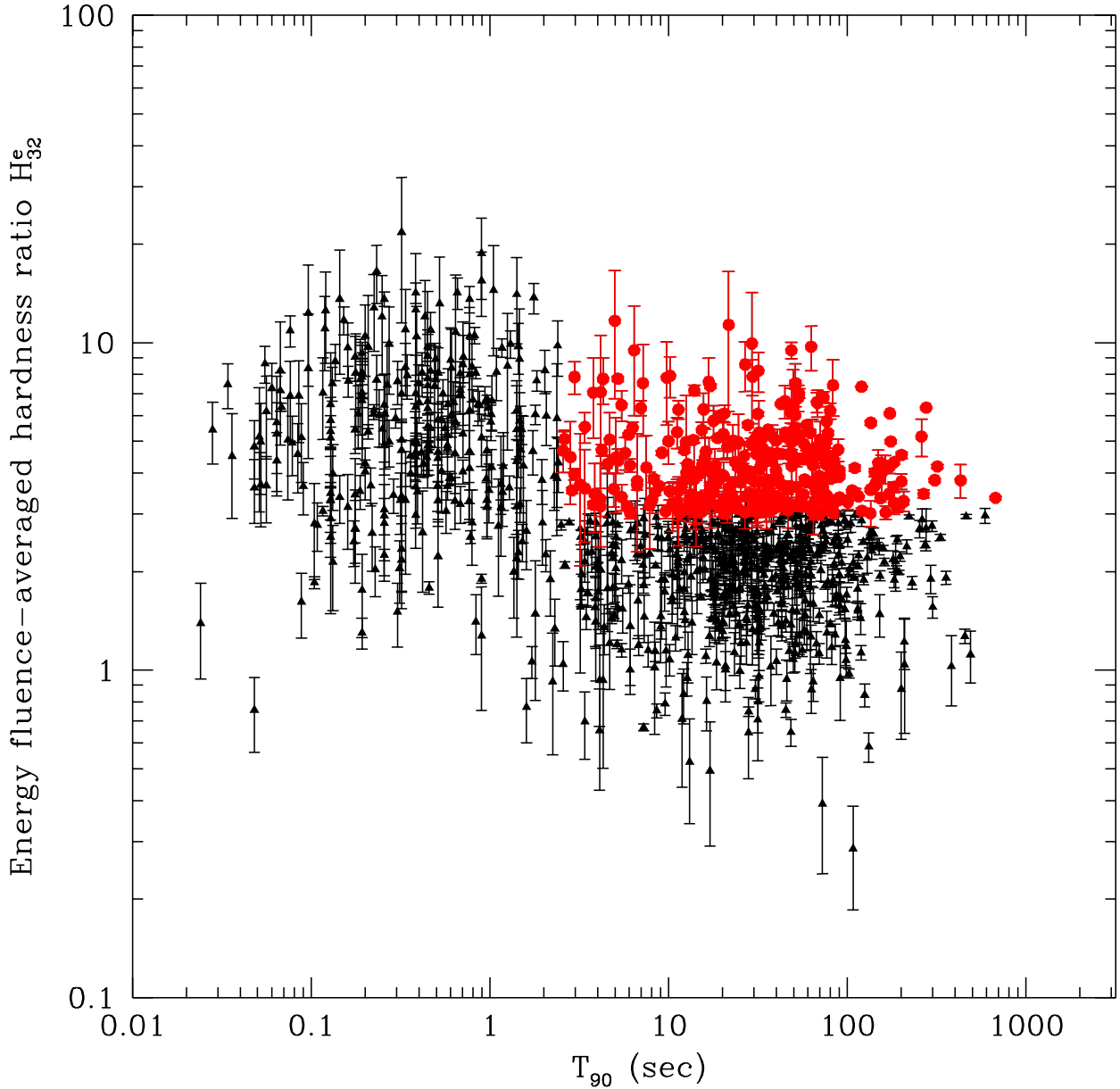} 
\caption{The energy fluence-averaged hardness ratio for short ($T < 1$ s) and long ($T> 1$ s) GRBs are represented. Reproduced, by his kind permission, from Tavani \cite{t98} where the details are given.}
\label{tavani} 
\end{figure}

\begin{figure} 
\centering 
\includegraphics[width=\hsize,clip]{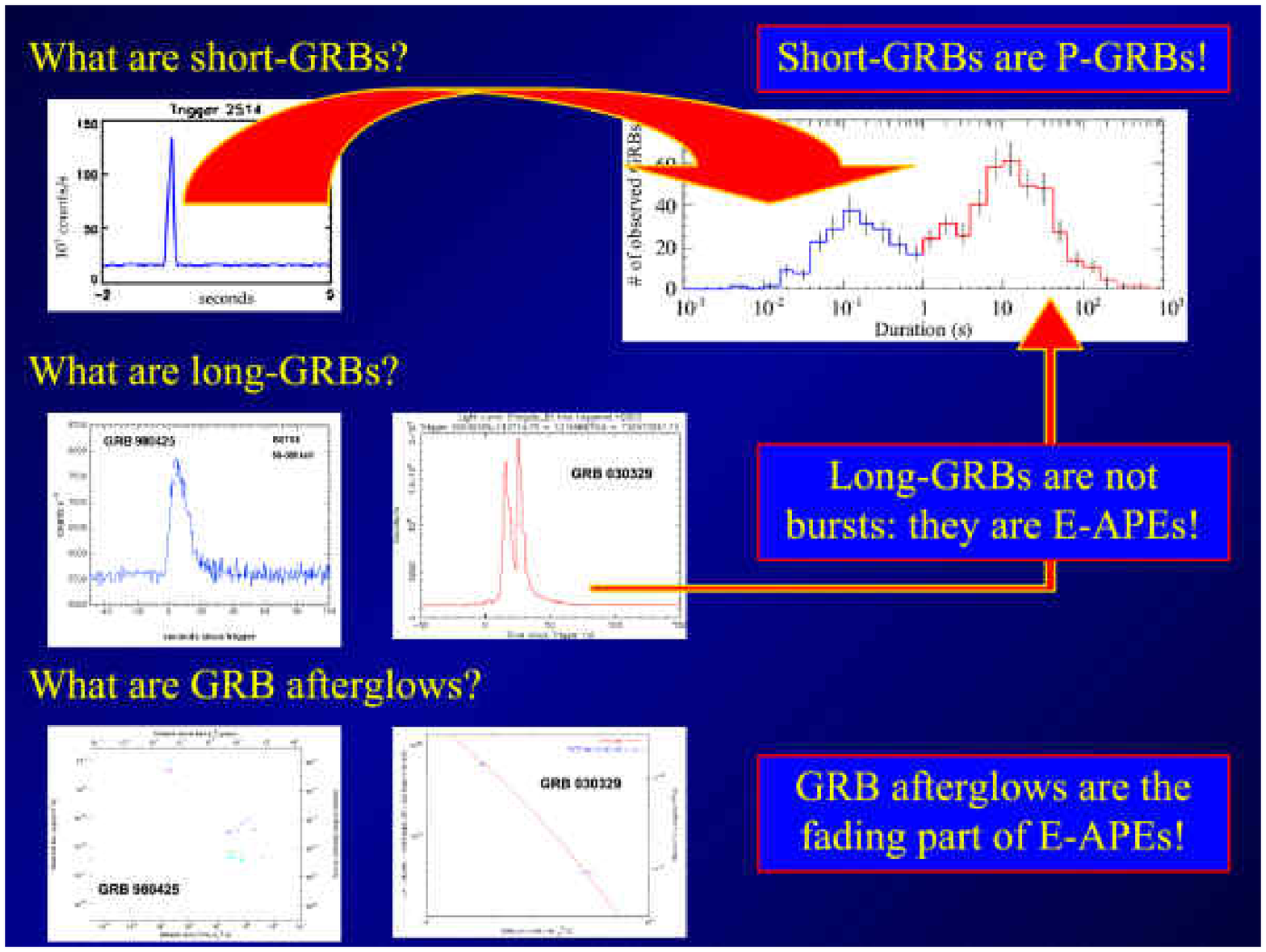} 
\caption{On the upper right part of the figure are plotted the number of the observed GRBs as a function of their duration. The bimodal distribution corresponding respectively to the short bursts, upper left figure, and the long bursts, middle figure, is quite evident. The structure of the long bursts as part of the afterglow phenomena of GRBs is illustrated in section \ref{eape}.}
\label{slb} 
\end{figure}

The mystery of these sources became more profound as the observations of the BATSE instrument on board of the Compton Gamma-Ray Observatory (CGRO) satellite\footnote{see http://cossc.gsfc.nasa.gov/batse/} over $9$ years proved the isotropy of these sources in the sky (See Fig. \ref{batse2k}). In addition to these data, the CGRO satellite gave an unprecedented number of details on the GRB structure, on their spectral properties and time variabilities which became encoded in the fourth BATSE catalog \cite{batse4b} (see e.g. Fig. \ref{grb_profiles_eng}). Out of the analysis of these BATSE sources it soon became clear (see e.g. Kouveliotou et al. \cite{ka93}, Tavani \cite{t98}) the existence of two distinct families of sources: the long bursts, lasting more then one second and softer in spectra, and the short bursts (see Fig. \ref{slb}), harder in spectra (see Fig. \ref{tavani}). We shall return shortly on this topic.

The situation drastically changed with the discovery of the afterglow by the Italian-Dutch satellite BeppoSAX (Costa et al. \cite{ca97}) and the possibility which led to the optical identification of the GRBs by the largest telescopes in the world, including the Hubble Space Telescope, the Keck Telescope in Hawaii and the VLT in Chile, and allowed as well the identification in the radio band of these sources. The outcome of this collaboration between complementary observational technique has led to the possibility of identifying in 1997 the distance of these sources from the Earth and their tremendous energy of the order up to $10^{54}$ erg/second during the burst. It is interesting, as we will show in the following, that an energetics of this magnitude for the GRBs had previously been predicted out of first principles already in 1974 by Damour and Ruffini \cite{dr75}.

The resonance between the X- and gamma ray astronomy from the satellites and the optical and radio astronomy from the ground, had already marked in the seventies the great success and development of the astrophysics of binary X-ray sources (see e.g. Giacconi \& Ruffini \cite{gr78}). This resonance is re-proposed here for GRBs on a much larger scale. The use of much larger satellites, like Chandra and XMM-Newton, and dedicated space missions, like HETE-2 and, in the near future, Swift, and the very fortunate circumstance of the coming of age of the development of unprecedented optical technologies for the telescopes offers opportunities without precedence in the history of mankind. In parallel, the enormous scientific interest on the nature of GRB sources and the exploration, not only of new regimes, but also of totally novel conceptual physical process of the blackholic energy, make the knowledge of GRBs an authentic new frontier in the scientific knowledge.

\section{The complexity an self-consistency of GRB modeling} 

The study of GRBs is very likely ``the'' most extensive computational and theoretical investigation ever done in physics and astrophysics. There are at least three different fields of research which underlie the foundation of the theoretical understanding of GRBs. All three, for different reasons, are very difficult.

The first field of research is the field of special relativity. As I always mention to my students in the course of theoretical physics, this field is paradoxically very difficult since it is extremely simple. In approaching special relativistic phenomena the extremely simple and clear procedures expressed by Einstein in his 1905 classic paper \cite{e05} are often ignored. Einstein makes use in his work of very few physical assumptions, an almost elementary mathematical framework and gives constant attention to a proper operational definition of all observable quantities. Those who work on GRBs use at times very intricate, complex and often wrong theoretical approaches lacking the necessary self-consistency. This is well demonstrated in the current literature on GRBs.

The second field of research essential for understanding the energetics of GRBs deals with quantum electrodynamics and the relativistic process of pair creation in overcritical electromagnetic fields. This topic is also very difficult but for a quite different conceptual reason: the process of pair creation, expressed in the classic works of Heisenberg-Euler-Schwinger \cite{he35,s51} later developed by many others, is based on a very powerful theoretical framework but has never been verified by experimental data. The quest for creating electron-positron pairs by vacuum polarization processes in heavy ion collisions or in lasers has not yet been successfully achieved in Earth-bound experiments (see e.g. Ruffini, Vitagliano, Xue \cite{rvx05}). As we will show here, there is the tantalizing possibility of observing this phenomenon, for the first time, in the astrophysical setting of GRBs on a more grandiose scale.

There is a third field which is essential for the understanding of the GRB phenomenon: general relativity. In this case, contrary to the case of special relativity, the field is indeed very difficult, since it is very difficult both from a conceptual, technical and mathematical point of view. The physical assumptions are indeed complex. The entire concept of geometrization of physics needs a  new conceptual approach to the field. The mathematical complexity of the pseudo-Riemannian geometry contrasts now with the simple structure of the pseudo-Euclidean Minkowski space. The operational definition of the observable quantities has to take into account the intrinsic geometrical properties and also the cosmological settings of the source. With GRBs we have the possibility to follow, from a safe position in an asymptotically flat space at large distance, the formation of the horizon of a black hole with all the associated relativistic phenomena of light bending and time dilatation. Most important, as we will show in details in this presentation, general relativity in connection with quantum phenomena offers, with the blackholic energy, the explanation of the tremendous GRB energy sources.

For these reasons GRBs offer an authentic new frontier in the field of physics and astrophysics. It is appropriate to mention some of the goals of such a new frontier in the above three fields. We recall in the special relativity field, for the first time, we observe phenomena occurring at Lorentz gamma factors of approximately $300$. In the field of relativistic quantum electro-dynamics we see for the first time the interchange between classical fields and the created quantum matter-antimatter pairs. In  the field of general relativity also for the first time we can test the blackholic energy which is the basic energetic physical variable underlying the entire GRB phenomenon.

The most appealing aspect of this work is that, if indeed these three different fields are treated and approached with the necessary technical and scientific maturity, the model which results has a very large redundancy built-in. The approach requires an unprecedented level of self-consistency. Any departures from the correct theoretical treatment in this very complex system lead to exponential departures from the correct solution and from the correct fit of the observations. 

It is so that, as the model is being properly developed and verified, its solution will have existence and uniqueness.

\subsection{GRBs and special relativity} 

The ongoing dialogue between our work and the one of the workers on GRBs, rests still on some elementary considerations presented by Einstein in his classic article of 1905 \cite{e05}. These considerations are quite general and even precede Einstein's derivation, out of first principles, of the Lorentz transformations. We recall here Einstein's words: ``We might, of course, content ourselves with time values determined by an observer stationed together with the watch at the origin of the co-ordinates, and co-ordinating the corresponding positions of the hands with light signals, given out by every event to be timed, and reaching him through empty space. But this co-ordination has the disadvantage that it is not independent of the standpoint of the observer with the watch or clock, as we know from experience''. 

\begin{figure} 
\centering 
\includegraphics[width=\hsize,clip]{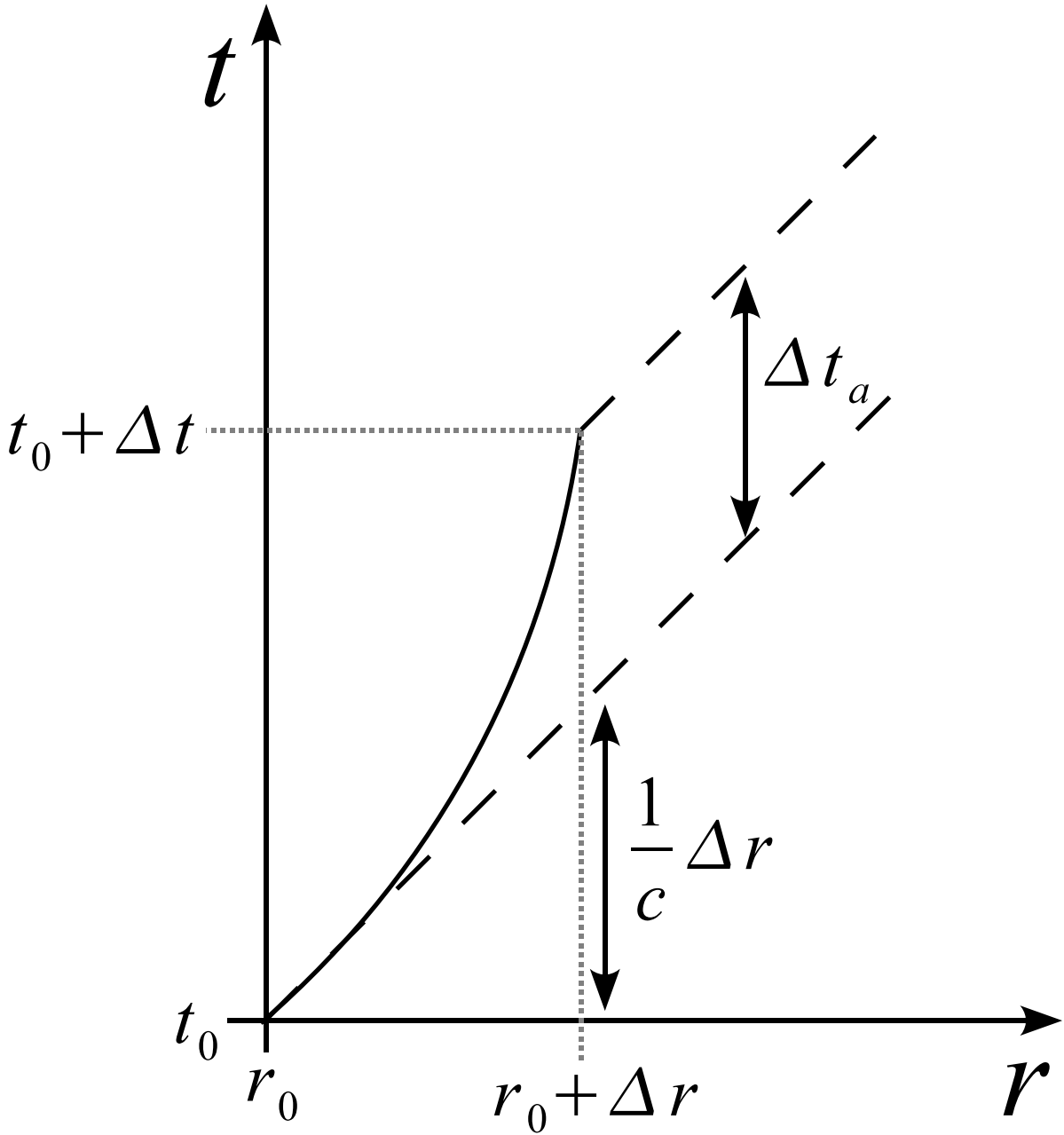} 
\caption{Relation between the arrival time $t_a$ and the laboratory time $t$. Details in Ruffini et al. \cite{lett1,Brasile}.} 
\label{ttasch_new_bn} 
\end{figure} 

The message by Einstein is simply illustrated in Fig. \ref{ttasch_new_bn}. If we consider in an inertial frame a source (solid line) moving with high speed and emitting light signals (dashed lines) along the direction of its motion, a far away observer will measure a delay $\Delta t_a$ between the arrival time of two signals emitted at the origin and after a time interval  $\Delta t$ in the laboratory frame. The real velocity of the source is given by: 
\begin{equation} 
v = \frac{\Delta r}{\Delta t} 
\label{v} 
\end{equation} 
and the apparent velocity is given by: 
\begin{equation} 
v_{app} = \frac{\Delta r}{\Delta t_a}\, , 
\label{vapp} 
\end{equation} 
As pointed out by Einstein the adoption of coordinating light signals simply by their arrival time as in Eq.(\ref{vapp}), without an adequate definition of synchronization, is incorrect and leads to unsurmountable difficulties as well as to apparently ``superluminal'' velocities as soon as motions close to the speed of light are considered.

The use of $\Delta t_a$ as a time coordinate, often tacitly adopted by astronomers, should be done, if at all, with proper care. The relation between $\Delta t_a$ and the correct time parameterization in the laboratory frame has to be taken into account:
\begin{equation} 
\Delta t_a = \Delta t - \frac{\Delta r}{c} = \Delta t - 
\frac{1}{c}\int_{t_\circ}^{t_\circ + \Delta t}{v\left(t'\right) dt'}\, . 
\label{tadef} 
\end{equation} 
In other words, the relation between the arrival time and the laboratory time cannot be done without a knowledge of the speed along the entire world-line of the source. In the case of GRBs, such a worldline starts at the moment of gravitational collapse. It is of course clear that the parameterization in the laboratory frame has to take into account the cosmological redshift $z$ of the source. We then have, at the detector:
\begin{equation}
\Delta t_a^d = \left(1+z\right) \Delta t_a\, .
\label{taddef}
\end{equation}

In the current GRB literature, Eq.(\ref{tadef}) has been systematically neglected by addressing only the afterglow description neglecting the previous history of the source. Often the integral equation has been approximated by a clearly incorrect instantaneous value: 
\begin{equation} 
\Delta t_a \simeq \frac{\Delta t}{2\gamma^2}\, . 
\label{taapp} 
\end{equation}
The attitude has been adopted that it should be possible to consider separately the afterglow part of the GRB phenomenon, without the knowledge of the entire equation of motion of the source. 

This point of view has reached its most extreme expression in the works reviewed by Piran \cite{p99,p00}, where the so-called ``prompt radiation'', lasting on the order of $10^2$ s, is considered as a burst emitted by the prolonged activity of an ``inner engine''. In these models, generally referred to as the ``internal shock model'', the emission of the afterglow is assumed to follow the ``prompt radiation'' phase \cite{rm94,px94,sp97,f99,fcrsyn99}.

As we outline in the following, such an extreme point of view originates from the inability of obtaining the time scale of the ``prompt radiation'' from a burst structure. These authors consequently appeal to the existence of an ``ad hoc'' inner engine in the GRB source to solve this problem.

We show in the following how this difficulty has been overcome in our approach by interpreting the ``prompt radiation'' as an integral part of the afterglow and {\em not} as a burst. This explanation can be reached only through a relativistically correct theoretical description of the entire afterglow (see section \ref{aft}). Within the framework of special relativity we show that it is not possible to describe a GRB phenomenon by disregarding the knowledge of the entire past worldline of the source. We show that at $10^2$ seconds the emission occurs from a region of dimensions of approximately $10^{16}$ cm, well within the region of activity of the afterglow. This point was not appreciated in the current literature due to the neglect of the apparent superluminal effects implied by the use of the ``pathological'' parametrization of the GRB phenomenon by the arrival time of light signals.

An additional difference between our treatment and the ones in the current literature relates to the assumption of the existence of scaling laws in the afterglow phase: the power law dependence of the Lorentz gamma factor on the radial coordinate is usually systematically assumed. From the proper use of the relativistic transformations and by the direct numerical and analytic integration of the special relativistic equations of motion we demonstrate (see section \ref{eqaft}) that no simple power-law relation can be derived for the equations of motion of the system. This situation is not new for workers in relativistic theories: scaling laws exist in the extreme ultrarelativistic regimes and in the Newtonian ones but not in the intermediate fully relativistic regimes (see e.g. Ruffini \cite{r70}).

\subsection{GRBs and general relativity}\label{genrel}

\begin{figure} 
\centering 
\includegraphics[width=\hsize,clip]{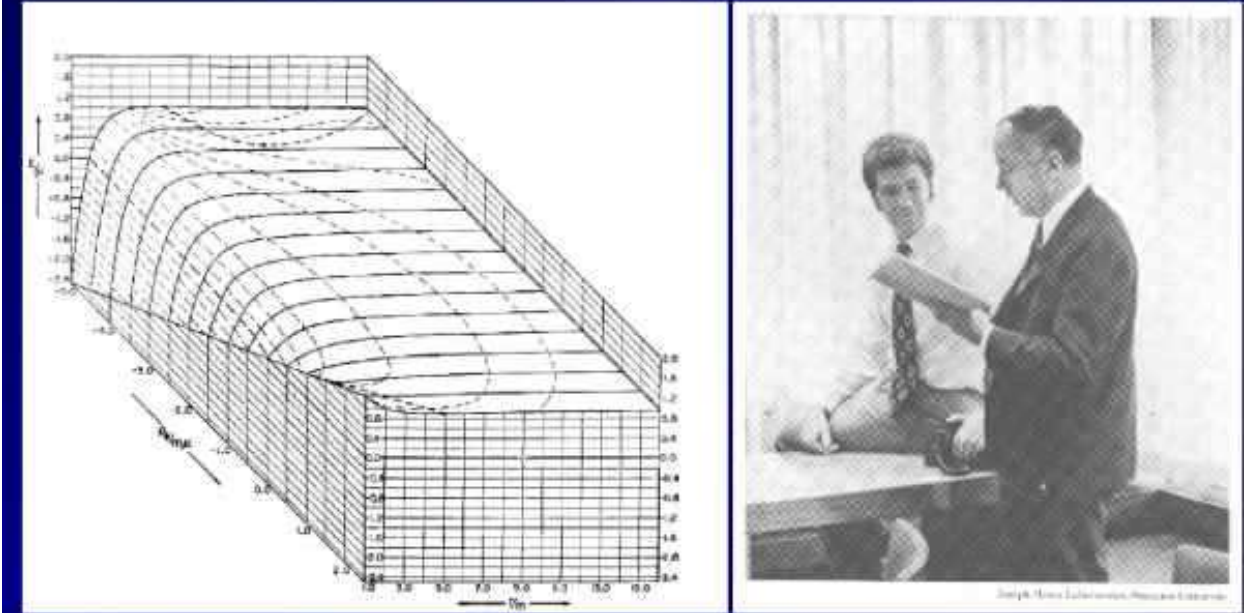} 
\caption{The effective potential corresponding to the circular orbits in the equatorial plane of a black hole is given as a function of the angular momentum of the test particle. This digram was originally derived by Ruffini and Wheeler (right picture). For details see Landau and Lifshitz \cite{ll2} and Rees, Ruffini and Wheeler \cite{rrw}.}
\label{effp} 
\end{figure}

Three of the most important works in the field of general relativity have certainly been the discovery of the Kerr solution \cite{kerr}, its generalization to the charged case (Newman et al. \cite{newman}) and the formulation by Brandon Carter \cite{carter} of the Hamilton-Jacobi equations for a charged test particle in the metric and electromagnetic field of a Kerr-Newman solution (see e.g. Landau and Lifshitz \cite{ll2}). The equations of motion, which are generally second order differential equations, were reduced by Carter to a set of first order differential equations which were then integrated by using an effective potential technique by Ruffini and Wheeler for the Kerr metric (see e.g. Landau and Lifshitz \cite{ll2}) and by Ruffini for the Reissner-Nordstr\"om geometry (Ruffini \cite{r70}, see Fig. \ref{effp}).

All the above mathematical results were essential for understanding the new physics of gravitationally collapsed objects and allowed the publication of a very popular article: ``Introducing the black hole'' (Ruffini and Wheeler \cite{rw71}). In that paper, we advanced the ansatz that the most general black hole is a solution of the Einstein-Maxwell equations, asymptotically flat and with a regular horizon: the Kerr-Newman solution, characterized only by three parameters: the mass $M$, the charge $Q$ and the angular momentum $L$. This ansatz of the ``black hole uniqueness theorem'' still today after thirty years presents challenges to the mathematical aspects of its complete proof (see e.g. Carter \cite{ckf} and Bini et al. \cite{bcjr}). In addition to these mathematical difficulties, in the field of physics this ansatz contains the most profound consequences. The fact that, among all the possible highly nonlinear terms characterizing the gravitationally collapsed objects, only the ones corresponding solely to the Einstein Maxwell equations survive the formation of the horizon has, indeed, extremely profound physical implications. Any departure from such a minimal configuration either collapses on the horizon or is radiated away during the collapse process. This ansatz is crucial in identifying precisely the process of gravitational collapse leading to the formation of the black hole and the emission of GRBs. Indeed, in this specific case, the Born-like nonlinear \cite{b33} term of the Heisenberg-Euler-Schwinger \cite{he35,s51} Lagrangian are radiated away prior to the formation of the horizon of the black hole (see e.g. Ruffini et al. \cite{rvx05}). Only the nonlinearity corresponding solely to the classical Einstein-Maxwell theory is left as the outcome of the gravitational collapse process.

The same effective potential technique (see Landau and Lifshitz \cite{ll2}), which allowed the analysis of circular orbits around the black hole, was crucial in reaching the equally interesting discovery of the reversible and irreversible transformations of black holes by Christodoulou and Ruffini \cite{cr71}, which in turn led to the mass-energy formula of the black hole:
\begin{equation} 
E_{BH}^2 = M^2c^4 = \left(M_{\rm ir}c^2 + \frac{Q^2}{2\rho_+}\right)^2+\frac{L^2c^2}{\rho_+^2}\, ,
\label{em} 
\end{equation} 
with 
\begin{equation} 
\frac{1}{\rho_+^4}\left(\frac{G^2}{c^8}\right)\left(Q^4+4L^2c^2\right)\leq 1\, , 
\label{s1}
\end{equation} 
where 
\begin{equation} 
S=4\pi\rho_+^2=4\pi(r_+^2+\frac{L^2}{c^2M^2})=16\pi\left(\frac{G^2}{c^4}\right) M^2_{\rm ir}\, ,
\label{sa} 
\end{equation} 
is the horizon surface area, $M_{\rm ir}$ is the irreducible mass, $r_{+}$ is the horizon radius and $\rho_+$ is the quasi-spheroidal cylindrical coordinate of the horizon evaluated at the equatorial plane. Extreme black holes satisfy the equality in Eq.(\ref{s1}).

From Eq.(\ref{em}) follows that the total energy of the black hole $E_{BH}$ can be split into three different parts: rest mass, Coulomb energy and rotational energy. In principle both Coulomb energy and rotational energy can be extracted from the black hole (Christodoulou and Ruffini \cite{cr71}). The maximum extractable rotational energy is 29\% and the maximum extractable Coulomb energy is 50\% of the total energy, as clearly follows from the upper limit for the existence of a black hole, given by Eq.(\ref{s1}). We refer in the following to both these extractable energies as the blackholic energy.

The existence of the black hole and the basic correctness of the circular orbits has been proven by the observations of Cygnus-X1 (see e.g. Giacconi and Ruffini \cite{gr78}). However, in binary X-ray sources, the black hole uniquely acts passively by generating the deep potential well in which the accretion process occurs. It has become tantalizing to look for astrophysical objects in order to verify the other fundamental prediction of general relativity that the blackholic energy is the largest energy extractable from any physical object.

As we shall see in the next section, the feasibility of the extraction of the blackholic energy has been made possible by the quantum processes of creating, out of classical fields, a plasma of electron-positron pairs in the field of black holes. The manifestation of such process of energy extraction from the black hole is astrophysically manifested by the occurrence of GRBs.

\subsection{GRBs and quantum electro-dynamics}\label{dyadosphere}

That a static electromagnetic field stronger than a critical value: 
\begin{equation} 
E_c = \frac{m_e^2c^3}{\hbar e} 
\label{ec} 
\end{equation} 
can polarize the vacuum and create electron-positron pairs was clearly evidenced by Heisenberg and Euler \cite{he35}. The major effort in verifying the correctness of this theoretical prediction has been directed in the analysis of heavy ion collisions (see Ruffini et al. \cite{rvx05} and references therein). From an order-of-magnitude estimate, it appears that around a nucleus with a charge: 
\begin{equation} 
Z_c \simeq \frac{\hbar c}{e^2} \simeq 137 
\label{zc} 
\end{equation} 
the electric field can be stronger than the electric field polarizing the vacuum. A more accurate detailed analysis taking into account the bound states levels around a nucleus brings to a value of
\begin{equation} 
Z_c \simeq 173
\label{zc2} 
\end{equation}  
for the nuclear charge leading to the existence of a critical field. From the Heisenberg uncertainty principle it follows that, in order to create a pair, the existence of the critical field should last a time
\begin{equation} 
\Delta t \sim \frac{\hbar}{m_e c^2} \simeq 10^{-18}\, \mathrm{s}\, ,
\label{dt} 
\end{equation} 
which is much longer then the typical confinement time in heavy ion collisions which is 
\begin{equation} 
\Delta t \sim \frac{\hbar}{m_p c^2} \simeq 10^{-21}\, \mathrm{s}\, .
\label{dt2} 
\end{equation} 
This is certainly a reason why no evidence for pair creation in heavy ion collisions has been obtained although remarkable effort has been spent in various accelerators worldwide. Similar experiments involving laser beams encounter analogous difficulties (see e.g. Ruffini et al. \cite{rvx05} and references therein).

The alternative idea was advanced in 1975 \cite{dr75} that the critical field condition given in Eq.(\ref{ec}) could be reached easily, and for a time much larger than the one given by Eq.(\ref{dt}), in the field of a Kerr-Newman black hole in a range of masses $3.2M_\odot \le M_{BH} \le 7.2\times 10^6M_\odot$. In that paper we have generalized to the curved Kerr-Newman geometry the fundamental theoretical framework developed in Minkowski space by Heisenberg-Euler \cite{he35} and Schwinger \cite{s51}. This result was made possible by the work on the structure of the Kerr-Newman spacetime previously done by Carter \cite{carter} and by the remarkable mathematical craftsmanship of Thibault Damour then working with me as a post-doc in Princeton.

The maximum energy extractable in such a process of creating a vast amount of electron-positron pairs around a black hole is given by:
\begin{equation} 
E_{max} = 1.8\times 10^{54} \left(M_{BH}/M_\odot\right)\, \mathrm{erg}\, \mathrm{.} 
\label{emax} 
\end{equation} 
We concluded in that paper that such a process ``naturally leads to a most simple model for the explanation of the recently discovered $\gamma$-rays bursts''.

At that time, GRBs had not yet been optically identified and nothing was known about their distance and consequently about their energetics. Literally thousands of theories existed in order to explain them and it was impossible to establish a rational dialogue with such an enormous number of alternative theories. We did not pursue further our model until the results of the BeppoSAX mission, which clearly pointed to the cosmological origin of GRBs, implying for the typical magnitude of their energy precisely the one predicted by our model.

It is interesting that the idea of using an electron-positron plasma as a basis of a GRB model was independently introduced years later in a set of papers by Cavallo and Rees \cite{cr78}, Cavallo and Horstman \cite{ch81} and Horstman and Cavallo \cite{hc83}. These authors did not address the issue of the physical origin of their energy source. They reach their conclusions considering the pair creation and annihilation process occurring in the confinement of a large amount of energy in a region of dimension $\sim 10$ km typical of a neutron star. No relation to the physics of black holes nor to the energy extraction process from a black hole was envisaged in their interesting considerations, mainly directed to the study of the opacity and the consequent dynamics of such an electron-positron plasma.

\begin{figure} 
\centering 
\includegraphics[width=\hsize,clip]{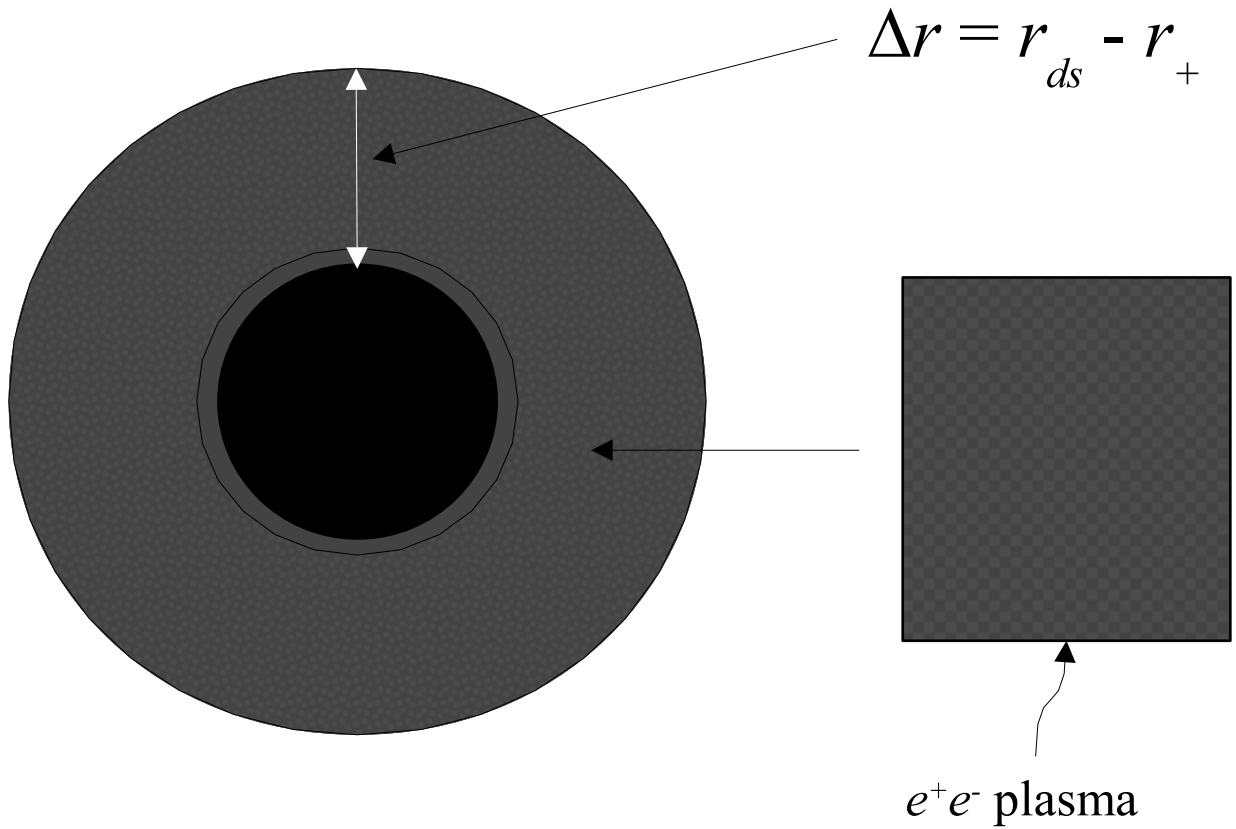} 
\caption{The dyadosphere is comprised between the horizon radius and the radius of the dyadosphere. All this region is filled with electron-positron pairs and photons in thermal equilibrium. Details in Ruffini \cite{rukyoto}, Preparata et al. \cite{prx98}, Ruffini et al. \cite{rvx03a}.} 
\label{dya} 
\end{figure} 

After the discovery of the afterglows and the optical identification of GRBs at cosmological distances, implying exactly the energetics predicted in Eq.(\ref{emax}), we returned to the analysis of the vacuum polarization process around a black hole and precisely identified the region around the black hole in which the vacuum polarization process and the consequent creation of electron-positron pairs occur. We defined this region, using the Greek name dyad for pairs ($\delta\upsilon\alpha\varsigma$, $\delta\upsilon\alpha\delta o \varsigma$), to be the ``dyadosphere'' of the black hole, bounded  by the black hole horizon and the dyadosphere radius $r_{ds}$ given by (see Ruffini \cite{rukyoto}, Preparata et al. \cite{prx98} and Fig.\ref{dya}):
\begin{equation} 
r_{ds}=\left(\frac{\hbar}{mc}\right)^\frac{1}{2}\left(\frac{GM}{ 
c^2}\right)^\frac{1}{2} \left(\frac{m_{\rm p}}{m}\right)^\frac{1}{2}\left(\frac{e}{q_{\rm p}}\right)^\frac{1}{2}\left(\frac{Q}{\sqrt{G}M}\right)^\frac{1}{2}=1.12\cdot 10^8\sqrt{\mu\xi} \, {\rm cm}, 
\label{rc} 
\end{equation} 
where we have introduced the dimensionless mass and charge parameters $\mu={M_{BH}/M_{\odot}}$, $\xi={Q/(M_{BH}\sqrt{G})}\le 1$. 

The analysis of the dyadosphere was developed, at that time, around an already formed black hole. In recent months we have been developing the dynamical formation of the black hole and correspondingly of the dyadosphere during the process of gravitational collapse, reaching some specific signatures which may be detectable in the structure of the short and long GRBs (Cherubini et al. \cite{crv02}, Ruffini and Vitagliano \cite{rv02a,rv02b}, Ruffini et al. \cite{rvx03a,rvx03b,rfvx05}).

\section{The dynamical phases following the dyadosphere formation}

Many details of this topic have been presented in great details in Ruffini et al. \cite{Brasile}.

After the vacuum polarization process around a black hole, one of the topics of the greatest scientific interest is the analysis of the dynamics of the electron-positron plasma formed in the dyadosphere. This issue was addressed by us in a very effective collaboration with Jim Wilson at Livermore. The numerical simulations of this problem were developed at Livermore, while the semi-analytic approach was developed in Rome (Ruffini et al. \cite{rswx99}).

The corresponding treatment in the framework of the Cavallo et al. analysis was performed by Piran et al. \cite{psn93} also using a numerical approach, by Bisnovaty-Kogan and Murzina \cite{bm95} using an analytic approach and by M\'esz\'aros, Laguna and Rees \cite{mlr93} using a numerical and semi-analytic approach.

Although some analogies exists between these treatments, they are significantly different in the theoretical details and in the final results. Since the final result of the GRB model is extremely sensitive to any departure from the correct treatment, it is indeed very important to detect at every step the appearance of possible fatal errors.

A conclusion common to all these treatments is that the electron-positron plasma is initially optically thick and expands till transparency reaching very high values of the Lorentz gamma 
factor. A second point, which is common, is the discovery of a new clear feature: the plasma shell expands but the Lorentz contraction is such that its width in the laboratory frame appears to be constant.

\begin{figure}
\centering
\includegraphics[width=\hsize,clip]{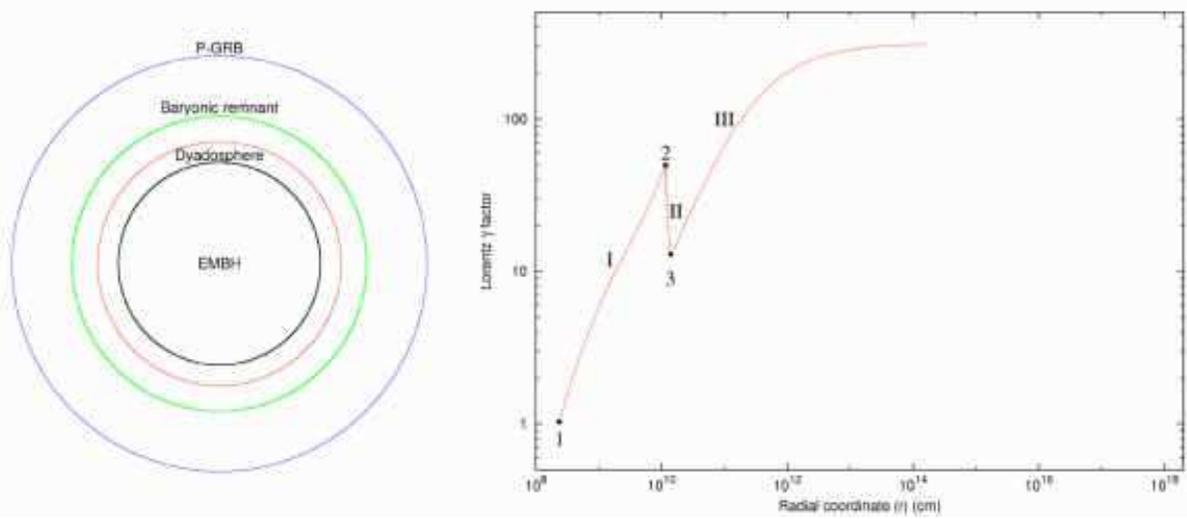}
\caption{The optically thick phase of our model are qualitatively represented in this diagram. There are clearly recognizable 1) the PEM pulse phase, 2) the impact on the baryonic remnant, 3) the PEMB pulse phase and the final approach to transparency with the emission of the P-GRB (see Fig. \ref{cip3}). Details in Ruffini et al. \cite{Brasile}.}
\label{cip2}
\end{figure}

There is however a major difference between our approach and the ones of Piran, M\'esz\'aros and Rees, in that the dyadosphere is assumed by us to be filled uniquely with an electron-positron plasma. Such a plasma expands in substantial agreement with the results presented in the work of Bisnovati-Kogan and Murzina \cite{bm95}. In our model the pulse of electron-positron pairs and photons (PEM Pulse, see Ruffini et al. \cite{rswx99}) evolves and at a radius on the order of $10^{10}$ cm it encounters the remnant of the star progenitor of the newly formed black hole. The PEM pulse is then loaded with baryons. A new pulse is  formed of electron-positron-photons and baryons (PEMB Pulse, see Ruffini et al. \cite{rswx00}) which expands all the way until transparency is reached. At transparency the emitted photons give origin to what we define as the Proper-GRB (see Ruffini et al. \cite{lett2} and Fig. \ref{cip2}).

In our approach, the baryon loading is measured by a dimensionless quantity
\begin{equation}
B = \frac{M_B c^2}{E_{dya}}\, ,
\label{Bdef}
\end{equation}
which gives direct information about the mass $M_B$ of the remnant. The corresponding treatment done by Piran and collaborators (Shemi \& Piran \cite{sp90}, Piran et al. \cite{psn93}) and by M\'esz\'aros, Laguna and Rees \cite{mlr93} differs in one important respect: the baryonic loading is assumed to occur since the beginning of the electron-positron pair formation and no relation to the mass of the remnant of the collapsed progenitor star is attributed to it.

A marked difference also exists between our description of the rate equation for the electron-positron pairs and the ones by those authors. While our results are comparable with the ones obtained by Piran under the same initial conditions, the set of approximations adopted by M\'esz\'aros, Laguna and Rees \cite{mlr93} appears to be too radical and leads to very different results violating energy and momentum conservation (see Bianco et al. \cite{bfrvx}).

From our analysis (Ruffini et al. \cite{rswx00}) it also becomes clear that such expanding dynamical evolution can only occur for values of $B < 10^{-2}$. This prediction, as we will show shortly in the three GRB sources considered here, is very satisfactorily confirmed by observations.

From the value of the $B$ parameter, related to the mass of the remnant, it therefore follows that the collapse to a black hole leading to a GRB is drastically different from the collapse to a neutron star. While in the case of a neutron star collapse a very large amount of matter is expelled, in many instances well above the mass of the neutron star itself, in the case of black holes leading to a GRB only a very small fraction of the initial mass ($\sim 10^{-2}$ or less) is expelled. The collapse to a black hole giving rise to a GRB appears to be much smoother than any collapse process considered until today: almost 99.9\% of the star has to be collapsing simultaneously!

\begin{figure} 
\centering 
\includegraphics[width=\hsize,clip]{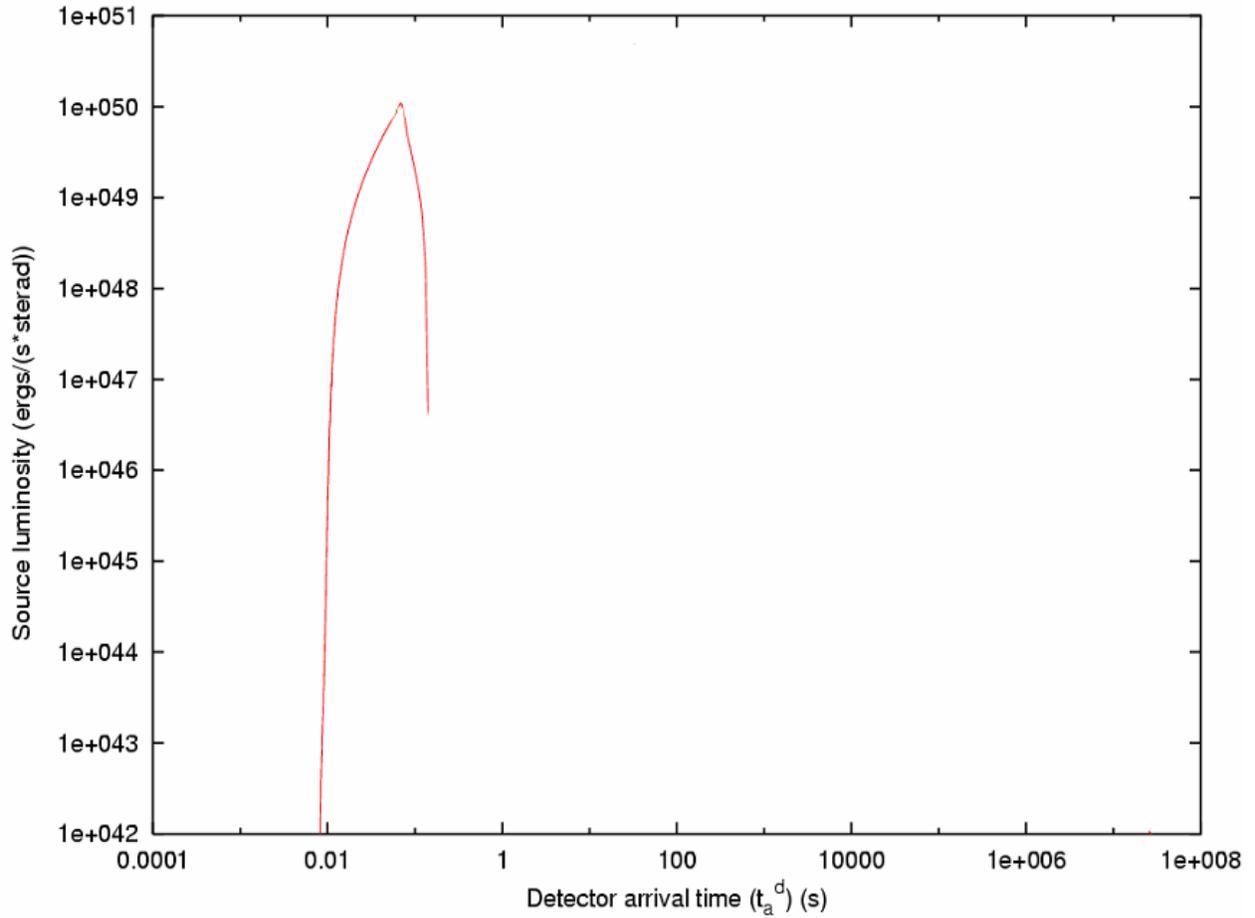} 
\caption{The P-GRB emitted at the transparency point at a time of arrival $t_a^d$ which has been computed following the prescriptions of Eq.(\ref{tadef}). Details in Ruffini et al. \cite{lett2,Brasile}.} 
\label{cip3} 
\end{figure}

We summarize in Figs. \ref{cip2}--\ref{cip3} the optically thick phase of GRBs in our model: we start from a given dyadosphere of energy $E_{dya}$; the pair-electromagnetic pulse (PEM pulse) self-accelerates outward typically reaching Lorentz gamma factors $\gamma \sim 200$ at $r \sim 10^{10}$ cm; at this point the collision of the PEM pulse with the remnant of the progenitor star occurs with an abrupt decrease in the value of the Lorentz gamma factor; a new pair-electromagnetic-baryon pulse (PEMB pulse) is formed which self-accelerates outward until the system becomes transparent.

The photon emission at this transparency point is the Proper-GRB (P-GRB). An accelerated beam of baryons with an initial Lorentz gamma factor $\gamma_\circ$ starts to interact with the interstellar medium at typical distances from the black hole of $r_\circ \sim 10^{14}$ cm and at a photon arrival time at the detector on the Earth surface of $t_a^d \sim 0.1$ s. These values determine the initial conditions of the afterglow.

\section{The description of the afterglow}\label{aft}

\begin{figure} 
\centering 
\includegraphics[width=\hsize,clip]{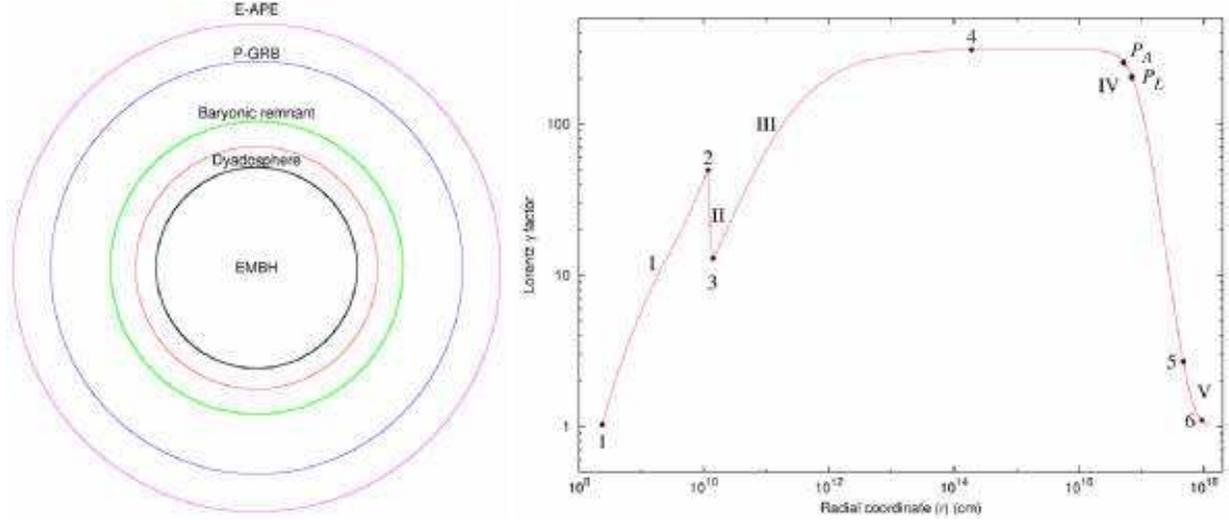} 
\caption{The GRB afterglow phase is here represented together with the optically thick phase (see Fig. \ref{cip2}). The value of the Lorentz gamma factor is here given from the transparency point all the way to the ultrarelativisitc, relativistic and non relativistic regimes. Details in Ruffini et al. \cite{Brasile}.}
\label{cip_tot} 
\end{figure}

After reaching transparency and the emission of the P-GRB, the accelerated baryonic matter (the ABM pulse) interacts with the interstellar medium (ISM) and gives rise to the afterglow (see Fig. \ref{cip_tot}). Also in the descriptions of this last phase many differences exist between our treatment and the other ones in the current literature. 

\subsection{The initial value problem}

The initial conditions of the afterglow era are determined at the end of the optically thick era when the P-GRB is emitted. As recalled in the last section, the transparency condition is determined by a time of arrival $t_a^d$, a value of the gamma Lorentz factor $\gamma_\circ$, a value of the radial coordinate $r_\circ$, an amount of baryonic matter $M_B$ which are only functions of the two parameters $E_{dya}$ and $B$ (see Eq.(\ref{Bdef})). It is appropriate here to emphasize again that, in order to have the expansion leading to an observed GRB, one must have $B < 10^{-2}$.

This connection to the optically thick era is missing in the current approach in the literature which attributes the origin of the ``prompt radiation'' to an unspecified inner engine activity (see Piran \cite{p99} and references therein). The initial conditions at the beginning of the afterglow era are obtained by a best fit of the later parts of the afterglow. This approach is quite unsatisfactory since, as we will explicitly show, the theoretical treatments currently adopted in the description of the afterglow are not correct. The fit using an incorrect theoretical treatment leads necessarily to the wrong conclusions as well as, in turn, to the determination of incorrect initial conditions.

\subsection{The equations of the afterglow dynamics}\label{eqaft}

Let us first summarize the commonalities between our approach and the ones in the current literature. In both cases (see Piran \cite{p99}, Chiang \& Dermer \cite{cd99} and Ruffini et al. \cite{Brasile}) a thin shell approximation is used to describe the collision between the ABM pulse and the ISM: 
\begin{subequations}\label{Taub_Eq} 
\begin{eqnarray} 
dE_{\mathrm{int}} &=& \left(\gamma - 1\right) dM_{\mathrm{ism}} c^2 
\label{Eint}\, ,\\ 
d\gamma &=& - \textstyle\frac{{\gamma}^2 - 1}{M} dM_{\mathrm{ism}}\, , 
\label{gammadecel}\\ 
dM &=& 
\textstyle\frac{1-\varepsilon}{c^2}dE_{\mathrm{int}}+dM_\mathrm{ism}\, 
,\label{dm}\\ 
dM_\mathrm{ism} &=& 4\pi m_p n_\mathrm{ism} r^2 dr \, , \label{dmism} 
\end{eqnarray} 
\end{subequations} 
where $E_{\mathrm{int}}$, $\gamma$ and $M$ are respectively the internal energy, the Lorentz factor and the mass-energy of the expanding pulse, $n_\mathrm{ism}$ is the ISM number density which is assumed to be constant, $m_p$ is the proton mass, $\varepsilon$ is the emitted fraction of the energy developed in the collision with the ISM and $M_\mathrm{ism}$ is the amount of ISM mass swept up within the radius $r$: $M_\mathrm{ism}=(4/3)\pi(r^3-{r_\circ}^3)m_pn_\mathrm{ism}$, where $r_\circ$ is the starting radius of the shock front. In general, an additional condition is needed in order to determine $\varepsilon$ as a function of the radial coordinate. In the following, $\varepsilon$ is 
assumed to be constant and such an approximation appears to be correct in the GRB context.

In both our work and in the current literature (see Piran \cite{p99}, Chiang \& Dermer \cite{cd99} and Ruffini et al. \cite{Brasile}) a first integral of these equations has been found, leading to expressions for the Lorentz gamma factor as a function of the radial coordinate. In the ``fully adiabatic condition'' (i.e. $\varepsilon = 0$) we have:
\begin{equation} 
\gamma^2=\frac{\gamma_\circ^2+2\gamma_\circ\left(M_\mathrm{ism}/M_B\right) 
+\left(M_\mathrm{ism}/M_B\right)^2}{1+2\gamma_\circ\left(M_\mathrm{ism}/M_B\right)+\left(M_\mathrm{ism}/M_B\right)^2}\, , 
\label{gamma_ad} 
\end{equation} 
while in the ``fully radiative condition'' (i.e. $\varepsilon = 1$) we have: 
\begin{equation} 
\gamma=\frac{1+\left(M_\mathrm{ism}/M_B\right)\left(1+\gamma_\circ^{-1}\right)\left[1+\left(1/2\right)\left(M_\mathrm{ism}/M_B\right)\right]}{\gamma_\circ^{-1}+\left(M_\mathrm{ism}/M_B\right)\left(1+\gamma_\circ^{-1}
\right)\left[1+\left(1/2\right)\left(M_\mathrm{ism}/M_B\right)\right]}\, , 
\label{gamma_rad} 
\end{equation} 
where $\gamma_\circ$ and $M_B$ are respectively the values of the Lorentz gamma factor and of the mass of the accelerated baryons at the beginning of the afterglow phase and $r_\circ$ is the value of the radius $r$ at the beginning of the afterglow phase. 

A major difference between our treatment and the other ones in the current literature is that we have integrated the above equations analytically, obtaining the explicit analytic form of the equations of motion for the expanding shell in the afterglow for a constant ISM density. For the fully radiative case we have explicitly integrated the differential equation for $r\left(t\right)$ in Eq.\eqref{gamma_rad}, recalling that $\gamma^{-2}=1-\left[dr/\left(cdt\right)\right]^2$, where $t$ is the time in the laboratory reference frame. We have then obtained a new explicit analytic solution of the equations of motion for the relativistic shell in the entire range from the ultra-relativistic to the non-relativistic regimes:
\begin{equation} 
\begin{split} 
& t = \tfrac{M_B  - m_i^\circ}{2c\sqrt C }\left( {r - r_\circ } \right) 
+ \tfrac{{r_\circ \sqrt C }}{{12cm_i^\circ A^2 }} \ln \left\{ 
{\tfrac{{\left[ {A + \left(r/r_\circ\right)} \right]^3 \left(A^3  + 
1\right)}}{{\left[A^3  + \left( r/r_\circ \right)^3\right] \left( {A + 1} 
\right)^3}}} \right\} - \tfrac{m_i^\circ r_\circ }{8c\sqrt C }\\ 
& + t_\circ + \tfrac{m_i^\circ r_\circ }{8c\sqrt C } \left( 
{\tfrac{r}{{r_\circ }}} \right)^4 + \tfrac{{r_\circ \sqrt{3C}}}{{6 c 
m_i^\circ A^2 }} \left[\arctan \tfrac{{2\left(r/r_\circ\right) - 
A}}{{A\sqrt 3 }} - \arctan \tfrac{{2 - A}}{{A\sqrt 3 }}\right] 
\end{split} 
\label{analsol} 
\end{equation} 
where $A=\sqrt[3]{\left(M_B-m_i^\circ\right)/m_i^\circ}$, $C={M_B}^2(\gamma_\circ-1)/(\gamma_\circ +1)$ and $m_i^\circ = \left(4/3\right)\pi m_p n_{\mathrm{ism}} r_\circ^3$.

Correspondingly, in the adiabatic case we have: 
\begin{equation} 
t = \left(\gamma_\circ-\tfrac{m_i^\circ}{M_B}\right)\tfrac{r-r_\circ}{c\sqrt{\gamma_\circ^2-1}} 
+ \tfrac{m_i^\circ}{4M_Br_\circ^3}\tfrac{r^4-r_\circ^4}{c\sqrt{\gamma_\circ^2-1}} 
+ t_\circ\, . 
\label{analsol_ad} 
\end{equation} 

In the current literature, following Blandford and McKee \cite{bm76}, a so-called ``ultrarelativistic'' approximation $\gamma_\circ \gg \gamma \gg 1$ has been widely adopted by many authors to solve Eqs.\eqref{Taub_Eq} (see e.g. Sari \cite{s97,s98}, Waxman \cite{w97}, Rees \& M\'esz\'aros \cite{rm98}, Granot et al. \cite{gps99}, Panaitescu \& M\'esz\'aros \cite{pm98c}, Piran \cite{p99}, Gruzinov \& Waxman \cite{gw99}, van Paradijs et al. \cite{vpkw00}, M\'esz\'aros \cite{m02} and references therein). This leads to simple constant-index power-law relations: 
\begin{subequations}\label{gr0-rct} 
\begin{equation} 
\gamma\propto r^{-a}\, , 
\label{gr0} 
\end{equation} 
with $a=3$ in the fully radiative case and $a=3/2$ in the fully adiabatic case. This simple relation is in stark contrast to the complexity of Eq.\eqref{gamma_rad} and Eq.\eqref{gamma_ad} respectively. In the same spirit, instead of Eq.\eqref{analsol} and Eq.\eqref{analsol_ad}, some authors have assumed the following much simpler approximation for the relation between the time and the radial coordinate of the expanding shell, both in the fully radiative and in the fully adiabatic cases:
\begin{equation} 
ct=r\, , 
\label{rct} 
\end{equation} 
while others, like e.g. Panaitescu \& M\'esz\'aros \cite{pm98c}, have integrated the approximate Eq.\eqref{gr0}, obtaining: 
\begin{equation} 
ct=r\left[1+\left(4a+2\right)^{-1}\gamma^{-2}\left(r\right)\right]\, . 
\label{t_app_pm98c} 
\end{equation} 
\end{subequations} 
Again, it is appropriate here to emphasize the stark contrast between Eqs.\eqref{rct},\eqref{t_app_pm98c} and the exact analytic solutions of Eqs.\eqref{Taub_Eq}, expressed in Eqs.\eqref{analsol},\eqref{analsol_ad}.

\subsection{The equitemporal surfaces (EQTSs)}\label{eqts}

As pointed out long ago by Couderc \cite{c39}, in all relativistic expansion the crucial geometrical quantities with respect to a physical observer are the ``equitemporal surfaces'' (EQTSs), namely the locus of source points of the signals arriving at the observer at the same time.

\begin{figure}
\centering
\includegraphics[width=\hsize,clip]{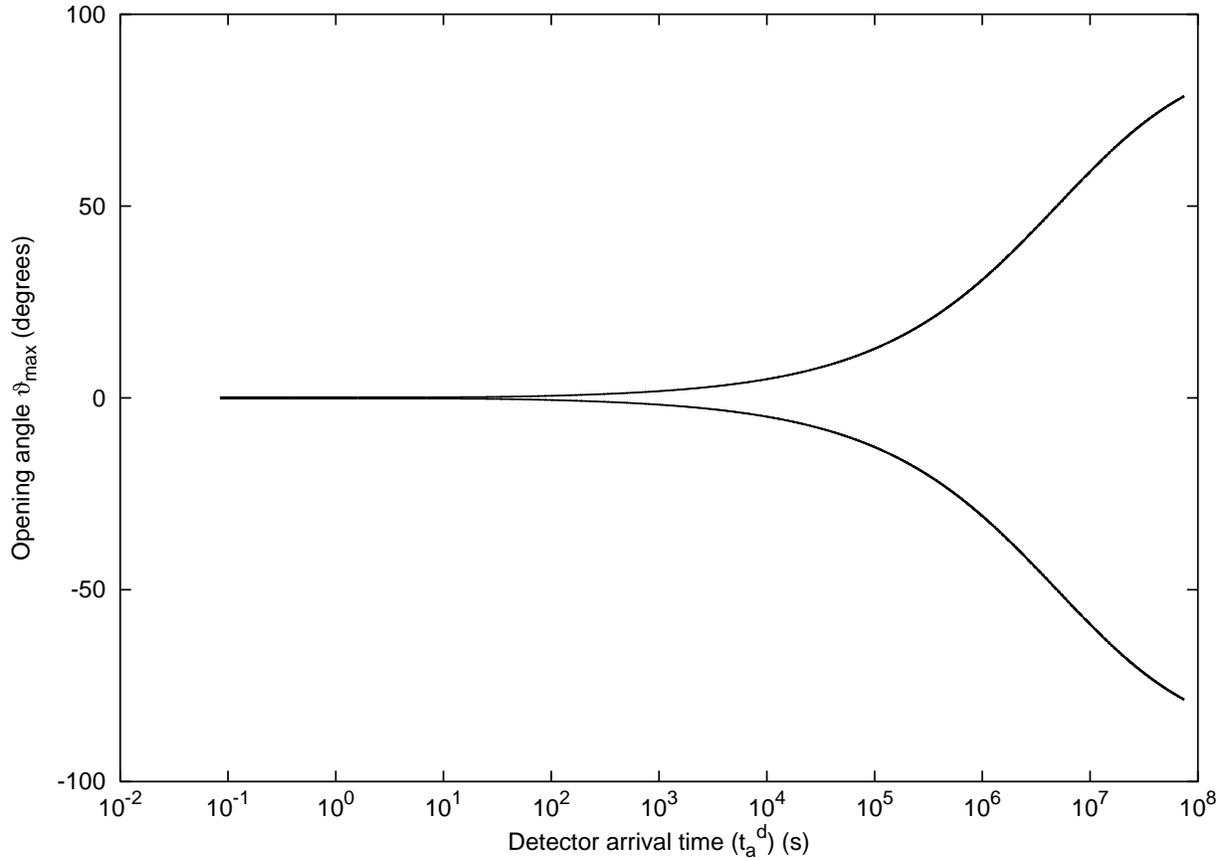}
\caption{Not all values of $\vartheta$ are allowed. Only photons emitted at an angle such that $\cos\vartheta \ge \left(v/c\right)$ can be viewed by the observer. Thus the maximum allowed $\vartheta$ value $\vartheta_{max}$ corresponds to $\cos\vartheta_{max} = (v/c)$. In this figure we show $\vartheta_{max}$ (i.e. the angular amplitude of the visible area of the ABM pulse) in degrees as a function of the arrival time at the detector for the photons emitted along the line of sight (see text). In the earliest GRB phases $v\sim c$ and so $\vartheta_{max}\sim 0$. On the other hand, in the latest phases of the afterglow the ABM pulse velocity decreases and $\vartheta_{max}$ tends to the maximum possible value, i.e. $90^\circ$. Details in Ruffini et al. \cite{rbcfx02_letter,Brasile}}
\label{openang}
\end{figure}

\begin{figure}
\centering
\includegraphics[width=\hsize,clip]{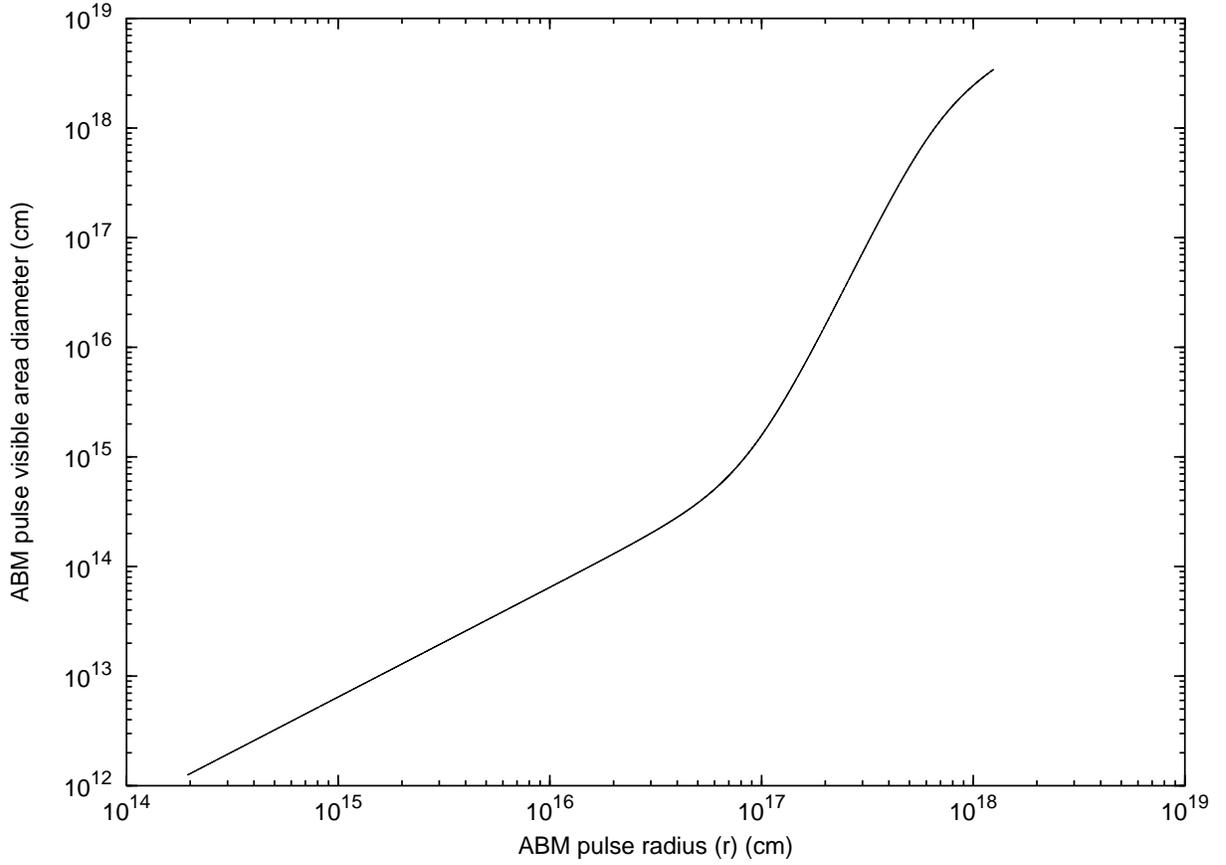}
\caption{The diameter of the visible area is represented as a function of the ABM pulse radius. In the earliest expansion phases ($\gamma\sim 310$) $\vartheta_{max}$ is very small (see left pane and Fig. \ref{opening}), so the visible area is just a small fraction of the total ABM pulse surface. On the other hand, in the final expansion phases $\vartheta_{max} \to 90^\circ$ and almost all the ABM pulse surface becomes visible. Details in Ruffini et al. \cite{rbcfx02_letter,Brasile}}
\label{opensrad}
\end{figure}

\begin{figure}
\centering
\includegraphics[width=8.0cm,clip]{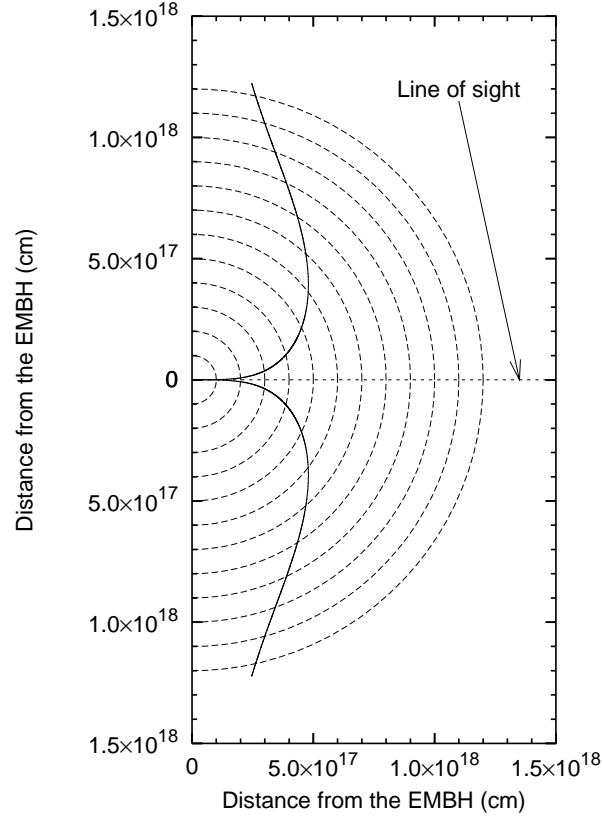}
\caption{This figure shows the temporal evolution of the visible area of the ABM pulse. The dashed half-circles are the expanding ABM pulse at radii corresponding to different laboratory times. The black curve marks the boundary of the visible region. The black hole is located at position (0,0) in this plot. Again, in the earliest GRB phases the visible region is squeezed along the line of sight, while in the final part of the afterglow phase almost all the emitted photons reach the observer. This time evolution of the visible area is crucial to the explanation of the GRB temporal structure. Details in Ruffini et al. \cite{rbcfx02_letter,Brasile}}
\label{opening}
\end{figure}

For a relativistically expanding spherically symmetric source the EQTSs are surfaces of revolution about the line of sight. The general expression for their profile, in the form $\vartheta = \vartheta(r)$, corresponding to an arrival time $t_a$ of the photons at the detector, can be obtained from (see e.g. Ruffini et al. \cite{Brasile}, Bianco and Ruffini \cite{EQTS_ApJL,EQTS_ApJL2} and Figs. \ref{openang}--\ref{opening}):
\begin{equation} 
ct_a = ct\left(r\right) - r\cos \vartheta  + r^\star\, , 
\label{ta_g} 
\end{equation} 
where $r^\star$ is the initial size of the expanding source, $\vartheta$ is the angle between the radial expansion velocity of a point on its surface and the line of sight, and $t = t(r)$ is its equation of motion, expressed in the laboratory frame, obtained by the integration of Eqs.(\ref{Taub_Eq}). From the definition of the Lorentz gamma factor $\gamma^{-2}=1-(dr/cdt)^2$, we have in fact:
\begin{equation} 
ct\left(r\right)=\int_0^r\left[1-\gamma^{-2}\left(r'\right)\right]^{-1/2}dr'\, , 
\label{tdir} 
\end{equation} 
where $\gamma(r)$ comes from the integration of Eqs.(\ref{Taub_Eq}). 

\begin{figure}
\centering
\includegraphics[width=\hsize,clip]{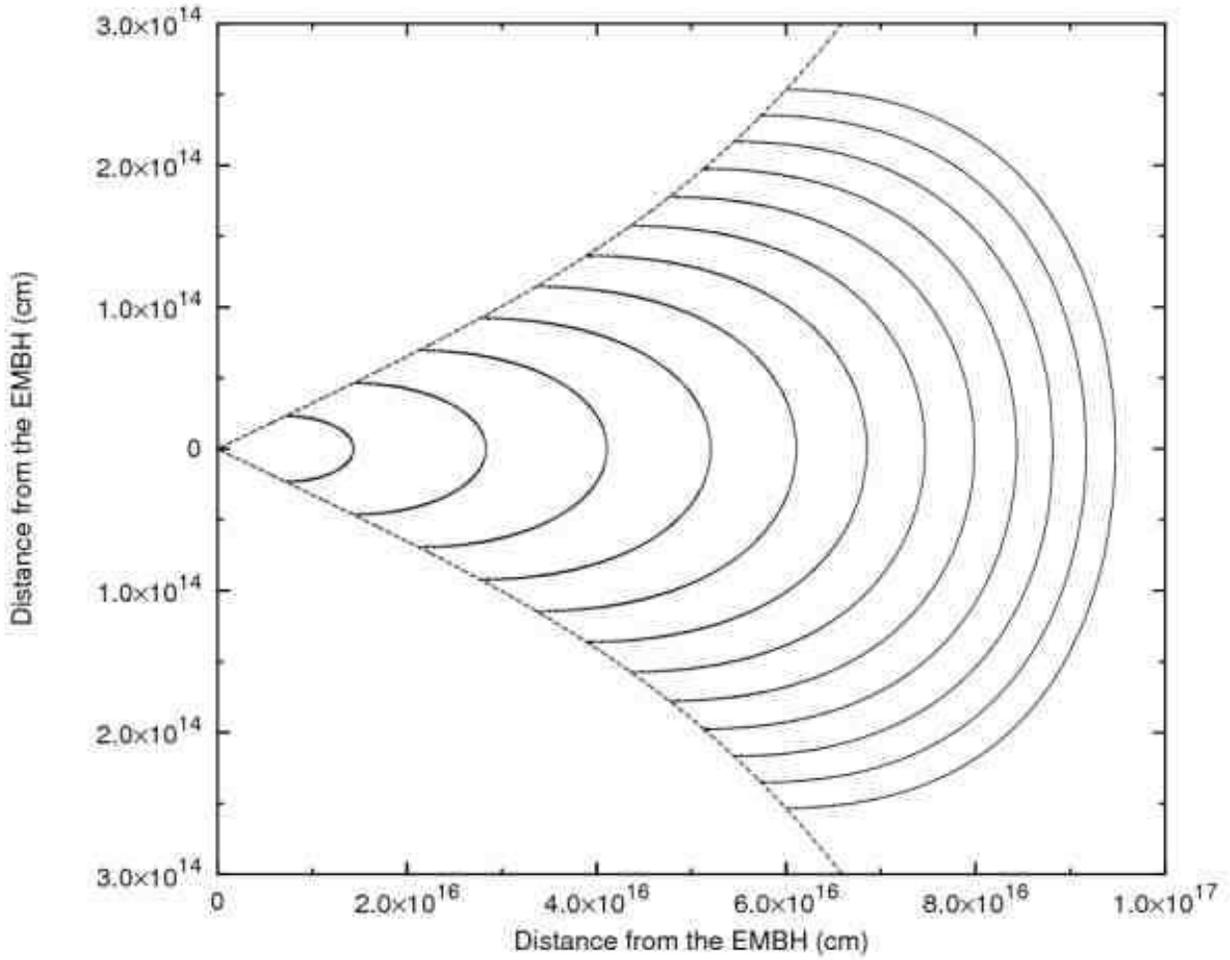}
\caption{Due to the extremely high and extremely varying Lorentz gamma factor, photons reaching the detector on the Earth at the same arrival time are actually emitted at very different times and positions. We represent here the surfaces of photon emission corresponding to selected values of the photon arrival time at the detector: the {\em equitemporal surfaces} (EQTS). Such surfaces differ from the ellipsoids described by Rees in the context of the expanding radio sources with typical Lorentz factor $\gamma\sim 4$ and constant. In fact, in GRB~991216 the Lorentz gamma factor ranges from $310$ to $1$. The EQTSs represented here (solid lines) correspond respectively to values of the arrival time ranging from $5\, s$ (the smallest surface on the left of the plot) to $60\, s$ (the largest one on the right). Each surface differs from the previous one by $5\, s$. To each EQTS contributes emission processes occurring at different values of the Lorentz gamma factor. The dashed lines are the boundaries of the  visible area of the ABM pulse and the black hole is located at position $(0,0)$ in this plot. Note the different scales on the two axes, indicating the very high EQTS ``effective eccentricity''. The time interval from $5\, {\rm s}$ to $60\, {\rm s}$ has been chosen to encompass the E-APE emission, ranging from $\gamma=308.8$ to $\gamma=56.84$. Details in Ruffini et al. \cite{rbcfx02_letter,Brasile}}
\label{ETSNCF}
\end{figure}

\begin{figure}
\centering
\includegraphics[width=8.2cm,clip]{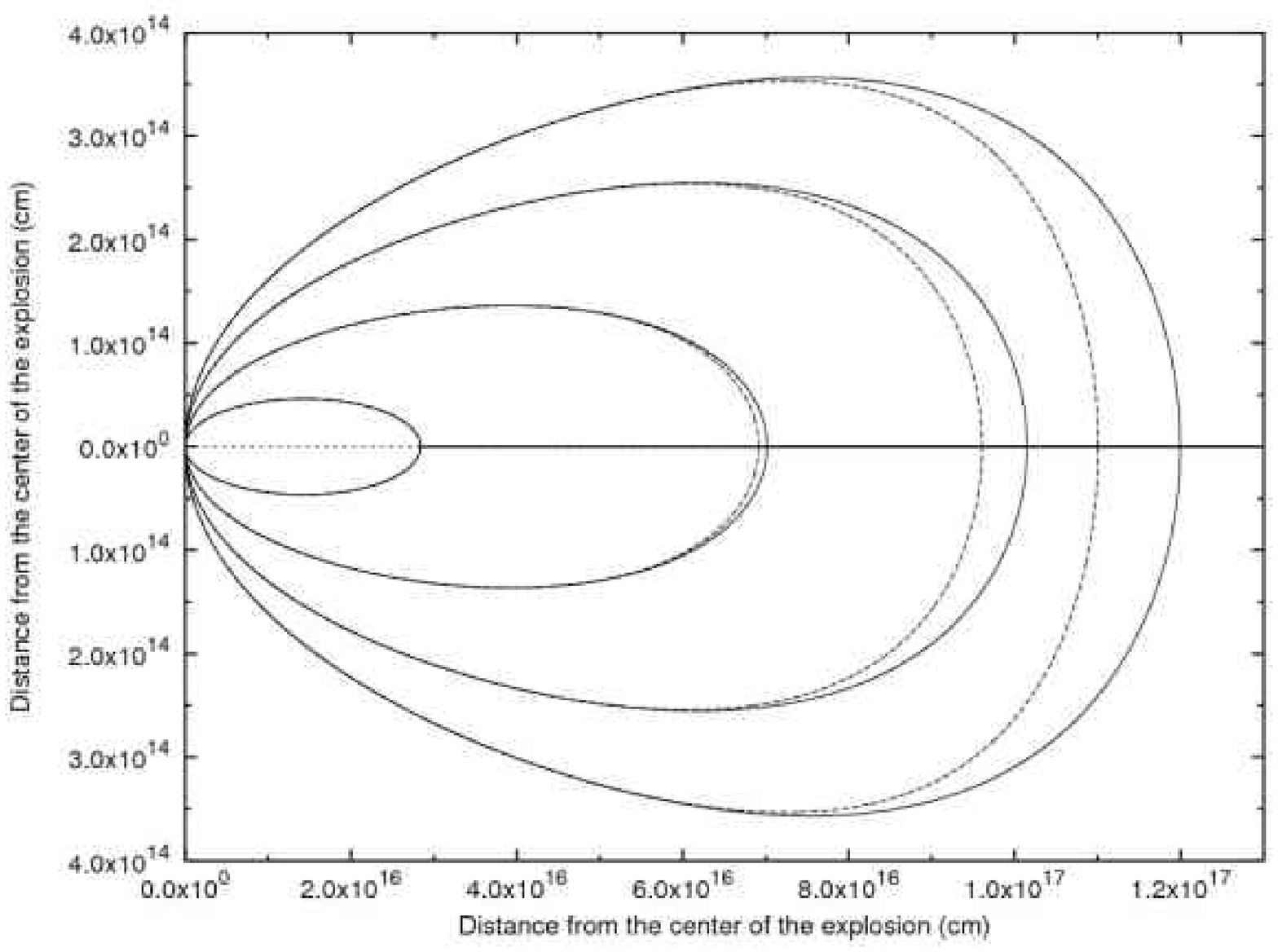}
\includegraphics[width=8.2cm,clip]{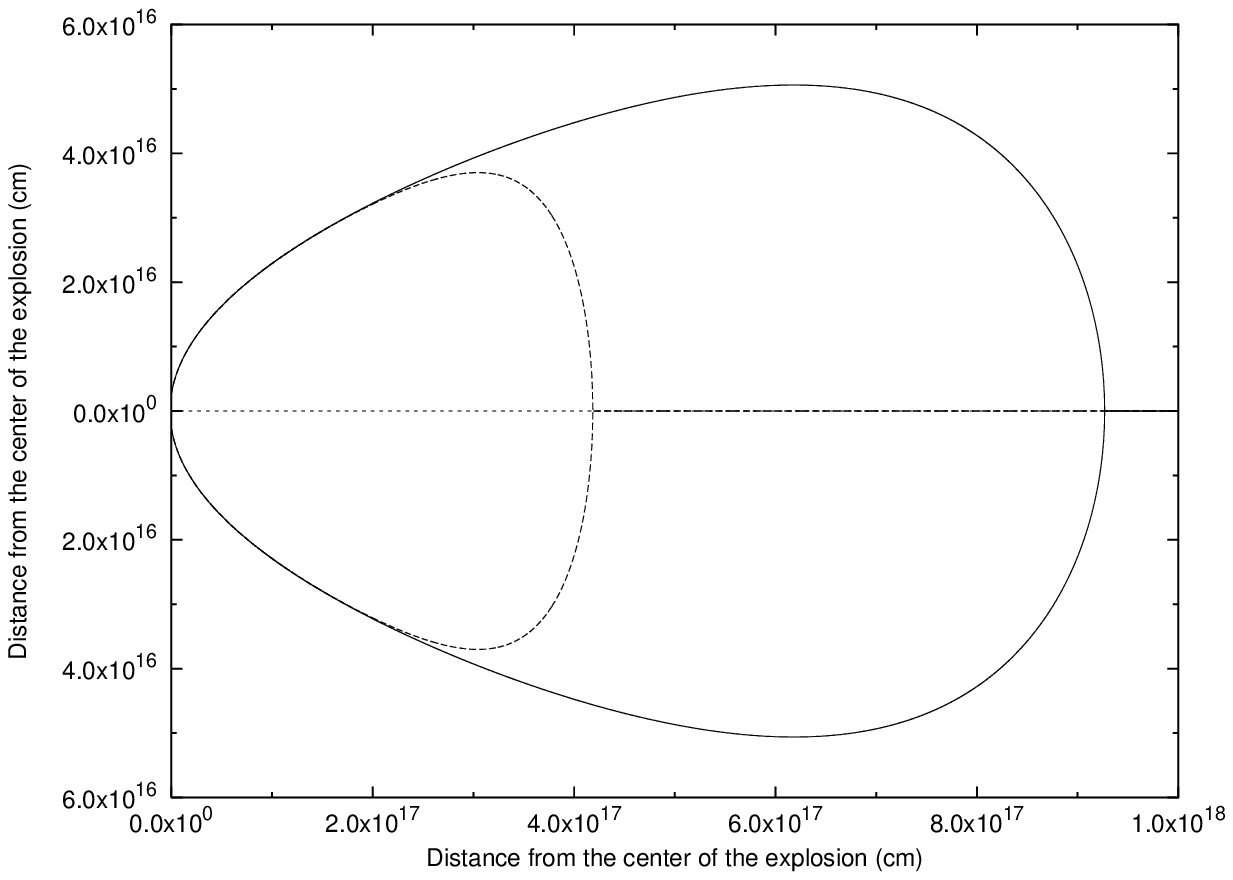} 
\caption{Comparison between EQTSs in the adiabatic regime (solid lines) and in the fully radiative regime (dashed lines). The left plot shows the EQTSs for $t_a=5$ s, $t_a=15$ s, $t_a=30$ s and $t_a=45$ s, respectively from the inner to the outer one. The right plot shows the EQTS at an arrival time of 2 days. Details in Bianco and Ruffini \cite{EQTS_ApJL2}.}
\label{eqts_comp} 
\end{figure}

We have obtained the expressions in the adiabatic case and in the fully radiative cases respectively (see Bianco and Ruffini \cite{EQTS_ApJL2}):
\begin{equation} 
\begin{split} 
\cos\vartheta & = 
\frac{m_i^\circ}{4M_B\sqrt{\gamma_\circ^2-1}}\left[\left(\frac{r}{r_\circ}\right)^3 
  - \frac{r_\circ}{r}\right] + \frac{ct_\circ}{r} \\[6pt] & - 
\frac{ct_a}{r} + \frac{r^\star}{r} - 
\frac{\gamma_\circ-\left(m_i^\circ/M_B\right)}{\sqrt{\gamma_\circ^2-1}}\left[\frac{r_\circ}{r} 
- 1\right]\, . 
\end{split} 
\label{eqts_g_dopo_ad} 
\end{equation} 
\begin{equation} 
\begin{split} 
&\cos\vartheta=\frac{M_B  - m_i^\circ}{2r\sqrt{C}}\left( {r - r_\circ } 
\right) +\frac{m_i^\circ r_\circ }{8r\sqrt{C}}\left[ {\left( 
{\frac{r}{{r_\circ }}} \right)^4  - 1} \right] \\[6pt] 
&+\frac{{r_\circ \sqrt{C} }}{{12rm_i^\circ A^2 }} \ln \left\{ 
{\frac{{\left[ {A + \left(r/r_\circ\right)} \right]^3 \left(A^3  + 
1\right)}}{{\left[A^3  + \left( r/r_\circ \right)^3\right] \left( {A + 1} 
\right)^3}}} \right\} +\frac{ct_\circ}{r}-\frac{ct_a}{r} \\[6pt] & + 
\frac{r^\star}{r} +\frac{{r_\circ \sqrt{3C} }}{{6rm_i^\circ A^2 }} \left[ 
\arctan \frac{{2\left(r/r_\circ\right) - A}}{{A\sqrt{3} }} - \arctan 
\frac{{2 - A}}{{A\sqrt{3} }}\right]\, . 
\end{split} 
\label{eqts_g_dopo} 
\end{equation} 
The two EQTSs are represented at selected values of the arrival time $t_a$ in Fig. \ref{eqts_comp}, where the illustrative case of GRB~991216 has been used as a prototype. The initial conditions at the beginning of the afterglow era are in this case given by $\gamma_\circ = 310.131$, $r_\circ = 1.943 \times 10^{14}$ cm, $t_\circ = 6.481 \times 10^{3}$ s, $r^\star = 2.354 \times 10^8$ cm (see Ruffini et al. \cite{lett1,lett2,rbcfx02_letter,Brasile}).

\subsection{The bolometric luminosity of the source} 

We assume that the internal energy due to kinetic collision is instantly radiated away and that the corresponding emission is isotropic. As in section \ref{eqaft}, let $\Delta \varepsilon$ be the internal energy density developed in the collision. In the comoving frame the energy per unit of volume and per solid angle is simply
\begin{equation} 
\left(\frac{dE}{dV d\Omega}\right)_{\circ}  =  \frac{\Delta \varepsilon}{4 
\pi} 
\label{dEo} 
\end{equation} 
due to the fact that the emission is isotropic in this frame. The total number of photons emitted is an invariant quantity independent of the frame used. Thus we can compute this quantity as seen by an observer in the comoving frame (which we denote with the subscript ``$\circ$'') and by an observer in the laboratory frame (which we denote with no subscripts). Doing this we find:
\begin{equation} 
\frac{dN_\gamma}{dt d \Omega d \Sigma}= \left(\frac{dN_\gamma}{dt d \Omega 
d \Sigma} \right)_{\circ} \Lambda^{-3} 
\cos \vartheta 
\, , 
\end{equation} 
where $\cos\vartheta$ comes from the projection of the elementary surface of the shell on the direction of propagation and $\Lambda = \gamma ( 1 - \beta \cos \vartheta )$ is the Doppler factor introduced in the two following differential transformation
\begin{equation} 
d \Omega_{\circ} = d \Omega \times \Lambda^{-2} 
\end{equation} 
for the solid angle transformation and 
\begin{equation} 
d t_{\circ} = d t \times \Lambda^{-1} 
\end{equation} 
for the time transformation. The integration in $d \Sigma$ is performed over the visible area of the ABM pulse at laboratory time $t$, namely with $0\le\vartheta\le\vartheta_{max}$ and $\vartheta_{max}$ defined in section \ref{eqts} (see Figs. \ref{openang}--\ref{opening}). An extra $\Lambda$ factor comes from the energy transformation:
\begin{equation} 
E_{\circ} = E \times \Lambda\, . 
\end{equation} 
See also Chiang and Dermer \cite{cd99}. Thus finally we obtain:
\begin{equation} 
\frac{dE}{dt d \Omega d \Sigma} = \left(\frac{dE}{dt d \Omega d \Sigma} 
\right)_{\circ} \Lambda^{-4} \cos \vartheta \, . 
\end{equation} 
Doing this we clearly identify  $  \left(\frac{dE}{dt d \Omega d \Sigma} \right)_{\circ} $ 
as the energy density in the comoving frame up to a factor $\frac{v}{4\pi}$ (see Eq.(\ref{dEo})). Then we have: 
\begin{equation} 
\frac{dE}{dt d \Omega } = \int_{shell} \frac{\Delta \varepsilon}{4 \pi} \; 
v \; \cos \vartheta \; \Lambda^{-4} \; d \Sigma\, , 
\label{fluxlab} 
\end{equation} 
where the integration in $d \Sigma$ is performed over the ABM pulse visible area at laboratory time $t$, namely with $0\le\vartheta\le\vartheta_{max}$ and $\vartheta_{max}$ defined in section \ref{eqts}. Eq.(\ref{fluxlab}) gives us the energy emitted toward the observer per unit solid angle and per unit laboratory time $t$ in the laboratory frame.

What we really need is the energy emitted per unit solid angle and per unit detector arrival time $t_a^d$, so we must use the complete relation between $t_a^d$ and $t$ given in Eq.(\ref{ta_g}). First we have to multiply the integrand in Eq.(\ref{fluxlab}) by the factor $\left(dt/dt_a^d\right)$ to transform the energy density generated per unit of laboratory time $t$ into the energy density generated per unit arrival time $t_a^d$. Then we have to integrate with respect to $d \Sigma$ over the {\em equitemporal surface} (EQTS, see section \ref{eqts}) of constant arrival time $t_a^d$ instead of the ABM pulse visible area at laboratory time $t$. The analog of Eq.(\ref{fluxlab}) for the source luminosity in detector arrival time is then:
\begin{equation} 
\frac{dE_\gamma}{dt_a^d d \Omega } = \int_{EQTS} \frac{\Delta 
\varepsilon}{4 \pi} \; v \; \cos \vartheta \; \Lambda^{-4} \; 
\frac{dt}{dt_a^d} d \Sigma\, . 
\label{fluxarr} 
\end{equation} 
It is important to note that, in the present case of GRB 991216, the Doppler factor $\Lambda^{-4}$ in Eq.(\ref{fluxarr}) enhances the apparent luminosity of the burst, as compared to the intrinsic luminosity, by a factor which at the peak of the afterglow is in the range between $10^{10}$ and $10^{12}$!
\begin{figure}
\centering
\includegraphics[width=\hsize,clip]{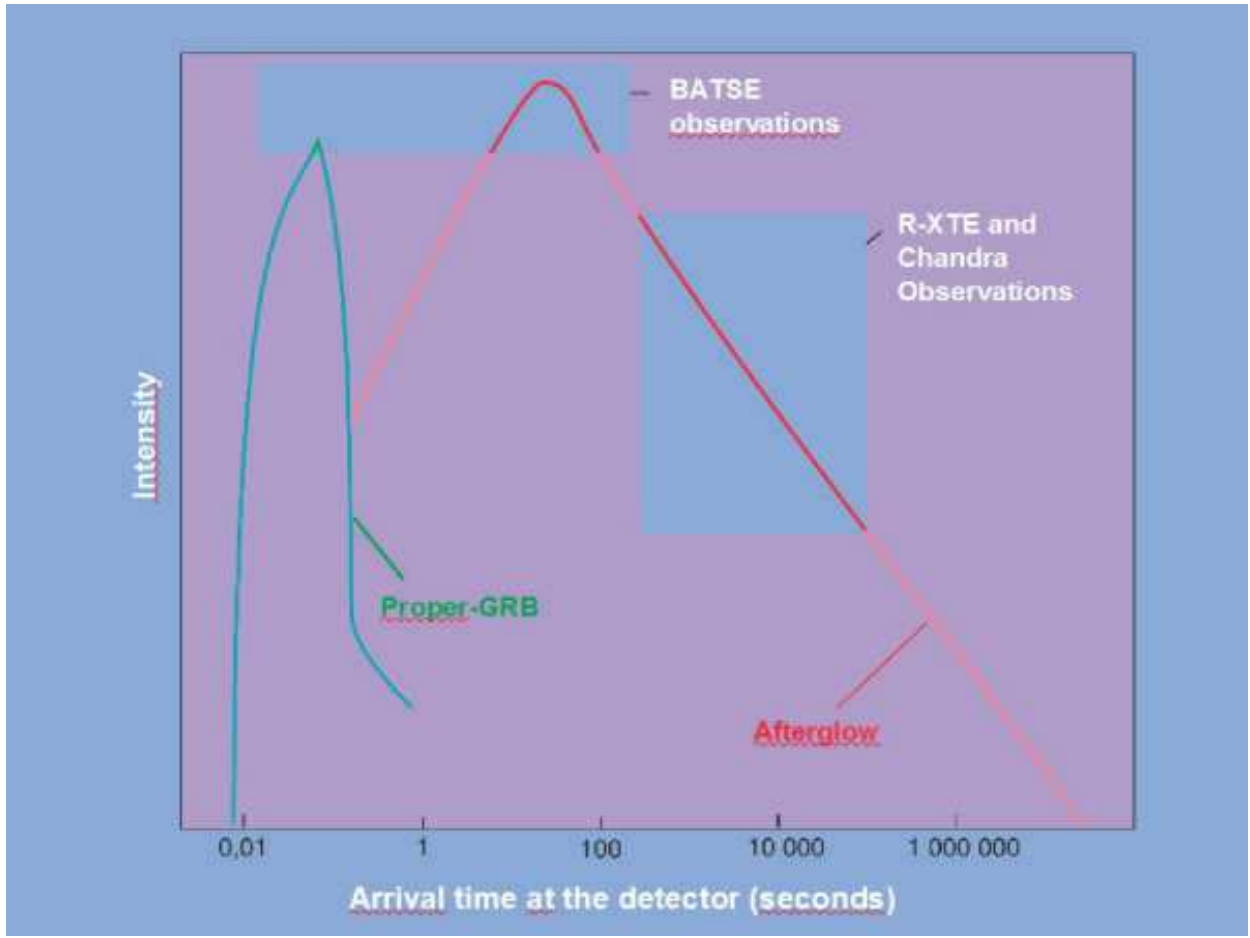}
\caption{Bolometric luminosity of P-GRB and afterglow as a function of the arrival time. Details in Ruffini et al. \cite{Brasile}. Reproduced and adapted from Ruffini et al. \cite{pls} with the kind permission of the publisher.}
\label{bolum}
\end{figure}

We are now able to reproduce in Fig. \ref{bolum} the general behavior of the luminosity starting from the P-GRB to the latest phases of the afterglow as a function of the arrival time. It is generally agreed that the GRB afterglow originates from an ultrarelativistic shell of baryons with an initial Lorentz factor $\gamma_\circ\sim 200$--$300$ with respect to the interstellar medium (see e.g. Ruffini et al. \cite{Brasile}, Bianco \& Ruffini \cite{EQTS_ApJL} and references therein). Using GRB 991216 as a prototype, in Ruffini et al. \cite{lett1,lett2} we have shown how from the time varying bolometric intensity of the afterglow it is possible to infer the average density $\left<n_{ism}\right>=1$ particle/cm$^3$ of the InterStellar Medium (ISM) in a region of approximately $10^{17}$ cm surrounding the black hole giving rise to the GRB phenomenon.

It was shown in Ruffini et al. \cite{rbcfx02_letter} that the theoretical interpretation of the intensity variations in the prompt phase in the afterglow implies the presence in the ISM of inhomogeneities of typical scale $10^{15}$ cm. Such inhomogeneities were there represented for simplicity as spherically symmetric over-dense regions with $\left<n_{ism}^{od}\right> \simeq 10^2\left<n_{ism}\right>$ separated by under-dense regions with $\left<n_{ism}^{ud}\right> \simeq 10^{-2}\left<n_{ism}\right>$ also of typical scale $\sim 10^{15}$ cm in order to keep $\left<n_{ism}\right>$ constant.

\begin{figure}
\centering
\includegraphics[width=\hsize,clip]{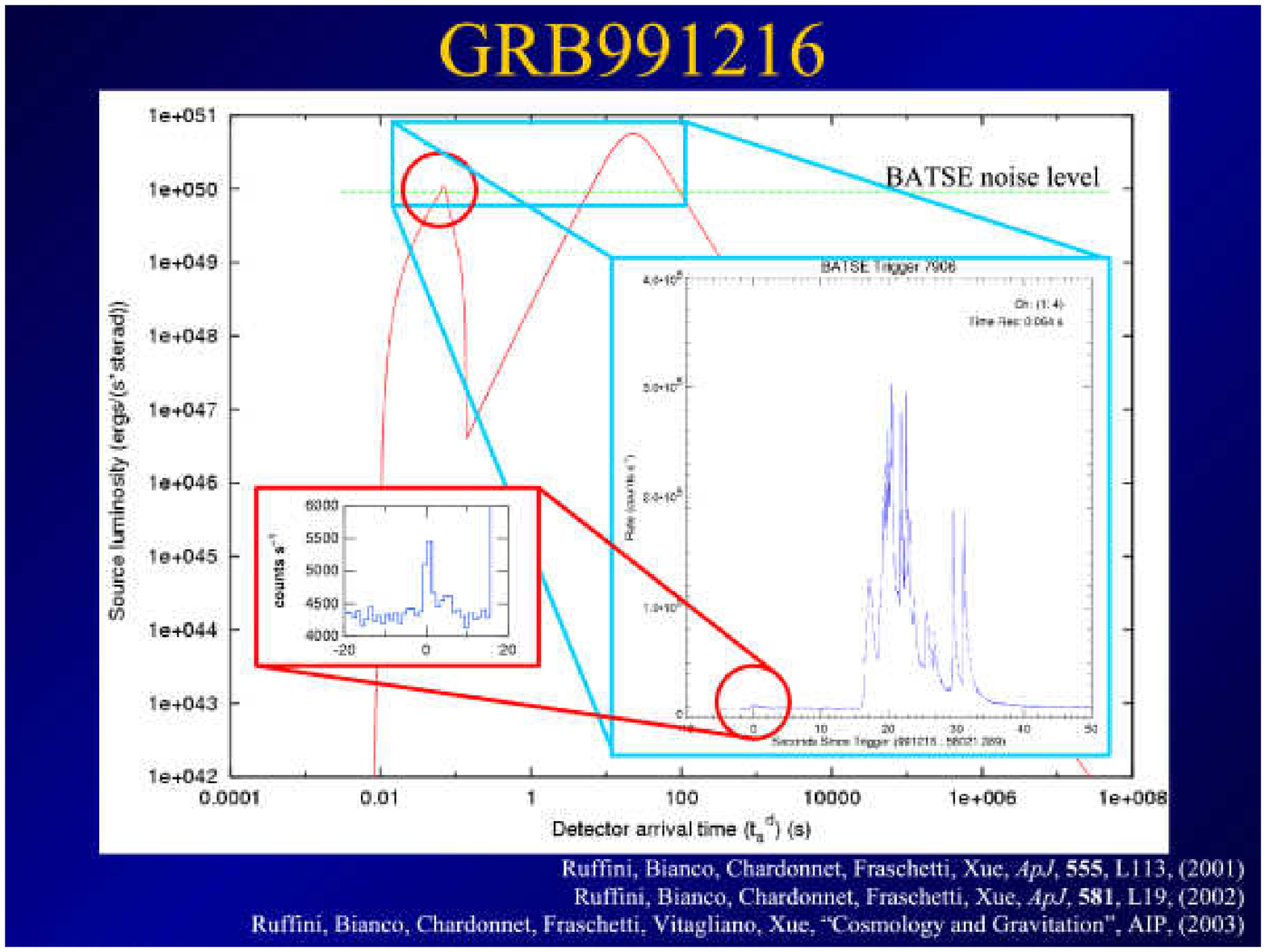}
\caption{The detailed features of GRB 991216 evidenced by our theoretical models are here reproduced. The P-GRB, the ``prompt radiation'' and what is generally called the afterglow. It is clear that the prompt radiation coincides with the extended afterglow peak emission (E-APE) and has been considered as a burst only as a consequence of the high noise threshold in the observations. Details in Ruffini et al. \cite{rbcfx02_letter,Brasile}.}
\label{grb991216}
\end{figure}

The summary of these general results are shown in Fig. \ref{grb991216}, where the P-GRB, the emission at the peak of the afterglow in relation to the ``prompt emission'' and the latest part of the afterglow are clearly identified for the source GRB 991216. Details in Ruffini et al. \cite{Brasile}.

\section{The theory of the luminosity in fixed energy bands and spectra of the afterglow} 

Having obtained a general agreement between the observed luminosity variability and our treatment of the bolometric luminosity, we have further developed the model in order to explain\\ 
a) the details  of the observed luminosity in fixed energy bands, which are the ones actually measured by the detectors on the satellites\\
b) the instantaneous as well as the average spectral distribution in the 
entire afterglow and\\ 
c) the observed hard to soft drift observed in GRB spectra. 

In order to do so we have developed (Ruffini et al. \cite{Spectr2}) a more detailed theory of the structure of the shock front giving rise to the afterglow. We have modeled the interaction between the ultrarelativistic shell of baryons and the ISM by a shock front with three well-defined layers (see e.g. secs. 85--89, 135 of Landau \& Lifshitz \cite{ll6}, ch. 2 and sec. 13--15 of Zel'dovich \& Rayzer \cite{zr66} and sec. IV, 11--13 of Sedov \cite{sedov}). From the back end to the leading edge of this shock front there is:\\ 
{\bf a)} A compressed high-temperature layer, of thickness $\Delta'$, in front of the relativistic baryonic shell, created by the accumulated material swept up in the ISM.\\ 
{\bf b)} A thin shock front, with a jump $\Delta T$ in the temperature which has been traditionally estimated in the comoving frame by the Rankine-Hugoniot adiabatic equations: 
\begin{equation} 
\Delta T \simeq (3/16) m_p\delta v^2/k \simeq 1.5\times 10^{11} 
\left[\delta v/(10^5 km s^{-1})\right]^2 K\,, 
\label{EqT} 
\end{equation} 
where $\delta v$ is the velocity jump, $m_p$ is the proton mass and $k$ is Boltzmann's constant. Of course such a treatment, valid for $\gamma \sim 1$, has to be modified (see below) in our novel treatment for the $\gamma \sim 200$ case relevant to GRBs.\\
{\bf c)} A pre-shock layer of ISM swept-up matter at much lower density and temperature, both of which change abruptly at the thin shock front behind it.

At larger distances ahead of the expanding fireball the ISM is at still smaller densities. The upper limit to the temperature jump at the thin shock front, given in Eq.(\ref{EqT}), is due to the transformation of kinetic energy to thermal energy, since the particle mean free path is assumed to be less then the thickness of the layer (a). The thermal emission of the observed X- and gamma ray radiation, which as seen from the observations reveals a high level of stability, is emitted in the above region (a) due to the sharp temperature gradient at the thin shock front described in the above region (b).

The optical and radio emission comes in our model from the extended region (c). The description of such a region, unlike the sharp and well-defined temperature gradient occurring in region (b), requires magnetohydrodynamic simulations of the evolution of the electron energy distribution of the synchrotron emission. Such analysis has been performed using 3-D Eulerian MHD codes for the particle acceleration models to produce the energy spectrum of cosmic rays at supernova envelope fronts (see e.g. McKee and Cowie \cite{mc75}, Tenorio-Tagle et al. \cite{tt91}, Stone and Norman \cite{sn92}, Jun \& Jones \cite{j99}). Other challenges are the magnetic field and the instabilities. We mention two key phenomena: first, the importance of the development of Kelvin-Helmholtz and Rayleigh-Taylor instabilities ahead of the thin shock front. The second is the dual effect that the shock front has on the ISM initial magnetic field, first through the compression of the swept-up matter containing the field and secondly the amplification of the radial magnetic field component due to the Rayleigh-Taylor instability. Simulations of both effects (see e.g. Jun and Jones \cite{j99} and references therein), modeling the synchrotron radio emission for an expanding supernova shell at various initial magnetic field and ISM parameter values, shows for example that the presence of an initial tangential magnetic field component may essentially affect the resulting magnetic field configuration and hence the outgoing radio flux and spectrum. Among the additional effects to be taken into account are the initial inhomogeneity of the ISM and the contribution of magnetohydrodynamic turbulence.

In our approach we focus uniquely on the X- and gamma ray radiation, which appears to be conceptually much simpler than the optical and radio emission. It is perfectly predictable by a set of constitutive equations (see next section), which leads to directly verifiable and very stable features in the spectral distribution of the observed GRB afterglows. In line with the observations of GRB 991216 and other GRB sources, we assume in the following that the X- and gamma ray luminosity represents approximately 90\% of the energy flux of the afterglow, while the optical and radio emission represents only the remaining 10\%. 

This approach differs significantly from the other ones in the current literature, where attempts are made to explain at once all the multi-wavelength emission in the radio, optical, X and gamma ray as coming from a common origin which is linked to boosted synchrotron emission. Such an approach has been shown to have a variety of difficulties (Ghirlanda et al. \cite{gcg02}, Preece et al. \cite{pa98}) and cannot anyway have the instantaneous variability needed to explain the structure in the ``prompt radiation'' in an external shock scenario, which is indeed confirmed by our model.

\subsection{The equations determining the luminosity in fixed energy bands} 

Here the fundamental new assumption is adopted (see also Ruffini et al. \cite{Spectr1}) that the X- and gamma ray radiation during the entire afterglow phase has a thermal spectrum in the co-moving frame. The temperature is then given by:
\begin{equation} 
T_s=\left[\Delta E_{\rm int}/\left(4\pi r^2 \Delta \tau \sigma 
\mathcal{R}\right)\right]^{1/4}\, , 
\label{TdiR} 
\end{equation} 
where $\Delta E_{\rm int}$ is the internal energy developed in the collision with the ISM in a time interval $\Delta \tau$ in the co-moving frame, $\sigma$ is the Stefan-Boltzmann constant and
\begin{equation} 
\mathcal{R}=A_{eff}/A\, , 
\label{Rdef} 
\end{equation} 
is the ratio between the ``effective emitting area'' of the afterglow and the surface area of radius $r$. In GRB 991216 such a factor is observed to be decreasing during the afterglow between: $3.01\times 10^{-8} \ge \mathcal{R} \ge 5.01 \times 10^{-12}$ (Ruffini et al. \cite{Spectr1}). 

\begin{figure}
\centering
\includegraphics[width=\hsize,clip]{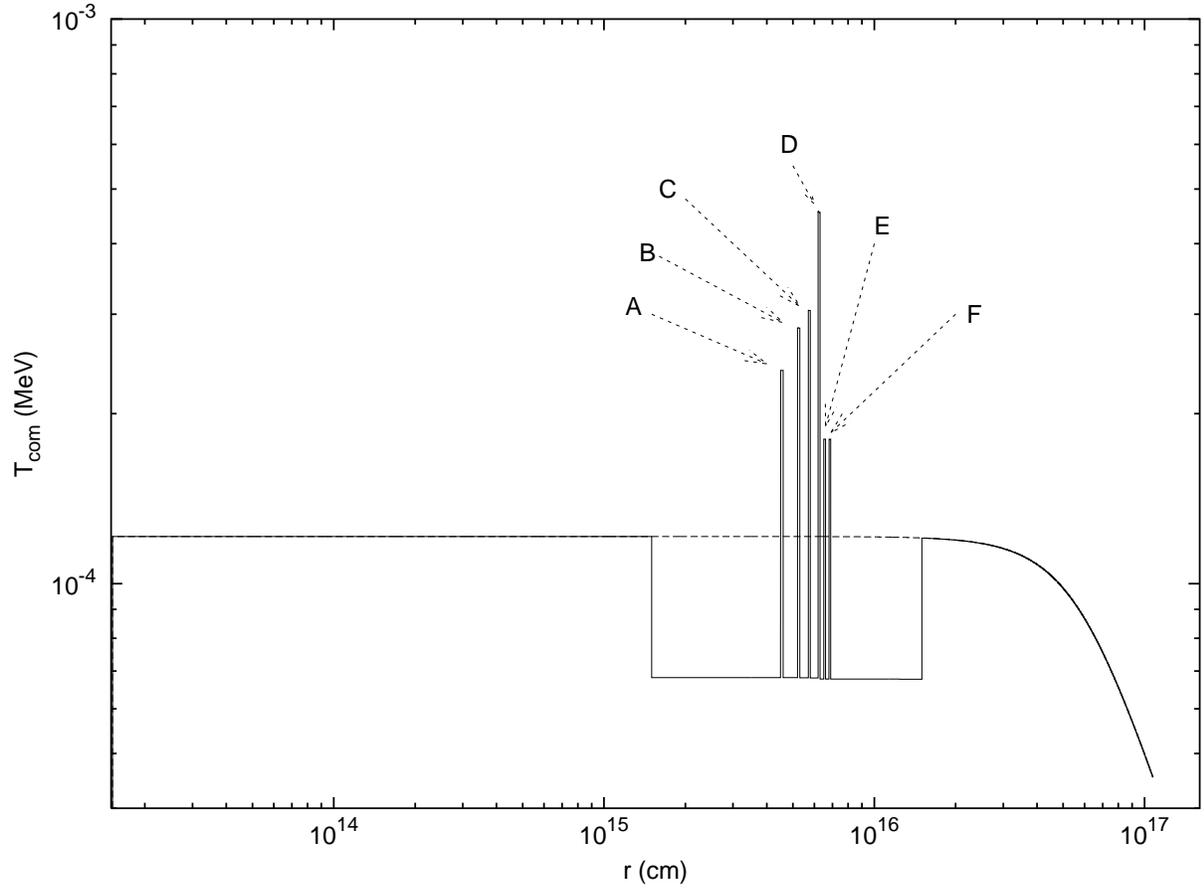}
\caption{The temperature in the comoving frame of the shock front corresponding to the density distribution with the six spikes A,B,C,D,E,F presented in Ruffini et al.$^5$. The dashed line corresponds to an homogeneous distribution with $n_{ism}=1$. Details in Ruffini et al. \cite{Spectr2}.}
\label{tcom_fig}
\end{figure}

The temperature in the comoving frame corresponding to the density distribution described in Ruffini et al. \cite{rbcfx02_letter} is shown in Fig. \ref{tcom_fig}. 

We are now ready to evaluate the source luminosity in a given energy band. The source luminosity at a detector arrival time $t_a^d$, per unit solid angle $d\Omega$ and in the energy band $\left[\nu_1,\nu_2\right]$ is given by (see Ruffini et al. \cite{Brasile,Spectr1}): 
\begin{equation} 
\frac{dE_\gamma^{\left[\nu_1,\nu_2\right]}}{dt_a^d d \Omega } = 
\int_{EQTS} \frac{\Delta \varepsilon}{4 \pi} \; v \; \cos \vartheta \; 
\Lambda^{-4} \; \frac{dt}{dt_a^d} W\left(\nu_1,\nu_2,T_{arr}\right) d 
\Sigma\, , 
\label{fluxarrnu} 
\end{equation} 
where $\Delta \varepsilon=\Delta E_{int}/V$ is the energy density released in the interaction of the ABM pulse with the ISM inhomogeneities measured in the comoving frame, $\Lambda=\gamma(1-(v/c)\cos\vartheta)$ is the Doppler factor, $W\left(\nu_1,\nu_2,T_{arr}\right)$ is an ``effective weight'' required to evaluate only the contributions in the energy band $\left[\nu_1,\nu_2\right]$, $d\Sigma$ is the surface element of the EQTS at detector arrival time $t_a^d$ on which the integration is performed (see also Ruffini et al. \cite{rbcfx02_letter}) and $T_{arr}$ is the observed temperature of the radiation emitted from $d\Sigma$: 
\begin{equation} 
T_{arr}=T_s/\left[\gamma 
\left(1-(v/c)\cos\vartheta\right)\left(1+z\right)\right]\, . 
\label{Tarr} 
\end{equation} 

The ``effective weight'' $W\left(\nu_1,\nu_2,T_{arr}\right)$ is given by the ratio of the integral over the given energy band of a Planckian distribution at a temperature $T_{arr}$ to the total integral $aT_{arr}^4$: 
\begin{equation} 
W\left(\nu_1,\nu_2,T_{arr}\right)=\frac{1}{aT_{arr}^4}\int_{\nu_1}^{\nu_2}\rho\left(T_{arr},\nu\right)d\left(\frac{h\nu}{c}\right)^3\, , 
\label{effweig} 
\end{equation} 
where $\rho\left(T_{arr},\nu\right)$ is the Planckian distribution at temperature $T_{arr}$: 
\begin{equation} 
\rho\left(T_{arr},\nu\right)=\left(2/h^3\right)h\nu/\left(e^{h\nu/\left(kT_{arr}\right)}-1\right) 
\label{rhodef} 
\end{equation} 

\section{On the time integrated spectra and the hard-to-soft spectral transition} 

\begin{figure}
\centering
\includegraphics{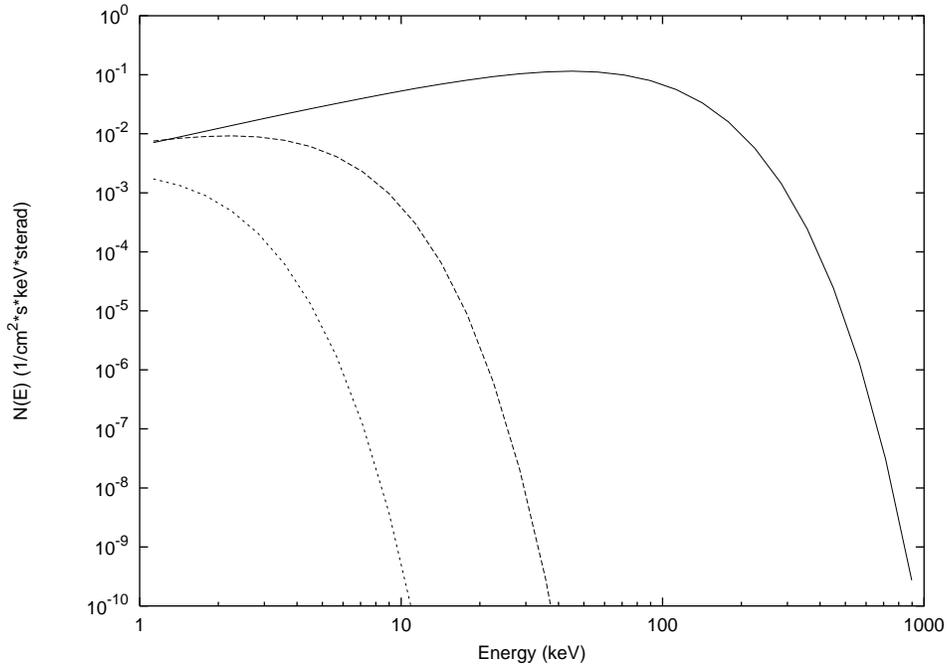} 
\caption{The instantaneous spectra of the radiation observed in GRB~991216 at three different EQTS respectively, from top to bottom, for $t_a^d=10$ s, $t_a^d=10^4$ s and $t_a^d=1.45\times10^5$ s. These diagrams have been computed assuming a constant $\left<n_{ism}\right>\simeq 1$ particle/cm$^3$ and clearly explains the often quoted hard-to-soft spectral evolution in GRBs. Details in Ruffini et al. \cite{Spectr1}.}
\label{spectrum} 
\end{figure}

We turn now to the much debated issue of the origin of the observed hard-to-soft spectral transition during the GRB observations (see e.g. Frontera et al. \cite{fa00}, Ghirlanda et al. \cite{gcg02}, Piran \cite{p99}, Piro et al. \cite{p99b}). We consider the instantaneous spectral distribution of the observed radiation for three different EQTSs:
\begin{itemize} 
\item $t_a^d=10$ s, in the early radiation phase near the peak of the luminosity, 
\item $t_a^d=1.45\times 10^5$ s, in the last observation of the afterglow by the Chandra satellite, and 
\item $t_a^d=10^4$ s, chosen in between the other two (see Fig. \ref{spectrum}). 
\end{itemize} 
The observed hard-to-soft spectral transition is then explained and traced back to: 
\begin{enumerate} 
\item a time decreasing temperature of the thermal spectrum measured in the comoving frame, 
\item the GRB equations of motion, 
\item the corresponding infinite set of relativistic transformations. 
\end{enumerate} 
A clear signature of our model is the existence of a common low-energy behavior of the instantaneous spectrum represented by a power-law with index $\alpha = +0.9$. This prediction will be possibly verified in future observations.

\begin{figure}
\centering
\includegraphics{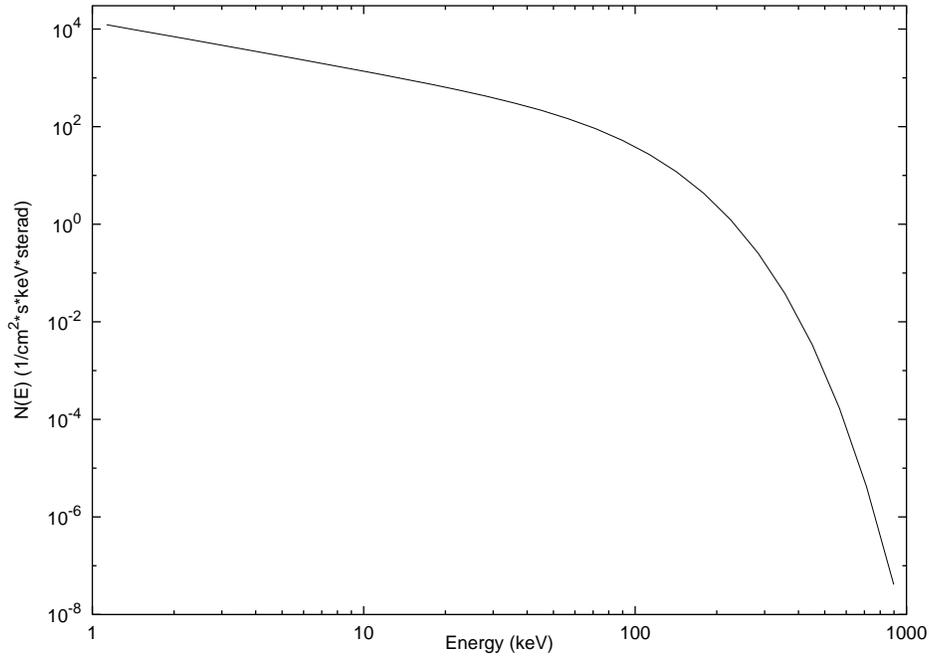} 
\caption{The time-integrated spectrum of the radiation observed in GRB~991216. The low energy part of the curve below $10$ keV is fit by a power-law with index $\alpha = -1.05$ and the high energy part above $500$ keV is fit by a power-law with an index $\beta < -16$. Details in Ruffini et al. \cite{Spectr1}.}
\label{spectband} 
\end{figure}

Starting from these instantaneous values, we integrate the spectra in arrival time obtaining what is usually fit in the literature by the ``Band relation'' (Band et al. \cite{b93}). Indeed we find for our integrated spectra a low energy spectral index $\alpha=-1.05$ and an high energy spectral index $\beta < -16$ when interpreted within the framework of a Band relation (see Fig. \ref{spectband}). This theoretical result can be submitted to a direct confrontation with the observations of GRB 991216 and, most importantly, the entire theoretical framework which we have developed can now be applied to any GRB source. The theoretical predictions on the luminosity in fixed energy bands so obtained can be then straightforwardly confronted with the observational data.

\section{The three paradigms for the interpretation of GRBs}\label{fp}

Having outlined the main features of our model and shown its application to GRB 991216 used as a prototype, before addressing the two new sources which are going to be the focus of this presentation, we recall the three paradigms for the interpretation of GRBs we had previously introduced.

The first paradigm, the relative space-time transformation (RSTT) paradigm (Ruffini et al. \cite{lett1}) emphasizes the importance of a global analysis of the GRB phenomenon encompassing both the optically thick and the afterglow phases. Since all the data are received in the detector arrival time it is essential to know the equations of motion of all relativistic phases with $\gamma > 1$ of the GRB sources in order to reconstruct the time coordinate in the laboratory frame, see Eq.(\ref{tadef}). Contrary to other phenomena in nonrelativistic physics or astrophysics, where every phase can be examined separately from the others, in the case of GRBs all the phases are inter-related by their signals received in arrival time $t_a^d$. There is the need, in order to describe the physics of the source, to derive the laboratory time $t$ as a function of the arrival time $t_a^d$ along the entire past worldline of the source using Eq.(\ref{taddef}).

\begin{figure}
\centering
\includegraphics[width=\hsize,clip]{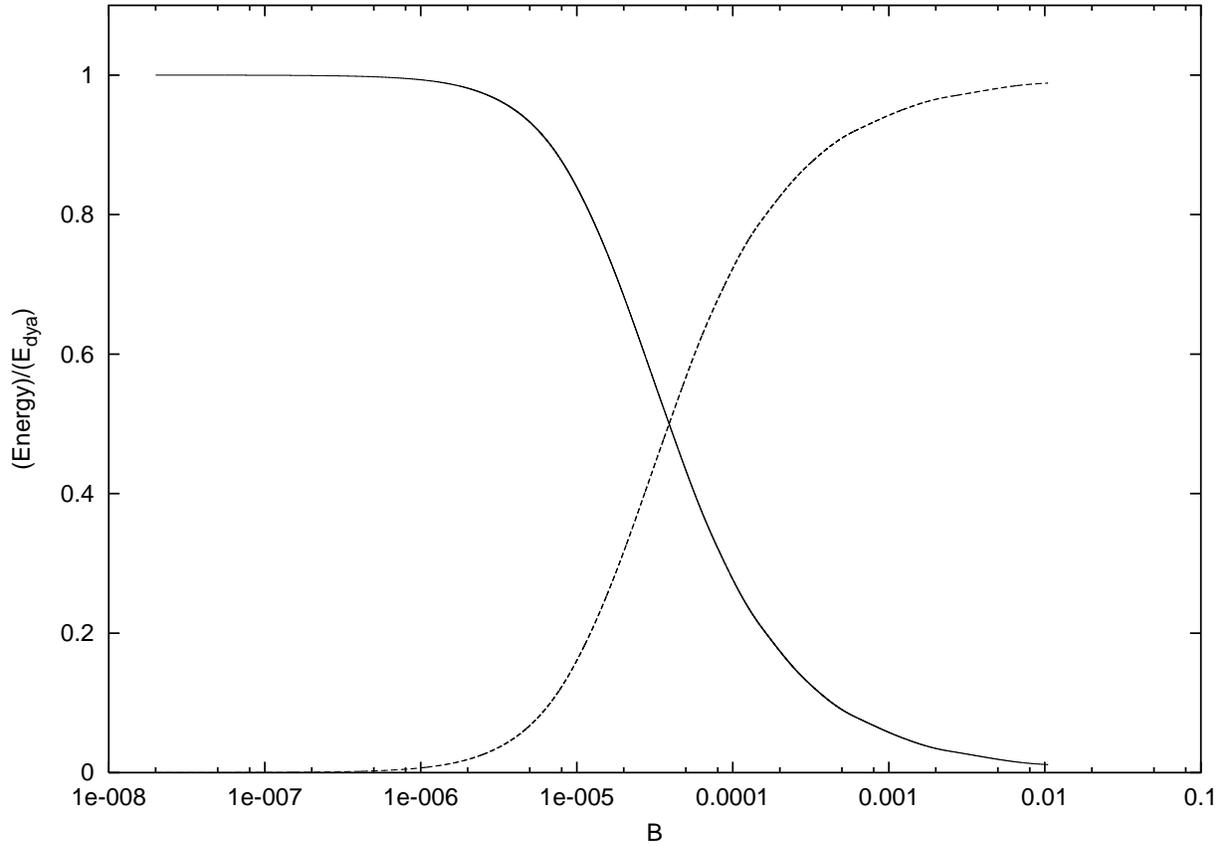}
\caption{The energy radiated in the P-GRB (the solid line) and in the afterglow (the dashed line), in units of the total energy of the dyadosphere ($E_{dya}$), are plotted as functions of the $B$ parameter.}
\label{bcross}
\end{figure}

The second paradigm, the interpretation of the burst structure (IBS) paradigm (Ruffini et al. \cite{lett2}) covers three fundamental issues:\\ 
a) the existence, in the general GRB, of two different components: the P-GRB and the afterglow related by precise equations determining their relative amplitude and temporal sequence (see Ruffini et al. \cite{Brasile});\\ 
b) what in the literature has been addressed as the ``prompt emission'' and considered as a burst, in our model is not a burst at all --- instead it is just the emission from the peak of the afterglow (see Fig. \ref{grb991216});\\
c) the crucial role of the parameter $B$ in determining the relative amplitude of the P-GRB to the afterglow and discriminating between the short and the long bursts (see Fig. \ref{bcross}). Both short and long bursts arise from the same physical phenomena: the dyadosphere. The absence of baryonic matter in the remnant leads to the short bursts and no afterglow. The presence of baryonic matter with $B < 10^{-2}$ leads to the afterglow and consequently to its peak emission which gives origin to the so-called long bursts.

The third paradigm, the GRB-Supernova Time Sequence (GSTS) paradigm (Ruffini et al. \cite{lett3}), deals with the relation of the GRB and the associated supernova process, and acquires a special meaning in relation to the sources GRB 980425 and GRB 030329 as we will show in the following.

We now shortly illustrate some consequences of these three paradigms.

\subsection{Long bursts are E-APEs}\label{eape}

The order of magnitude estimate usually quoted for the the characteristic time scale to be expected for a burst emitted by a GRB at the moment of transparency at the end of the expansion of the optically thick phase is given by $\tau \sim GM/c^3$, which for a $10M_\odot$ black hole will give $\sim 10^{-3}$ s. There are reasons today not to take seriously such an order of magnitude estimate (see e.g. Ruffini et al. \cite{rfvx05}). In any case this time is much shorter then the ones typically observed in ``prompt radiation'' of the long bursts, from a few seconds all the way to $10^2$ s. In the current literature (see e.g. Piran \cite{p99} and references therein), in order to explain the ``prompt radiation'' and overcome the above difficulty it has been generally assumed that its origin should be related to a prolonged ``inner engine'' activity preceding the afterglow which is not well identified.

To us this explanation has always appeared logically inconsistent since there remain to be explained not one but two very different mechanisms, independent of each other, of similar and extremely large energetics. This approach has generated an additional very negative result: it has distracted everybody working in the field from the earlier very interesting work on the optically thick phase of GRBs.

The way out of this dichotomy in our model is drastically different: 1) indeed the optically thick phase exists, is crucial to the GRB phenomenon and terminates with a burst: the P-GRB; 2) the ``prompt radiation'' follows the P-GRB; 3) the ``prompt radiation'' is not a burst: it is actually the temporally extended peak emission of the afterglow (E-APE). The observed structures of the prompt radiation can all be traced back to inhomogeneities in the interstellar medium (see Fig. \ref{grb991216} and Ruffini et al. \cite{rbcfx02_letter}).

\subsection{Short bursts are P-GRBs} 

The fundamental diagram determining the relative intensity of the P-GRB and the afterglow as a function of the dimensionless parameter $B$ has been shown in Fig. \ref{bcross}. The underlying machine generating the short and the long GRBs is identical: in both cases is the dyadosphere. The main difference relates to the amount of baryonic matter engulfed by the electron-positron plasma in their optically thick phase prior to transparency. In the limit of small $B < 10^{-5}$ the intensity of the P-GRB is larger and dominates the afterglow. This corresponds to the short bursts. For $10^{-5} < B < 10^{-2}$ the  afterglow dominates the GRBs and we have the so-called ``long bursts''. For $B > 10^{-2}$ we may observe a third class of ``bursts'', eventually related to a turbulent process occurring prior to transparency (Ruffini et al. \cite{rswx00}). This third family should be characterized  by  smaller values of the Lorentz gamma factors than in the case of the short or long bursts.

\subsection{The trigger of multiple gravitational collapses}

The relation between the GRBs and the supernovae is one of the most complex aspects to be addressed by our model, which needs the understanding of new fields of general relativistic physics in relation to yet unexplored many-body solutions in a substantially new astrophysical scenario.

As we will show in the two systems GRB980425/SN1998bw and GRB030329/SN2003dh which we are going to discuss next, there is in each one the possibility of an astrophysical ``triptych''\footnote{A picture or carving in three panels side by side; {\em esp}: an altarpiece with a central panel and two flanking panels half its size that fold over it [Webster's New collegiate dictionary, G. \& C. Merriam Co. (Springfield, Massachussets, U.S.A., 1977)]} formed by:\\
1) the formation of the black hole and the emission of the GRB,\\ 
2) the gravitational collapse of an evolved companion star, leading to a supernova,\\ 
3) a clearly identified URCA source whose nature appears to be of the greatest interest.

This new astrophysical scenario presents new challenges:\\
a) The identification of the physical reasons of the instability leading to the gravitational collapse of a $\sim 10M_\odot$ star, giving origin to the black hole. Such an implosion must occur radially with negligible mass of the remnant ($B < 10^{-2}$).\\
b) The identification of the physical reasons for the instability leading to the gravitational collapse of an evolved companion star, giving origin to the supernova.\\
c) The theoretical issues related to the URCA sources, which range today in many possible directions: from the physics of black holes, to the physical processes occurring in the expanding supernova remnants, and finally to the very exciting possibility that we are observing for the first time a newly born neutron star. The main effort in the next sections is to show that the detailed understanding we have reached for the GRB phenomenon and its afterglow allows us to state, convincingly, that the URCA source, contrary to what established in the current literature, is not part of the GRB nor of its afterglow.

We will draw in the conclusions some considerations on the possible nature of the URCA sources.

\section{Applications} 

We illustrate the application of our GRB model to two different systems, which are quite different in the energetics but are both related to supernovae: GRB 980425 and GRB 030329. We will let the gradual theoretical understanding of the system to unveil the underlying astrophysical scenario.

\subsection{GRB 980425 / SN 1998bw}

Approaches in the current literature have always attempted to explain both the supernova and the GRB as two aspects of a single phenomenon assuming that the GRB takes his origin from a specially strong and yet unobserved supernova process: a hypernova (see Paczy\'nski \cite{p98}, Kulkarni \cite{ka98}, Iwamoto \cite{ia98}).

We have taken a very different approach, following Cicero's classic aphorism ``divide et impera'', which was adopted as the motto of the Roman empire: ``divide and conquer''. In this specific case of GRBs, which are indeed a very complex system, we plan to divide and identify the truly independent physical constituents and conquer the understanding of the underlying astrophysical process. As we will see, this approach will lead to an unexpected and much richer scenario.

\begin{figure}
\centering 
\includegraphics[width=\hsize]{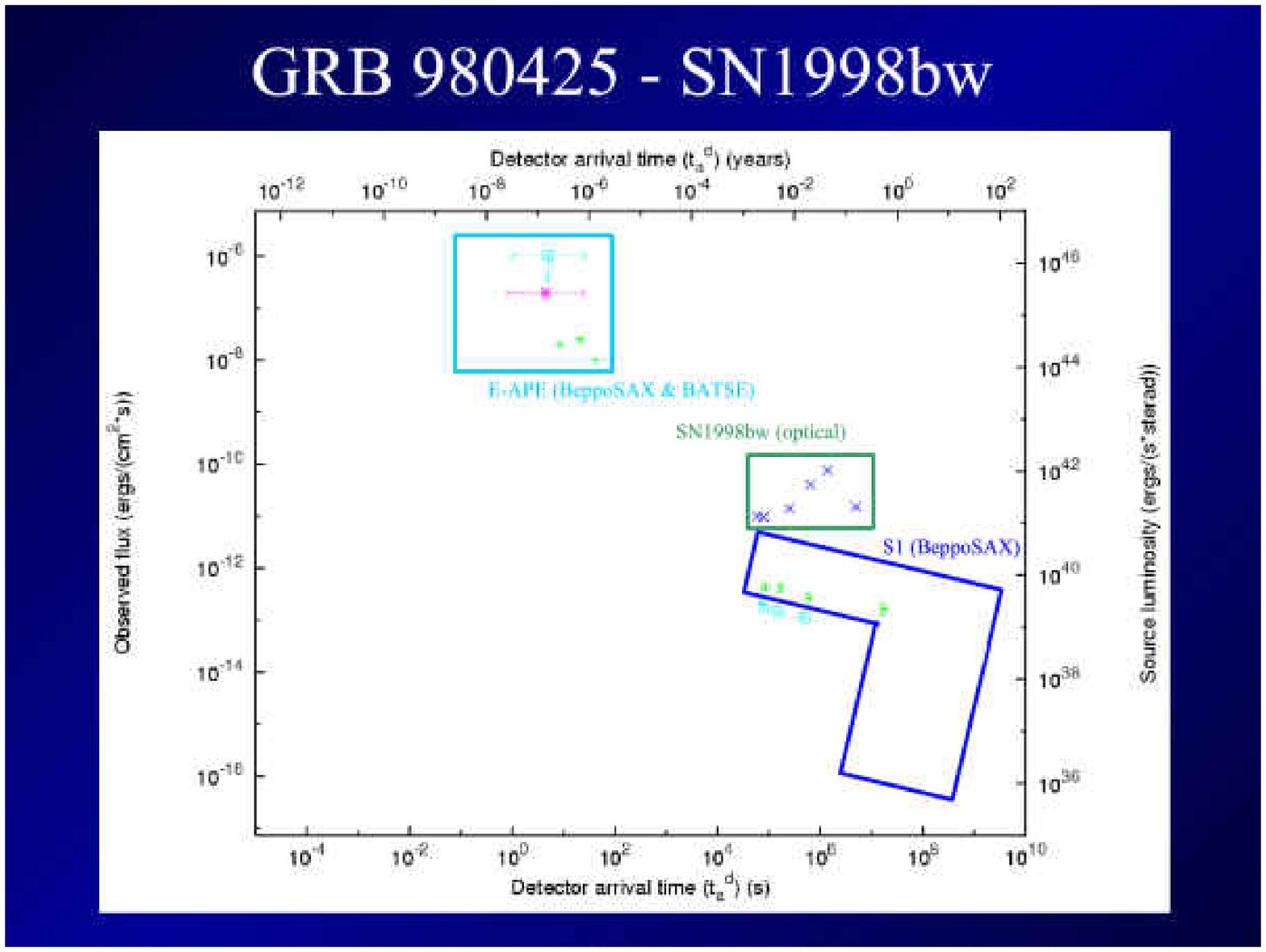}
\caption{The ``divide et impera'' concept applied to the system GRB 980425 / SN 1998bw. Four different components are identified: the GRB 980425, the SN 1998bw, and the two sources S1 and S2.}
\label{d14}
\end{figure}

In addition to the source GRB 980425 and the supernova SN1998bw, two X-ray sources have been found by BeppoSAX in the error box for the location of GRB 980425: a source {\em S1} and a source {\em S2} (Pian et al. \cite{pian00}), which have been traditionally interpreted either as a background source or as a part of the GRB afterglow. See Fig. \ref{d14}. Our approach has been: to  first comprehend the entire afterglow of GRB 980425 within our theory. This allows the computation of the luminosity in given energy bands, the spectra, the Lorentz gamma factors, and more generally of all the dynamical aspects of the source. Having characterized the features of GRB 980425, we can gradually approach the remaining part of the scenario, disentangling the GRB observations from those of the supernova and then disentangling both the GRB and the supernova observations from those of the sources {\em S1} and {\em S2}. This leads to a natural identification of distinct events and to their autonomous astrophysical characterization.

\begin{figure}
\centering 
\includegraphics[width=\hsize]{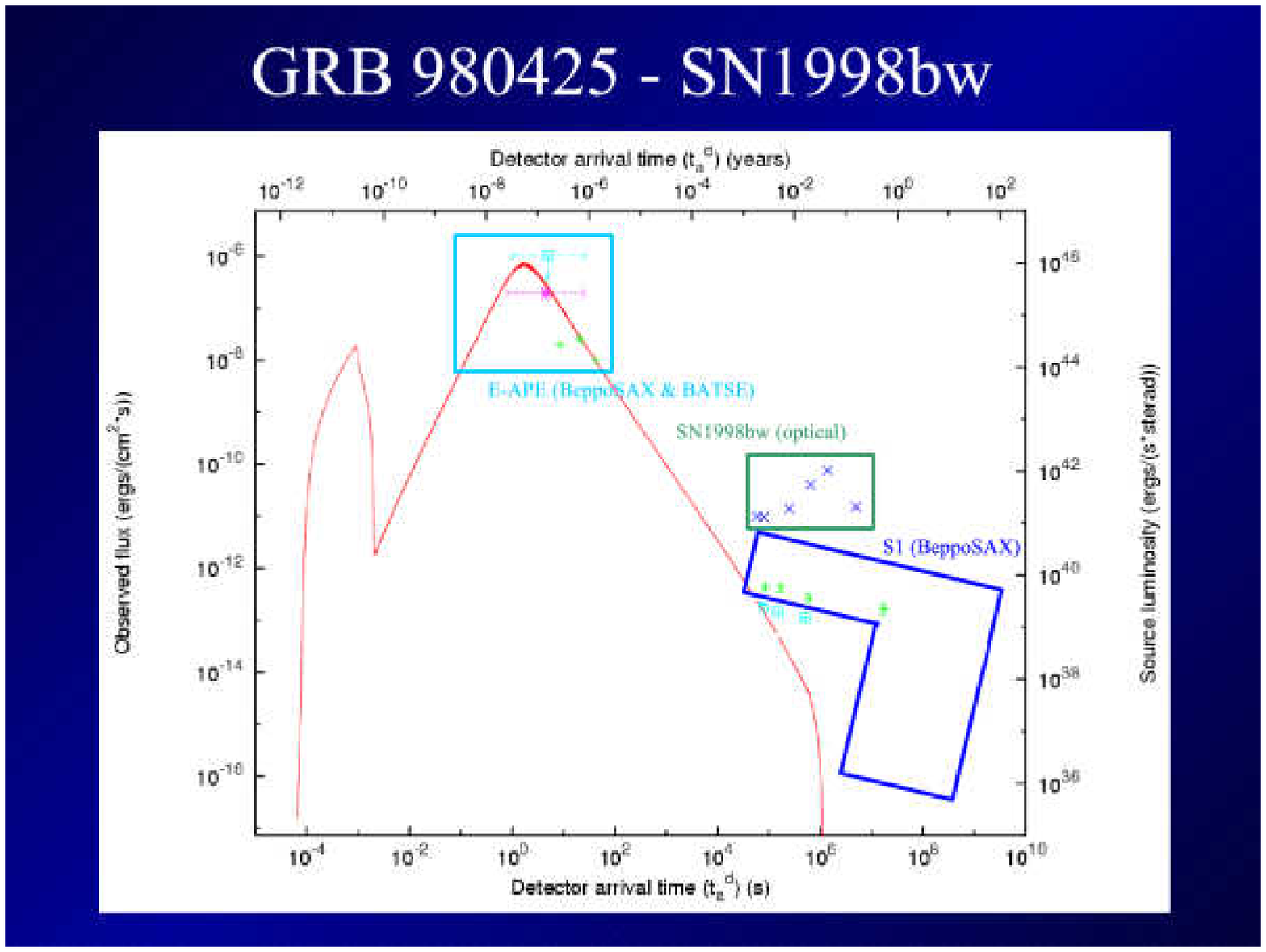}
\caption{The bolometric luminosity as a function of the arrival time. The peak of the P-GRB is just below the observational noise level.}
\label{d15} 
\end{figure}

\begin{figure}
\centering 
\includegraphics[width=\hsize]{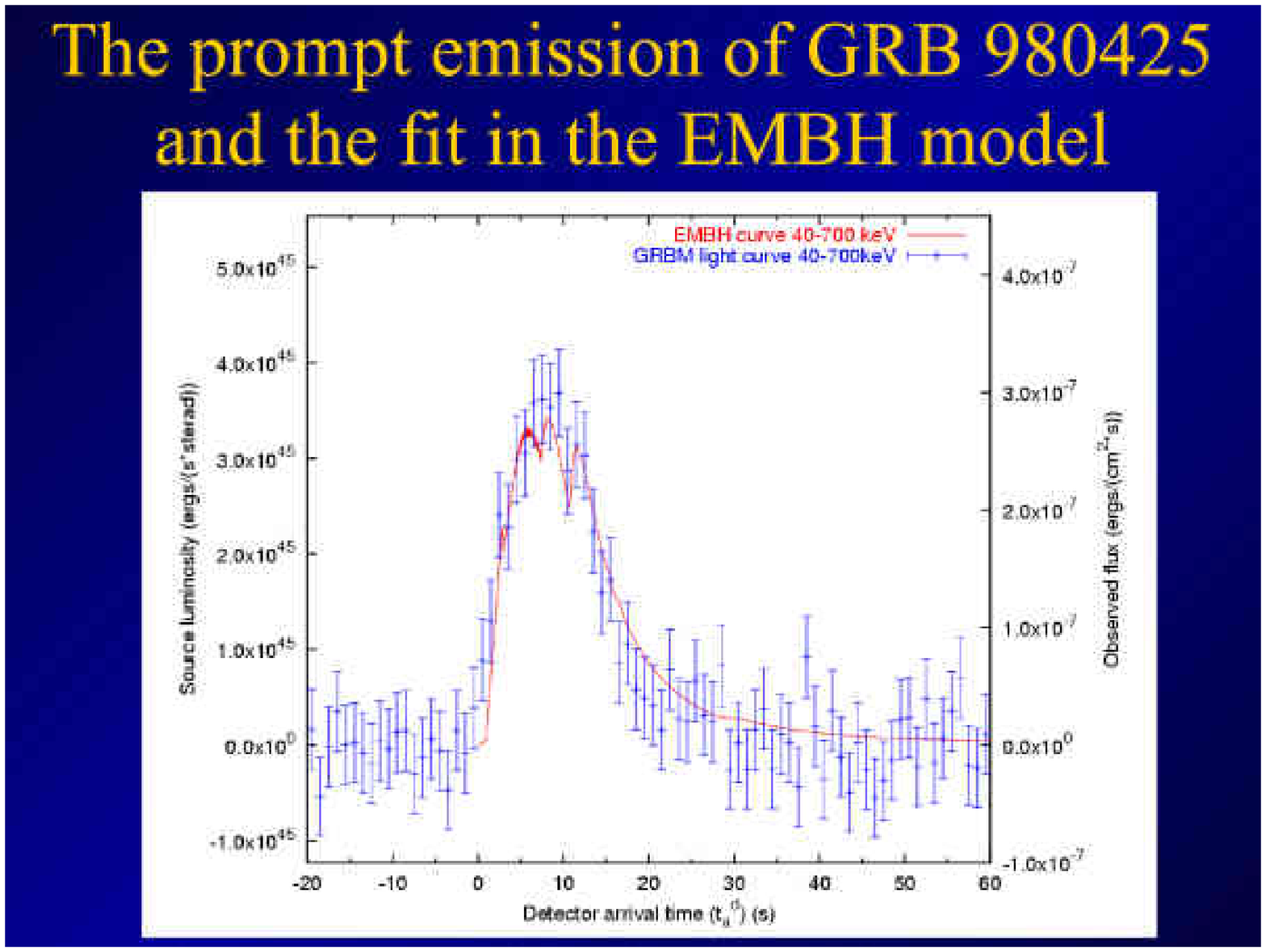}
\caption{The observation by BeppoSAX of the peak of the afterglow in the $40$--$700$ keV energy band is fitted by our model.}
\label{d17} 
\end{figure}

Our best fit for GRB 980425 corresponds to $E_{dya}=1.1\times 10^{48}\, {\rm ergs}$, $B=7\times 10^{-3}$ and the ISM average density is found to be $\left<n_{ism}\right>=0.02~{\rm particle}/{\rm cm}^3$. The plasma temperature and the total number of pairs in the dyadosphere are respectively $T=1.028\, {\rm MeV}$ and $N_{e^\pm}=5.3274\times10^{53}$. The light curve of the GRB is shown in Figs. \ref{d15}--\ref{d17}. The P-GRB is under the threshold and in the case of this source is not observable (see Ruffini et al. \cite{cospar02}, Fraschetti et al. \cite{f03mg10}).

\begin{figure} 
\centering
\includegraphics[width=\hsize]{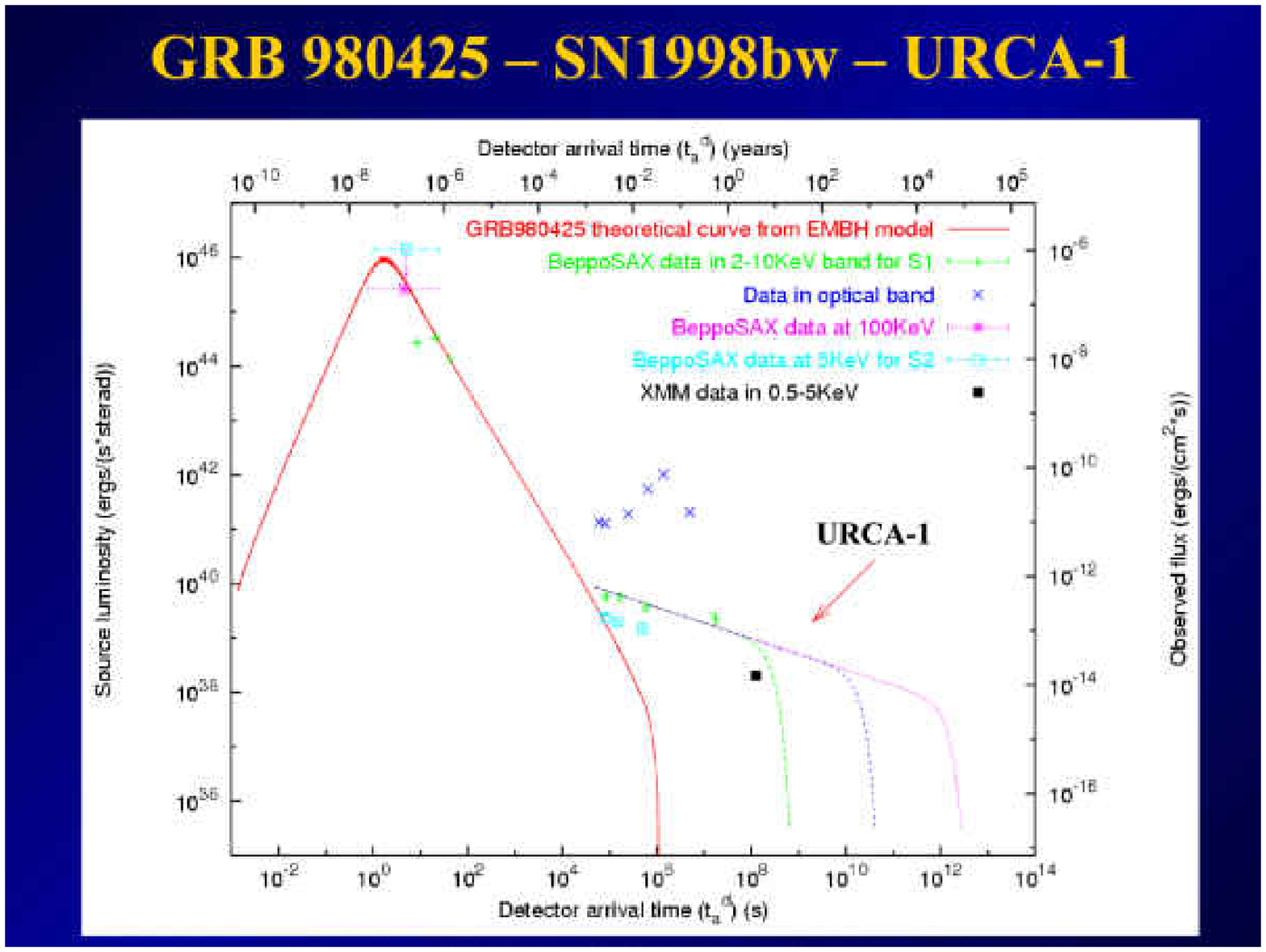} 
\caption{The bolometric light curves is reported as well as the BeppoSAX MECS observations in the $2$--$10$ keV band of S1 and S2 (Pian et al. \cite{pian00}) and the optical data of SN 1998bw (Iwamoto \cite{i99}). Here we also report some very qualitative curves to be expected for the URCA-1 luminosity. Details in Ruffini et al. \cite{cospar02}.}
\label{d19} 
\end{figure}

The characteristic parameter $\mathcal{R}$, defining the filamentary structure of the ISM, monotonically decreases from $4.81\times 10^{-10}$ to $2.65\times 10^{-12}$). The results are given in Fig. \ref{d19} where the bolometric luminosity is represented together with the optical data of SN1998bw, the source {\em S1} and the source {\em S2}. It is then clear that GRB 980425 is separated both from the supernova data and from the sources {\em S1} and {\em S2}. 

While the occurrence of the supernova in relation to the GRB has already been discussed within the GRB-Supernova Time Sequence (GSTS) paradigm (Ruffini et al. \cite{lett3}), we like to address here a different fundamental issue: the nature of the source {\em S1} which we have named, in celebration of the work of Gamow and Shoenberg, URCA-1. It is clear, from the theoretical predictions of the afterglow luminosity, that the URCA-1 cannot be part of the afterglow (see Figs. \ref{d15}, \ref{d19}). There are three different possibilities for the explanation of such source:\\
1) Its possible relation to the black hole formed during the process of gravitational collapse leading to the GRB emission.\\
2) Its possible relation to emission originating in the early phases of the expansion of the supernova remnant.\\
3) The very exciting possibility that for the first time we are observing a newly born neutron star out of the supernova phenomenon.

While some general considerations will be discussed in the conclusions, we would like to stress here the paramount importance of following the further time history of URCA-1 and of the source {\em S2}. If, as we propose, {\em S2} is a background source, its flux should be practically constant in time and this source has nothing to do with the GRB 980425 / SN1998bw system. The drastic behavior of the URCA-1 luminosity reported in the talk by Elena Pian in this meeting, showing the latest URCA-1 observations by the XMM and Chandra satellites, is crucial for the understanding of the nature of this source. Some very qualitative luminosity curves are sketched in Fig. \ref{d19}, illustrating the possible time evolution of URCA-1. They are still very undetermined today due to a lack of attention to these observational data and, consequently, to the lack of a detailed theoretical model of the phenomenon. We therefore propose to have a dedicated attention to the astrophysical ``triptych'' GRB 980425 / SN 1998bw / URCA-1.

\subsection{GRB 030329 / SN 2003dh}

We have adopted for our modeling of GRB030329 a spherically symmetric distribution for the source and, as initial conditions at $t = 10^{-21}$ s, an $e^+$-$e^-$-photon neutral plasma lying between the radii $r_1 = 2.9\times 10^6$ cm and $r_2 = 9.0\times 10^7$ cm. The temperature of such a plasma is 2.1 MeV, the total energy $E_{tot} = 2.1\times 10^{52}$ erg and the total number of pairs $N_{e^+e^-} = 1.1\times 10^{57}$. The baryonic matter component $M_B$ is the second free parameter of our theory: $B = 4.8 \times 10^{-3}$. At the emission of the P-GRB, the Lorentz gamma factor is $\gamma_\circ = 183.6$ and the radial coordinate is $r_\circ = 5.3 \times 10^{13}$ cm. The ISM average density is best fit by $<n_{ism}> = 1$ particle/cm$^3$. The third free parameter of our theory is given by $1.1\times 10^{-7} < \mathcal{R} < 5.0 \times 10^{-11}$.

\begin{figure} 
\centering
\includegraphics[width=\hsize]{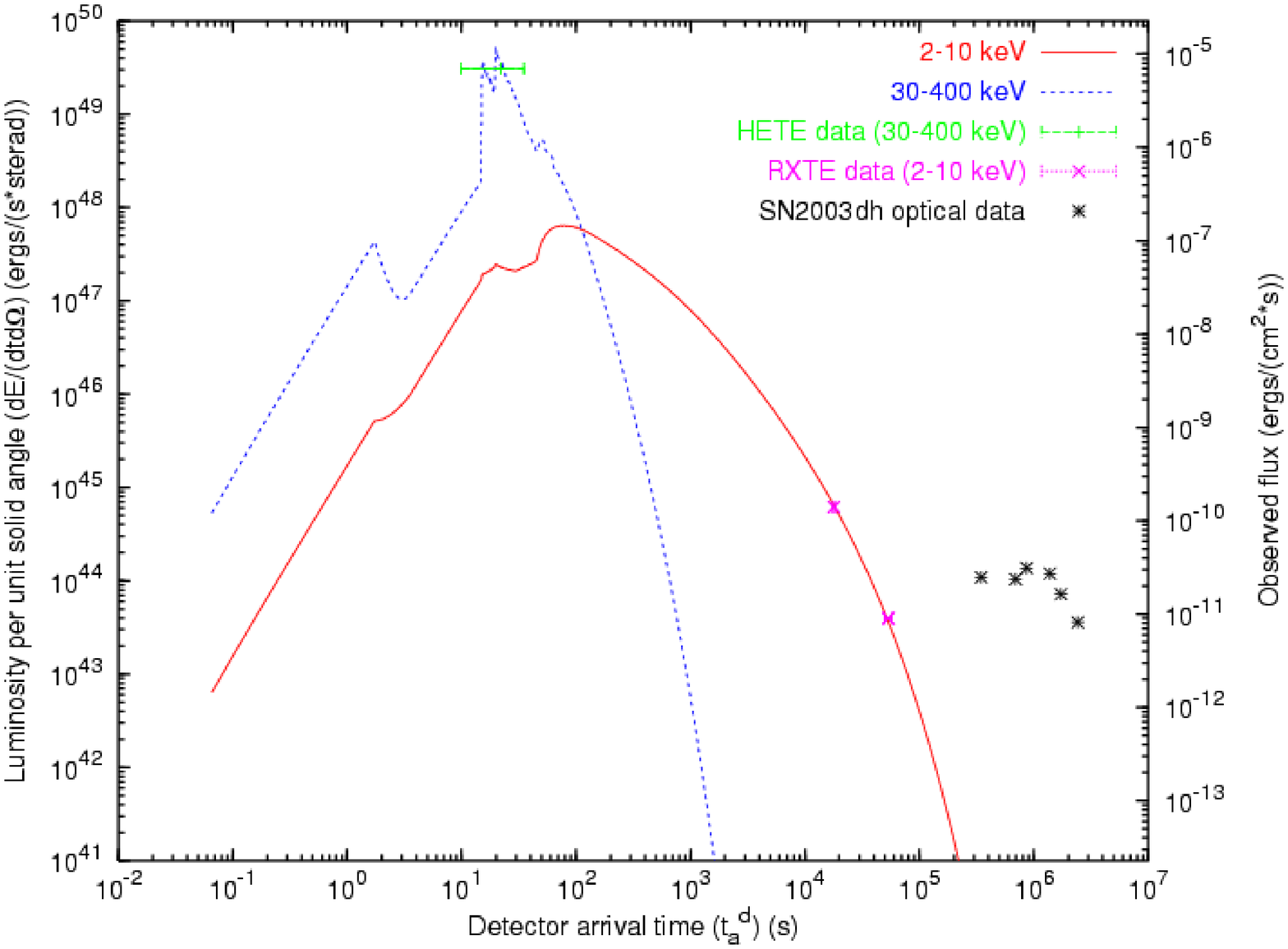} 
\caption{The luminosity in the $2$--$10$ keV and in the $30$--$400$ keV energy bands predicted by our model are fitted to the data of R-XTE (GCN Circ. 1996 \cite{gcn1996}) and HETE-2 (GCN Circ. 1997 \cite{gcn1997}) respectively. The SN 2003dh optical luminosity is given by the crosses (Hjorth et al. \cite{ha03}). Details in Bernardini et al. \cite{b03mg10}.}
\label{030329global} 
\end{figure}

\begin{figure} 
\centering
\includegraphics[width=\hsize]{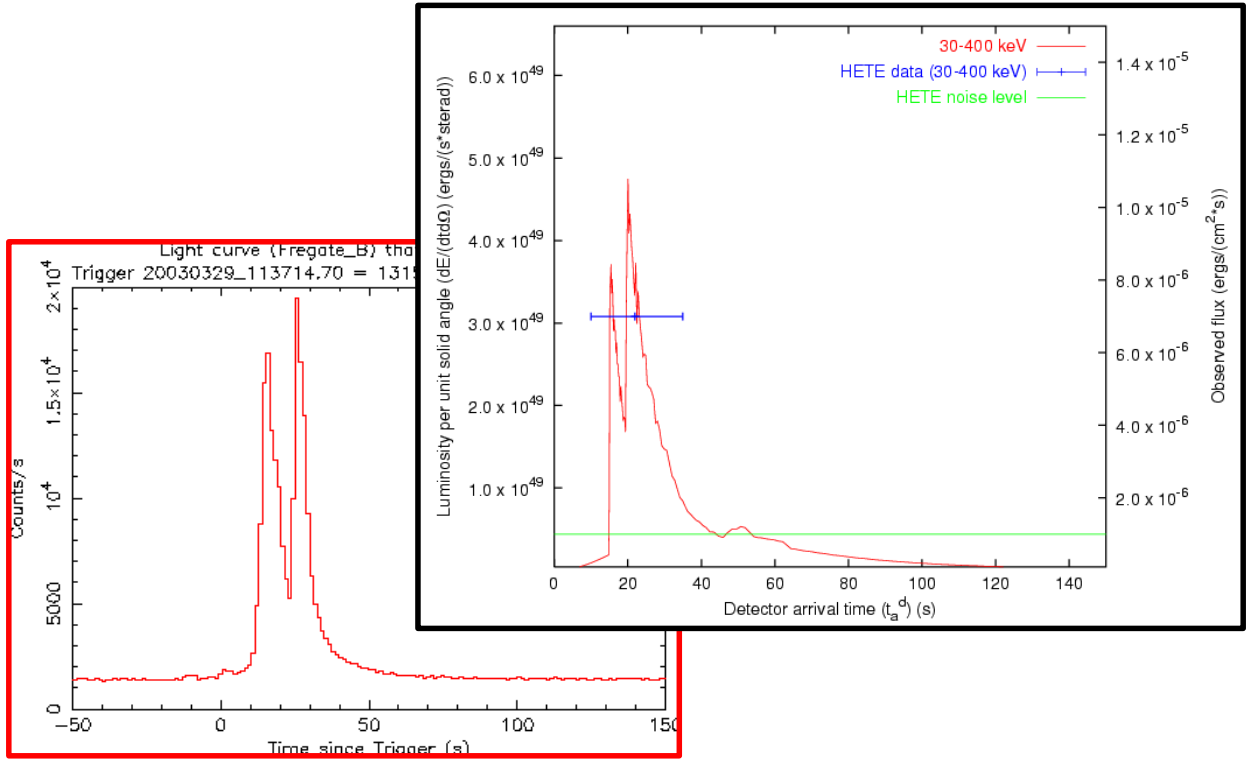} 
\caption{The details of the theoretical fit of the prompt radiation of GRB 030329 have been reproduced by the filamentary structure in the ISM in our model. Details in Bernardini et al. \cite{b03mg10}.} 
\label{030329sptp} 
\end{figure}

\begin{figure} 
\centering
\includegraphics[width=\hsize]{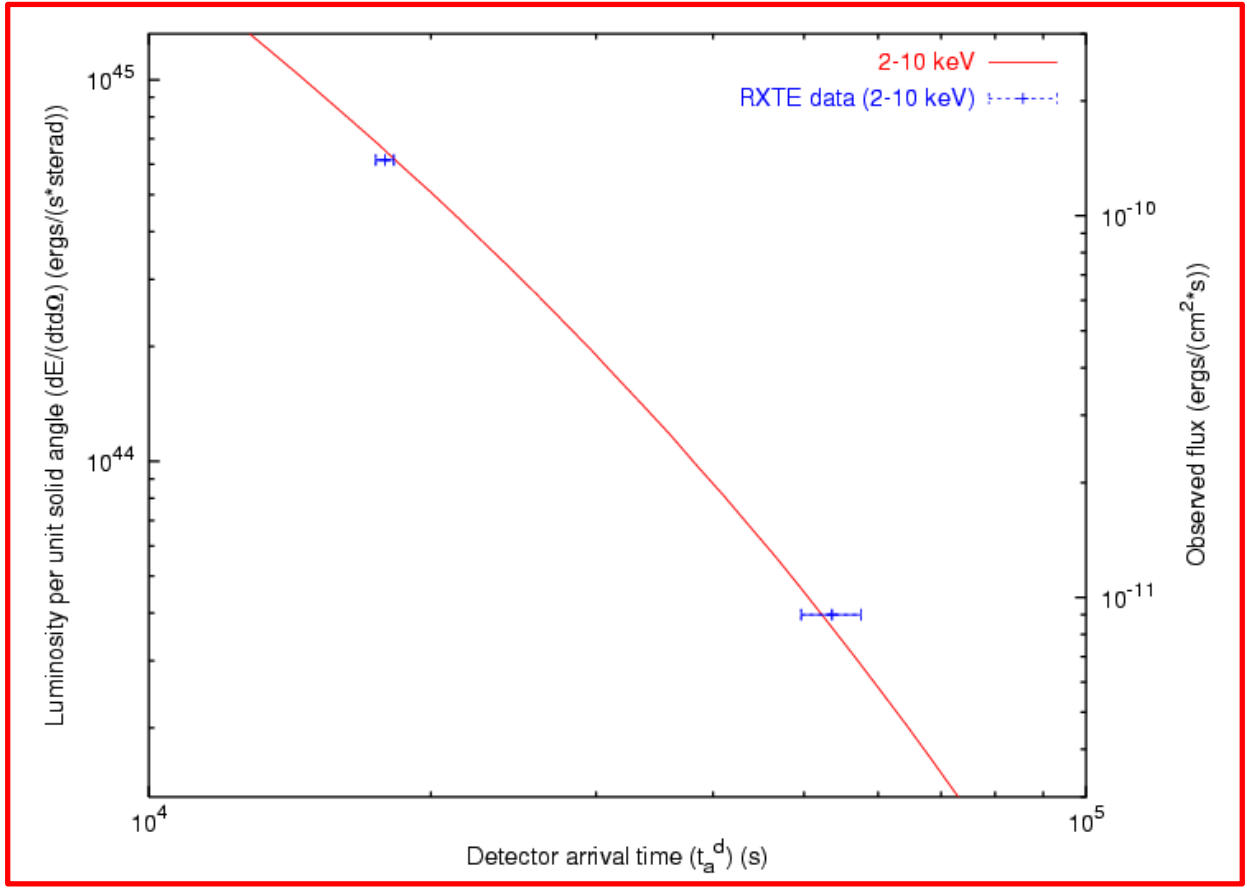} 
\caption{The perfect fit of the late part of the afterglow of our theoretical model for the $2$--$10$ keV energy bands. The data refers to the R-XTE observations (GCN Circ. 1996 \cite{gcn1996}). Details in Bernardini et al. \cite{b03mg10}.}
\label{030329af} 
\end{figure}

\begin{figure} 
\centering
\includegraphics[width=\hsize]{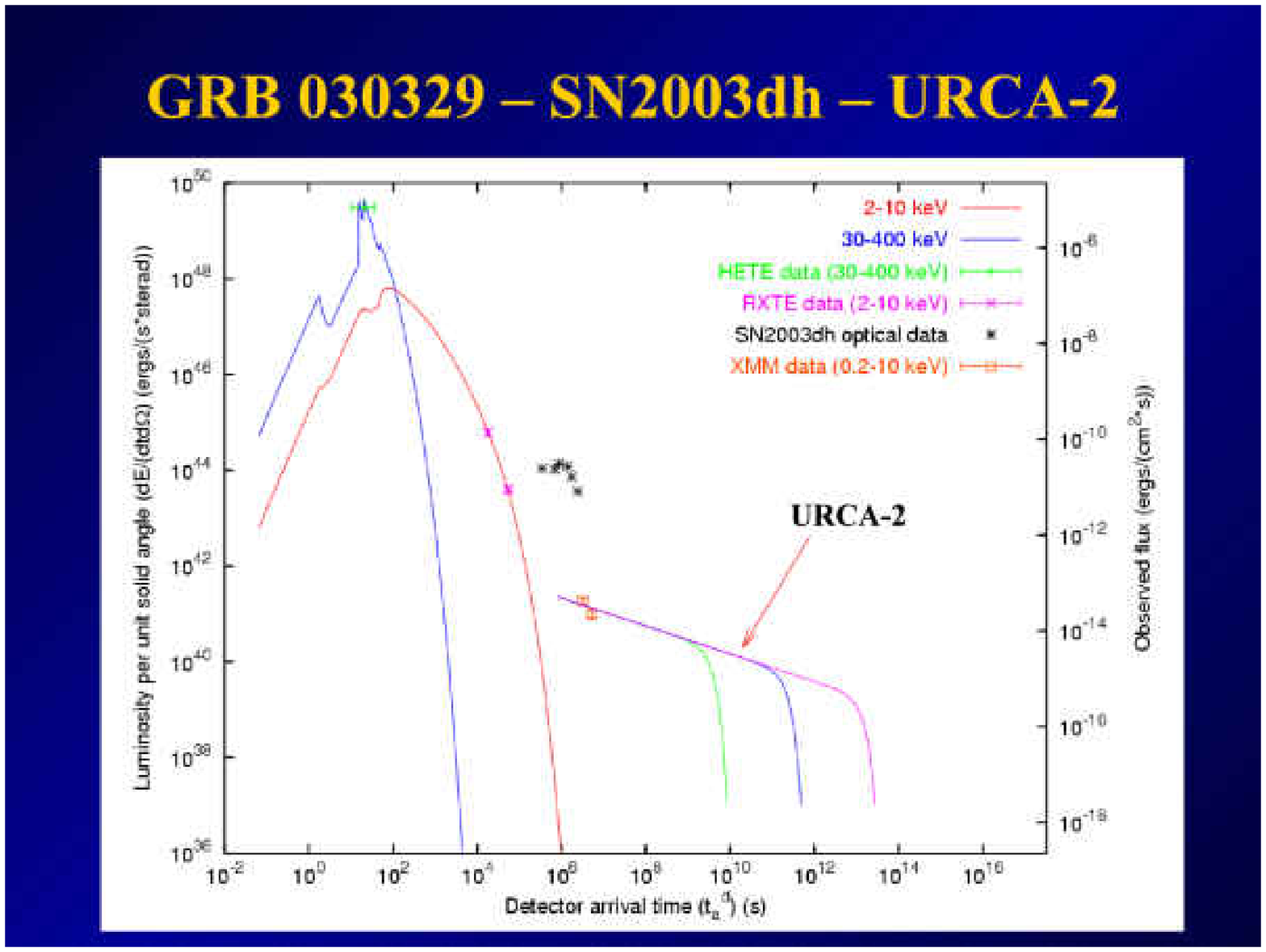} 
\caption{The dotted line represents our theoretically predicted GRB030329 light curve in $\gamma$-rays (30-400 keV) with the horizontal bar corresponding to the mean peak flux from HETE-2 (GCN Circ. 1997 \cite{gcn1997}). The solid line represents the corresponding one in X-rays (2-10 keV) with the experimental data obtained by R-XTE (GCN Circ. 1996 \cite{gcn1996}). The remaining points refer respectively to the optical VLT data (Hjorth et al. \cite{ha03}) of SN2003bw and to the X-ray XMM data (Tiengo et al. \cite{ta03}) of URCA-2. The dash-dotted lines corresponds to qualitative luminosity curves expected for URCA-2. It is interesting to compare and contrast these results with the ones for GRB980425/SN1998bw (see Fig. 3 in Ruffini et al. \cite{cospar02}). Details in Ruffini et al. \cite{030329}.}
\label{fig1} 
\end{figure}

We then obtain (see also Bernardini et al. \cite{b03mg10}) for the GRB 030329 the luminosities in given energy bands, computed in the range $2$--$400$ keV with very high accuracy. Figs. \ref{030329global}--\ref{fig1} shows the results for the luminosities in the $30$--$400$ keV and $2$--$10$ keV bands, including the ``prompt radiation''. Subsequently, the theoretically predicted GRB spectra have been evaluated at selected values of the arrival time (Ruffini et al. \cite{030329}). 

The splendid news received the evening before the presentation of this talk is graphically represented by the XMM observations shown in Fig. \ref{fig1}. Again, the XMM observations, like the corresponding ones of GRB 980425, occur after the decaying part of the afterglow and, in analogy to the one occurring in the system GRB 980425 / SN 1998bw / URCA-1, we call this source URCA-2. Further observations by XMM are highly recommended to follow the URCA-2 temporal evolution. Also in this system we are dealing with an astrophysical ``triptych'': GRB 030329 / SN 2003dh / URCA-2.

\section{On the short GRBs}\label{Short GRBs}

By the analysis of the first and second BATSE catalogs\footnote{see http://cossc.gsfc.nasa.gov/batse/} Tavani in 1998 \cite{t98} (see Fig. \ref{tavani}) confirmed the previous results by Kouveliotou et al. \cite{ka93} on the existence of two families of GRBs: the so-called ``long-bursts'' with a soft spectrum and duration $\Delta t > 2.5$sec and the ``short-bursts'' with harder spectrum and duration $\Delta t < 2.5$sec. In 2001 we have proposed the theory \cite{lett2} that both short-bursts and long-bursts originate from the same underlying physical process: the vacuum polarization of electromagnetic overcritical gravitational collapse leading to the creation of $e^+-e^-$ pairs at the expenses of the ``blackholic'' energy \cite{cr71}. The difference between the short-bursts and long-bursts in this theory is mainly due to the amount of baryonic matter, described by the dimensionless parameter $B$ previously mentioned, encountered by the $e^+e^-$ pairs in their relativistic expansion (see Fig. \ref{LongShort}). Short-bursts occur in a range of B:
\begin{figure} 
\centering
\includegraphics[width=\hsize]{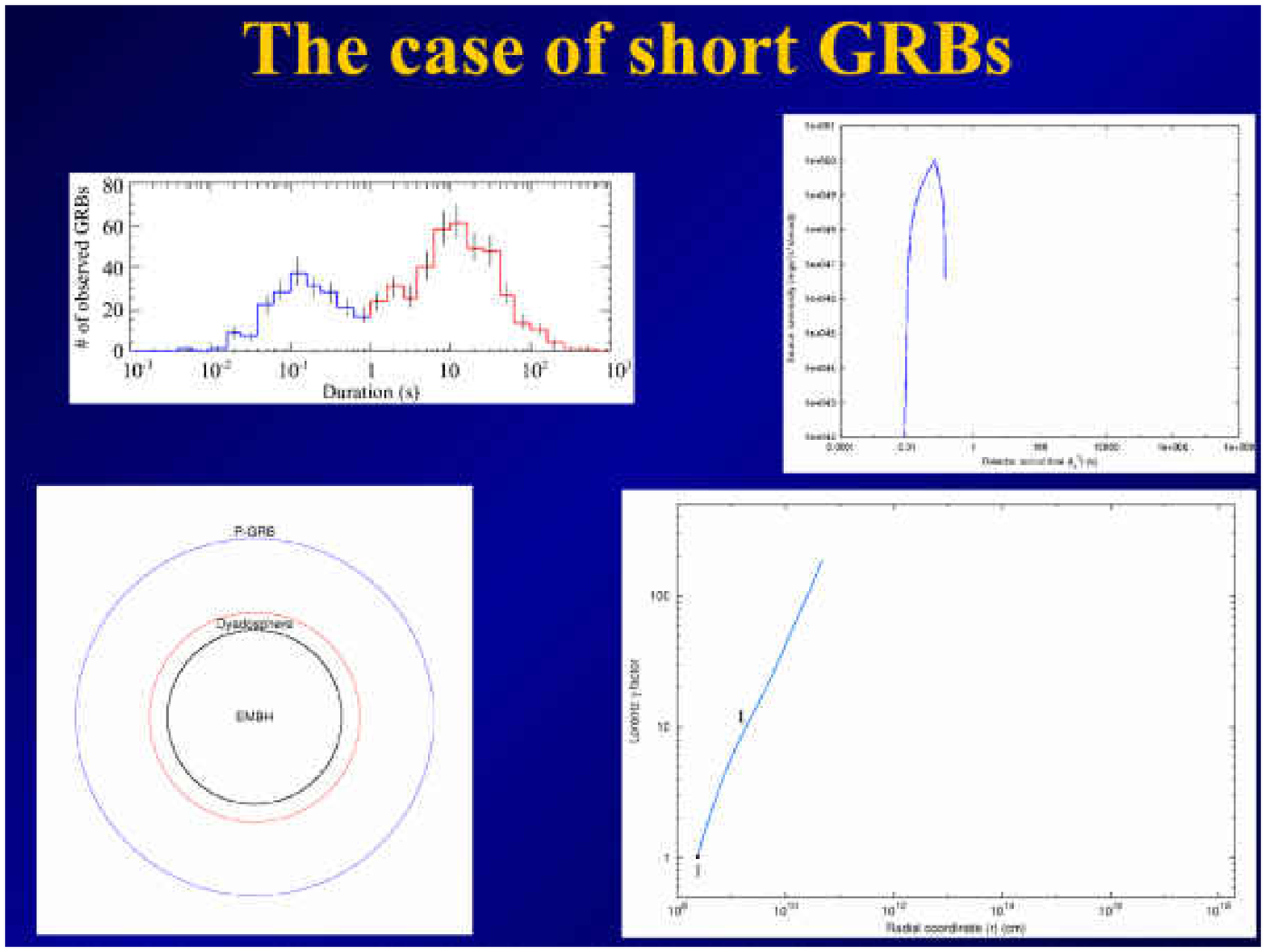} 
\caption{In our theory the short GRBs originate by the same basic process leading also to the long GRBs (see Fig. \ref{cip_tot}) in the limiting case of no baryonic matter present ($B \to 0$). The optically thick electron-positron pairs expand to ultrarelativistic Lorentz gamma factors reaching, then, transparency. The emission at the transparency point originates the short GRBs.}
\label{LongShort} 
\end{figure} 
\begin{equation}
0 < B < 10^{-5}\, ,
\end{equation}
and the long-bursts occur for:
\begin{equation}
10^{-5} < B < 10^{-2}\, .
\end{equation}
Compare and contrast Fig. \ref{cip_tot} with Fig. \ref{LongShort}.

An indirect support of our theory was given by Schmidt \cite{s01} who has shown that short-bursts and long-bursts have the same isotropic-equivalent characteristic peak luminosity.

In recent work we have systematically developed the theoretical background of the process
of gravitational collapse of matter involving an electromagnetic field with field strength higher than the critical value for $e^+e^-$ pair creation \cite{crv02,rv02a,rv02b,rvx03b,rvx03,rvx03a}. The goal has been to clarify the
physical nature of the process of extracting the ``blackholic'' energy by the creation of $e^+e^-$ matter pairs \cite{cr71} and to analyze the electromagnetic radiation emission process during the transient dynamical phases of the gravitational collapse, leading to the final formation of the black hole. 

All the considerations presented in the description of the long GRBs were based on a dyadosphere of an already formed black hole, presented in section \ref{dyadosphere}. This approximate treatment is very satisfactory in estimating the general dependence of the energy of the P-GRB, the kinetic energy of the baryonic matter pulse generating the afterglow and consequently the intensity of the afterglow itself. In particular it is possible to obtain the overall time structure of the GRB and especially the time of the release of the P-GRB in respect to the moment of gravitational collapse and its relative intensity with respect to the afterglow. If, however, we address the issue of the detailed temporal structure of the P-GRB and its detailed spectral distribution, the dynamical considerations on the dyadosphere formation, which we are going to present in the following sections, are needed (see also \cite{rvx03b}). In turn, this detailed analysis is needed in order to describe the general relativistic effects close to the horizon formation. As expressed already in section \ref{dyadosphere}, all general relativistic quantum field theory effects are encoded in the fine structure of the P-GRB. As emphasized in section \ref{fp}, the only way to differentiate between solutions with same $E_{dya}$ but different black hole mass and charge is to observe the P-GRBs in the limit $B \to 0$, namely, to observe the short GRBs (see Fig. \ref{LongShort}).

\section{Some propaedeutic analysis for the dynamical formation of the Black Hole}\label{luca}

While the formation in time of the dyadosphere is the fundamental phenomena we are interested in, we can get an insight on the issue of gravitational collapse of an electrically charged star core studying in details a simplified model, namely a thin shell of charged dust.

\subsection{On the collapsing charged shell in general relativity}

In \cite{i66,idlc67} it is shown that the problem of a collapsing charged shell in general relativity can be reduced to a set of ordinary differential equations. We reconsider here the following relativistic system: a spherical shell of electrically charged dust which is moving radially in the Reissner-Nordstr\"{o}m background of an already formed nonrotating black hole of mass $M_{1}$ and charge $Q_{1}$, with $Q_{1}\leq M_{1}$.

The world surface spanned by the shell divides the space-time into two regions: an internal one $\mathcal{M}_{-}$ and an external one $\mathcal{M}_{+}$. The line element in Schwarzschild like coordinate is \cite{crv02}
\begin{equation}
ds^{2}=\left\{
\begin{array}[c]{l}
-f_{+}dt_{+}^{2}+f_{+}^{-1}dr^{2}+r^{2}d\Omega^{2}\qquad\text{in } \mathcal{M}_{+}\\
-f_{-}dt_{-}^{2}+f_{-}^{-1}dr^{2}+r^{2}d\Omega^{2}\qquad\text{in } \mathcal{M}_{-}
\end{array}
\right.  , \label{E0}
\end{equation}
where $f_{+}=1-\tfrac{2M}{r}+\tfrac{Q^{2}}{r^{2}}$, $f_{-}=1-\tfrac{2M_{1}} {r}+\tfrac{Q_{1}^{2}}{r^{2}}$ and $t_{-}$ and $t_{+}$ are the Schwarzschild-like time coordinates in $\mathcal{M}_{-}$ and $\mathcal{M}_{+}$ respectively. $M$ is the total mass-energy of the system formed by the shell and the black hole, measured by an observer at rest at infinity and $Q=Q_{0}+Q_{1}$ is the total charge: sum of the charge $Q_{0}$ of the shell and the charge $Q_{1}$ of the internal black hole.

Indicating by $R$ the radius of the shell and by $T_{\pm}$ its time coordinate, the equations of motion of the shell become \cite{rv02a}
\begin{align}
\left(  \tfrac{dR}{d\tau}\right)  ^{2}  &  =\tfrac{1}{M_{0}^{2}}\left(M-M_{1}+\tfrac{M_{0}^{2}}{2R}-\tfrac{Q_{0}^{2}}{2R}-\tfrac{Q_{1}Q_{0}}{R}\right)  ^{2}-f_{-}\left(  R\right) \nonumber\\
&  =\tfrac{1}{M_{0}^{2}}\left(  M-M_{1}-\tfrac{M_{0}^{2}}{2R}-\tfrac{Q_{0}^{2}}{2R}-\tfrac{Q_{1}Q_{0}}{R}\right)  ^{2}-f_{+}\left(  R\right),\label{EQUYa}\\
\tfrac{dT_{\pm}}{d\tau}  &  =\tfrac{1}{M_{0}f_{\pm}\left(  R\right)  }\left(M-M_{1}\mp\tfrac{M_{0}^{2}}{2R}-\tfrac{Q_{0}^{2}}{2R}-\tfrac{Q_{1}Q_{0}}{R}\right)  , \label{EQUYb}
\end{align}
where $M_{0}$ is the rest mass of the shell and $\tau$ is its proper time. Eqs.(\ref{EQUYa},\ref{EQUYb}) (together with Eq.(\ref{E0})) completely describe a 5-parameter ($M$, $Q$, $M_{1}$, $Q_{1}$, $M_{0}$) family of solutions of the Einstein-Maxwell equations. Note that Eqs.(\ref{EQUYa},\ref{EQUYb}) imply that
\begin{equation}
M-M_{1}-\tfrac{Q_{0}^{2}}{2R}-\tfrac{Q_{1}Q_{0}}{R}>0
\label{Constraint}
\end{equation}
holds for $R>M+\sqrt{M^{2}-Q^{2}}$ if $Q<M$ and for $R>M_{1}+\sqrt{M_{1}^{2}-Q_{1}^{2}}$ if $Q>M$.

For astrophysical applications \cite{rvx03b} the trajectory of the shell $R=R\left(  T_{+}\right)  $ is obtained as a function of the time coordinate $T_{+}$ relative to the space-time region $\mathcal{M}_{+}$. In the following we drop the $+$ index from $T_{+}$. From Eqs.(\ref{EQUYa},\ref{EQUYb}) we have
\begin{equation}
\tfrac{dR}{dT}=\tfrac{dR}{d\tau}\tfrac{d\tau}{dT}=\pm\tfrac{F}{\Omega} \sqrt{\Omega^{2}-F},
\label{EQUAISRDLC}
\end{equation}
where
\begin{align}
F  & \equiv f_{+}\left(  R\right)  =1-\tfrac{2M}{R}+\tfrac{Q^{2}}{R^{2}},\\
\Omega & \equiv\Gamma-\tfrac{M_{0}^{2}+Q^{2}-Q_{1}^{2}}{2M_{0}R},\\
\Gamma & \equiv\tfrac{M-M_{1}}{M_{0}}.
\end{align}
Since we are interested in an imploding shell, only the minus sign case in (\ref{EQUAISRDLC}) will be studied. We can give the following physical interpretation of $\Gamma$. If $M-M_{1}\geq M_{0}$, $\Gamma$ coincides with the Lorentz $\gamma$ factor of the imploding shell at infinity; from Eq.(\ref{EQUAISRDLC}) it satisfies
\begin{equation}
\Gamma=\tfrac{1}{\sqrt{1-\left(  \frac{dR}{dT}\right)  _{R=\infty}^{2}}}\geq1.
\end{equation}
When $M-M_{1}<M_{0}$ then there is a \emph{turning point} $R^{\ast}$, defined by $\left.  \tfrac{dR}{dT}\right|  _{R=R^{\ast}}=0$. In this case $\Gamma$ coincides with the ``effective potential'' at $R^{\ast}$ :
\begin{equation}
\Gamma=\sqrt{f_{-}\left(  R^{\ast}\right)  }+M_{0}^{-1}\left(  -\tfrac{M_{0}^{2}}{2R^{\ast}}+\tfrac{Q_{0}^{2}}{2R^{\ast}}+\tfrac{Q_{1}Q_{0}}{R^{\ast}}\right)
\leq1.
\end{equation}
The solution of the differential equation (\ref{EQUAISRDLC}) is given by:
\begin{equation}
\int dT=-\int\tfrac{\Omega}{F\sqrt{\Omega^{2}-F}}dR.
\label{GRYD}
\end{equation}
The functional form of the integral (\ref{GRYD}) crucially depends on the degree of the polynomial $P\left(  R\right)  =R^{2}\left(  \Omega^{2}-F\right)  $, which is generically two, but in special cases has lower values. We therefore distinguish the following cases:

\begin{enumerate}
\item {\boldmath$M=M_{0}+M_{1}$}; {\boldmath$Q_{1}=M_{1}$}; {\boldmath $Q=M$}: $P\left(  R\right)  $ is equal to $0$, we simply have
\begin{equation}
R(T)=\mathrm{{const}.}
\end{equation}
\item {\boldmath$M=M_{0}+M_{1}$}; {\boldmath$M^{2}-Q^{2}=M_{1}^{2}-Q_{1}^{2}$}; {\boldmath$Q\neq M$}: $P\left(  R\right)  $ is a constant, we have
\begin{align}
T  & =\mathrm{const}+\tfrac{1}{2\sqrt{M^{2}-Q^{2}}}\left[  \left(
R+2M\right)  R\right.  \nonumber\\
& \left.  +r_{+}^{2}\log\left(  \tfrac{R-r_{+}}{M}\right)  +r_{-}^{2}%
\log\left(  \tfrac{R-r_{-}}{M}\right)  \right]  .\label{CASO1}%
\end{align}
\item {\boldmath$M=M_{0}+M_{1}$}; {\boldmath$M^{2}-Q^{2}\neq M_{1}^{2}-Q_{1}^{2}$}: $P\left(  R\right)  $ is a first order polynomial and
\begin{align}
T &  =\mathrm{const}+2R\sqrt{\Omega^{2}-F}\left[  \tfrac{M_{0}R}{3\left(
M^{2}-Q^{2}-M_{1}^{2}+Q_{1}^{2}\right)  }\right.  \nonumber\\
&  \left.  +\tfrac{\left(  M_{0}^{2}+Q^{2}-Q_{1}^{2}\right)  ^{2}%
-9MM_{0}\left(  M_{0}^{2}+Q^{2}-Q_{1}^{2}\right)  +12M^{2}M_{0}^{2}%
+2Q^{2}M_{0}^{2}}{3\left(  M^{2}-Q^{2}-M_{1}^{2}+Q_{1}^{2}\right)  ^{2}%
}\right]  \nonumber\\
&  -\tfrac{1}{\sqrt{M^{2}-Q^{2}}}\left[  r_{+}^{2}\mathrm{arctanh}\left(
\tfrac{R}{r_{+}}\tfrac{\sqrt{\Omega^{2}-F}}{\Omega_{+}}\right)  \right.
\nonumber\\
&  \left.  -r_{-}^{2}\mathrm{arctanh}\left(  \tfrac{R}{r_{-}}\tfrac
{\sqrt{\Omega^{2}-F}}{\Omega_{-}}\right)  \right]  ,\label{CASO2}%
\end{align}

where $\Omega_{\pm}\equiv\Omega\left(  r_{\pm}\right)  $.

\item {\boldmath$M\neq M_{0}+M_{1}$}: $P\left(  R\right)  $ is a second order polynomial and
\begin{align}
T &  =\mathrm{const}-\tfrac{1}{2\sqrt{M^{2}-Q^{2}}}\left\{  \tfrac
{2\Gamma\sqrt{M^{2}-Q^{2}}}{\Gamma^{2}-1}R\sqrt{\Omega^{2}-F}\right.
\nonumber\\
&  +r_{+}^{2}\log\left[  \tfrac{R\sqrt{\Omega^{2}-F}}{R-r_{+}}+\tfrac
{R^{2}\left(  \Omega^{2}-F\right)  +r_{+}^{2}\Omega_{+}^{2}-\left(  \Gamma
^{2}-1\right)  \left(  R-r_{+}\right)  ^{2}}{2\left(  R-r_{+}\right)
R\sqrt{\Omega^{2}-F}}\right]  \nonumber\\
&  -r_{-}^{2}\log\left[  \tfrac{R\sqrt{\Omega^{2}-F}}{R-r_{-}}+\tfrac
{R^{2}\left(  \Omega^{2}-F\right)  +r_{-}^{2}\Omega_{-}^{2}-\left(  \Gamma
^{2}-1\right)  \left(  R-r_{-}\right)  ^{2}}{2\left(  R-r_{-}\right)
R\sqrt{\Omega^{2}-F}}\right]  \nonumber\\
&  -\tfrac{\left[  2MM_{0}\left(  2\Gamma^{3}-3\Gamma\right)  +M_{0}^{2}%
+Q^{2}-Q_{1}^{2}\right]  \sqrt{M^{2}-Q^{2}}}{M_{0}\left(  \Gamma^{2}-1\right)
^{3/2}}\log\left[  \tfrac{R\sqrt{\Omega^{2}-F}}{M}\right.  \nonumber\\
&  \left.  \left.  +\tfrac{2M_{0}\left(  \Gamma^{2}-1\right)  R-\left(
M_{0}^{2}+Q^{2}-Q_{1}^{2}\right)  \Gamma+2M_{0}M}{2M_{0}M\sqrt{\Gamma^{2}-1}%
}\right]  \right\}  .\label{CASO3}%
\end{align}
\end{enumerate}

\begin{figure}
\centering
\includegraphics[width=8.5cm,clip]{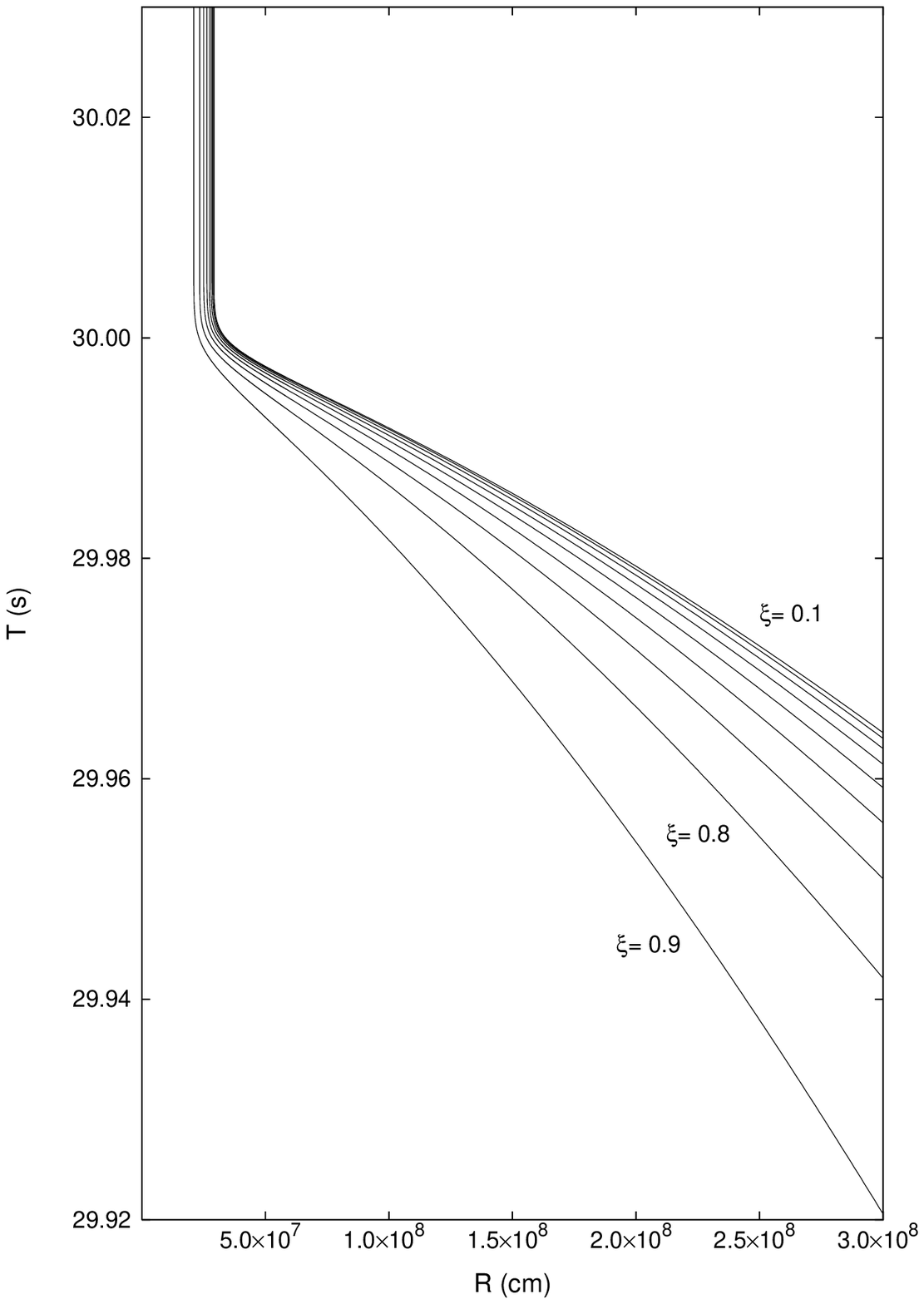}
\includegraphics[width=8.5cm,clip]{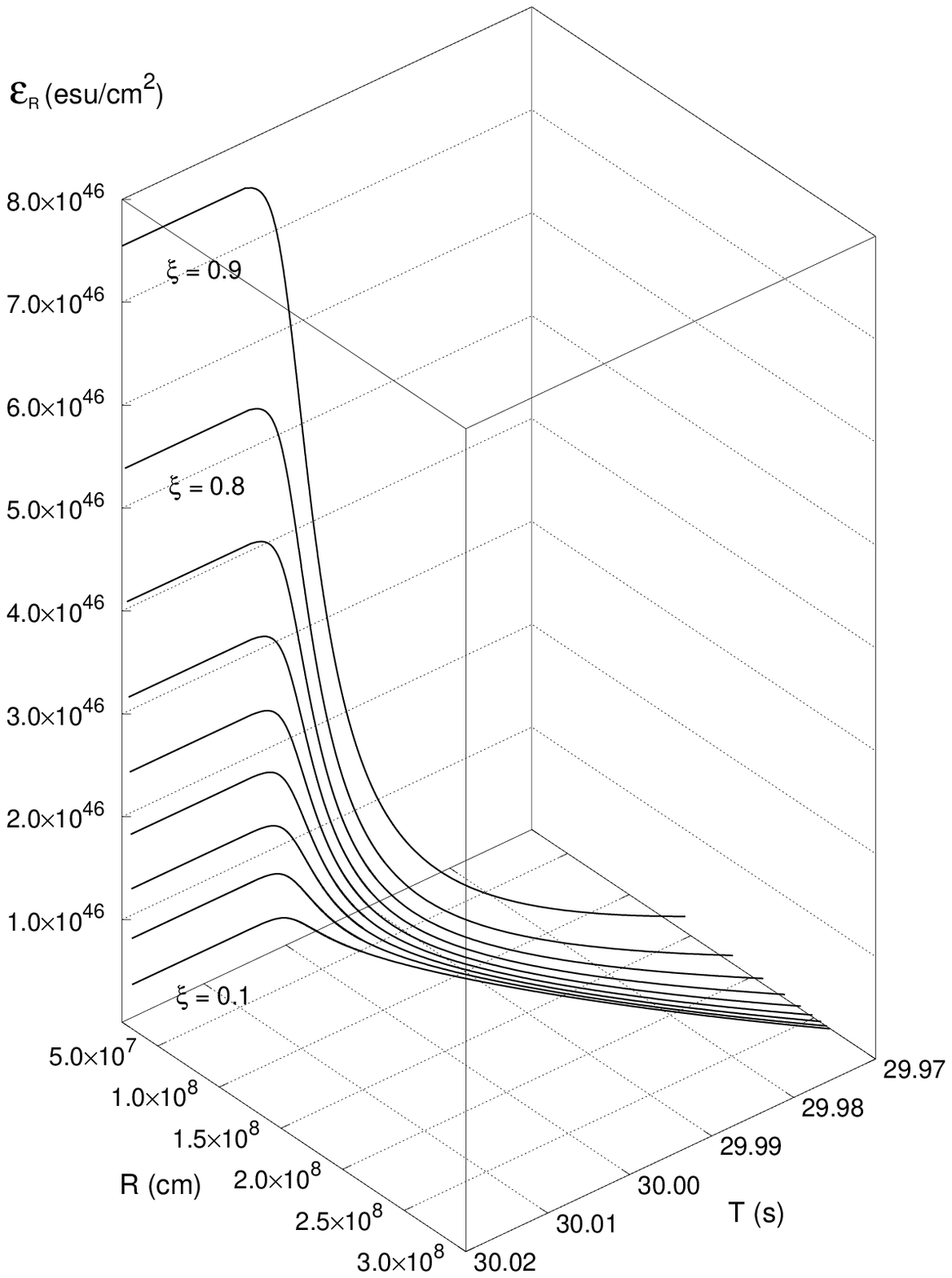}
\caption{{\bf Left)} Collapse curves in the plane $(T,R)$ for $M=20M_{\odot}$ and for different values of the parameter $\xi$. The asymptotic behavior is the clear manifestation of general relativistic effects as the horizon of the black hole is approached. {\bf Right)} Electric field behavior at the surface of the shell for $M=20M_{\odot}$ and for different values of the parameter $\xi$. The asymptotic behavior is the clear manifestation of general relativistic effects as the horizon of the black hole is approached.}
\label{elec3d}
\end{figure}

Of particular interest is the time varying electric field $\mathcal{E}%
_{R}=\tfrac{Q}{R^{2}}$ on the external surface of the shell. In order to study
the variability of $\mathcal{E}_{R}$ with time it is useful to consider in the
tridimensional space of parameters $(R,T,\mathcal{E}_{R})$ the parametric
curve $\mathcal{C}:\left(  R=\lambda,\quad T=T(\lambda),\quad\mathcal{E}%
_{R}=\tfrac{Q}{\lambda^{2}}\right)  $. In astrophysical applications
\cite{rvx03b} we are specially interested in the family of solutions such that
$\frac{dR}{dT}$ is 0 when $R=\infty$ which implies that $\Gamma=1$. In Fig. 
\ref{elec3d} we plot the collapse curves in the plane $(T,R)$ for different
values of the parameter $\xi\equiv\frac{Q}{M}$, $0<\xi<1$. The initial data
$\left(  T_{0},R_{0}\right)  $ are chosen so that the integration constant in
equation (\ref{CASO2}) is equal to 0. In all the cases we can follow the
details of the approach to the horizon which is reached in an infinite
Schwarzschild time coordinate. In Fig. \ref{elec3d} we plot the parametric
curves $\mathcal{C}$ in the space $(R,T,\mathcal{E}_{R})$ for different values
of $\xi$. Again we can follow the exact asymptotic behavior of the curves
$\mathcal{C}$, $\mathcal{E}_{R}$ reaching the asymptotic value $\frac{Q}%
{r_{+}^{2}}$. The detailed knowledge of this asymptotic behavior is of great
relevance for the observational properties of the black hole formation (see e.g. \cite{rv02a}).

In the case of a shell falling in a flat background ($M_{1}=Q_{1}=0$) Eq.(\ref{EQUYa}) reduces to
\begin{equation}
\left(  \tfrac{dR}{d\tau}\right)  ^{2}=\tfrac{1}{M_{0}^{2}}\left(
M+\tfrac{M_{0}^{2}}{2R}-\tfrac{Q^{2}}{2R}\right)  ^{2}-1. \label{EQUY2}%
\end{equation}
Introducing the total radial momentum $P\equiv M_{0}u^{r}=M_{0}\tfrac
{dR}{d\tau}$ of the shell, we can express the kinetic energy of the shell as
measured by static observers in $\mathcal{M}_{-}$ as $T\equiv-M_{0}u_{\mu}%
\xi_{-}^{\mu}-M_{0}=\sqrt{P^{2}+M_{0}^{2}}-M_{0}$. Then from equation
(\ref{EQUY2}) we have
\begin{equation}
M=-\tfrac{M_{0}^{2}}{2R}+\tfrac{Q^{2}}{2R}+\sqrt{P^{2}+M_{0}^{2}}%
=M_{0}+T-\tfrac{M_{0}^{2}}{2R}+\tfrac{Q^{2}}{2R}. \label{EQC}%
\end{equation}
where we choose the positive root solution due to the constraint
(\ref{Constraint}). Eq.(\ref{EQC}) is the \emph{mass formula} of the shell,
which depends on the time-dependent radial coordinate $R$ and kinetic energy
$T$. If $M\geq Q$, a black hole is formed and we have
\begin{equation}
M=M_{0}+T_{+}-\tfrac{M_{0}^{2}}{2r_{+}}+\tfrac{Q^{2}}{2r_{+}}\,, \label{EQL}%
\end{equation}
where $T_{+}\equiv T\left(  r_{+}\right)  $ and $r_{+}=M+\sqrt{M^{2}-Q^{2}}$
is the radius of external horizon of the black hole. 

\subsection{On the physical origin of the terms in mass formula of the black hole}

We know from the
Christodoulou-Ruffini black hole mass formula that
\begin{equation}
M=M_{\mathrm{irr}}+\tfrac{Q^{2}}{2r_{+}}, \label{irrmass}%
\end{equation}
so it follows that {}
\begin{equation}
M_{\mathrm{irr}}=M_{0}-\tfrac{M_{0}^{2}}{2r_{+}}+T_{+}, \label{EQM}
\end{equation}
namely that $M_{\mathrm{irr}}$ is the sum of only three contributions: the rest
mass $M_{0}$, the gravitational potential energy and the kinetic energy of the
rest mass evaluated at the horizon. $M_{\mathrm{irr}}$ is independent of the
electromagnetic energy, a fact noticed by Bekenstein \cite{b71}. We have taken
one further step here by identifying the independent physical contributions to
$M_{\mathrm{irr}}$.

Next we consider the physical interpretation of the electromagnetic term
$\tfrac{Q^{2}}{2R}$, which can be obtained by evaluating the conserved Killing
integral
\begin{align}
\int_{\Sigma_{t}^{+}}\xi_{+}^{\mu}T_{\mu\nu}^{\mathrm{(em)}}d\Sigma^{\nu}  &
=\int_{R}^{\infty}r^{2}dr\int_{0}^{1}d\cos\theta\int_{0}^{2\pi}d\phi
\ T^{\mathrm{(em)}}{}{}_{0}{}^{0}\nonumber\\
& =\tfrac{Q^{2}}{2R}\,,\label{EQR}%
\end{align}
where $\Sigma_{t}^{+}$ is the space-like hypersurface in $\mathcal{M}_{+}$
described by the equation $t_{+}=t=\mathrm{const}$, with $d\Sigma^{\nu}$ as
its surface element vector and where $T_{\mu\nu}^{\mathrm{(em)}}=-\tfrac
{1}{4\pi}\left(  F_{\mu}{}^{\rho}F_{\rho\nu}+\tfrac{1}{4}g_{\mu\nu}%
F^{\rho\sigma}F_{\rho\sigma}\right)  $ is the energy-momentum tensor of the
electromagnetic field. The quantity in Eq.(\ref{EQR}) differs from the purely
electromagnetic energy
\[
\int_{\Sigma_{t}^{+}}n_{+}^{\mu}T_{\mu\nu}^{\mathrm{(em)}}d\Sigma^{\nu}%
=\tfrac{1}{2}\int_{R}^{\infty}dr\sqrt{g_{rr}}\tfrac{Q^{2}}{r^{2}},
\]
where $n_{+}^{\mu}=f_{+}^{-1/2}\xi_{+}^{\mu}$ is the unit normal to the
integration hypersurface and $g_{rr}=f_{+}$. This is similar to the analogous
situation for the total energy of a static spherical star of energy density
$\epsilon$ within a radius $R$, $m\left(  R\right)  =4\pi\int_{0}^{R}%
dr\ r^{2}\epsilon$, which differs from the pure matter energy $m_{\mathrm{p}%
}\left(  R\right)  $ $=4\pi\int_{0}^{R}dr\sqrt{g_{rr}}r^{2}\epsilon$ by the
gravitational energy (see \cite{mtw73}). Therefore the term $\tfrac{Q^{2}}%
{2R}$ in the mass formula (\ref{EQC}) is the \emph{total} energy of the
electromagnetic field and includes its own gravitational binding energy. This
energy is stored throughout the region $\Sigma_{t}^{+}$, extending from $R$ to infinity.

\subsection{On the energy extraction process of blackholic energy}

We now turn to the problem of extracting the blackholic energy from a black hole (see \cite{cr71}). We can distinguish between two conceptually physically
different processes, depending on whether the electric field strength
$\mathcal{E}=\frac{Q}{r^{2}}$ is smaller or greater than the critical value
$\mathcal{E}_{\mathrm{c}}=\tfrac{m_{e}^{2}c^{3}}{e\hbar}$. Here $m_{e}$ and
$e$ are the mass and the charge of the electron. As already mentioned in this
paper an electric field $\mathcal{E}>\mathcal{E}_{\mathrm{c}}$ polarizes the
vacuum creating electron-positron pairs (see \cite{he35}). The maximum value
$\mathcal{E}_{+}=\tfrac{Q}{r_{+}^{2}}$ of the electric field around a black hole is
reached at the horizon. We then have the following:

\begin{enumerate}
\item  For $\mathcal{E}_{+}<\mathcal{E}_{\mathrm{c}}$ the leading energy
extraction mechanism consists of a sequence of discrete elementary decay
processes of a particle into two oppositely charged particles. The condition
$\mathcal{E}_{+}<\mathcal{E}_{\mathrm{c}}$ implies
\begin{align}
\xi & \equiv\tfrac{Q}{\sqrt{G}M}\nonumber\\
& \lesssim\left\{
\begin{array}
[c]{r}%
\tfrac{GM/c^{2}}{\lambda_{\mathrm{C}}}\tfrac{\sqrt{G}m_{e}}{e}\sim
10^{-6}\tfrac{M}{M_{\odot}}\quad\text{if }\tfrac{M}{M_{\odot}}\leq10^{6}\\
1\quad\quad\quad\quad\quad\quad\quad\quad\text{if }\tfrac{M}{M_{\odot}}>10^{6}%
\end{array}
\right.  ,\label{critical3}%
\end{align}
where $\lambda_{\mathrm{C}}$ is the Compton wavelength of the electron.
\cite{dr73} and \cite{dhr74} have defined as the \emph{effective ergosphere} the region around a black hole
where the energy extraction processes occur. This region extends from the
horizon $r_{+}$ up to a radius
\begin{align}
r_{\mathrm{Eerg}}  & =\tfrac{GM}{c^{2}}\left[  1+\sqrt{1-\xi^{2}\left(
1-\tfrac{e^{2}}{G{m_{e}^{2}}}\right)  }\right]  \nonumber\\
& \simeq\tfrac{e}{m_{e}}\tfrac{Q}{c^{2}}\,.\label{EffErg}%
\end{align}
The energy extraction occurs in a finite number $N_{\mathrm{PD}}$ of such
discrete elementary processes, each one corresponding to a decrease of the black hole
charge. We have
\begin{equation}
N_{\mathrm{PD}}\simeq\tfrac{Q}{e}\,.
\end{equation}
Since the total extracted energy is (see Eq.(\ref{irrmass})) $E^{\mathrm{tot}%
}=\tfrac{Q^{2}}{2r_{+}}$, we obtain for the mean energy per accelerated
particle $\left\langle E\right\rangle _{\mathrm{PD}}=\tfrac{E^{\mathrm{tot}}%
}{N_{\mathrm{PD}}}$
\begin{equation}
\left\langle E\right\rangle _{\mathrm{PD}}=\tfrac{Qe}{2r_{+}}=\tfrac{1}%
{2}\tfrac{\xi}{1+\sqrt{1-\xi^{2}}}\tfrac{e}{\sqrt{G}m_{e}}\ m_{e}c^{2}%
\simeq\tfrac{1}{2}\xi\tfrac{e}{\sqrt{G}m_{e}}\ m_{e}c^{2},
\end{equation}
which gives
\begin{equation}
\left\langle E\right\rangle _{\mathrm{PD}}\lesssim\left\{
\begin{array}
[c]{r}%
\left(  \tfrac{M}{M_{\odot}}\right)  \times10^{21}eV\quad\text{if }\tfrac
{M}{M_{\odot}}\leq10^{6}\\
10^{27}eV\quad\quad\text{if }\tfrac{M}{M_{\odot}}>10^{6}%
\end{array}
\right.  . \label{UHECR}%
\end{equation}
One of the crucial aspects of the energy extraction process from a black hole is
its back reaction on the irreducible mass expressed in \cite{cr71}. Although
the energy extraction processes can occur in the entire effective ergosphere
defined by Eq. (\ref{EffErg}), only the limiting processes occurring on the
horizon with zero kinetic energy can reach the maximum efficiency while
approaching the condition of total reversibility (see Fig. 2 in \cite{cr71} for details). 
The farther from the horizon that a decay occurs, the more it
increases the irreducible mass and loses efficiency. Only in the complete
reversibility limit \cite{cr71} can the energy extraction process from an
extreme black hole reach the upper value of $50\%$ of the total black hole energy.

\item  For $\mathcal{E}_{+}\geq\mathcal{E}_{\mathrm{c}}$ the leading
extraction process is a \emph{collective} process based on an
electron-positron plasma generated by the vacuum polarization, (see Fig. \ref{dya}) 
as discussed in section III in \cite{Brasile} The condition $\mathcal{E}_{+}%
\geq\mathcal{E}_{\mathrm{c}}$ implies
\begin{equation}
\tfrac{GM/c^{2}}{\lambda_{\mathrm{C}}}\left(  \tfrac{e}{\sqrt{G}m_{e}}\right)
^{-1}\simeq2\cdot10^{-6}\tfrac{M}{M_{\odot}}\leq\xi\leq1\,.
\end{equation}
This vacuum polarization process can occur only for a black hole with mass smaller
than $2\cdot10^{6}M_{\odot}$. The electron-positron pairs are now produced in
the dyadosphere of the black hole, (note that the dyadosphere is a subregion of the
effective ergosphere) whose radius $r_{ds}$ is given in Eq.(\ref{rc}).
We have $r_{ds}\ll r_{\mathrm{Eerg}}$. The number of particles
created and the total energy stored in dyadosphere are given in Eqs.(17,18) of Ref. \cite{rv02a} respectively and we have approximately
\begin{align}
N^\circ_{e^+e^-}  & \simeq\left(  \tfrac{r_{ds}}%
{\lambda_{\mathrm{C}}}\right)  \tfrac{Q}{e}\,,\label{numdya}\\
E_{dya}  & \simeq\tfrac{Q^{2}}{2r_{+}}\,
\end{align}
The mean energy per particle produced in the dyadosphere $\left\langle
E\right\rangle _{\mathrm{ds}}=\tfrac{E_{dya}}{N^\circ_{e^+e^-}}$ is
then
\begin{equation}
\left\langle E\right\rangle _{\mathrm{ds}}\simeq\tfrac{3}{8}\left(
\tfrac{\lambda_{\mathrm{C}}}{r_{ds}}\right)  \tfrac{Qe}{r_{+}%
}\,,\label{meanenedya}%
\end{equation}
which can be also rewritten as
\begin{equation}
\left\langle E\right\rangle _{\mathrm{ds}}\simeq\tfrac{1}{2}\left(
\tfrac{r_{\mathrm{ds}}}{r_{+}}\right)  \ m_{e}c^{2}\sim\sqrt{\tfrac{\xi
}{M/M_{\odot}}}10^{5}keV\,.\label{GRB}%
\end{equation}
Such a process of vacuum polarization, occurring not at the horizon but in the
extended dyadosphere region ($r_{+}\leq r\leq r_{\mathrm{ds}}$) around a
black hole, has been observed to reach the maximum efficiency limit of $50\%$ of the
total mass-energy of an extreme black hole (see e.g. \cite{prx98}). The conceptual
justification of this result follows from the present work: the $e^{+}e^{-}$
creation process occurs at the expense of the Coulomb energy given by Eq.
(\ref{EQR}) and does not affect the irreducible mass given by Eq. (\ref{EQM}),
which indeed, as we have proved, does not depend of the electromagnetic
energy. In this sense, $\delta M_{\mathrm{irr}}=0$ and the transformation is
fully reversible. This result will be further validated by the study of the
dynamical formation of the dyadosphere, which we have obtained using the
present work and \cite{crv02} (see \cite{rvx03a,rvx03b}).
\end{enumerate}

Let us now compare and contrast these two processes. We have
\begin{align}
r_{\mathrm{Eerg}}  & \simeq\left(  \tfrac{r_{ds}}{\lambda
_{\mathrm{C}}}\right)  r\\
N_{\mathrm{dya}}  & \simeq\left(  \tfrac{r_{ds}}{\lambda
_{\mathrm{C}}}\right)  N_{\mathrm{PD}},\\
\left\langle E\right\rangle _{\mathrm{dya}}  & \simeq\left(  \tfrac
{\lambda_{\mathrm{C}}}{r_{ds}}\right)  \left\langle E\right\rangle
_{\mathrm{PD}}.
\end{align}
Moreover we see (Eqs. (\ref{UHECR}), (\ref{GRB})) that $\left\langle
E\right\rangle _{\mathrm{PD}}$ is in the range of energies of UHECR, while for
$\xi\sim0.1$ and $M\sim10M_{\odot}$, $\left\langle E\right\rangle
_{\mathrm{ds}}$ is in the gamma ray range. In other words, the discrete
particle decay process involves a small number of particles with ultra high
energies ($\sim10^{21}eV$), while vacuum polarization involves a much larger
number of particles with lower mean energies ($\sim10MeV$).

\begin{figure}
\centering
\includegraphics[width=10cm,clip]{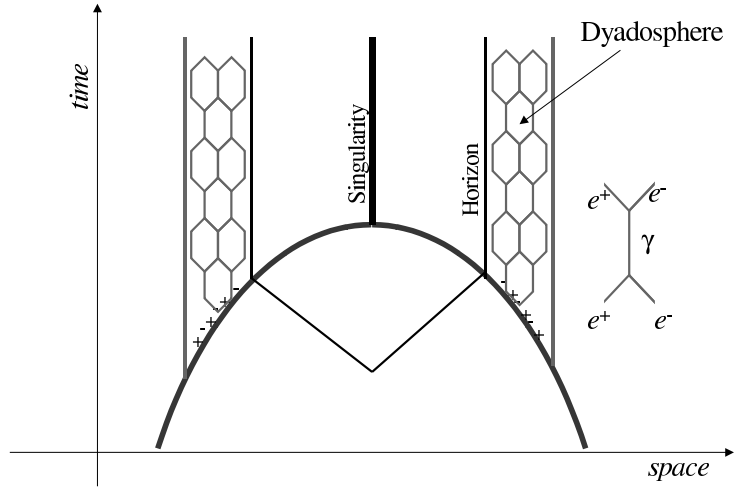}
\caption{Space-time diagram of the collapse process leading to the formation of the dyadosphere. As the collapsing core crosses the dyadosphere radius the pair creation process starts, and the pairs thermalize in a neutral plasma configuration. Then also the horizon is crossed and the singularity is formed.}
\label{dyaform}
\end{figure}

Having so established and clarified the basic conceptual processes of the energetic of the black hole, we are now ready to approach, using the new analytic solution obtained, the dynamical process of vacuum polarization occurring during the formation of a black hole as qualitatively represented in Fig. \ref{dyaform}. The study of the dyadosphere dynamical formation as well as of the electron-positron plasma dynamical evolution will lead to the first possibility of directly observing the general relativistic effects approaching the black hole horizon.

\section{Contribution of the short GRBs to the black hole theory}\label{luca2}

\subsection{On the gravitational binding energy of white dwarf and neutron stars}

The aim of this section is to point out how the knowledge obtained from the black hole model is of relevance also for the basic theory of black holes and further how very high precision verification of general relativistic effects in the very strong field near the formation of the horizon should be expected in the near future.

We shall first see how Eq.(\ref{EQM}) for $M_{\mathrm{irr}}$,
\begin{equation}
M_{\mathrm{irr}}=M_{0}-\tfrac{M_{0}^{2}}{2r_{+}}+T_{+}\, ,
\end{equation}
leads to a deeper physical understanding of the role of the gravitational interaction in the maximum energy extraction process of a black hole. This formula can also be of assistance in clarifying some long lasting epistemological issue on the role of general relativity, quantum theory and thermodynamics.

It is well known that if a spherically symmetric mass distribution without any electromagnetic structure undergoes free gravitational collapse, its total mass-energy $M$ is conserved according to the Birkhoff theorem: the increase in the kinetic energy of implosion is balanced by the increase in the gravitational energy of the system. If one considers the possibility that part of the kinetic energy of implosion is extracted then the situation is very different: configurations of smaller mass-energy and greater density can be attained without violating Birkhoff theorem.

We illustrate our considerations with two examples: one has found confirmation from astrophysical observations, the other promises to be of relevance for gamma ray bursts (GRBs) (see \cite{rv02a}). Concerning the first example, it is well known from the work of \cite{l32} that at the endpoint of thermonuclear evolution, the gravitational collapse of a spherically symmetric star can be stopped by the Fermi pressure of the degenerate electron gas (white dwarf). A configuration of equilibrium can be found all the way up to the critical number of particles
\begin{equation}
N_{\mathrm{crit}}=0.775\tfrac{m_{Pl}^{3}}{m_{0}^{3}},
\end{equation}
where the factor $0.775$ comes from the coefficient $\tfrac{3.098}{\mu^{2}}$ of the solution of the Lane-Emden equation with polytropic index $n=3$, and $m_{Pl}=\sqrt{\tfrac{\hbar c}{G}}$ is the Planck mass, $m_{0}$ is the nucleon mass and $\mu$ the average number of electrons per nucleon. As the kinetic energy of implosion is carried away by radiation the star settles down to a configuration of mass
\begin{equation}
M=N_{\mathrm{crit}}m_{0}-U, \label{BE}%
\end{equation}
where the gravitational binding energy $U$ can be as high as $5.72\times 10^{-4}N_{\mathrm{crit}}m_{0}$.

Similarly Gamov (see \cite{g51}) has shown that a gravitational collapse process to still higher densities can be stopped by the Fermi pressure of the neutrons (neutron star) and Oppenheimer \cite{ov39} has shown that, if the effects of strong interactions are neglected, a configuration of equilibrium exists also in this case all the way up to a critical number of particles
\begin{equation}
N_{\mathrm{crit}}=0.398\tfrac{m_{Pl}^{3}}{m_{0}^{3}},
\end{equation}
where the factor $0.398$ comes now from the integration of the
Tolman-Oppenheimer-Volkoff equation (see e.g. \cite{htww65}). If the kinetic energy of implosion is again carried away by radiation of photons or neutrinos and antineutrinos the final configuration is characterized by the formula (\ref{BE}) with $U\lesssim2.48\times10^{-2}N_{\mathrm{crit}}m_{0}$. These considerations and the existence of such large values of the gravitational binding energy have been at the heart of the explanation of astrophysical phenomena such as red-giant stars and supernovae: the corresponding measurements of the masses of neutron stars and white dwarfs have been carried out with unprecedented accuracy in binary systems \cite{gr75}.

\subsection{On the minimum value of the reducible mass of a black hole formed in a spherically symmetric gravitational collapse}

From a theoretical physics point of view it is still an open question how far such a sequence can go: using causality nonviolating interactions, can one find a sequence of braking and energy extraction processes by which the density and the gravitational binding energy can increase indefinitely and the mass-energy of the collapsed object be reduced at will? This question can also be formulated in the mass-formula language of a black hole given in \cite{cr71} (see also \cite{rv02a}): given a collapsing core of nucleons with a given rest mass-energy $M_{0}$, what is the minimum irreducible mass of the black hole which is formed?

Following \cite{crv02} and \cite{rv02a}, consider a spherical shell of rest mass $M_{0}$ collapsing in a flat space-time. In the neutral case the irreducible mass of the final black hole satisfies the equation (see \cite{rv02a})
\begin{equation}
M_{\mathrm{irr}}=M=M_{0}-\tfrac{M_{0}^{2}}{2r_{+}}+T_{+}, \label{Mirr2}%
\end{equation}
where $M$ is the total energy of the collapsing shell and $T_{+}$ the kinetic energy at the horizon $r_{+}$. Recall that the area $S$ of the horizon is \cite{cr71}
\begin{equation}
S=4\pi r_{+}^{2}=16\pi M_{\mathrm{irr}}^{2} \label{Sbis}%
\end{equation}
where $r_{+}=2M_{\mathrm{irr}}$ is the horizon radius. The minimum irreducible mass $M_{\mathrm{irr}}^{\left(  {\mathrm{min}}\right)  }$ is obtained when the kinetic energy at the horizon $T_{+}$ is $0$, that is when the entire kinetic energy $T_{+}$ has been extracted. We then obtain the simple result
\begin{equation}
M_{\mathrm{irr}}^{\left(  \mathrm{min}\right)  }=\tfrac{M_{0}}{2}.
\label{Mirrmin}%
\end{equation}
We conclude that in the gravitational collapse of a spherical shell of rest mass $M_{0}$ at rest at infinity (initial energy $M_{\mathrm{i}}=M_{0}$), an energy up to $50\%$ of $M_{0}c^{2}$ can in principle be extracted, by braking processes of the kinetic energy. In this limiting case the shell crosses the horizon with $T_{+}=0$. The limit $\tfrac{M_{0}}{2}$ in the extractable kinetic energy can further increase if the collapsing shell is endowed with kinetic energy at infinity, since all that kinetic energy is in principle extractable.

\begin{figure}
\centering
\includegraphics[width=10cm,clip]{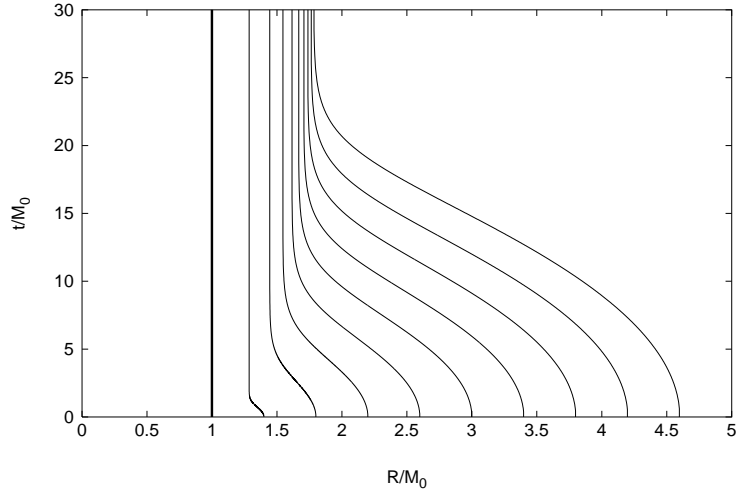}
\caption{Collapse curves for neutral shells with rest mass $M_{0}$ starting at rest at selected radii $R^{\ast}$ computed by using the exact solutions given in \cite{crv02}. A different value of $M_{\mathrm{irr}}$ (and therefore of $r_{+}$) corresponds to each curve. The time parameter is the Schwarzschild time coordinate $t$ and the asymptotic behavior at the respective horizons is evident. The limiting configuration $M_{\mathrm{irr}}=\tfrac{M_{0}}{2}$ (solid line) corresponds to the case in which the shell is trapped, at the very beginning of its motion, by the formation of the horizon.}
\label{fig1l2}
\end{figure}

\begin{figure}
\centering
\includegraphics[width=10cm,clip]{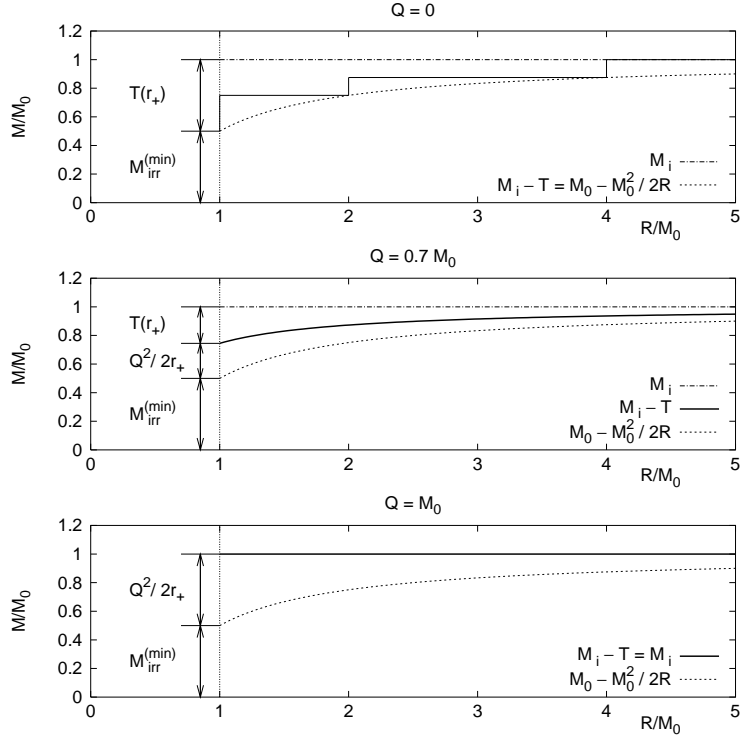}
\caption{Energetics of a shell such that $M_{\mathrm{i}}=M_{0}$, for selected values of the charge. In the first diagram $Q=0$; the dashed line represents the total energy for a gravitational collapse without any braking process as a function of the radius $R$ of the shell; the solid, stepwise line represents a collapse with suitable braking of the kinetic energy of implosion at selected radii; the dotted line represents the rest mass energy plus the gravitational binding energy. In the second and third diagram $Q/M_{0}=0.7$, $Q/M_{0}=1$ respectively; the dashed and the dotted lines have the same meaning as above; the solid lines represent the total energy minus the kinetic energy. The region between the solid line and the dotted line corresponds to the stored electromagnetic energy. The region between the dashed line and the solid line corresponds to the kinetic energy of collapse. In all the cases the sum of the kinetic energy and the electromagnetic energy at the horizon is 50\% of $M_{0}$. Both the electromagnetic and the kinetic energy are extractable. It is most remarkable that the same underlying process occurs in the three cases: the role of the electromagnetic interaction is twofold: a) to reduce the kinetic energy of implosion by the Coulomb repulsion of the shell; b) to store such an energy in the region around the black hole. The stored electromagnetic energy is extractable as shown in \cite{rv02a}.}
\label{fig3l2}
\end{figure}

In order to illustrate the physical reasons for this result, using the formulas of \cite{crv02}, we have represented in Fig. \ref{fig1l2} the world lines of spherical shells of the same rest mass $M_{0}$, starting their gravitational collapse at rest at selected radii $R^{\ast}$. These initial conditions can be implemented by performing suitable braking of the collapsing shell and concurrent kinetic energy extraction processes at progressively smaller radii (see also Fig. \ref{fig3l2}). The reason for the existence of the minimum (\ref{Mirrmin}) in the black hole mass is the ``self closure'' occurring by the formation of a horizon in the initial configuration (thick line in Fig. \ref{fig1l2}).

Is the limit $M_{\mathrm{irr}}\rightarrow\tfrac{M_{0}}{2}$ actually attainable without violating causality? Let us consider a collapsing shell with charge $Q$. If $M\geq Q$ a black hole is formed. As pointed out in \cite{rv02a} the irreducible mass of the final black hole does not depend on the charge $Q$. Therefore Eqs.(\ref{Mirr2}) and (\ref{Mirrmin}) still hold in the charged case with $r_{+}=M+\sqrt{M^{2}-Q^{2}}$. In Fig. \ref{fig3l2} we consider the special case in which the shell is initially at rest at infinity, i.e. has initial energy $M_{\mathrm{i}}=M_{0}$, for three different values of the charge $Q$. We plot the initial energy $M_{i}$, the energy of the system when all the kinetic energy of implosion has been extracted as well as the sum of the rest mass energy and the gravitational binding energy $-\tfrac{M_{0}^{2}}{2R}$ of the system (here $R$ is the radius of the shell). In the extreme case $Q=M_{0}$, the shell is in equilibrium at all radii (see \cite{crv02}) and the kinetic energy is identically zero. In all three cases, the sum of the extractable kinetic energy $T$ and the electromagnetic energy $\tfrac{Q^{2}}{2R}$ reaches $50\%$ of the rest mass energy at the horizon, according to Eq.(\ref{Mirrmin}).

What is the role of the electromagnetic field here? If we consider the case of a charged shell with $Q\simeq M_{0}$, the electromagnetic repulsion implements the braking process and the extractable energy is entirely stored in the electromagnetic field surrounding the black hole (see \cite{rv02a}). In \cite{rv02a} we have outlined two different processes of electromagnetic energy extraction. We emphasize here that the extraction of $50\%$ of the mass-energy of a black hole is not specifically linked to the electromagnetic field but depends on three factors: a) the increase of the gravitational energy during the collapse, b) the formation of a horizon, c) the reduction of the kinetic energy of implosion. Such conditions are naturally met during the formation of an extreme black hole but are more general and can indeed occur in a variety of different situations, e.g. during the formation of a Schwarzschild black hole by a suitable extraction of the kinetic energy of implosion (see Fig. \ref{fig1l2} and Fig. \ref{fig3l2}).

\subsection{On the Bekenstein-Hawking consideration of incompatibility between general relativity and thermodynamics}

Now consider a test particle of mass $m$ in the gravitational field of an already formed Schwarzschild black hole of mass $M$ and go through such a sequence of braking and energy extraction processes. Kaplan \cite{k49} found for the energy $E$ of the particle as a function of the radius $r$
\begin{equation}
E=m\sqrt{1-\tfrac{2M}{r}}.\label{pointtest}
\end{equation}
It would appear from this formula that the entire energy of a particle could be extracted in the limit $r\rightarrow2M$. Such $100\%$ efficiency of energy extraction has often been quoted as evidence for incompatibility between General Relativity and the second principle of Thermodynamics (see \cite{b73} and references therein). J. Bekenstein and S. Hawking have gone as far as to consider General Relativity not to be a complete theory and to conclude that in order to avoid inconsistencies with thermodynamics, the theory should be implemented through a quantum description \cite{b73,h74}. Einstein himself often expressed the opposite point of view (see e.g. \cite{d02}).

The analytic treatment presented in \cite{crv02} can clarify this fundamental issue. It allows to express the energy increase $E$ of a black hole of mass $M_{1}$ through the accretion of a shell of mass $M_{0}$ starting its motion at rest at a radius $R$ in the following formula which generalizes Eq.(\ref{pointtest}):
\begin{equation}
E\equiv M-M_{1}=-\tfrac{M_{0}^{2}}{2R}+M_{0}\sqrt{1-\tfrac{2M_{1}}{R}},
\end{equation}
where $M=M_{1}+E$ is clearly the mass-energy of the final black hole. This formula differs from the Kaplan formula (\ref{pointtest}) in three respects: a) it takes into account the increase of the horizon area due to the accretion of the shell; b) it shows the role of the gravitational self energy of the imploding shell; c) it expresses the combined effects of a) and b) in an exact closed formula.

The minimum value $E_{\mathrm{\min}}$ of $E$ is attained for the minimum value of the radius $R=2M$: the horizon of the final black hole. This corresponds to the maximum efficiency of the energy extraction. We have
\begin{equation}
E_{\min}=-\tfrac{M_{0}^{2}}{4M}+M_{0}\sqrt{1-\tfrac{M_{1}}{M}}=-\tfrac
{M_{0}^{2}}{4(M_{1}+E_{\min})}+M_{0}\sqrt{1-\tfrac{M_{1}}{M_{1}+E_{\min}}},
\end{equation}
or solving the quadratic equation and choosing the positive solution for physical reasons
\begin{equation}
E_{\min}=\tfrac{1}{2}\left(  \sqrt{M_{1}^{2}+M_{0}^{2}}-M_{1}\right)  .
\end{equation}
The corresponding efficiency of energy extraction is
\begin{equation}
\eta_{\max}=\tfrac{M_{0}-E_{\min}}{M_{0}}=1-\tfrac{1}{2}\tfrac{M_{1}}{M_{0}%
}\left(  \sqrt{1+\tfrac{M_{0}^{2}}{M_{1}^{2}}}-1\right)  , \label{efficiency}%
\end{equation}
which is strictly \emph{smaller than} 100\% for \emph{any} given $M_{0}\neq0$. It is interesting that this analytic formula, in the limit $M_{1}\ll M_{0}$, properly reproduces the result of equation (\ref{Mirrmin}), corresponding to an efficiency of $50\%$. In the opposite limit $M_{1}\gg M_{0}$ we have
\begin{equation}
\eta_{\max}\simeq1-\tfrac{1}{4}\tfrac{M_{0}}{M_{1}}.
\end{equation}
Only for $M_{0}\rightarrow0$, Eq.(\ref{efficiency}) corresponds to an efficiency of 100\% and correctly represents the limiting reversible transformations introduced in \cite{cr71}. It seems that the difficulties of reconciling General Relativity and Thermodynamics are ascribable not to an incompleteness of General Relativity but to the use of the Kaplan formula in a regime in which it is not valid. The generalization of the above results to stationary black holes is being considered.

\section{On a Separatrix in the Gravitational Collapse to an Overcritical Electromagnetic Black Hole}\label{separatrix}

We are now ready to analyze the dynamical properties of an electron--positron--photon plasma created by the vacuum polarization process occurring around a charged gravitationally collapsing core of an initially neutral star are examined within the framework of General Relativity and Quantum Field Theory. The Reissner--Nordstr\"{o}m geometry is assumed to apply between the collapsing core and the oppositely charged remnant of the star. The appearance of a separatrix at radius $\bar{R}$, well outside the asymptotic approach to the horizon, is evidenced. The neutral electron--positron--photon plasma created at radii $r>\bar{R}$ self-propels outwards to infinity, following the classical PEM--pulse analysis \cite{rswx99,rswx00}. The plasma created at $r<\bar{R}$ remains trapped and follows the gravitational collapse of the core only contributing to the reduction of the electromagnetic energy of the black hole and to the increase of its irreducible mass. This phenomenon has consequences for the observational properties of gamma--ray bursts and is especially relevant for the theoretical prediction of the temporal and spectral structure of the short bursts.

The formulation of the physics of the \emph{dyadosphere} of an electromagnetic
black hole (black hole) has been until now approached by assuming the vacuum
polarization process \emph{\`{a} l\`{a}} Sauter--Heisenberg--Euler--Schwinger
\cite{s31,he35,s51} in the field of an already formed Kerr--Newmann
\cite{dr75} or Reissner--Nordstr\"{o}m black hole \cite{prx98,rv02a}. This
acausal approach is certainly valid in order to describe the overall
energetics and the time development of the gamma--ray bursts (GRBs) reaching a
remarkable agreement between the observations and the theoretical prediction,
in particular with respect to: a) the existence of a proper gamma--ray burst
(P--GRB) \cite{lett1}, b) the afterglow detailed luminosity function and
spectral properties \cite{rbcfx03a,Brasile,rbcfx02_letter} and c) the relative
intensity of the P-GRB to the afterglow \cite{lett2,rbcfx03a,Brasile}.

This acausal approach has to be improved by taking into account the causal
dynamical process of the formation of the dyadosphere as soon as the detailed
description on timescales of $10^{-4}-10^{-3}$s of the P--GRB are considered.
Such a description leads to theoretical predictions on the time variability of
the P--GRB spectra which may become soon testable by a new class of specially
conceived space missions.

We report progress in this theoretically
challenging process which is marked by distinctive and precise quantum and
general relativistic effects. These new results have been made possible by the
recent progress in Refs.~\cite{crv02}, \cite{rv02a} and especially
\cite{rvx03}. There it was demonstrated the intrinsic stability of the
gravitational amplification of the electromagnetic field at the surface of a
charged star core collapsing to a black hole. The $e^{+}e^{-}$ plasma generated by
the vacuum polarization process around the core is entangled in the
electromagnetic field \cite{rvx03a}. The $e^{+}e^{-}$ pairs do thermalize in
an electron--positron--photon plasma on a time scale $10^{2}-10^{4}$ times
larger than $\hbar/m_{e}c$ \cite{rvx03}, where $c$ is the speed of light and
$m_{e}$ the electron mass. As soon as the thermalization has occurred, a
dynamical phase of this electrically neutral plasma starts following the
considerations already discussed in \cite{rswx99,rswx00}. While the temporal
evolution of the $e^{+}e^{-}\gamma$ plasma takes place, the gravitationally
collapsing core moves inwards, giving rise to a further amplified
supercritical field, which in turn generates a larger amount of $e^{+}e^{-}$
pairs leading to a yet higher temperature in the newly formed $e^{+}%
e^{-}\gamma$ plasma. We report, in the following, progress in the
understanding of this crucial dynamical process: the main difference from the
previous treatments is the fact that we do not consider an already formed black hole
but we follow the dynamical phase of the formation of dyadosphere and of the
asymptotic approach to the horizon by examining the time varying process at
the surface of the gravitationally collapsing core.

The space--time external to the surface of the spherically symmetric
collapsing core is described by the Reissner-Nordstr\"{o}m geometry \cite{P74}
with line element
\begin{equation}
ds^{2}=-\alpha^{2}dt^{2}+\alpha^{-2}dr^{2}+r^{2}d\Omega^{2},\label{ds}%
\end{equation}
with $d\Omega^{2}=d\theta^{2}+\sin^{2}\theta d\phi^{2}$, $\alpha^{2}%
=\alpha^{2}\left(  r\right)  =1-2M/r+Q^{2}/r^{2}$, where $M$ and $Q$ are the
total energy and charge of the core as measured at infinity. On the core
surface, which at the time $t_{0}$ has radial coordinate $r_{0}$, the
electromagnetic field strength is $\mathcal{E}=\mathcal{E}\left(
r_{0}\right)  =Q/r_{0}^{2}$. The equation of core's collapse is (see
\cite{crv02}):
\begin{equation}
\tfrac{dr_{0}}{dt_{0}}=-\tfrac{\alpha^{2}\left(  r_{0}\right)  }{H\left(
r_{0}\right)  }\sqrt{H^{2}\left(  r_{0}\right)  -\alpha^{2}\left(
r_{0}\right)  }\label{Motion}%
\end{equation}
where $H\left(  r_{0}\right)  =\tfrac{M}{M_{0}}-\tfrac{M_{0}^{2}+Q^{2}}%
{2M_{0}r_{0}}$ and $M_{0}$ is the core rest mass. Analytic expressions for the
solution of Eq.(\ref{Motion}) were given in \cite{crv02}. We here recall that
the dyadosphere radius is defined by $\mathcal{E}\left(  r_{\mathrm{ds}%
}\right)  =\mathcal{E}_{\mathrm{c}}=$ $m_{e}^{2}c^{3}/e\hbar$ \cite{prx98} as
$r_{\mathrm{ds}}=\sqrt{eQ\hbar/m_{e}^{2}c^{3}}$, where $e$ is the electron
charge. In the following we assume that the dyadosphere starts to be formed at
the instant $t_{\mathrm{ds}}=t_{0}\left(  r_{\mathrm{ds}}\right)  =0$.

Having formulated the core collapse in General Relativity in Eq.(\ref{Motion}%
), in order to describe the quantum phenomena, we consider, at each value of
$r_{0}$ and $t_{0}$, a slab of constant coordinate thickness $\Delta r$ small
in comparison with $r_{\mathrm{ds}}$ and larger than $\hbar/m_{e}c^{2}$. All
the results will be shown to be independent on the choice of the value of
$\Delta r$. In each slab the process of vacuum polarization leading to
$e^{+}e^{-}$ pair creation is considered. As shown in \cite{rvx03,rvx03a} the
pairs created oscillate \cite{KESCM91,KESCM92,CEKMS93,BMP...99} with
ultrarelativistic velocities and partially annihilate into photons; the
electric field oscillates around zero and the amplitude of such oscillations
decreases with a characteristic time of the order of $10^{2}-10^{4}$
$\hbar/m_{e}c^{2}$. The electric field is effectively screened to the critical
value $\mathcal{E}_{\mathrm{c}}$ and the pairs thermalize to an $e^{+}%
e^{-}\gamma$ plasma. While the average of the electric field $\mathcal{E}$
over one oscillation is $0$, the average of $\mathcal{E}^{2}$ is of the order
of $\mathcal{E}_{c}^{2}$, therefore the energy density in the pairs and
photons, as a function of $r_{0}$, is given by \cite{rv02a}
\begin{equation}
\epsilon_{0}\left(  r_{0}\right)  =\tfrac{1}{8\pi}\left[  \mathcal{E}%
^{2}\left(  r_{0}\right)  -\mathcal{E}_{c}^{2}\right]  =\tfrac{\mathcal{E}%
_{c}^{2}}{8\pi}\left[  \left(  \tfrac{r_{\mathrm{ds}}}{r_{0}}\right)
^{4}-1\right]  .\label{eps0}%
\end{equation}
For the number densities of $e^{+}e^{-}$ pairs and photons at thermal
equilibrium we have $n_{e^{+}e^{-}}\simeq n_{\gamma}$; correspondingly the
equilibrium temperature $T_{0}$, which is clearly a function of $r_{0}$ and is
different for each slab, is such that
\begin{equation}
\epsilon\left(  T_{0}\right)  \equiv\epsilon_{\gamma}\left(  T_{0}\right)
+\epsilon_{e^{+}}\left(  T_{0}\right)  +\epsilon_{e^{-}}\left(  T_{0}\right)
=\epsilon_{0},\label{eq0}%
\end{equation}
with $\epsilon$ and $n$ given by Fermi (Bose) integrals (with zero chemical
potential):
\begin{align}
\epsilon_{e^{+}e^{-}}\left(  T_{0}\right)   &  =\tfrac{2}{\pi^{2}\hbar^{3}%
}\int_{m_{e}}^{\infty}\tfrac{\left(  E^{2}-m_{e}^{2}\right)  ^{1/2}}%
{\exp\left(  E/kT_{0}\right)  +1}E^{2}dE,\quad\epsilon_{\gamma}\left(
T_{0}\right)  =\tfrac{\pi^{2}}{15\hbar^{3}}\left(  kT_{0}\right)
^{4},\label{Integrals1}\\
n_{e^{+}e^{-}}\left(  T_{0}\right)   &  =\tfrac{1}{\pi^{2}\hbar^{3}}%
\int_{m_{e}}^{\infty}\tfrac{\left(  E^{2}-m_{e}^{2}\right)  ^{1/2}}%
{\exp\left(  E/kT_{0}\right)  +1}EdE,\quad n_{\gamma}\left(  T_{0}\right)
=\tfrac{2\zeta\left(  3\right)  }{\hbar^{3}}\left(  kT_{0}\right)
^{3},\label{Integrals2}%
\end{align}
where $k$ is the Boltzmann constant. From the conditions set by Eqs.(\ref{eq0}%
), (\ref{Integrals1}), (\ref{Integrals2}), we can now turn to the dynamical
evolution of the $e^{+}e^{-}\gamma$ plasma in each slab. We use the covariant
conservation of energy momentum and the rate equation for the number of pairs
in the Reissner--Nordstr\"{o}m geometry external to the star core:
\begin{equation}
\nabla_{a}T^{ab}=0,\quad\nabla_{a}\left(  n_{e^{+}e^{-}}u^{a}\right)
=\overline{\sigma v}\left[  n_{e^{+}e^{-}}^{2}\left(  T\right)  -n_{e^{+}%
e^{-}}^{2}\right]  ,\label{na}%
\end{equation}
where $T^{ab}=\left(  \epsilon+p\right)  u^{a}u^{b}+pg^{ab}$ is the
energy--momentum tensor of the plasma with proper energy density $\epsilon$
and proper pressure $p$, $u^{a}$ is the fluid $4-$velocity, $n_{e^{+}e^{-}}$
is the number of pairs, $n_{e^{+}e^{-}}\left(  T\right)  $ is the equilibrium
number of pairs and $\overline{\sigma v}$ is the mean of the product of the
$e^{+}e^{-}$ annihilation cross-section and the thermal velocity of pairs. We
follow closely the treatment which we developed for the consideration of a
plasma generated in the dyadosphere of an already formed black hole
\cite{rswx99,rswx00}. It was shown in \cite{rswx99,rswx00} that the plasma
expands as a pair--electromagnetic pulse (PEM pulse) of constant thickness in
the laboratory frame. Since the expansion, hydrodynamical timescale is much
larger than the pair creation ($\hbar/m_{e}c^{2}$) and the thermalization
($10^{2}-10^{4}\hbar/m_{e}c^{2}$) time-scales, in each slab the plasma remains
at thermal equilibrium in the initial phase of the expansion and the right
hand side of the rate Eq.(\ref{na}) is effectively $0$, see Fig. 24 (second
panel) of \cite{Brasile} for details.

If we denote by $\xi^{a}$ the static Killing vector field normalized at unity
at spacial infinity and by $\left\{  \Sigma_{t}\right\}  _{t}$ the family of
space-like hypersurfaces orthogonal to $\xi^{a}$ ($t$ being the Killing time)
in the Reissner--Nordstr\"{o}m geometry, from Eqs.(\ref{na}), the following
integral conservation laws can be derived (see for instance \cite{D79,S60})
\begin{equation}
\int_{\Sigma_{t}}\xi_{a}T^{ab}d\Sigma_{b}=E,\quad\int_{\Sigma_{t}}%
n_{e^{+}e^{-}}u^{b}d\Sigma_{b}=N_{e^{+}e^{-}},\label{Ne}%
\end{equation}
where $d\Sigma_{b}=\alpha^{-2}\xi_{b}r^{2}\sin\theta drd\theta d\phi$ is the
vector surface element, $E$ the total energy and $N_{e^{+}e^{-}}$ the total
number of pairs which remain constant in each slab. We then have
\begin{equation}
\left[  \left(  \epsilon+p\right)  \gamma^{2}-p\right]  r^{2}=\mathfrak
{E},\quad n_{e^{+}e^{-}}\gamma\alpha^{-1}r^{2}=\mathfrak{N}_{e^{+}e^{-}%
},\label{ne}%
\end{equation}
where $\mathfrak{E}$ and $\mathfrak{N}_{e^{+}e^{-}}$ are constants and
\begin{equation}
\gamma\equiv\alpha^{-1}u^{a}\xi_{a}=\left[  1-\alpha^{-4}\left(  \tfrac
{dr}{dt}\right)  ^{2}\right]  ^{-1/2}%
\end{equation}
is the Lorentz $\gamma$ factor of the slab as measured by static observers. We
can rewrite Eqs.(\ref{Ne}) for each slab as
\begin{align}
\left(  \tfrac{dr}{dt}\right)  ^{2} &  =\alpha^{4}f_{r_{0}},\label{eq17}\\
\left(  \tfrac{r}{r_{0}}\right)  ^{2} &  =\left(  \tfrac{\epsilon+p}%
{\epsilon_{0}}\right)  \left(  \tfrac{n_{e^{+}e^{-}0}}{n_{e^{+}e^{-}}}\right)
^{2}\left(  \tfrac{\alpha}{\alpha_{0}}\right)  ^{2}-\tfrac{p}{\epsilon_{0}%
}\left(  \tfrac{r}{r_{0}}\right)  ^{4},\label{eq18}\\
f_{r_{0}} &  =1-\left(  \tfrac{n_{e^{+}e^{-}}}{n_{e^{+}e^{-}0}}\right)
^{2}\left(  \tfrac{\alpha_{0}}{\alpha}\right)  ^{2}\left(  \tfrac{r}{r_{0}%
}\right)  ^{4}\label{eq19}%
\end{align}
where pedex $_{0}$ refers to quantities evaluated at selected initial times
$t_{0}>0$, having assumed $r\left(  t_{0}\right)  =r_{0}$, $\left.
dr/dt\right|  _{t=t_{0}}=0$, $T\left(  t_{0}\right)  =T_{0}$.

Eq.(\ref{eq17}) is only meaningful when $f_{r_{0}}\left(  r\right)  \geq0$.
From the structural analysis of such equation it is clearly identifiable a
critical radius $\bar{R}$ such that:

\begin{itemize}
\item  for any slab initially located at $r_{0}>\bar{R}$ we have $f_{r_{0}%
}\left(  r\right)  \geq0$ for any value of $r\geq r_{0}$ and $f_{r_{0}}\left(
r\right)  <0$ for $r\lesssim r_{0}$; therefore a slab initially located at a
radial coordinate $r_{0}>\bar{R}$ moves outwards,

\item  for any slab initially located at $r_{0}<\bar{R}$ we have $f_{r_{0}%
}\left(  r\right)  \geq0$ for any value of $r_{+}<r\leq r_{0}$ and $f_{r_{0}%
}\left(  r\right)  <0$ for $r\gtrsim r_{0}$; therefore a slab initially
located at a radial coordinate $r_{0}<\bar{R}$ moves inwards and is trapped by
the gravitational field of the collapsing core.
\end{itemize}

We define the surface $r=\bar{R}$, the \emph{dyadosphere trapping surface
}(DTS). The radius $\bar{R}$ of DTS is generally evaluated by the condition
$\left.  \tfrac{df_{\bar{R}}}{dr}\right|  _{r=\bar{R}}=0$.
$\bar{R}$ is so close to the horizon value $r_{+}$ that the initial
temperature $T_{0}$ satisfies $kT_{0}\gg m_{e}c^{2}$ and we can obtain for
$\bar{R}$ an analytical expression. Namely the ultrarelativistic approximation
of all Fermi integrals, Eqs.(\ref{Integrals1}) and (\ref{Integrals2}), is
justified and we have $n_{e^{+}e^{-}}\left(  T\right)  \propto T^{3}$ and
therefore $f_{r_{0}}\simeq1-\left(  T/T_{0}\right)  ^{6}\left(  \alpha
_{0}/\alpha\right)  ^{2}\left(  r/r_{0}\right)  ^{4}$ ($r\leq\bar{R}$).
The defining equation of $\bar{R}$, together with (\ref{eq19}), then gives
\begin{equation}
\bar{R}=2M\left[  1+\left(  1-3Q^{2}/4M^{2}\right)  ^{1/2}\right]  >r_{+}.
\end{equation}

\begin{figure}
\centering
\includegraphics[width=10cm]{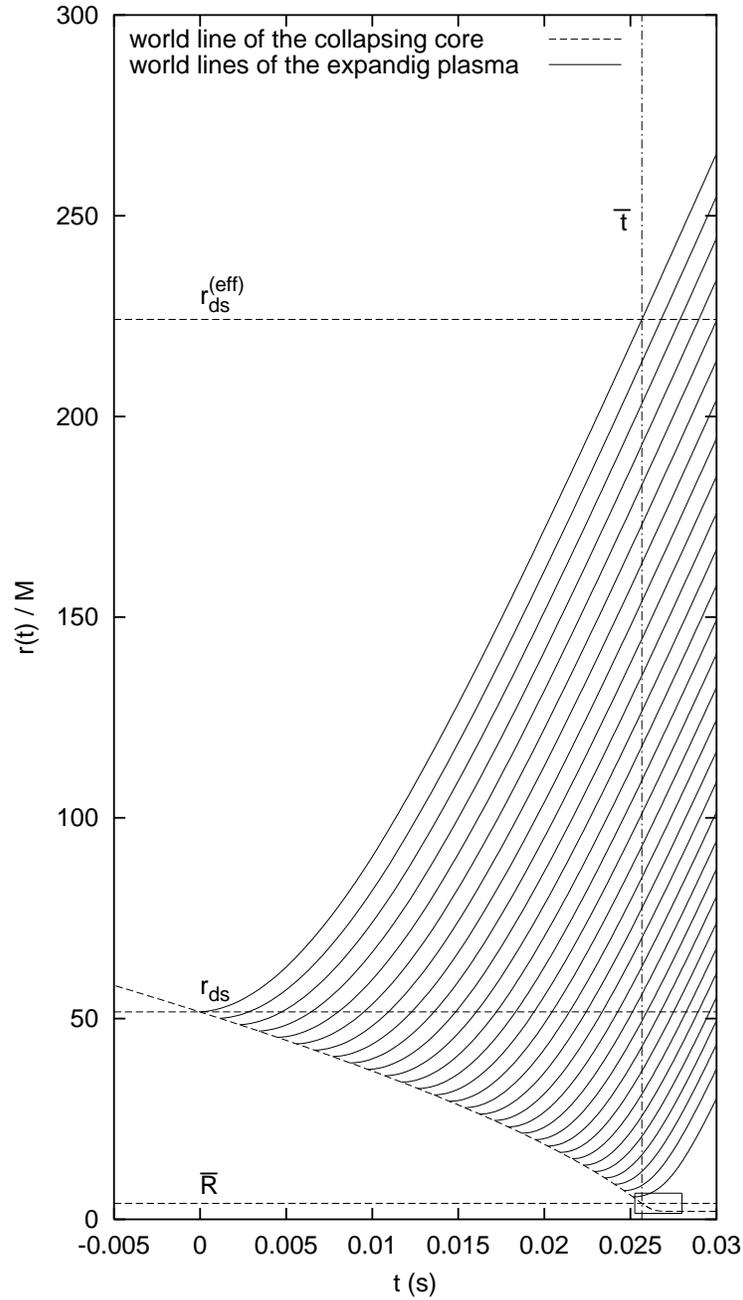}
\caption{World line of the collapsing charged core (dashed line) as derived
from Eq.(\ref{Motion}) for a black hole with $M=20M_{\odot}$, $Q=0.1M$; world
lines of slabs of plasma for selected radii $r_{0}$ in the interval $\left(
\bar{R},r_{\mathrm{ds}}\right)  $. At time $\bar{t}$ the expanding plasma
extends over a region which is almost one order of magnitude larger than the
dyadosphere. The small rectangle in the right bottom is enlarged in
Fig. \ref{f2}.}
\label{f1}
\end{figure}

\begin{figure}
\centering
\includegraphics[width=13cm]{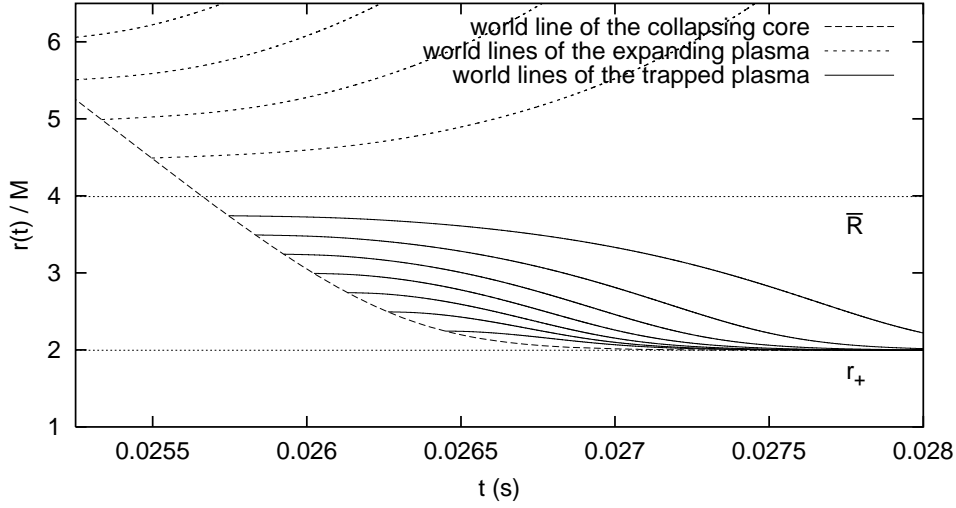}
\caption{Enlargement of the small rectangle in the right bottom of
Fig. \ref{f1}. World--lines of slabs of plasma for selected radii $r_{0}$ in
the interval $\left(  r_{+,}\bar{R}\right)  $.}
\label{f2}
\end{figure}

In the case of a black hole with $M=20M_{\odot}$, $Q=0.1M$, we compute:

\begin{itemize}
\item  the fraction of energy trapped in DTS:
\begin{equation}
\bar{E}=\int_{r_{+}<r<\bar{R}}\alpha\epsilon_{0}d\Sigma\simeq0.53\int
_{r_{+}<r<r_{\mathrm{ds}}}\alpha\epsilon_{0}d\Sigma;
\end{equation}

\item  the world--lines of slabs of plasma for selected $r_{0}$ in the
interval $\left(  \bar{R},r_{\mathrm{ds}}\right)  $ (see Fig. \ref{f1});

\item  the world--lines of slabs of plasma for selected $r_{0}$ in the
interval $\left(  r_{+},\bar{R}\right)  $ (see Fig. \ref{f2}).
\end{itemize}

\begin{figure}
\centering
\includegraphics[width=13cm]{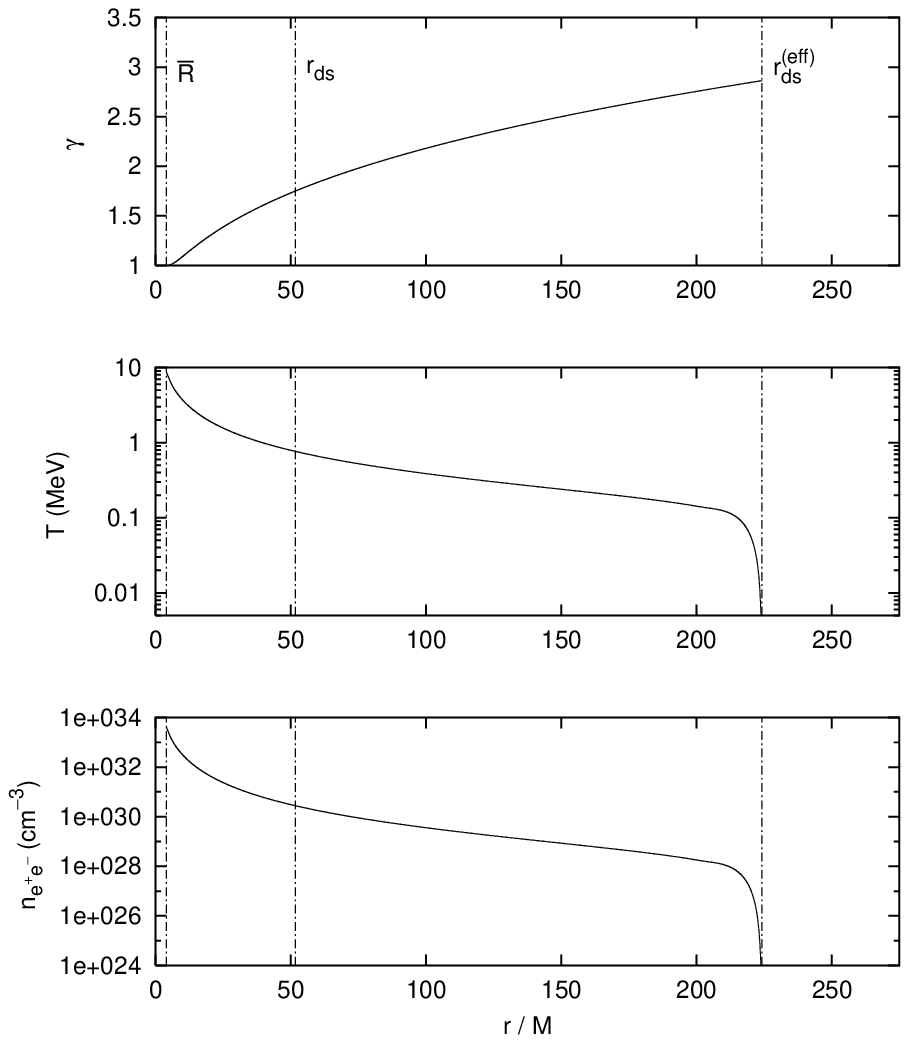}
\caption{Physical parameters in the effective dyadosphere: Lorentz $\gamma$
factor, proper temperature and proper $e^{+}e^{-}$ number density as functions
at time $\bar{t}$ for a black hole with $M=20M_{\odot}$ and $Q=0.1M$.}
\label{f3}
\end{figure}

At time $\bar{t}\equiv t_{0}\left(  \bar{R}\right)  $ when the DTS is formed,
the plasma extends over a region of space which is almost one order of
magnitude larger than the dyadosphere and which we define as the
\emph{effective dyadosphere}. The values of the Lorentz $\gamma$ factor, the
temperature and $e^{+}e^{-}$ number density in the effective dyadosphere are
given in Fig. \ref{f3}.

In conclusion we see how the causal description of the dyadosphere formation
can carry important messages on the time variability and spectral distribution
of the P--GRB due to quantum effects as well as precise signature of General Relativity.

\section{Description of the electron-positron plasma oscillations by generalized Vlasov-Boltzmann-Maxwell equation in the PEM pulse phase}\label{plasma}

We describe the creation and
evolution of electron-positron pairs in a strong electric field as well as the
pairs annihilation into photons. The formalism is based on generalized Vlasov equations, which are numerically integrated. We recover previous results about the oscillations of the charges,
discuss the electric field screening and the relaxation of the
system to a thermal equilibrium configuration. The timescale of the thermalization is estimated to be $\sim 10^{3}-10^{4}\ \hbar /m_{e}c^{2}$.

\subsection{On the observability of electron-positron pairs created in vacuum polarization in Earth bound experiment and in astrophysics}

Three different earth-bound experiments and one astrophysical observation have
been proposed for identifying the polarization of the electronic vacuum due to
a supercritical electric field ($\mathcal{E}>\mathcal{E}_{\mathrm{c}}\equiv
m_{e}^{2}c^{3}/e\hbar$, where $m_{e}$ and $e$ are the electron mass and
charge) postulated by Sauter-Heisenberg-Euler-Schwinger \cite{s31}:

\begin{enumerate}
\item  In central collisions of heavy ions near the Coulomb barrier, as first
proposed in \cite{GZ69,GZ70} (see also \cite{PR71,P72,ZP72}). Despite some
apparently encouraging results \cite{S...83}, such efforts have failed so far
due to the small contact time of the colliding ions
\cite{A...95,G...96,L...97,B...95,H...98}. Typically the electromagnetic
energy involved in the collisions of heavy ions with impact parameter
$l_{1}\sim10^{-12}$cm is $E_{1}\sim10^{-6}$erg and the lifetime of the
diatomic system is $t_{1}\sim10^{-22}$s.

\item  In collisions of an electron beam with optical laser pulses: a signal
of positrons above background has been observed in collisions of a 46.6 GeV
electron beam with terawatt pulses of optical laser in an experiment at the
Final Focus Test Beam at SLAC \cite{B...97}; it is not clear if this
experimental result is an evidence for the vacuum polarization phenomenon.
The energy of the laser pulses was $E_{2}\sim10^{7}$erg, concentrated in a
space-time region of spacial linear extension (focal length) $l_{2}\sim
10^{-3}$cm and temporal extension (pulse duration) $t_{2}\sim10^{-12}$s
\cite{B...97}.

\item  At the focus of an X-ray free electron laser (XFEL) (see
\cite{R01,AHRSV01,RSV02} and references therein). Proposals for this
experiment exist at the TESLA collider at DESY and at the LCLS facility at
SLAC \cite{R01}. Typically the electromagnetic energy at the focus of an XFEL
can be $E_{3}\sim10^{6}$erg, concentrated in a space-time region of spacial
linear extension (spot radius) $l_{3}\sim10^{-8}$cm and temporal extension
(coherent spike length) $t_{3}\sim10^{-13}$s \cite{R01}.
\end{enumerate}

and from astrophysics

\begin{enumerate}
\item  around an electromagnetic black hole (black hole) \cite{dr75,prx98,prx02},
giving rise to the observed phenomenon of gamma-ray bursts (GRB)
\cite{lett1,lett2,lett3,rbcfx02_letter}. The electromagnetic energy of an
black hole of mass $M\sim10M_{\odot}$ and charge $Q\sim0.1M/\sqrt{G}$ is $E_{4}%
\sim10^{54}$erg and it is deposited in a space-time region of spacial linear extension $l_{4}%
\sim10^{8}$cm \cite{prx98,rv02a} and temporal extension (collapse time) $t_{4}\sim10^{-2}$s
\cite{rvx03}.
\end{enumerate}

\subsection{On the role of transparency condition in the electron-positron plasma}

In addition to their marked quantitative difference in testing the same basic
physical phenomenon, there is a very important conceptual difference among
these processes: the first three occur in a transparency condition in which the
created electron-positron pairs and, possibly, photons freely propagate to
infinity, while the one in the black hole occurs in an opacity condition
\cite{rswx00}. Under the opacity condition a thermalization effect occurs and
a final equipartition between the $e^{+}e^{-}$ and $\gamma$ is reached. Far
from being just an academic issue, this process and its characteristic
timescale is of the greatest importance in physics and astrophysics. It has
been shown by a numerical simulation done in Livermore and an analytic work
done in Rome \cite{rswx00}, that, as soon as the thermalization of $e^{+}%
e^{-}$ and $\gamma$ created around a black hole has been reached, the plasma self
propels outwards and this process is at the very heart of the gamma-ray burst
(GRB) phenomenon. A critical step was missing up to now: how to bridge the gap
between the creation of pairs in the supercritical field of the black hole and the
thermalization of the system to a plasma configuration. We report
some progress on this topic with special attention to the timescale needed for
the thermalization of the newly created $e^{+}e^{-}$ pairs in the background
field. The comparison of the thermalization timescale to the one of
gravitational collapse, which occurs on general relativistic timescale, is at
the very ground of the comprehension of GRBs \cite{rvx03}.

The evolution of a system of particle-antiparticle pairs created by the
Schwinger process has been often described by a transport Vlasov equation
(see, for example, \cite{KM85,GKM87}). More recently it has been showed that
such an equation can be derived from quantum field theory
\cite{SRS...97,KME98,SBR...98}. In the homogeneous case, the equations have
been numerically integrated taking into account the back reaction on the
external electric field \cite{KESCM91,KESCM92,CEKMS93,BMP...99}. In many
papers (see \cite{V...01} and references therein) a phenomenological term
describing equilibrating collisions is introduced in the transport equation
which is parameterized by an effective relaxation time $\tau$. In \cite{V...01}
one further step is taken by allowing time variability of $\tau$; the
ignorance on the collision term is then parameterized by a free dimensionless
constant. The introduction of a relaxation time corresponds to the assumption
that the system rapidly evolves towards thermal equilibrium. In this paper we
focus on the evolution of a system of $e^{+}e^{-}$ pairs, explicitly taking
into account the scattering processes $e^{+}e^{-}\rightleftarrows\gamma\gamma
$. Since we are mainly interested in a system in which the electric field
varies on macroscopic length scale ($l\sim10^{8}$cm, above), we
can limit ourselves to a homogeneous electric field. Also, we will use
transport equations for electrons, positrons and photons, with collision
terms, coupled to Maxwell equations. There is no free parameter here: the
collision terms can be exactly computed, since the QED cross sections are
known. Starting from a regime which is far from thermal equilibrium, we find
that collisions do not prevent plasma oscillations in the initial phase of the
evolution and analyse the issue of the timescale of the approach to a
$e^{+}e^{-}\gamma$ plasma equilibrium configuration, which is the most
relevant quantity in the process of gravitational collapse \cite{rvx03}.

\subsection{The Vlasov-Boltzmann-Maxwell equations and their solutions}

The motion of positrons (electrons) is the resultant of three contributions:
the pair creation, the electric acceleration and the annihilation damping. The
homogeneous system consisting of electric field, electrons, positrons and
photons can be described by the equations
\begin{align}
\partial_{t}f_{e}+e\mathbf{E}\partial_{\mathbf{p}}f_{e} &  =\mathcal{S}\left(
\mathbf{E},\mathbf{p}\right)  -\tfrac{1}{\left(  2\pi\right)  ^{5}}%
\epsilon_{\mathbf{p}}^{-1}\mathcal{C}_{e}\left(  t,\mathbf{p}\right)
,\label{pairs}\\
\partial_{t}f_{\gamma} &  =\tfrac{2}{\left(  2\pi\right)  ^{5}}\epsilon
_{\mathbf{k}}^{-1}\mathcal{C}_{\gamma}\left(  t,\mathbf{k}\right)
,\label{photons}\\
\partial_{t}\mathbf{E} &  =-\mathbf{j}_{p}\left(  \mathbf{E}\right)
-\mathbf{j}_{c}\left(  t\right)  ,\label{Maxwell}%
\end{align}
where $f_{e}=f_{e}\left(  t,\mathbf{p}\right)  $ is the distribution function
in the phase-space of positrons (electrons), $f_{\gamma}=f_{\gamma}\left(
t,\mathbf{k}\right)  $ is the distribution function in the phase-space of
photons, $\mathbf{E}$ is the electric field, $\epsilon_{\mathbf{p}}=\left(
\mathbf{p}\cdot\mathbf{p}+m_{e}^{2}\right)  ^{1/2}$ is the energy of an
electron of 3-momentum $\mathbf{p}$ ($m_{e}$ is the mass of the electron) and
$\epsilon_{\mathbf{k}}=\left(  \mathbf{k}\cdot\mathbf{k}\right)  ^{1/2}$ is
the energy of a photon of 3-momentum $\mathbf{k}$. $f_{e}$ and $f_{\gamma}$
are normalized so that $\int\tfrac{d^{3}\mathbf{p}}{\left(  2\pi\right)  ^{3}%
}\ f_{e}\left(  t,\mathbf{p}\right)  =n_{e}\left(  t\right)  $, $\int
\tfrac{d^{3}\mathbf{k}}{\left(  2\pi\right)  ^{3}}\ f_{\gamma}\left(
t,\mathbf{k}\right)  =n_{\gamma}\left(  t\right)  $ , where $n_{e}$ and
$n_{\gamma}$ are number densities of positrons (electrons) and photons,
respectively. The term
\begin{equation}
\mathcal{S}\left(  \mathbf{E},\mathbf{p}\right)  =\left(  2\pi\right)
^{3}\tfrac{dN}{dtd^{3}\mathbf{x}d^{3}\mathbf{p}}=-\left|  e\mathbf{E}\right|
\log\left[  1-\exp\left(  -\tfrac{\pi(m_{e}^{2}+\mathbf{p}_{\perp}^{2}%
)}{\left|  e\mathbf{E}\right|  }\right)  \right]  \delta(p_{\parallel
})\label{S}%
\end{equation}
is the Schwinger source for pair creation (see \cite{KESCM91,KESCM92}):
$p_{\parallel}$ and $\mathbf{p}_{\perp}$ are the components of the 3-momentum
$\mathbf{p}$ parallel and orthogonal to $\mathbf{E}$. We assume that the pairs
are produced at rest in the direction parallel to the electric field
\cite{KESCM91,KESCM92}. We also have, in Eqs. (\ref{pairs}), (\ref{photons}) and
(\ref{Maxwell}),
\begin{align}
\mathcal{C}_{e}\left(  t,\mathbf{p}\right)   &  \simeq\int\tfrac
{d^{3}\mathbf{p}_{1}}{\epsilon_{\mathbf{p}_{1}}}\tfrac{d^{3}\mathbf{k}_{1}%
}{\epsilon_{\mathbf{k}_{1}}}\tfrac{d^{3}\mathbf{k}_{2}}{\epsilon
_{\mathbf{k}_{2}}}\delta^{\left(  4\right)  }\left(  p+p_{1}-k_{1}%
-k_{2}\right)  \nonumber\\
&  \times\left|  \mathcal{M}\right|  ^{2}\left[  f_{e}\left(  \mathbf{p}%
\right)  f_{e}\left(  \mathbf{p}_{1}\right)  -f_{\gamma}\left(  \mathbf{k}%
_{1}\right)  f_{\gamma}\left(  \mathbf{k}_{2}\right)  \right]  ,\label{Ce}\\
\mathcal{C}_{\gamma}\left(  t,\mathbf{k}\right)   &  \simeq\int\tfrac
{d^{3}\mathbf{p}_{1}}{\epsilon_{\mathbf{p}_{1}}}\tfrac{d^{3}\mathbf{p}_{2}%
}{\epsilon_{\mathbf{p}_{2}}}\tfrac{d^{3}\mathbf{k}_{1}}{\epsilon
_{\mathbf{k}_{1}}}\delta^{\left(  4\right)  }\left(  p_{1}+p_{2}%
-k-k_{1}\right)  \nonumber\\
\times &  \left|  \mathcal{M}\right|  ^{2}\left[  f_{e}\left(  \mathbf{p}%
_{1}\right)  f_{e}\left(  \mathbf{p}_{2}\right)  -f_{\gamma}\left(
\mathbf{k}\right)  f_{\gamma}\left(  \mathbf{k}_{1}\right)  \right]
,\label{Cf}%
\end{align}
which describe probability rates for pair creation by photons and pair
annihilation into photons, $\mathcal{M}=\mathcal{M}_{e^{+}\left(
\mathbf{p}_{1}\right)  e^{-}\left(  \mathbf{p}_{2}\right)  \rightleftarrows
\gamma\left(  \mathbf{k}\right)  \gamma\left(  \mathbf{k}_{1}\right)  }$ being
the matrix element for the process $e^{+}\left(  \mathbf{p}_{1}\right)
e^{-}\left(  \mathbf{p}_{2}\right)  \rightarrow\gamma\left(  \mathbf{k}%
\right)  \gamma\left(  \mathbf{k}_{1}\right)  $. Note that the collisional
terms (\ref{Ce}) and (\ref{Cf}) are either inapplicable or negligible in the
case of the above three earth-bound experiments where the
created pairs do not originate a dense plasma. They have been correctly
neglected in previous works (see e. g. \cite{RSV02}). Collisional terms have
also been considered in the different physical context of vacuum polarization by strong chromoelectric fields. Unlike the present QED case, where expressions for the cross sections
are known exactly, in the QCD case the cross sections are yet unknown and
such collisional terms are of a phenomenological type and useful uniquely near
the equilibrium regime \cite{V...01}. Finally $\mathbf{j}_{p}\left(
\mathbf{E}\right)  =2\tfrac{\mathbf{E}}{\mathbf{E}^{2}}\int\tfrac
{d^{3}\mathbf{p}}{\left(  2\pi\right)  ^{3}}\epsilon_{\mathbf{p}}%
\mathcal{S}\left(  \mathbf{E},\mathbf{p}\right)  $ and $\mathbf{j}_{c}\left(
t\right)  =2en_{e}\int\tfrac{d^{3}\mathbf{p}}{\left(  2\pi\right)  ^{3}}%
\tfrac{\mathbf{p}}{\epsilon_{\mathbf{p}}}f_{e}\left(  \mathbf{p}\right)  $ are
polarization and conduction current respectively (see \cite{GKM87}). In Eqs.
(\ref{Ce}) and (\ref{Cf}) we neglect, as a first approximation, Pauli blocking
and Bose enhancement (see e.g. \cite{KESCM92}). By suitably integrating
(\ref{pairs}) and (\ref{photons}) over the phase spaces of positrons
(electrons) and photons, we find the following exact equations for mean
values:
\begin{align}
\tfrac{d}{dt}n_{e} &  =S\left(  \mathbf{E}\right)  -n_{e}^{2}\left\langle
\sigma_{1}v^{\prime}\right\rangle _{e}+n_{\gamma}^{2}\left\langle \sigma
_{2}v^{\prime\prime}\right\rangle _{\gamma},\nonumber\\
\tfrac{d}{dt}n_{\gamma} &  =2n_{e}^{2}\left\langle \sigma_{1}v^{\prime
}\right\rangle _{e}-2n_{\gamma}^{2}\left\langle \sigma_{2}v^{\prime\prime
}\right\rangle _{\gamma},\nonumber\\
\tfrac{d}{dt}n_{e}\left\langle \epsilon_{\mathbf{p}}\right\rangle _{e} &
=en_{e}\mathbf{E}\cdot\left\langle \mathbf{v}\right\rangle _{e}+\tfrac{1}%
{2}\mathbf{E\cdot j}_{p}-n_{e}^{2}\left\langle \epsilon_{\mathbf{p}}\sigma
_{1}v^{\prime\prime}\right\rangle _{e}+n_{\gamma}^{2}\left\langle
\epsilon_{\mathbf{k}}\sigma_{2}v^{\prime\prime}\right\rangle _{\gamma
},\nonumber\\
\tfrac{d}{dt}n_{\gamma}\left\langle \epsilon_{\mathbf{k}}\right\rangle
_{\gamma} &  =2n_{e}^{2}\left\langle \epsilon_{\mathbf{p}}\sigma_{1}v^{\prime
}\right\rangle _{e}-2n_{\gamma}^{2}\left\langle \epsilon_{\mathbf{k}}%
\sigma_{2}v^{\prime\prime}\right\rangle _{\gamma},\nonumber\\
\tfrac{d}{dt}n_{e}\left\langle \mathbf{p}\right\rangle _{e} &  =en_{e}%
\mathbf{E}-n_{e}^{2}\left\langle \mathbf{p}\sigma_{1}v^{\prime}\right\rangle
_{e},\nonumber\\
\tfrac{d}{dt}\mathbf{E} &  =-2en_{e}\left\langle \mathbf{v}\right\rangle
_{e}-\mathbf{j}_{p}\left(  \mathbf{E}\right)  ,\label{System1}%
\end{align}
where, for any function of the momenta
\begin{align}
\left\langle F\left(  \mathbf{p}_{1},...,\mathbf{p}_{n}\right)  \right\rangle
_{e} &  \equiv n_{e}^{-n}\int\tfrac{d^{3}\mathbf{p}_{1}}{\left(  2\pi\right)
^{3}}...\tfrac{d^{3}\mathbf{p}_{n}}{\left(  2\pi\right)  ^{3}}\ F\left(
\mathbf{p}_{1},...,\mathbf{p}_{n}\right)  \cdot f_{e}\left(  \mathbf{p}%
_{1}\right)  \cdot...\cdot f_{e}\left(  \mathbf{p}_{n}\right)  ,\\
\left\langle G\left(  \mathbf{k}_{1},...,\mathbf{k}_{l}\right)  \right\rangle
_{\gamma} &  \equiv n_{\gamma}^{-l}\int\tfrac{d^{3}\mathbf{k}_{1}}{\left(
2\pi\right)  ^{3}}...\tfrac{d^{3}\mathbf{k}_{l}}{\left(  2\pi\right)  ^{3}%
}\ G\left(  \mathbf{k}_{1},...,\mathbf{k}_{l}\right)  \cdot f_{\gamma}\left(
\mathbf{k}_{1}\right)  \cdot...\cdot f_{\gamma}\left(  \mathbf{k}_{l}\right)
.
\end{align}
Furthermore $v^{\prime}$ is the relative velocity between electrons and
positrons, $v^{\prime\prime}$ is the relative velocity between photons,
$\sigma_{1}=\sigma_{1}\left(  \epsilon_{\mathbf{p}}^{\mathrm{CoM}}\right)  $
is the total cross section for the process $e^{+}e^{-}\rightarrow\gamma\gamma$
and $\sigma_{2}=\sigma_{2}\left(  \epsilon_{\mathbf{k}}^{\mathrm{CoM}}\right)
$ is the total cross section for the process $\gamma\gamma\rightarrow
e^{+}e^{-}$ (here $\epsilon^{\mathrm{CoM}}$ is the energy of a particle in the
reference frame of the center of mass).

In order to evaluate the mean values in system (\ref{System1}) we need some
further hypotheses on the distribution functions. Let us define $\bar
{p}_{\parallel}$, $\bar{\epsilon}_{\mathbf{p}}$ and $\mathbf{\bar{p}}_{\perp
}^{2}$ such that $\left\langle p_{\parallel}\right\rangle _{e}\equiv\bar
{p}_{\parallel},~\left\langle \epsilon_{\mathbf{p}}\right\rangle _{e}%
\equiv\bar{\epsilon}_{\mathbf{p}}\equiv(\bar{p}_{\parallel}^{2}+\mathbf{\bar
{p}}_{\perp}^{2}+\ m_{e}^{2})^{1/2}$. We assume
\begin{equation}
f_{e}\left(  t,\mathbf{p}\right)  \propto n_{e}\left(  t\right)  \delta\left(
p_{\parallel}-\bar{p}_{\parallel}\right)  \delta\left(  \mathbf{p}_{\perp}%
^{2}-\mathbf{\bar{p}}_{\perp}^{2}\right)  . \label{fe}%
\end{equation}
Since in the scattering $e^{+}e^{-}\rightarrow\gamma\gamma$ the coincidence of
the scattering direction with the incidence direction is statistically
favored, we also assume
\begin{equation}
f_{\gamma}\left(  t,\mathbf{k}\right)  \propto n_{\gamma}\left(  t\right)
\delta\left(  \mathbf{k}_{\perp}^{2}-\mathbf{\bar{k}}_{\perp}^{2}\right)
\left[  \delta\left(  k_{\parallel}-\bar{k}_{\parallel}\right)  +\delta\left(
k_{\parallel}+\bar{k}_{\parallel}\right)  \right]  , \label{fgamma}%
\end{equation}
where $k_{\parallel}$ and $\mathbf{k}_{\perp}$ have analogous meaning as
$p_{\parallel}$ and $\mathbf{p}_{\perp}$ and the terms $\delta\left(
k_{\parallel}-\bar{k}_{\parallel}\right)  $ and $\delta\left(  k_{\parallel
}+\bar{k}_{\parallel}\right)  $ account for the probability of producing,
respectively, forwardly scattered and backwardly scattered photons. Since the
Schwinger source term (\ref{S}) implies that the positrons (electrons) have
initially fixed $p_{\parallel}$, $p_{\parallel}=0$, assumption (\ref{fe})
((\ref{fgamma})) means that the distribution of $p_{\parallel}$ ($k_{\parallel
}$) does not spread too much with time and, analogously, that the distribution
of energies is sufficiently peaked to be describable by a $\delta-$function.
The dependence on the momentum of the distribution functions has been
discussed in \cite{KESCM92,KME98}. Approximations (\ref{fe}), (\ref{fgamma})
reduce Eqs. (\ref{System1}) to a system of ordinary differential equations. In
average, since the inertial reference frame we fix coincides with the center
of mass frame for the processes $e^{+}e^{-}\rightleftarrows\gamma\gamma$,
$\epsilon^{\mathrm{CoM}}\simeq\bar{\epsilon}$ for each species. Substituting
(\ref{fe}) and (\ref{fgamma}) into (\ref{System1}) we find
\begin{align}
\tfrac{d}{dt}n_{e}  &  =S\left(  \mathcal{E}\right)  -2n_{e}^{2}\sigma_{1}%
\rho_{e}^{-1}\left|  {\pi}_{e\parallel}\right|  +2n_{\gamma}^{2}\sigma
_{2},\nonumber\\
\tfrac{d}{dt}n_{\gamma}  &  =4n_{e}^{2}\sigma_{1}\rho_{e}^{-1}\left|  {\pi
}_{e\parallel}\right|  -4n_{\gamma}^{2}\sigma_{2},\nonumber\\
\tfrac{d}{dt}\rho_{e}  &  =en_{e}\mathcal{E}\rho_{e}^{-1}\left|  {\pi
}_{e\parallel}\right|  +\tfrac{1}{2}\mathcal{E}j_{p}-2n_{e}\rho_{e}\sigma
_{1}\rho_{e}^{-1}\left|  {\pi}_{e\parallel}\right|  +2n_{\gamma}\rho_{\gamma
}\sigma_{2},\nonumber\\
\tfrac{d}{dt}\rho_{\gamma}  &  =4n_{e}\rho_{e}\sigma_{1}\rho_{e}^{-1}\left|
{\pi}_{e\parallel}\right|  -4n_{\gamma}\rho_{\gamma}\sigma_{2},\nonumber\\
\tfrac{d}{dt}{\pi}_{e\parallel}  &  =en_{e}\mathcal{E}-2n_{e}{\pi}%
_{e\parallel}\sigma_{1}\rho_{e}^{-1}\left|  {\pi}_{e\parallel}\right|
,\nonumber\\
\tfrac{d}{dt}\mathcal{E}  &  =-2en_{e}\rho_{e}^{-1}\left|  {\pi}_{e\parallel
}\right|  -j_{p}\left(  \mathcal{E}\right)  , \label{System2}%
\end{align}
where $\rho_{e}=n_{e}\bar{\epsilon}_{\mathbf{p}}$, $\rho_{\gamma}=n_{\gamma
}\bar{\epsilon}_{\mathbf{k}}$, ${\pi}_{e\parallel}=n_{e}\bar{p}_{\parallel}$
are the energy density of positrons (electrons), the energy density of photons
and the density of ``parallel momentum'' of positrons (electrons),
$\mathcal{E}$ is the electric field strength and $j_{p}$ the unique component
of $\mathbf{j}_{p}$ parallel to $\mathbf{E}$. $\sigma_{1}$ and $\sigma_{2}$
are evaluated at $\epsilon^{\mathrm{CoM}}=\bar{\epsilon}$ for each species.
Note that Eqs.(\ref{System2}) are ``classical'' in the sense that the only quantum
information is encoded in the terms describing pair creation and scattering
probabilities. Eqs.(\ref{System2}) are consistent with energy density
conservation: $\tfrac{d}{dt}\left(  \rho_{e}+\rho_{\gamma}+\tfrac{1}%
{2}\mathcal{E}^{2}\right)  =0.$

\begin{figure}
\centering
\includegraphics[width=8.5cm]{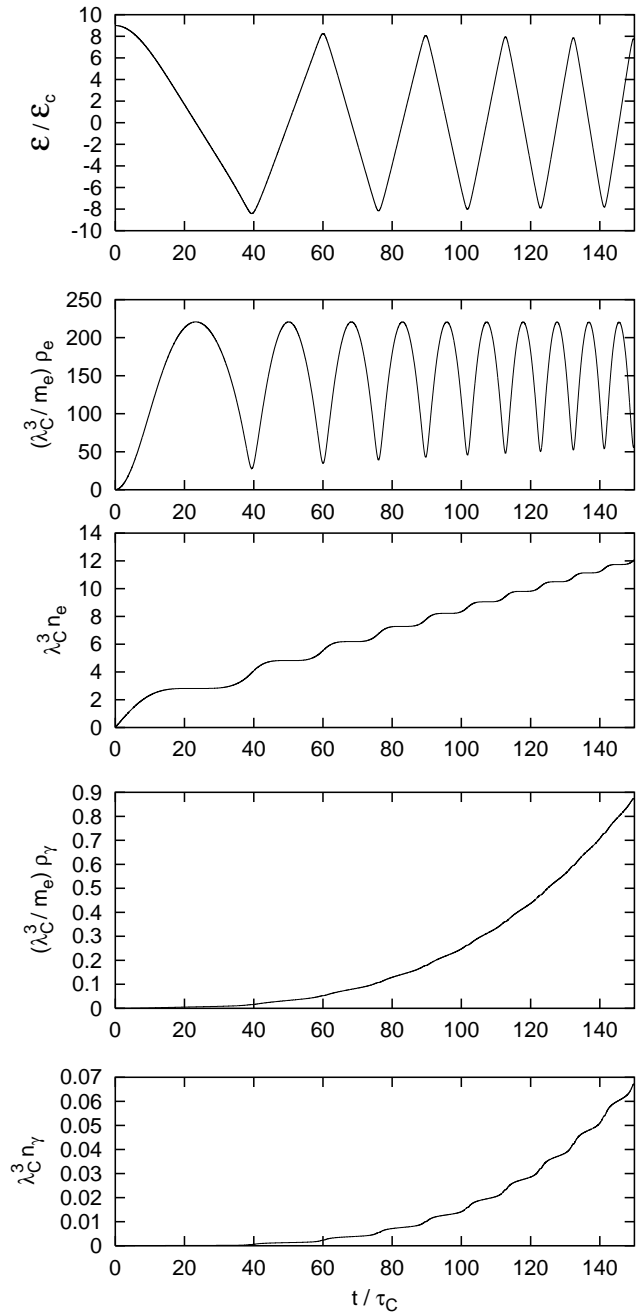}
\caption{Plasma oscillations. We set $\mathcal{E}_{0}=9\mathcal{E}%
_{\mathrm{c}}$, $t<150\tau_{\mathrm{C}}$ and plot: a) electromagnetic field
strength; b) electrons energy density; c) electrons number density; d) photons
energy density; e) photons number density as functions of time.}
\label{fig1luca}
\end{figure}

The initial conditions for Eqs.(\ref{System2}) are $n_{e}=n_{\gamma}=\rho
_{e}=\rho_{\gamma}=\pi_{e\parallel}=0,~\mathcal{E}=\mathcal{E}_{0}$. In Fig.
\ref{fig1luca} the results of the numerical integration for $\mathcal{E}%
_{0}=9\mathcal{E}_{\mathrm{c}}$ is showed. The integration stops at
$t=150\ \tau_{\mathrm{C}}$ (where $\tau_{\mathrm{C}}=\hbar/m_{e}c^{2}$). Each
variable is represented in units of $m_{e}$ and
$\lambda_{\mathrm{C}}=\hbar/m_{e}c$. The numerical integration confirms
\cite{KESCM91,KESCM92} that the system undergoes plasma oscillations: a) the
electric field oscillates with decreasing amplitude rather than abruptly
reaching the equilibrium value; b) electrons and positrons oscillates in the
electric field direction, reaching ultrarelativistic velocities; c) the role
of the $e^{+}e^{-}\rightleftarrows$ $\gamma\gamma$ scatterings is marginal in
the early time of the evolution, the electrons are too extremely relativistic
and consequently the density of photons builds up very slowly (see. details in
Fig. \ref{fig1luca}).

\begin{figure}
\centering
\includegraphics[width=8.5cm]{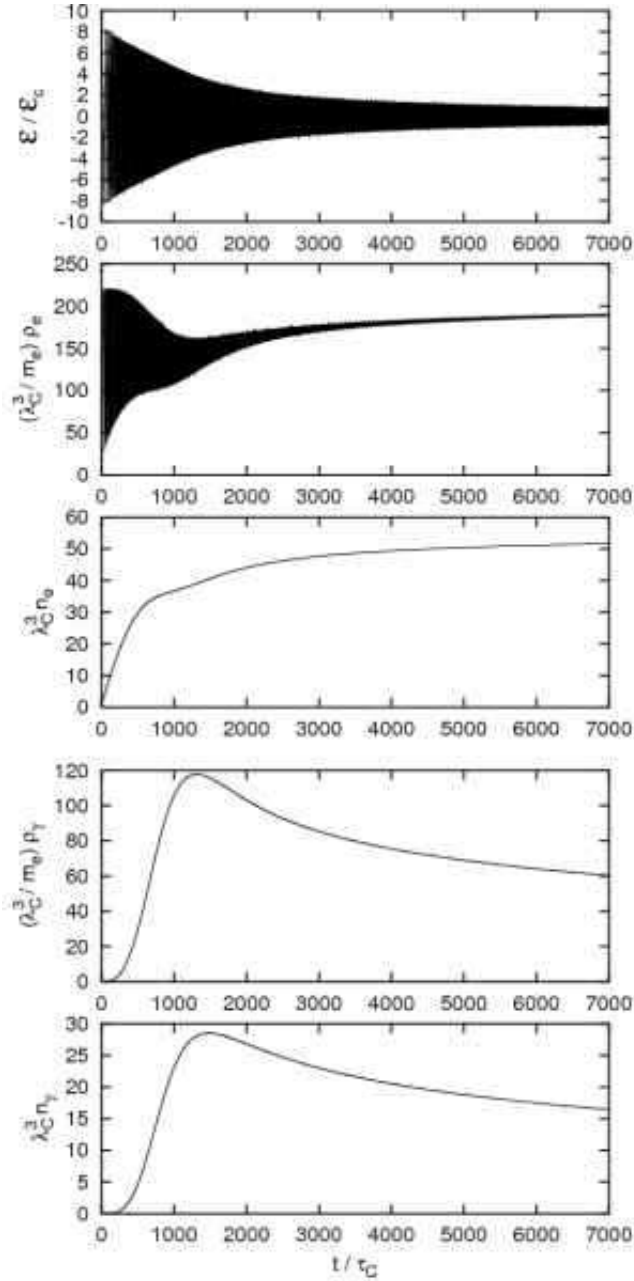}
\caption{Plasma oscillations. We set $\mathcal{E}_{0}=9\mathcal{E}%
_{\mathrm{c}}$, $t<7000\tau_{\mathrm{C}}$ and plot: a) electromagnetic field
strength; b) electrons energy density; c) electrons number density; d) photons
energy density; e) photons number density as functions of time - the
oscillation period is not resolved in these plots. The model used should have
a breakdown at a time much earlier than $7000\tau_{\mathrm{C}}$ and therefore
this plot contains no more than qualitative informations.}
\label{fig2}
\end{figure}

At late times the system is expected to relax to a plasma configuration of
thermal equilibrium and assumptions (\ref{fe}) and (\ref{fgamma}) have to be
generalized to take into account quantum spreading of the distribution
functions. It is nevertheless interesting to look at the solutions of
Eqs.(\ref{System2}) in this regime. In Fig. \ref{fig2} we plot the numerical
solution of Eqs.(\ref{System2}) but the integration extends here all the way
up to $t=7000\ \tau_{\mathrm{C}}$ (the time scale of oscillations is not
resolved in these plots). It is interesting that the leading term recovers the
expected asymptotic behavior: a) the electric field is screened to about the
critical value: $\mathcal{E}\simeq\mathcal{E}_{\mathrm{c}}$ for $t\sim
10^{3}-10^{4}\tau_{\mathrm{C}}\gg\tau_{\mathrm{C}}$; b) the initial
electromagnetic energy density is distributed over electron-positron pairs and
photons, indicating energy equipartition; c) photons and electron-positron
pairs number densities are asymptotically comparable, indicating number
equipartition. At such late times a regime of thermalized
electrons-positrons-photons plasma begins and the system is describable by
hydrodynamic equations \cite{rvx03,rswx00}.

We provided a very simple formalism apt to describe simultaneously the
creation of electron-positron pairs by a strong electric field $\mathcal{E}%
\gtrsim\mathcal{E}_{c}$ and the pairs annihilation into photons. As discussed
in literature, we find plasma oscillations. In particular the collisions do
not prevent such a feature. This is because the momentum of electrons
(positrons) is very high, therefore the cross section for the process
$e^{+}e^{-}\rightarrow\gamma\gamma$ is small and the annihilation into photons
is negligible in the very first phase of the evolution. As a result, the
system takes some time ($t\sim10^{3}-10^{4}\tau_{\mathrm{C}}$) to thermalize
to a $e^{+}e^{-}\gamma$ plasma equilibrium configuration. We finally remark
that, at least in the case of electromagnetic Schwinger mechanism, the picture
could be quite different from the one previously depicted in literature, where
the system is assumed to thermalize in a very short time (see \cite{V...01}
and references therein).

It is conceivable that in the race to first identify the vacuum polarization
process \emph{\`a la} Sauter-Euler-Heisenberg-Schwinger, the astrophysical observations will reach a positive result before earth-bound experiments, much like in the case of the discovery of lines in the Sun chromosphere by J. N. Lockyer in 1869, later identified with the Helium spectral lines by W. Ramsay in 1895 \cite{G03}.

\quad

\section{Observational signatures of an electromagnetic overcritical gravitational collapse and prediction of spectral evolution of short GRBs}\label{pgrb}\

We finally present theoretical predictions for the spectral, temporal and intensity signatures of the electromagnetic radiation emitted during the process of the gravitational collapse of a stellar core to a black hole, during which electromagnetic field strengths rise over the critical value for $e^+e^-$ pair creation. The last phases of this gravitational collapse are studied, using the result presented in the previous sections, leading to the formation of a black hole with a subcritical electromagnetic field, likely with zero charge, and an outgoing pulse of initially optically thick $e^+e^-$-photon plasma. Such a pulse reaches transparency at Lorentz gamma factors of $10^2$--$10^4$. We find a clear signature in the outgoing electromagnetic signal, drifting from a soft to a hard spectrum, on very precise time-scales and with a very specific intensity modulation. 

We outline finally the relevance of these theoretical results for the understanding of short gamma-ray bursts.

\subsection{The model}

The dynamics of the collapse of an electrically-charged stellar core, separating
itself from an oppositely charged remnant in an initially neutral star, was
first modeled by an exact solution of the Einstein-Maxwell equations
corresponding to a shell of charged matter in Ref.~\cite{crv02}. The
fundamental dynamical equations and their analytic solutions were obtained, 
revealing the amplification of the electromagnetic field strength
during the process of collapse and the asymptotic approach to the final static
configuration. The results, which properly account for general
relativistic effects, are summarized in Fig. 1 and Fig. 2 of
Ref.~\cite{crv02}.

A first step toward the understanding of the process of extracting energy
from a black hole was obtained in Ref.~\cite{rv02a}, where it
was shown how the extractable electromagnetic energy is not stored behind the
horizon but is actually distributed all around the black hole. Such a stored
energy is in principle extractable, very efficiently, on time-scales
$\sim\hbar/m_{e}c^{2}$, by a vacuum polarization process \emph{\`{a} l\`{a}}
Sauter-Heisenberg-Euler-Schwinger \cite{s31,he35,s51}. Such a process occurs
if the electromagnetic field becomes larger than the critical field strength
$\mathcal{E}_{\mathrm{c}}$ for $e^+e^-$ pair creation. In
Ref.~\cite{rv02a} we followed the approach of Damour and Ruffini \cite{dr75} in
order to evaluate the energy density and the temperature of the created
$e^+e^-$-photon plasma. As a byproduct, a formula for the
irreducible mass of a black hole was also derived solely in terms of the
gravitational, kinetic and rest mass energies of the collapsing core. This
surprising result allowed us in Ref.~\cite{rv02b} to obtain a deeper
understanding of the maximum limit for the extractable energy during the process of
gravitational collapse, namely 50\% of the initial energy of
the star: the well known result of a 50\% maximum efficiency for energy extraction in
the case of a Reissner-Nordstr\"{o}m black hole \cite{cr71} then becomes a particular
case of a process of much more general validity.

The crucial issue
of the survival of the electric charge of the collapsing core in the presence of a
copious process of $e^+e^-$ pair creation was addressed in
Refs.~\cite{rvx03a,rvx03b}. By using theoretical techniques borrowed from
plasma physics and statistical mechanics
\cite{GKM87,KESCM91,KESCM92,CEKMS93,KME98,SBR...98,BMP...99} based on a
generalized Vlasov equation, it was possible to show that while the core keeps
collapsing, the created $e^+e^-$ pairs are entangled in the
overcritical electric field. The electric field itself, due to the back
reaction of the created $e^+e^-$ pairs, undergoes damped
oscillations in sign finally settling down to the critical value
$\mathcal{E}_{\mathrm{c}}$. The pairs fully thermalize to an
$e^+e^-$-photon plasma on time-scales typically of the order of
$10^{2}$--$10^{4}\hbar/m_{e}c^{2}$. During this characteristic damping time,
which we recall is much larger than the pair creation time-scale
$\hbar/m_{e}c^{2}$, the core moves
inwards, collapsing with a speed $0.2$--$0.8c$,
further amplifying the electric field strength at its surface and
enhancing the pair creation process.

Turning now to the dynamical evolution of such an $e^+e^-$ plasma we recall that, after some original attempt to consider a steady state emission \cite{p86,p90}, the crucial progress was represented by the understanding that during the optically thick phase such a plasma expands as a thin shell. There exists a fundamental relation between the width of the expanding shell and the Lorentz gamma factor. The shell expands, but the Lorentz contraction is such that its width in laboratory frame appears to be constant. Such a result was found in \cite{psn93} on the basis of a numerical approach, further analyzed in Bisnovatyi-Kogan and Murzina \cite{bm95} on the basis of an analytic approach. Attention to the role of the rate equations governing the $e^+e^-$ annihilation were given in \cite{GW98}, where approximations to the full equation were introduced. These results were improved in two important respects in 1999 and 2000 \cite{rswx99,rswx00}: the initial conditions were made more accurate by the considerations of the dyadosphere as well as the dynamics of the shell was improved by the self-consistent solution of the hydrodynamical equation and the rate equation for the $e^+e^-$ plasma following both an analytic and numerical approach. 

We are now ready to report the result of using the approach in \cite{rswx99,rswx00} in this general framework describing the dynamical formation of the dyadosphere.

The first attempt to analyze the
expansion of the newly generated and thermalized $e^+e^-$-photon
plasma was made in Ref.~\cite{rvx03a}. The initial dynamical phases of the
expansion were analyzed, using the general relativistic equations of
Ref.~\cite{crv02} for the gravitational collapse of the core. A {\itshape separatrix} was found in the motion of the plasma
at a critical radius $\bar{R}$: the plasma created at radii larger than
$\bar{R}$ expands to infinity, while the one created at radii smaller than
$\bar{R}$ is trapped by the gravitational field of the collapsing core and
implodes towards the black hole. The value of $\bar{R}$ was found in
Ref.~\cite{rvx03a} to be 
$\bar{R}=2GM/c^{2}[1+\left(  1-3Q^{2}/4GM^{2}\right)^{1/2}]$, 
where $M$ and $Q$ are the mass and the charge of the core, respectively.

We now pursue further the evolution of such a
system, describing the dynamical phase of the expansion of the pulse of the
optically thick plasma all the way to the point where the transparency
condition is reached. Some pioneering work in this respect were presented in Goodman in 1986 \cite{G86}. In this process the pulse reaches ultrarelativistic
regimes with Lorentz factor $\gamma\sim10^{2}$--$10^{4}$. The spectra, the
luminosities and the time-sequences of the electromagnetic signals captured
by a far-away observer are analyzed here in detail for the first time. 
The relevance of these theoretical results for short-bursts is then discussed.

\subsection{The expansion of the $e^{+}e^{-}\gamma$ plasma as a discrete set of
elementary slabs}

We discretize the gravitational collapse of a spherically symmetric
core of mass $M$ and charge $Q$
by considering a set of events along the world line of a point of fixed
angular position on the collapsing core surface. Between each of
these events we consider a spherical shell slab of plasma of constant coordinate
thickness $\Delta r$ so that:

\begin{enumerate}
\item $\Delta r$ is assumed to be a constant which is 
small with respect to the core radius;

\item $\Delta r$ is assumed to be large with respect to the mean free path of
the particles so that the statistical description of the $e^{+}e^{-}\gamma$
plasma can be used;

\item  There is no overlap among the slabs and their union describes the
entirety of the process.
\end{enumerate}

We check that the final results are independent of the special value of the
chosen $\Delta r$.

In order to describe the dynamics of the
expanding plasma pulse the energy-momentum conservation law and the rate equation
for the number of pairs in the Reissner-Nordstr\"{o}m geometry external to
the collapsing core have to be integrated:
\begin{align}
T^{\mu\nu}{}_{;\mu}  &  =0,\label{Tab}\\
\left(  n_{e^{+}e^{-}}u^{\mu}\right)  _{;\mu}  &  =\overline{\sigma v}\left[
n_{e^{+}e^{-}}^{2}\left(  T\right)  -n_{e^{+}e^{-}}^{2}\right]  , \label{nabis}%
\end{align}
where $T^{\mu\nu}=\left(  \epsilon+p\right)  u^{\mu}u^{\nu}+pg^{\mu\nu}$ is
the energy-momentum tensor of the plasma with proper energy density
$\epsilon$ and proper pressure $p$, $u^{a}$ is the fluid 4-velocity,
$n_{e^{+}e^{-}}$ is the pair number density, $n_{e^{+}e^{-}}\left(  T\right)
$ is the equilibrium pair number density at the temperature $T$ of the plasma
and $\overline{\sigma v}$ is the mean of the product of the $e^{+}e^{-}$
annihilation cross-section and the thermal velocity of the pairs. We use
Eqs.(\ref{Tab}) and (\ref{nabis}) to study the expansion of each slab, following
closely the treatment developed in Refs~\cite{rswx99,rswx00} where it was
shown how a homogeneous slab of plasma expands as a pair-electromagnetic
pulse (PEM pulse) of constant thickness in the laboratory frame. Two regimes
can be identified in the expansion of the slabs:

\begin{enumerate}
\item  In the initial phase of expansion the plasma experiences the strong
gravitational field of the core and a fully general relativistic description of
its motion is needed. The plasma is sufficiently hot in this first phase that
the $e^{+}e^{-}$ pairs and the photons remain at thermal equilibrium in it. As
shown in Ref.~\cite{rvx03a}, under these circumstances, the right hand side of
Eqs.(\ref{nabis}) is effectively $0$ and Eqs.(\ref{Tab}) and (\ref{nabis}) 
are equivalent to:
\begin{equation}
\left.
\begin{array}
[c]{c}%
\left(  \tfrac{dr}{cdt}\right)  ^{2}=\alpha^{4}\left[  1-\left(
\tfrac{n_{e^{+}e^{-}}}{n_{e^{+}e^{-}0}}\right)  ^{2}\left(  \tfrac{\alpha_{0}%
}{\alpha}\right)  ^{2}\left(  \tfrac{r}{r_{0}}\right)  ^{4}\right]  ,\\
\left(  \tfrac{r}{r_{0}}\right)  ^{2}=\left(  \tfrac{\epsilon+p}{\epsilon_{0}%
}\right)  \left(  \tfrac{n_{e^{+}e^{-}0}}{n_{e^{+}e^{-}}}\right)  ^{2}\left(
\tfrac{\alpha}{\alpha_{0}}\right)  ^{2}-\tfrac{p}{\epsilon_{0}}\left(
\tfrac{r}{r_{0}}\right)  ^{4},
\end{array}
\right.  \label{Eq2}%
\end{equation}
where $r$ is the radial coordinate of a slab of plasma, $\alpha=\left(
1-2MG/c^{2}r\right.$ $\left.+Q^{2}G/c^{4}r^{2}\right)^{1/2}$ is
the gravitational redshift factor and the subscript 
``$\scriptstyle{0}$" refers to quantities evaluated at the initial time.

\item  At asymptotically late times the temperature of the plasma drops below
an equivalent energy of $0.5$ MeV 
and the $e^{+}e^{-}$ pairs and the photons can no longer be
considered to be
in equilibrium: the full rate equation for pair
annihilation needs to be used. However, the plasma is so far from the central
core that gravitational effects can be neglected. In this new regime, as
shown in Ref.~\cite{rswx99}, Eqs.(\ref{Tab}) and (\ref{nabis}) reduce to:
\begin{align}
\tfrac{\epsilon_{0}}{\epsilon}  &  =\left(  \tfrac{\gamma\mathcal{V}}%
{\gamma_{0}\mathcal{V}_{0}}\right)  ^{\Gamma},\nonumber\\
\tfrac{\gamma}{\gamma_{0}}  &  =\sqrt{\tfrac{\epsilon_{0}\mathcal{V}_{0}%
}{\epsilon\mathcal{V}}},\label{Eq3}\\
\tfrac{\partial}{\partial t}N_{e^{+}e^{-}}  &  =-N_{e^{+}e^{-}}\tfrac
{1}{\mathcal{V}}\tfrac{\partial\mathcal{V}}{\partial t}+\overline{\sigma
v}\tfrac{1}{\gamma^{2}}\left[  N_{e^{+}e^{-}}^{2}\left(  T\right)
-N_{e^{+}e^{-}}^{2}\right]  ,\nonumber
\end{align}
where $\Gamma=1+p/\epsilon$, $\mathcal{V}$ is the volume of a single
slab as measured in the laboratory frame by an observer at rest with the black hole, $N_{e^{+}e^{-}}=\gamma n_{e^{+}e^{-}}$ is the pair number density as
measured in the laboratory frame by an observer at rest with the black hole,
and $N_{e^{+}e^{-}}\left(  T\right)  $ is the equilibrium laboratory pair
number density.
\end{enumerate}

\subsection{The reaching of transparency and the signature of the outgoing
gamma ray signal}

\begin{figure}
\centering
\includegraphics[width=10cm]{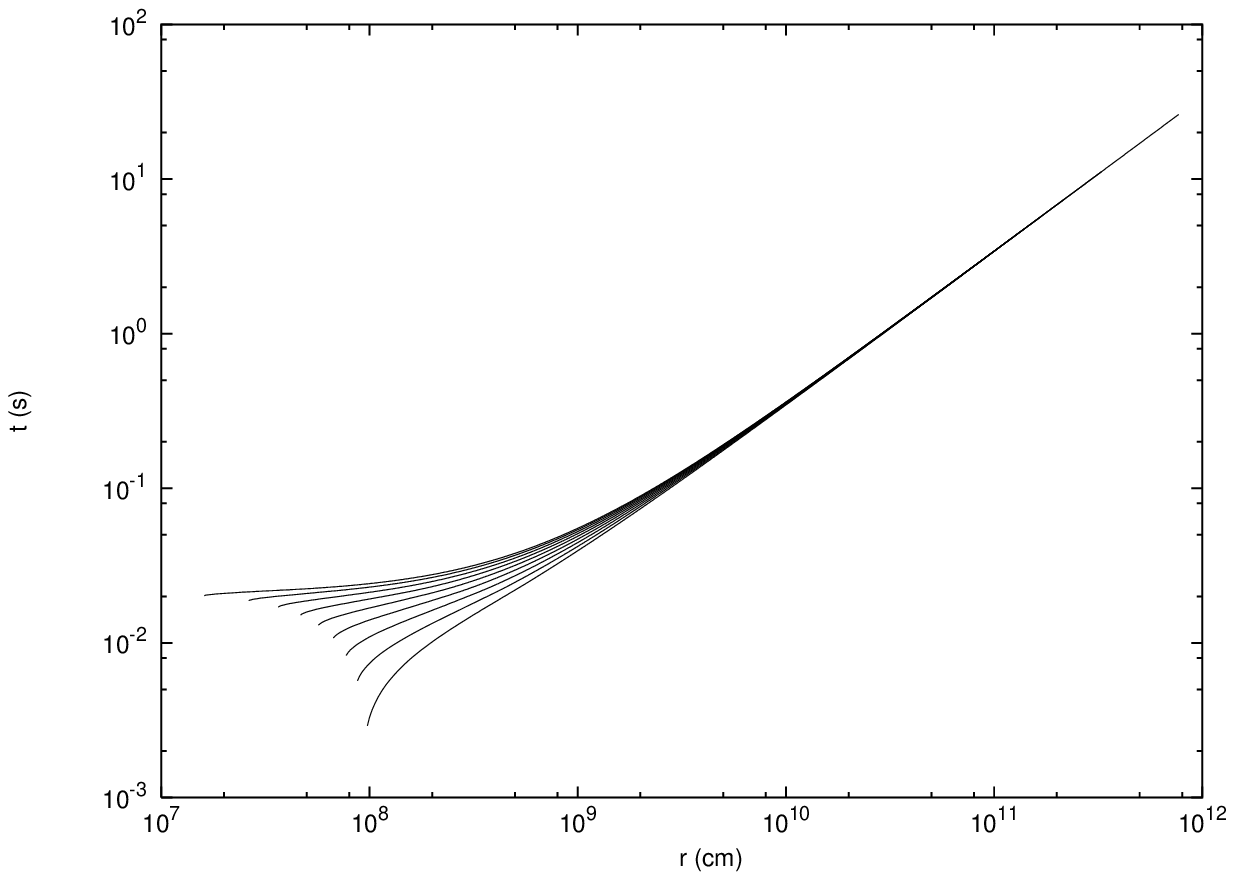} \includegraphics
[width=10cm]{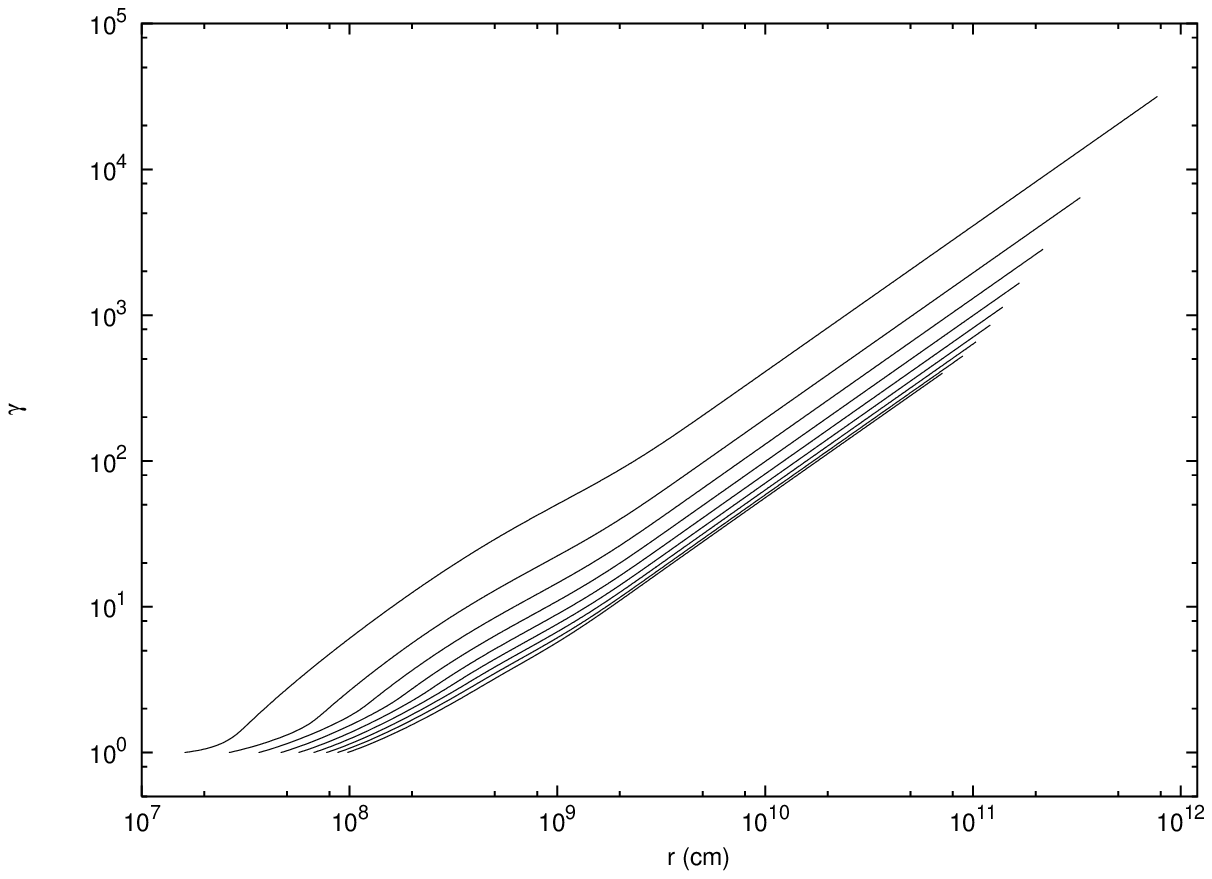}
\caption{Expansion of the plasma created around an overcritical collapsing
stellar core with $M=10M_{\odot}$ and $Q=0.1\sqrt{G}M$. Upper diagram: world
lines of the plasma. Lower diagram: Lorentz $\gamma$ factor as a function of
the radial coordinate $r$.}%
\label{ff1}%
\end{figure}

Eqs.(\ref{Eq2}) and (\ref{Eq3}) must be separately integrated 
and the solutions matched at
the transition between the two regimes. The integration stops when each slab
of plasma reaches the optical transparency condition given by
\begin{equation}
\int_0^{\Delta r}\sigma_{T}n_{e^{+}e^{-}}dr\sim1 \,,
\end{equation}
where $\sigma_{T}$ is the Thomson cross-section and the integral extends over
the radial thickness $\Delta r$ of the slab. The evolution of each slab occurs
without any collision or interaction with the other slabs; see the upper diagram
in Fig. \ref{ff1}. The outer layers are colder than the inner ones and
therefore reach transparency earlier; see the lower diagram in Fig. \ref{ff1}.

In Fig. \ref{ff1}, Eqs.(\ref{Eq2}) and (\ref{Eq3}) have been integrated for a
core with
\begin{equation}
M=10M_{\odot},\quad Q=0.1\sqrt{G}M;
\end{equation}
the upper diagram represents the world lines of the plasma as functions of the radius,
while the lower diagram shows the corresponding Lorentz $\gamma$ factors. The overall independence of the result of the dynamics
on the number $N$ of the slabs adopted in the discretization process or analogously on the value of $\Delta r$ has also
been checked. We have repeated the integration for $N=10$, $N=100$ reaching 
the same result to extremely good accuracy. The results in Fig. \ref{ff1}
correspond to the case $N=10$.

\begin{figure}
\centering
\includegraphics[width=10cm]{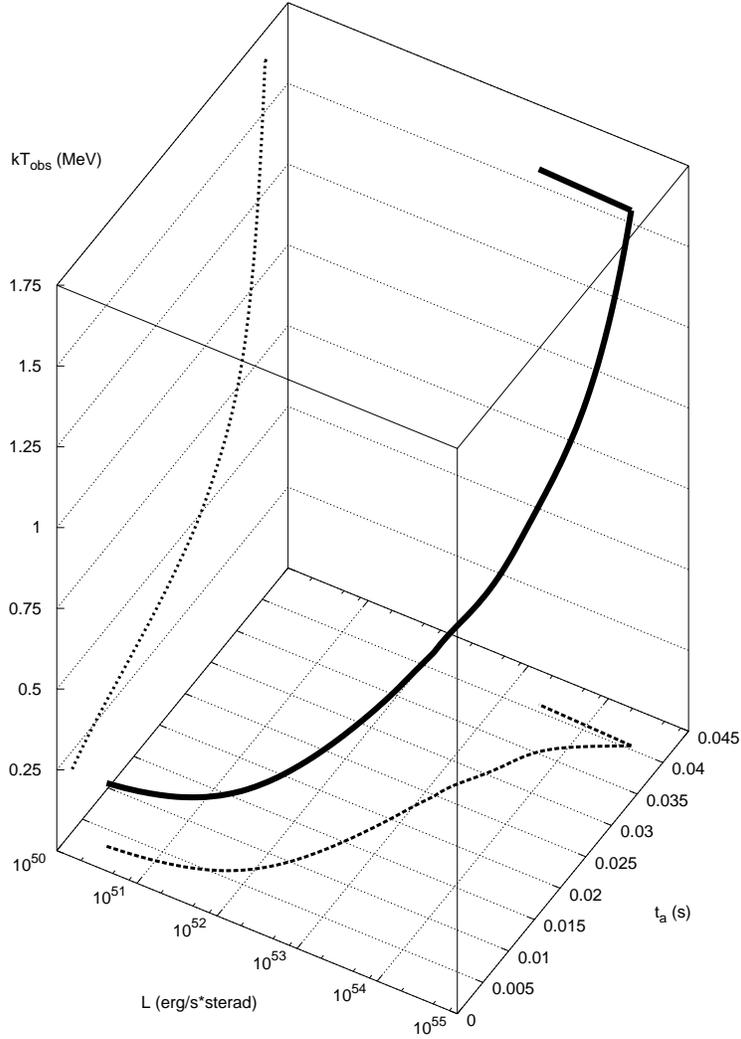}
\caption{Predicted observed luminosity and observed spectral hardness of the
electromagnetic signal from the gravitational collapse of a collapsing core
with $M=10M_{\odot}$, $Q=0.1\sqrt{G}M$ at $z=1$ as functions of the arrival
time $t_{a}$.}%
\label{ff4}%
\end{figure}

 We now turn to the results in
Fig. \ref{ff4}, where we plot both the theoretically predicted luminosity $L$
and the spectral hardness of the signal reaching a far-away observer as
functions of the arrival time $t_{a}$. Since all three of these quantities
depend in an essential way on the cosmological redshift factor $z$, see
Refs.~\cite{brx01,rbcfx03a}, we have adopted a cosmological
redshift $z=1$ for this figure.

\begin{figure}
\centering
\includegraphics[width=10cm]{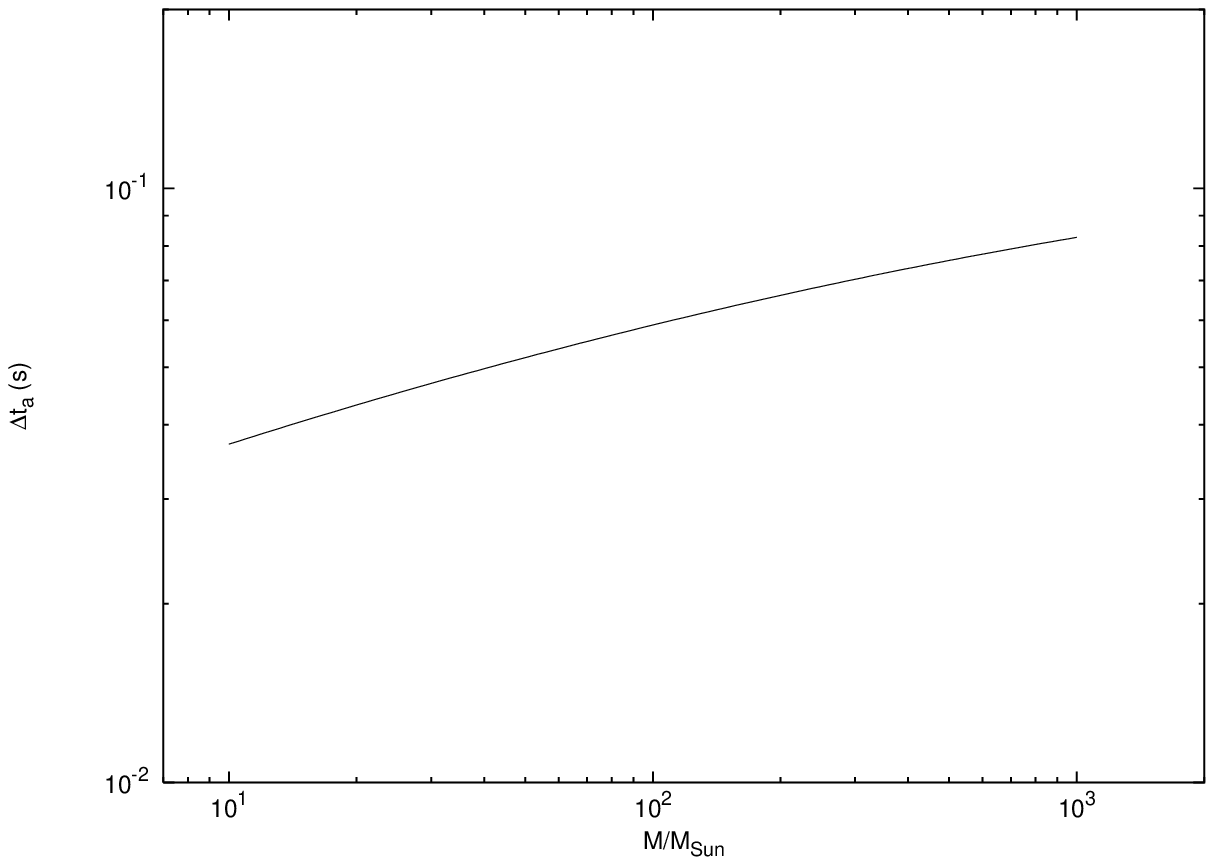}
\caption{Arrival time duration of the electromagnetic signal from the
gravitational collapse of a stellar core with charge $Q=0.1\sqrt{G}M$ as a
function of the mass $M$ of the core.}%
\label{ff5}%
\end{figure}

 As the plasma becomes transparent, gamma ray
photons are emitted. The energy $\hbar\omega$ of the observed photon is
$\hbar\omega=k\gamma T/\left(  1+z\right)  $, where $k$ is the Boltzmann
constant, $T$ is the temperature in the comoving frame of the pulse and
$\gamma$ is the Lorentz factor of the plasma at the transparency time. We also
recall that if the initial zero of time is chosen as the time when the first photon
is observed, then the arrival time $t_{a}$ of a photon at the detector in
spherical coordinates centered on the black hole is given by
\cite{brx01,rbcfx03a}:
\begin{equation}
t_{a}=\left(  1+z\right)  \left[  t+\tfrac{r_{0}}{c}-\tfrac{r\left(  t\right)
}{c}\cos\theta\right]
\end{equation}
where $\left(  t,r\left(  t\right)  ,\theta,\phi\right)  $ labels the
laboratory emission event along the world line of the emitting slab and
$r_{0}$ is the initial position of the slab. The projection of the plot in
Fig. \ref{ff4} onto the $t_{a}$-$L$ plane gives the total luminosity as the
sum of the partial luminosities of the single slabs. The sudden decrease of
the intensity at the time $t=0.040466$ s corresponds to the creation of the
{\itshape separatrix} introduced in Ref.~\cite{rvx03b}. We find that the duration of the
electromagnetic signal emitted by the relativistically expanding pulse is
given in arrival time by
\begin{equation}
\Delta t_{a}\sim5\times10^{-2}\mathrm{s}\label{ta1} \,.
\end{equation}
The projection of the plot in Fig. \ref{ff4} onto the $k T_{\mathrm{obs}%
}$, $t_{a}$ plane describes the temporal evolution of the spectral hardness.
We observe a precise soft-to-hard evolution of the spectrum of the gamma ray signal from $\sim10^{2}$ KeV monotonically increasing to $\sim1$ MeV. We
recall that $kT_{\mathrm{obs}}=k\gamma T/\left(  1+z\right)  $. 

 The
above quantities are clearly functions of the cosmological redshift $z$, of
the charge $Q$ and the mass $M$ of the collapsing core. We present in Fig. 3
the arrival time interval for $M$ ranging from $M\sim10M_{\odot}$ to
$10^{3}M_{\odot}$, keeping $Q=0.1\sqrt{G}M$. The arrival time interval is very
sensitive to the mass of the black hole:
\begin{equation}
\Delta t_{a}\sim10^{-2}-10^{-1}\mathrm{s} \,.
\label{ta2}
\end{equation}
Similarly the spectral hardness of the signal is sensitive to the ratio
$Q/\sqrt{G}M$ \cite{rfvx05}. Moreover the duration, the spectral hardness and
luminosity are all sensitive to the cosmological redshift $z$ (see
Ref.~\cite{rfvx05}).
All the above quantities can also be sensitive to a possible baryonic
contamination of the plasma due to the remnant of the progenitor star which
has undergone the process of gravitational collapse.

\section{Conclusions}

\subsection{On the GRB-Supernova connection}

We first stress some general considerations originating from comparing and contrasting the three GRB sources we have discussed:
\begin{enumerate}
\item The value of the $B$ parameter for all three sources occurs, as theoretically expected, in the allowed range (see Fig. \ref{bcross})
\begin{equation}
10^{-5} \le B \le 10^{-2}\, .
\label{brange}
\end{equation}
We have in fact:\\
\begin{center}
\begin{tabular}{ccc}
GRB 991216 & $B = 3.0\times 10^{-3}$ & $E_{dya} = 4.8 \times 10^{53}$ erg\\
GRB 980425 & $B = 7.0\times 10^{-3}$ & $E_{dya} = 1.1 \times 10^{48}$ erg\\
GRB 030329 & $B = 4.8\times 10^{-3}$ & $E_{dya} = 2.1 \times 10^{52}$ erg\\
\end{tabular}
\end{center}
\item The enormous difference in the GRB energy of the sources simply relates to the electromagnetic energy of the black hole given in Eq.(\ref{em}) which turns out to be smaller than the critical value given by Eq.(\ref{s1}). The fact that the theory is valid over $5$ orders of magnitude is indeed very satisfactory.
\item Also revealing is the fact that in both sources GRB 980425 and GRB 030329 the associated supernova energies are similar. We have, in fact, for both SN 1998bw and SN 2003dh an energy $\sim 10^{49}$ erg. Details in Fraschetti et al. \cite{f03mg10} and Bernardini et al. \cite{b03mg10}. The further comparison between the SN luminosity and the GRB intensity is crucial. In the case of GRB 980425 the GRB and the SN energies are comparable, and no dominance of one source over the other can be ascertained. In the case of GRB 030329 the energy of the GRB source is $10^3$ larger than the SN: in no way the GRB can originate from the SN event.
\end{enumerate}
The above stringent energetics considerations and the fact that GRBs occur also without an observed supernova give a strong evidence that GRBs cannot originate from supernovae.

\subsection{URCA-1 and URCA-2}

We turn now to the most exciting search for the nature of URCA-1 and URCA-2. We have already mentioned above that a variety of possibilities naturally appear. The first possibility is that the URCA sources are related to the black hole originating the GRB phenomenon. In order to probe such an hypothesis, it would be very important to find even a single case in which an URCA source occurs in association with a GRB and in absence of an associated supernova. Such a result, theoretically unexpected, would open an entire new problematic in relativistic astrophysics and in the physics of black holes.

If indeed, as we expect, the clear association between URCA sources and the supernovae occurring together with the GRBs, then it is clear that the analysis of the other two possibilities will be favored. Namely, an emission from processes occurring in the early phases of the expansion of the supernova remnant or the very exciting possibility that for the first time we are observing a newly born neutron star out of the supernova phenomenon. Of course, this last hypothesis is the most important one, since it would offer new fundamental information about the outcome of the gravitational collapse, about the equations of state at supranuclear densities and about a variety of fundamental issues of relativistic astrophysics of neutron stars. We shall focus in the following only on this last topic.

We have already recalled how the need for a rapid cooling process due to neutrino anti-neutrino emission in the process of gravitational collapse leading to the formation of a neutron star was considered for the first time by George Gamow and Mario Schoenberg in 1941 \cite{gs41}. It was Gamow who gave this process the name ``Urca process'', see \ref{gamow} and \ref{urca}. Since then, a systematic analysis of the theory of neutron star cooling was advanced by Tsuruta \cite{t64,t79}, Tsuruta and Cameron \cite{tc66}, Tsuruta et al. \cite{t02} and by Canuto \cite{c78}. The coming of age of X-ray observatories such as Einstein (1978-1981), EXOSAT (1983-1986), ROSAT (1990-1998), and the contemporary missions of Chandra and XMM-Newton since 1999 dramatically presented an observational situation establishing very embarrassing and stringent upper limits to the surface temperature of neutron stars in well known historical supernova remnants (see e.g. Romani \cite{r87}). It was so that, for some remnants, notably SN 1006 and the Tycho supernova, the upper limits to the surface temperatures were significantly lower than the temperatures given by standard cooling times (see e.g. Romani \cite{r87}). Much of the theoretical works has been mainly directed, therefore, to find theoretical arguments in order to explain such low surface temperature $T_s \sim 0.5$--$1.0\times 10^6$ K --- embarrassingly low, when compared to the initial hot ($\sim 10^{11}$ K) birth of a neutron star in a supernova explosion (see e.g. Romani \cite{r87}). Some very important steps in this direction of research have been represented by the works of Van Riper \cite{vr88,vr91}, Lattimer and his group \cite{bl86,lvrpp94} and by the most extensive work of Yakovlev and his group \cite{yp04}. The youngest neutron star to be searched for using its thermal emission in this context has been the pulsar PSR J0205+6449 in 3C 58 (see e.g. Yakovlev and Pethick \cite{yp04}), which is $820$ years old! Recently, evidence for the detection of thermal emission from the crab nebula pulsar was reported by Trumper \cite{t05} which is, again, $951$ years old.

In the case of URCA-1 and URCA-2, we are exploring a totally different regime: the X-ray emission possibly from a recently born neutron star in the first days -- months of its existence, where no observations have yet been performed and no embarrassing constraints upper limits on the surface temperature exist. The reason of approaching first the issue of the thermal emission from the neutron star surface is extremely important, since in principle it can give information on the equations of state in the core at supranuclear densities and on the detailed mechanism of the formation of the neutron star itself and the related neutrino emission. It is of course possible that the neutron star is initially fast rotating and its early emission is dominated by the magnetospheric emission or by accretion processes from the remnant which would overshadow the thermal emission. In that case a periodic signal related to the neutron star rotational period should in principle be observable in a close enough GRB source provided the suitable instrumentation from the Earth.

The literature on young born neutron star is relatively scarse today. There are some very interesting contributions which state: ``The time for a neutron star's center to cool by the direct URCA process to a temperature $T$ has been estimated to be $t = 20 \left[T/\left(10^9 K\right)\right]^{-4}$ s. The direct URCA process and all the exotic cooling mechanisms only occur at supranuclear densities. Matter at subnuclear densities in neutron star crust cools primarily by diffusion of heat to the interior. Thus the surface temperature remains high, in the vicinity of $10^6$ K or more, until the crust's heat reservoir is consumed. After this diffusion time, which is on the order of $1$--$100$ years, the surface temperature abruptly plunges to values below $5 \times 10^5$ K'' (Lattimer et al. \cite{lvrpp94}). ``Soon after a supernova explosion, the young neutron star has large temperature gradients in the inner part of the crust. While the powerful neutrino emission quickly cools the core, the crust stays hot. The heat gradually flows inward on a conduction time scale and the whole process can be thought of as a cooling wave propagation from the center toward the surface'' (Gnedin et al. \cite{ya01}).

The two considerations we have quoted above are developed in the case of spherical symmetry and we would like to keep the mind open, in this new astrophysical field, to additional factors, some more traditional than others, to be taken into account. Among the traditional ones we recall: 1) the presence of rotation and magnetic field which may affect the thermal conductivity and the structure of the surface, as well as the above mentioned magnetospheric emission; 2) there could be accretion of matter from the expanding nebula; and, among the nontraditional ones, we recall 3) some exciting theoretical possibilities advanced by Dyson on volcanoes on neutron stars \cite{d69} as well as iron helide on neutron star \cite{d71}, as well as the possibility of piconuclear reactions on neutron star surface discussed in Lai \& Salpeter \cite{ls97}.

All the above are just scientific arguments to attract attention on the abrupt fall in luminosity reported in this meeting on URCA-1 by Elena Pian which is therefore, in this light, of the greatest scientific interest and further analysis should be followed to check if a similar behavior will be found in future XMM and Chandra observations also in URCA-2.

\subsection{Astrophysical implications}

In addition to these very rich problematics in the field of theoretical physics and theoretical astrophysics, there are also more classical astronomical and astrophysical issues, which will need to be answered if indeed the observations of a young neutron star will be confirmed. An important issue to be addressed will be how the young neutron star can be observed, escaping from being buried under the expelled matter of the collapsing star. A possible explanation can originate from the binary nature of the newly born neutron star: the binary system being formed by the newly formed black hole and the triggered gravitational collapse of a companion evolved star leading, possibly, to a ``kick'' on and ejection of the newly born neutron star. Another possibility, also related to the binary nature of the system, is that the supernova progenitor star has been depleted of its outer layer by dynamic tidal effects.

In addition, there are other topics in which our scenario can open new research directions in fundamental physics and astrophysics:\\
1) The problem of the instability leading to the complete gravitational collapse of a $\sim 10M_\odot$ star needs the introduction of a new critical mass for gravitational collapse, which is quite different from the one for white dwarfs and neutron stars which has been widely discussed in the current literature (see e.g. Giacconi \& Ruffini \cite{gr78}).\\
2) The issue of the trigger of the instability of gravitational collapse induced by the GRB on the progenitor star of the supernova or, vice versa, by the supernova on the progenitor star of the GRB needs accurate timing and the considerations of new relativistic phenomena.\\
3) The general relativistic instability induced on a nearby star by the formation of a black hole needs some very basic new developments in the field of general relativity.

Only a very preliminary work exists on this subject, by Jim Wilson and his collaborators, see e.g. the paper by Mathews and Wilson \cite{mw}. The reason for the complexity in answering such a question is simply stated: unlike the majority of theoretical work on black holes, which deals mainly with one-body solutions, we have to address here a many-body problem in general relativity. We are starting in these days to reconsider, in this framework, some classic work by Fermi \cite{f21}, Hanni and Ruffini \cite{hr73}, Majumdar \cite{m47}, Papapetrou \cite{p47}, Parker et al. \cite{p73}, Bini et al. \cite{bgr04} which may lead to a new understanding of general relativistic effects relevant to these astrophysical ``triptychs''.

\subsection{The short GRBs}

After concluding the problematic of the long GRBs and their vast astrophysical implications, we have turned to the physics of short GRBs. We first report some progress in the understanding the dynamical phase of collapse, the mass-energy formula and the extraction of blackholic energy which have been motivated by the analysis of the short GRBs. In this context progress has also been accomplished on establishing an absolute lower limit to the irreducible mass of the black hole as well as on some critical considerations about the relations of general relativity and the second law of thermodynamics. We recall how this last issue has been one of the most debated in theoretical physics in the past thirty years due to the work of Bekenstein and Hawking. Following these conceptual progresses we analyze the vacuum polarization process around an overcritical collapsing shell. We evidence the existence of a separatrix and a dyadosphere trapping surface in the dynamics of the electron-positron plasma generated during the process of gravitational collapse. We then analyze, using recent progress in the solution of the Vlasov-Boltzmann-Maxwell system, the oscillation regime in the created electron-positron plasma and their rapid convergence to a thermalized spectrum. We conclude by making precise predictions for the spectra, the energy fluxes and characteristic time-scales of the radiation for short-bursts.

\subsection{Short GRBs as cosmological candles}

The characteristic spectra, time variabilities and luminosities of the electromagnetic signals from collapsing overcritical stellar cores have been derived from first principles, and they agree with preliminary observations of short-bursts \cite{batse4b}. Hopefully new space missions will be planned, with temporal resolution down to fractions of $\mu$s and higher collecting area and spectral resolution than at present, in order to verify the detailed agreement between our model and the observations. It is now clear that if our theoretical predictions will be confirmed, we will have a very powerful tool for cosmological observations: the independent information about luminosity, time-scale and spectrum can uniquely determine the mass, the electromagnetic structure and the distance from the observer of the collapsing core, see e.g. Fig. \ref{ff5} and Ref. \cite{rfvx05}. In that case short-bursts, in addition to give a detailed information on all general relativistic and relativistic field theory phenomena occurring in the approach to the horizon, may also become the best example of standard candles in cosmology \cite{r03tokyo}. We are currently analyzing the introduction of baryonic matter in the optically thick phase of the expansion of the $e^+e^-$ plasma, within this detailed time-varying description of the gravitational collapse, which may affect the structure of the Proper-GRB (P-GRB) \cite{lett1} as well as the structure of the long-bursts \cite{Brasile}.

\subsection{On the dyadosphere of Kerr-Newman black holes}

An interesting proposal was advanced in 2002 \cite{it02} that the $e^+e^-$ plasma may have a fundamental role as well in the physical process generating jets in the extragalactic radio sources. The concept of dyadosphere originally introduced in Reissner-Nordstr\"{o}m black hole in order to create the $e^+e^-$ plasma relevant for GRBs can also be generalized to the process of vacuum polarization originating in a Kerr-Newman black hole due to magneto-hydrodynamical process of energy extraction (see e.g. \cite{pu01} and references therein). The concept therefore introduced here becomes relevant for both the extraction of rotational and electromagnetic energy from the most general black hole \cite{cr71}.

\subsection*{* * *}

After the completion of these lectures, we have become aware that Ghirlanda et al. \cite{ggc03} have given evidence for the existence of an exponential cut off at high energies in the spectra of short bursts. We are currently comparing and contrasting these observational results with the predicted cut off in Fig. \ref{ff4} which results from the existence of the separatrix introduced in \cite{rvx03a}. The observational confirmation of the results presented in Fig. \ref{ff4} would lead for the first time to the identification of a process of gravitational collapse and its general relativistic self-closure as seen from an asymptotic observer.

\section*{Acknowledgments}
We are thankful to Charles Dermer, Hagen Kleinert, Tsvi Piran, Andreas Ringwald, Rashid Sunyaev, Lev Titarchuk, Jim Wilson and Dima Yakovlev for many interesting theoretical discussions, as well as to Lorenzo Amati, Lucio Angelo Antonelli, Enrico Costa, Filippo Frontera, Luciano Nicastro, Elena Pian, Luigi Piro, Marco Tavani and all the BeppoSAX team for assistance in the data analysis.

\appendix

\section{The numerical integration of the hydrodynamics and the rate equations in the Livermore code}\label{num_int1}

\subsection{The hydrodynamics and the rate equations for the plasma of $e^+e^-$-pairs}\label{hydro_pem}

The evolution of the $e^+e^-$-pair plasma generated in the dyadosphere has been treated in two papers \cite{rswx99,rswx00}. We recall here the basic governing equations in the most general case in which the plasma fluid is composed of $e^+e^-$-pairs, photons and baryonic matter. The plasma is described by the stress-energy tensor
\begin{equation}
T^{\mu\nu}=pg^{\mu\nu}+(p+\rho)U^\mu U^\nu\, ,
\label{tensor}
\end{equation}
where $\rho$ and $p$ are respectively the total proper energy density and pressure in the comoving frame of the plasma fluid and $U^\mu$ is its four-velocity, satisfying
\begin{equation}
g_{tt}(U^t)^2+g_{rr}(U^r)^2=-1 ~,
\label{tt}
\end{equation}
where $U^r$ and $U^t$ are the radial and temporal contravariant components of the 4-velocity and 
\begin{equation}
d^2s=g_{tt}(r)d^2t+g_{rr}(r)d^2r+r^2d^2\theta +r^2\sin^2\theta
d^2\phi ~,
\label{s}
\end{equation}
where $g_{tt}(r) \equiv - \alpha^2(r)$ and $g_{rr}(r)= \alpha^{-2}(r)$.

The conservation law for baryon number can be expressed in terms of the proper baryon number density $n_B$
\begin{eqnarray}
(n_B U^\mu)_{;\mu}&=& g^{-{1\over2}}(g^{1\over2}n_B
U^\nu)_{,\nu}\nonumber\\
&=&(n_BU^t)_{,t}+{1\over r^2}(r^2 n_BU^r)_{,r}=0 ~.
\label{contin}
\end{eqnarray}
The radial component of the energy-momentum conservation law of the plasma fluid reduces to 
\begin{equation}
{\partial p\over\partial r}+{\partial \over\partial t}\left((p+\rho)U^t U_r\right)+{1\over r^2} { \partial
\over \partial r}  \left(r^2(p+\rho)U^r U_r\right)
-{1\over2}(p+\rho)\left[{\partial g_{tt}
 \over\partial r}(U^t)^2+{\partial g_{rr}
 \over\partial r}(U^r)^2\right] =0 ~.
\label{cmom2}
\end{equation}
The component of the energy-momentum conservation law of the plasma fluid equation along a flow line is
\begin{eqnarray}
U_\mu(T^{\mu\nu})_{;\nu}&=&-(\rho U^\nu)_{;\nu}
-p(U^\nu)_{;\nu},\nonumber\\ &=&-g^{-{1\over2}}(g^{1\over2}\rho
U^\nu)_{,\nu} - pg^{-{1\over2}}(g^{1\over2} U^\nu)_{,\nu}\nonumber\\
&=&(\rho U^t)_{,t}+{1\over r^2}(r^2\rho
U^r)_{,r}\nonumber\\
&+&p\left[(U^t)_{,t}+{1\over r^2}(r^2U^r)_{,r}\right]=0 ~.
\label{conse1}
\end{eqnarray}

We define also the total proper internal energy density $\epsilon$ and the baryonic mass density $\rho_B$ in the comoving frame of the plasma fluid,
\begin{equation}
\epsilon \equiv \rho - \rho_B,\hskip0.5cm \rho_B\equiv n_Bmc^2 ~.
\label{cpp}
\end{equation} 

\subsection{The numerical integration}\label{num_int2}

A computer code \cite{wsm97,wsm98} has been used to
evolve the spherically symmetric general relativistic hydrodynamic equations starting from the dyadosphere \cite{rswx99}.

We define the generalized gamma factor $\gamma$ and the radial 3-velocity in the laboratory frame $V^r$
\begin{equation}
\gamma \equiv \sqrt{ 1 + U^r U_r},\hskip0.5cm V^r\equiv {U^r\over U^t}.
\label{asww}
\end{equation}
From Eqs.(\ref{s}, \ref{tt}), we then have
\begin{equation}
(U^t)^2=-{1\over g_{tt}}(1+g_{rr}(U^r)^2)={1\over\alpha^2}\gamma^2.
\label{rr}
\end{equation}
Following Eq.(\ref{cpp}), we also define 
\begin{equation}
E \equiv \epsilon \gamma,\hskip0.5cmD \equiv \rho_B \gamma,
\hskip0.3cm {\rm and}\hskip0.3cm\tilde\rho \equiv \rho\gamma
\label{cp}
\end{equation} 
so that the conservation law of baryon number (\ref{contin}) can then
be written as
\begin{equation}
{\partial D \over \partial t} = - {\alpha \over r^2} {
\partial \over \partial r} ({r^2 \over \alpha} D V^r).
\label{jay1}
\end{equation}
Eq.(\ref{conse1}) then takes the form,
\begin{equation}
{\partial E \over \partial t} = - {\alpha \over r^2} {
\partial \over \partial r} ({r^2 \over \alpha} E V^r) - p
\biggl[ {\partial \gamma \over \partial t} + {\alpha \over r^2}
{\partial \over \partial r} ({ r^2 \over \alpha} \gamma V^r)
\biggr].
\label{jay2}
\end{equation}
Defining the radial momentum density in the laboratory frame
\begin{equation}
S_r\equiv \alpha (p+\rho)U^tU_r = (D + \Gamma E) U_r,  
\label{mstate}
\end{equation}
we can express the radial component of the energy-momentum
conservation law given in Eq.(\ref{cmom2}) by
\begin{eqnarray}
{\partial S_r \over \partial t} &=& - {\alpha \over r^2} { \partial
\over \partial r} ({r^2 \over \alpha} S_r V^r) - \alpha {\partial p
\over \partial r}\nonumber\\ 
&-&{\alpha\over2}(p+\rho)\left[{\partial g_{tt}
\over\partial r}(U^t)^2+{\partial g_{rr} \over\partial
r}(U^r)^2\right]\nonumber\\ 
&=& - {\alpha \over r^2} { \partial \over
\partial r} ({r^2 \over \alpha} S_r V^r) - \alpha {\partial p \over
\partial r}\nonumber\\
&-& \alpha\left({M \over r^2}-{Q^2 \over r^3}\right)
\biggl({D + \Gamma E \over \gamma} \biggr) \biggl[ \left({\gamma \over
\alpha}\right)^2 + {(U^r)^2 \over \alpha^4 } \biggr] ~.
\label{jay3}
\end{eqnarray}

In order to determine the number-density of $e^+e^-$ pairs, we use the pair rate equation. 
We define the $e^+e^-$-pair density in the laboratory frame 
$N_{e^\pm} \equiv\gamma n_{e^\pm}$ and $N_{e^\pm}(T) \equiv\gamma
n_{e^\pm}(T)$, where $n_{e^{\pm}}(T)$ is the total proper number density of pairs in comoving frame at thermodynamic equilibrium with temperature $T$ in the process $e^+ + e^- \rightarrow \gamma + \gamma$ $\left(n_{e^-}(m, T) = n_{\gamma}(T)\right)$, $n_{e^{\pm}}$ is the total proper number density of pairs in comoving frame at a generic time before reaching the equilibrium. We write the rate equation in the form
\begin{equation}
{\partial N_{e^\pm} \over \partial t} = - {\alpha \over r^2} {
\partial \over \partial r} ({r^2 \over \alpha} N_{e^\pm} V^r) +
\overline{\sigma v} (N^2_{e^\pm} (T) - N^2_{e^\pm})/\gamma^2~,
\label{jay:E:ndiff}
\end{equation}
These equations are integrated starting from the dyadosphere distributions given in Fig. 17 (Right) in \cite{Brasile} and assuming as usual ingoing boundary conditions on the horizon of the black hole.

\section{The numerical integration of the hydrodynamics and the rate equations in the Rome code}\label{rome_code}

\subsection{Era I: expansion of PEM-pulse}\label{era1}\

After the explosion from the dyadosphere a thermal plasma of $e^+e^-$ pairs and photons optically thick with respect to scattering processes begins to expand at ultrarelativistic velocity. In this era the expansion takes place in a region of very low baryonic contamination.

Recalling that the local number density of electron and positron pairs created as a function of radius is given by
\begin{equation}
n_{e^+e^-}(r) = {Q\over 4\pi r^2\left({\hbar\over
mc}\right)e}\left[1-\left({r\over r^\star}\right)^2\right] ~,
\label{nd}
\end{equation}
the limit on such baryonic contamination, where $\rho_{B_c}$ is the mass-energy density of baryons, is given by 
\begin{equation}
\rho_{B_c}\ll m_pn_{e^+e^-}(r) = 3.2\cdot 10^8\left({r_{ds}\over r}\right)^2\left[1-\left({r\over r_{ds}}\right)^2\right](g/cm^3).
\label{nb} 
\end{equation}
Near the horizon $r\simeq r_+$, this gives
\begin{equation}
\rho_{B_c}\ll m_pn_{e^+e^-}(r) =1.86 \cdot 10^{14}\left({\xi\over\mu}\right)(g/cm^3)\, ,
\label{nb1} 
\end{equation}
and near the radius of the dyadosphere $r_{ds}$:
\begin{equation}
\rho_{B_c}\ll m_pn_{e^+e^-}(r) = 3.2\cdot 10^8\left[1-\left({r\over r_{ds}}\right)^2\right]_{r\rightarrow r_{ds}}(g/cm^3)\, .
\label{nb2} 
\end{equation}
Such conditions can be easily satisfied in the collapse to a black hole, but not necessarily in a collapse to a neutron star. 

\begin{figure}
\centering
\includegraphics[width=9cm,clip]{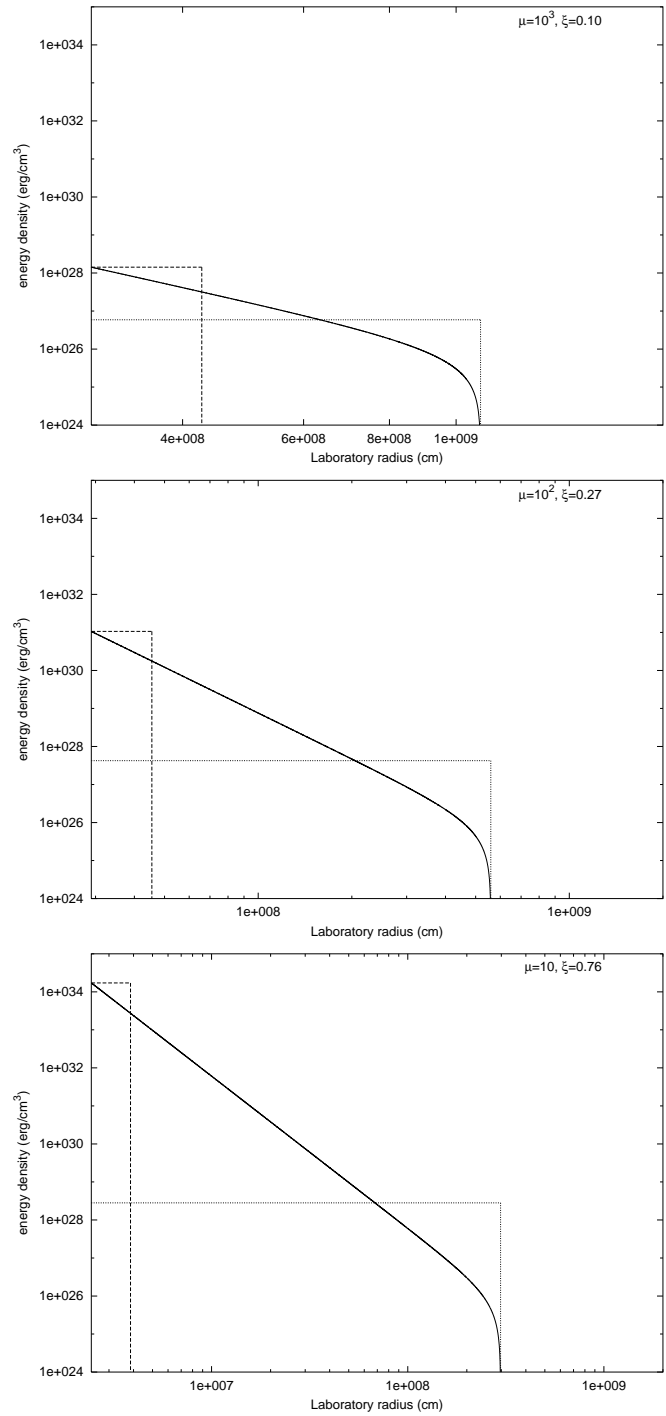}
\caption{{\itshape Three different dyadospheres corresponding to the same value and to different values of the two parameters $\mu$ and $\xi$ are given. The three different configurations are markedly different in their spatial extent as well as in their energy-density distribution (see text)}.}
\label{3dens}
\end{figure}

Consequently we have solved the equations governing a plasma composed solely of $e^+e^-$-pairs and electromagnetic radiation, starting at time zero from the dyadosphere configurations corresponding to constant density in Fig. \ref{3dens}.

The plasma of $e^+e^-$ pairs and photons is described by the covariant energy-momentum tensor:
\begin{equation}
T^{\mu\nu} = pg^{\mu\nu} + \left(p + \rho\right)U^{\mu}U^{\nu} + \Delta T^{\mu\nu}\,
\label{tensorenim}
\end{equation}
where $\rho$ and $p$ are respectively total proper energy density and pressure in the comoving system; $U^{\mu}$ are contravariant components of 4-velocity and $\Delta T^{\mu\nu}$ takes into account of dissipative effects due to heat conduction and viscosity, but in this treatment it has been neglected. In general we have $g_{\mu \nu}U^{\mu}U^{\nu} = -1$. For a spherically symmetric motion this reduces to $g_{tt}(U^t)^2 + g_{rr}(U^r)^2 = -1$, where $U^t$ and $U^r$ are respectively temporal and radial controvariant components of 4-velocity $U^{\mu}$.
 
It is assumed that the gravitational interaction with central black hole is negligible with respect to the total energy of PEM-pulse such that a fluid expansion with special relativistic equations can be considered.

Moreover it is assumed that photons remain trapped inside fireball until complete transparency, i.e. the emission of electromagnetic radiation is negligible during the first phases of expansion, being therefore adiabatic \cite{rswx99}. This assumption is valid until the photon mean free path is negligible with respect to the thickness of pulse.

The thermodynamic quantities used to describe the process are the total proper internal energy density of pulse $\epsilon$, given by $\epsilon = \epsilon_{e^+} + \epsilon_{e^-} +  \epsilon_{\gamma}$, where $\epsilon_{e^+}$ ($\epsilon_{e^-}$) is total proper internal energy density of electrons (positrons) and $\epsilon_{\gamma}$ 
of photons. The proper number density of pairs $n_{e^{\pm}}$, if the system is in thermodynamic equilibrium initially at temperature $T$ of order $T \sim MeV$, enough for $e^+e^-$ pair creation, equals the proper number density of photons $n_{\gamma}$. This is not valid at lower temperature \cite{brx01}. The pressure is $p={p_{e^+}}+{p_{e^-}}+{p_{\gamma}}$, where $p_{e^{\pm}}$ are electrons and positrons pressures and $p_{\gamma}$ is photons pressure. The system is highly relativistic, so the equation of state $p = {\epsilon} / {3}$ can be considered valid. This equation of state is represented with thermal index $\Gamma$:
\begin{equation}
\Gamma = 1 + { p\over \epsilon} ~.
\label{state}
\end{equation}

\subsection{Fermi integrals}    

Thermodynamical quantities before introduced are expressed in terms of integrals over Bose distribution for photons and Fermi distribution for $e^+e^-$ pairs with zero chemical potentials $\mu_{\gamma}$ and $\mu_{e^{\pm}}$. We begin from the reaction $e^+ + e^- \rightarrow \gamma + \gamma$. From statistical mechanics it is known that given a thermodynamic system at temperature $T$ kept inside a volume $V$ and made of a number of particle variable $N$, the thermodynamic equilibrium is expressed by the condition that the potential free energy of Helmholtz $F(T, V, N)$ is stationary with respect to $N$ variations:
\begin{equation}
\left({\partial F \over \partial N}\right)_{T,V} = 0 ;
\label{helmh}
\end{equation}
by definition chemical potential $\mu$ is given by
\begin{equation}
\mu = \left({\partial F \over \partial N}\right)_{T,V} ;
\label{potenziale}
\end{equation}
so that for a system made by a photon gas at equilibrium with matter with respect to creation and adsorption processes, we have $\mu_{\gamma}=0$ \cite{ll5}. Therefore the chemical potential of electrons and positrons associated to reaction $e^+ + e^- \rightarrow \gamma + \gamma$ is equal and opposite: $\mu_{e^-} = - \mu_{e^+} = \mu$; moreover also $\mu$ must be zero since the total electric charge of fireball is zero: if $Q$ is total electric charge of fireball, we have
\begin{equation}
Q = e\left[{n_{e^-}}\left(m, T, \mu\right) - {n_{e^+}}\left(m, T, -\mu\right)\right] =0
\label{charge}
\end{equation} 
where ${n_{e^-}}\left(m, T, \mu\right)$ is given by
\begin{equation}
{n_{e^-}}\left(m, T, \mu\right) = \frac{aT^3}{k} {7\over8} {1\over A} \int_0^{+\infty} \frac{z^2}{e^{\sqrt{z^2+(mc^2/kT)^2} + \frac{\mu}{kT}}+1} dz;
\label{num_pote}
\end{equation}
so $\mu = 0$.

\begin{figure}
\centering
\includegraphics[width=10cm]{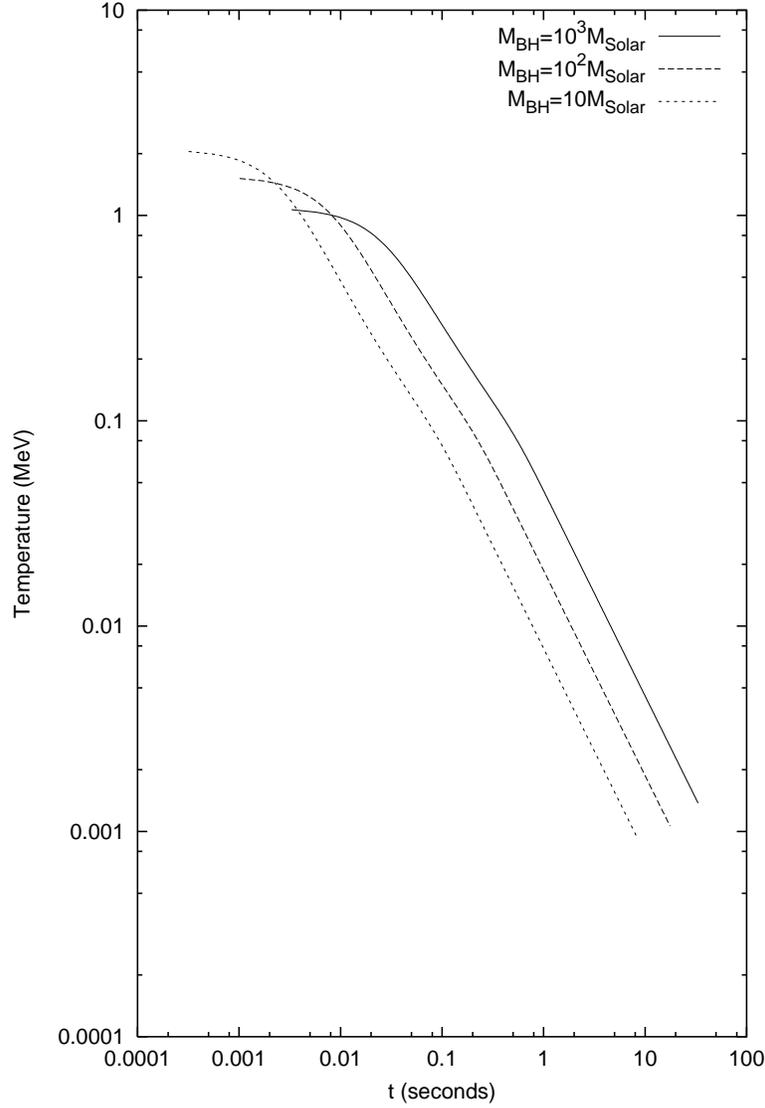}
\caption{{\itshape Temperature in comoving system as a function of emission time for different values of black hole mass $\mu$}.} 
\label{tem2}
\end{figure}

\begin{figure}
\centering
\includegraphics[width=10cm]{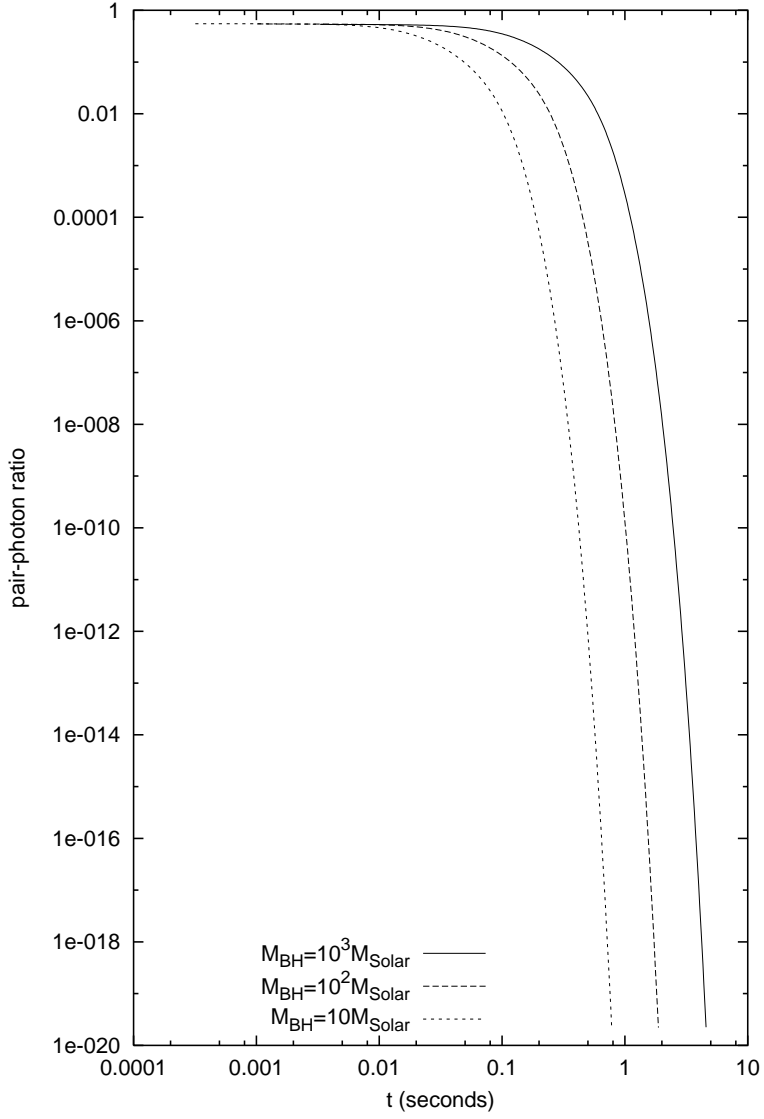}
\caption{{\itshape Ratio between number density of pairs $e^+e^-$ $n_{e^{\pm}}$ and number density of photons ${n_{\gamma}}\left(T\right)$ as a function of emission time for different values of black hole mass $\mu$}.} 
\label{pairs_fig}
\end{figure}

In the following the expressions of thermodynamical quantities as Fermi integrals are listed. The proper number density of electrons \cite{ws} is given by
\begin{eqnarray}
{n_{e^-}}\left(m, T, \mu_{e^-}\right) & = & \frac{2}{h^3} \int \frac{d^3 p}{e^{\frac{\sqrt{(pc)^2+(mc^2)^2}}{kT}}+1} = \nonumber \\
& = & \frac{8\pi}{h^3} \int_0^{+\infty} \frac{p^2}{e^{\frac{\sqrt{(pc)^2+(mc^2)^2}}{kT}}+1} dp = \nonumber \\ 
& = & \frac{aT^3}{k} {7\over8} {1\over A} \int_0^{+\infty} \frac{z^2}{e^{\sqrt{z^2+(mc^2/kT)^2}}+1} dz,
\label{nel1}
\end{eqnarray}
where $z=pc/kT$, $m$ is the electron mass, $T$[MeV] is the temperature of fireball in comoving frame, $a$ is a constant given by $a= {8\pi^5k^4} / {15h^3c^3} = 1.37 \cdot 10^{26} {erg} / {cm^3 MeV^4}$, $k$ is the Boltzmann constant and $A=(7/4)(\pi^4/15)$ is a numerical constant introduced for convenience.\\
Since the thermodynamic equilibrium is assumed and in all cases considered the initial temperature is larger than $e^+e^-$ pairs creation threshold ($T = 1$~MeV), the proper number density of electrons is roughly equal to that one of photons:
\begin{equation}
n_{e^{\pm}} \sim {n_{e^-}}\left(T\right) \sim {n_{\gamma}}\left(T\right); 
\label{equil_num}
\end{equation}
in these conditions the number of particles is conserved:
\begin{equation}
\left(n_{e^{\pm}}U^{\mu}\right);_{\mu} = 0.
\label{cons_equil}
\end{equation}
Later on, for $T \ll 1$MeV (see Fig. \ref{tem2}), $e^+e^-$ pairs go on in annihilation but can not be created anymore, therefore
\begin{equation}
{n_{\gamma}}\left(T\right) > n_{e^{\pm}} > n_{e^{\pm}}\left(T\right)\, 
\label{equil_num2}
\end{equation}
as shown in Fig. \ref{pairs_fig}.

The total proper internal energy density for photons is given by
\begin{equation}
\epsilon_{\gamma} = \frac{2}{h^3} \int \frac{h\nu}{e^{\frac{h\nu}{kT}}-1} d^3p = aT^4
\label{enpho}
\end{equation}
where $p=h\nu/c$. The total proper internal energy density for electrons is given by:
\begin{eqnarray}
\epsilon_{e^-} & = & \frac{2}{h^3} \int \frac{\sqrt{(pc)^2+(mc^2)^2}}{e^{\frac{\sqrt{(pc)^2+(mc^2)^2}}{kT}}+1} d^3 p = \nonumber \\ 
& = & \frac{8\pi}{h^3} \int_0^{+\infty} \frac{p^2\sqrt{(pc)^2+(mc^2)^2}}{e^{\frac{\sqrt{(pc)^2+(mc^2)^2}}{kT}}+1} dp = \nonumber \\ 
& = & aT^4 {7\over4} {1\over A} \int_0^{+\infty} \frac{z^2\sqrt{z^2+(mc^2/kT)^2}}{e^{\sqrt{z^2+(mc^2/kT)^2}}+1} dz
\label{enel1}
\end{eqnarray}
where $z=pc/kT$ and the integral is computed numerically. Therefore the total proper internal energy density of the PEM-pulse, summing up all the contributions of photons and $e^+e^-$ pairs, is given by
\begin{equation}
\epsilon_{tot} = aT^4 \left[ 1+ {7\over4} {2\over A} \int_0^{+\infty} \frac{z^2\sqrt{z^2+(mc^2/kT)^2}}{e^{\sqrt{z^2+(mc^2/kT)^2}}+1} dz \right]
\label{entot1}
\end{equation}  
where the factor $2$ in front of the integral takes into account of electrons and positrons.\\
About the pressure of the photons it holds
\begin{equation}
p_{\gamma} = {\epsilon_{\gamma} \over 3} = {aT^4 \over 3}; 
\label{ppho}
\end{equation}
and about the pressure of electrons
\begin{eqnarray}
p_{e^-} & = & \frac{2}{3h^3} \int \frac{1}{e^{\frac{\sqrt{(pc)^2+(mc^2)^2}}{kT}} + 1}\cdot \frac{(pc)^2}{\sqrt{(pc)^2+(mc^2)^2}} d^3 p = \nonumber \\ 
& = & \frac{8\pi}{3h^3} \int_0^{+\infty} \frac{p^2}{e^{\frac{\sqrt{(pc)^2+(mc^2)^2}}{kT}} + 1}\cdot \frac{(pc)^2}{\sqrt{(pc)^2+(mc^2)^2}} dp = \nonumber \\ 
& = & {aT^4\over3} {7\over4} {1\over A} \int_0^{+\infty} \frac{z^4}{e^{\sqrt{z^2+(mc^2/kT)^2}} + 1}\cdot \frac{1}{\sqrt{z^2+(mc^2/kT)^2}} dz.
\label{pel}
\end{eqnarray}
Therefore the total pressure of PEM-pulse is given by
\begin{equation}
p_{tot}= {\frac{aT^4}{3}} \left[ 1+ {7\over4} {2\over A} \int_0^{+\infty} \frac{z^4}{e^{\sqrt{z^2+(mc^2/kT)^2}} + 1}\cdot \frac{1}{\sqrt{z^2+(mc^2/kT)^2}} dz \right] .
\label{ptot}
\end{equation}

\subsection{Numerical code}    

In the following we recall a zeroth order approximation of the fully relativistic equations of the previous section \cite{rswx99}:\\
(i) Since we are mainly interested in the expansion of the $e^+e^-$ plasma away from the black hole, we neglect the gravitational interaction.\\
(ii) We describe the expanding plasma by a special relativistic set of equations.\\

In the PEM-pulse phase the expansion in vacuum is described by a set of equation expressing:
\begin{itemize} 
\item entropy conservation, because of the assumption that emission of electromagnetic radiation is negligible up to transparency;
\item energy conservation, because the increase of kinetic energy is compensated by a decrease of total internal energy.
\end{itemize} 
For the expansion of a single shell, the adiabaticity is given by
\begin{equation}
{d\left(V\epsilon\right)}+p{dV}={dE}+p{dV}=0 ~,
\label{volume}
\end{equation}
where $V$ is the volume of the shell in the comoving frame and $E = V\epsilon$ is the total proper internal energy of plasma. By using the equation of state \ref{state} we find
\begin{equation}
d{ln{\epsilon}} + \Gamma d{lnV} = 0\,
\label{cons_s_diff}
\end{equation}
and, by integrating, we find
\begin{equation}
\frac{\bar\epsilon_{\circ}}{\bar\epsilon} = \left(\frac{V}{V_{\circ}}\right)^{\Gamma};
\label{scale1}
\end{equation} 
recalling that the volume of the fireball in the comoving frame is given by $V = {\mathcal V} {\bar\gamma}$, where ${\mathcal V}$ is the volume in the laboratory frame, we find
\begin{eqnarray}
{\bar\epsilon_\circ\over \bar\epsilon} &=& 
\left({V\over V_\circ}\right)^\Gamma=
\left({ {\mathcal V}\over  {\mathcal V}_\circ}\right)^\Gamma\left({\bar\gamma
\over \bar\gamma_\circ}\right)^\Gamma.
\label{scale} 
\end{eqnarray}   

The total energy conservation of the shell implies \cite{rswx99}:
\begin{equation}
(\Gamma\bar\epsilon) {\mathcal V} \bar\gamma^2 = (\Gamma\bar\epsilon_\circ) {\mathcal V}_\circ \bar\gamma_\circ^2;
\label{cons_e2}
\end{equation}
and this gives the evolution for $\bar\gamma$:
\begin{equation}
\bar\gamma = \bar\gamma_\circ\sqrt{{\bar\epsilon_\circ{\mathcal V}_\circ
\over\bar\epsilon{\mathcal V}}} 
\label{gammavar}
\end{equation}
Substituting this expression for $\bar\gamma$ in (\ref{scale}) the final equation for proper internal energy density is found
\begin{equation}
\bar\epsilon = {\bar\epsilon_{\circ}} \left({{\mathcal V}_{\circ} \over {\mathcal V}}\right)^{\frac{\Gamma}{2 - \Gamma}}
\label{scale2}
\end{equation}
The evolution of a plasma of $e^+e^-$ pairs and photons should be treated by relativistic hydrodynamics equations describing the variation of the number of particles in the process. The 4-vector number density of pairs is defined $(n_{e^{\pm}}U^{\mu})$, which in the comoving frame reduces to the 4-vector $(n_{e^{\pm}}, 0, 0, 0)$. The law of number conservation for pairs is
\begin{eqnarray}
\left(n_{e^\pm}U^\mu\right)_{;\mu} &=& \frac{1}{\sqrt {-g}} \left({\sqrt {-g}} n_{e^{\pm}} U^\mu\right)_{,\mu} = \nonumber \\
& = & \left(n_{e^\pm}U^t\right)_{,t}+{1\over r^2}\left(r^2 n_{e^\pm}U^r\right)_{,r} = 0
\label{rate}
\end{eqnarray}
where $g = \parallel g^{\mu \nu} \parallel = -r^4 sin^2\theta $ is the determinant of Reissner-Nordstr{\o}m metric. In the system processes of creation and annihilation of particles occur due to collisions between particles. If the number of particles is conserved, it holds $\left(n_{e^\pm}U^\mu\right)_{;\mu} = 0$; if instead it is not conserved, in the assumptions that only binary collisions between particles occur and in the hypothesis of molecular caos, the Eq.\eqref{rate} becomes
\begin{equation}
\left(n_{e^\pm}U^\mu \right)_{;\mu} = {\overline{\sigma v}} \left[n_{e^-}(T)n_{e^+}(T) - n_{e^-}n_{e^+}\right]
\label{rate1}
\end{equation}
where $\sigma$ is the cross section for the process of creation and annihilation of pairs, given by {}
\begin{equation}
\sigma = {\frac{\pi {r_e}^2}{\alpha + 1 }} \left[{\frac{\alpha^2 + 4 \alpha + 1}{\alpha^2 - 1}}ln\left(\alpha + \sqrt{\alpha^2 - 1}\right) - {\frac{\alpha + 3}{\sqrt{\alpha^2 - 1}}}\right],
\label{cross}
\end{equation}
with $\alpha = \frac{E}{mc^2}$ and $E$ total energy of positrons in the laboratory frame, and $r_e = \frac{e^2}{mc^2}$ the classical radius of electron, $v$ is the sound velocity in the fireball:
\begin{equation}
v = c \sqrt{\frac{p_{tot}}{\epsilon_{tot}}},
\label{vsound}
\end{equation}
and $\overline{\sigma v}$ is the mean value of $\sigma v$; for $\sigma$ we use as a first approximation the Thomson cross section, $\sigma_T = 0.665\cdot 10^{-24} cm^2$; $n_{e^{\pm}}(T)$ is the total proper number density of electrons and positrons in comoving frame at thermodynamic equilibrium in the process $e^+ + e^- \rightarrow \gamma + \gamma$ $\left(n_{e^-}(m, T) = n_{\gamma}(T)\right)$, $n_{e^{\pm}}$ is the total proper number density of electrons and positrons in comoving frame at a generic time before reaching the equilibrium.\\
Using the approximation of special relativity, the 4-velocity is written $U^\mu = (\bar\gamma, \bar\gamma \frac{v}{c})$; substituting to $n_{e^\pm}(T)$ the $\bar n_{e^\pm}(T)$ and to $n_{e^\pm}$ the $\bar n_{e^\pm}$, Eq.\eqref{rate1} in hybrid form becomes
\begin{equation}
{\partial \left({\bar n_{e^{\pm}} \bar\gamma}\right) \over \partial t} = - {1\over r^2} {\partial \over \partial r} \left(r^2 \bar n_{e^\pm} \bar\gamma V^r\right) + \overline{\sigma v} \left(\bar n^2_{e^\pm} (T) - \bar n^2_{e^\pm}\right) ,
\label{paira}
\end{equation}
valid for electrons and positrons.\\
Now we have a complete set of equations for numerical integration: (\ref{scale2}), (\ref{gammavar}) e la (\ref{paira}).

If we now turn from a single shell to a finite distribution of shells, we can introduce the average values of the proper internal energy and pair number densities ($\bar\epsilon, \bar n_{e^\pm}$) for the PEM-pulse, where the average $\bar\gamma$-factor is defined by
\begin{equation}
\bar\gamma={1\over{\mathcal V}}\int_{\mathcal V}\gamma(r) d{\mathcal V},
\label{ga}
\end{equation}
and ${\mathcal V}$ is the total volume of the shell in the laboratory frame \cite{rswx99}.

\begin{figure}
\centering
\includegraphics[width=15cm]{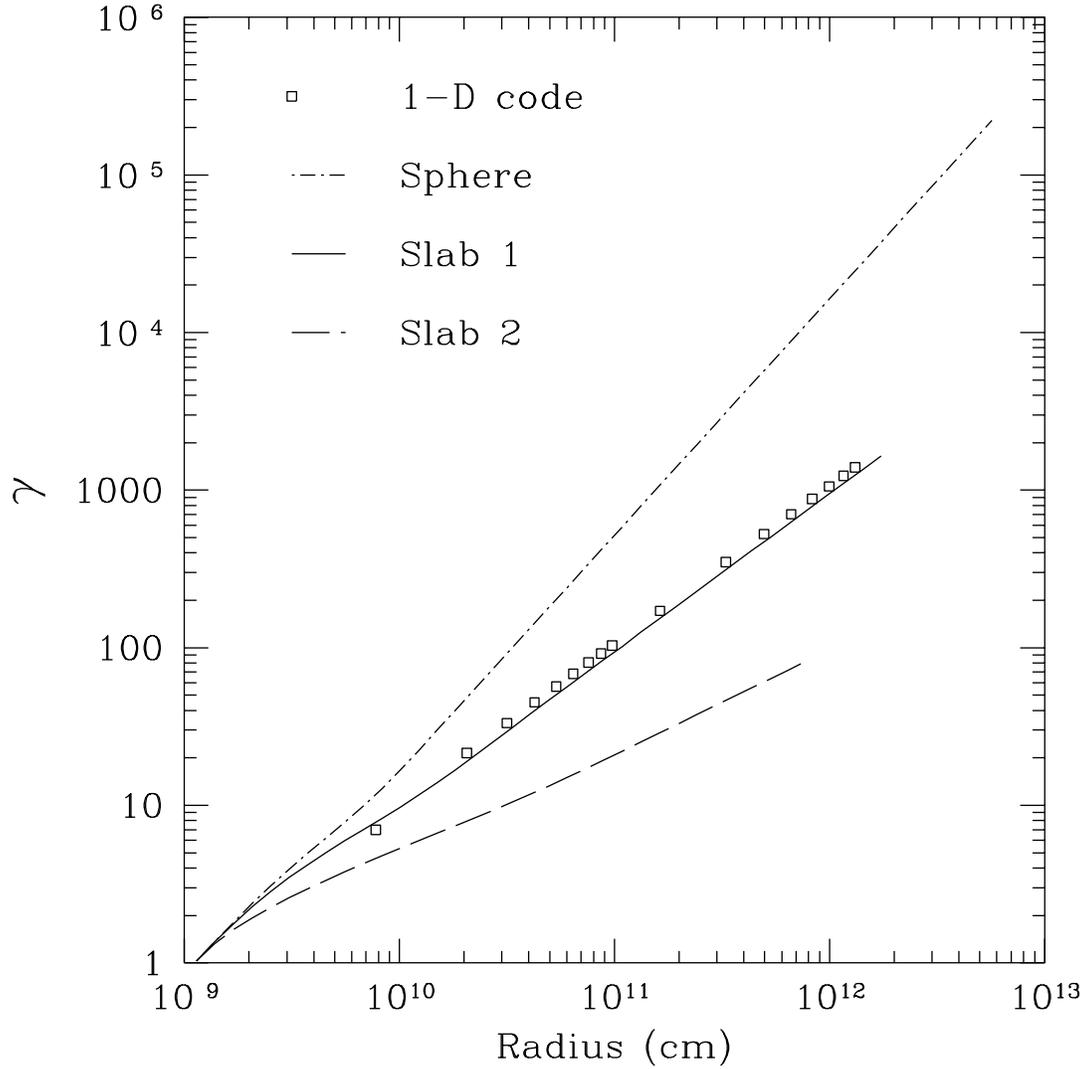}
\caption{{\itshape Lorentz $\bar\gamma$ factor as a function of radial coordinate. Three schemes of expansion of PEM-pulse (see text) are compared with solution of hydrodynamics relativistic equations numerically integrated for a black hole with $\mu = 10^3$ and $\xi = 0.1$. The result is in accordance with the scheme of a fireball with constant thickness in laboratory frame}.}
\label{shells}
\end{figure}

In principle we could have an infinite number of possible schemes to define geometry of the expanding shell. Three different possible schemes have been proposed \cite{rswx99}:
\begin{itemize}
\item Sphere. An expansion with radial component of 4-velocity proportional to the distance to the black hole $\displaystyle U_r (r) = U \frac{r}{{\mathcal R}(t)}$, where $U$ is the radial component of 4-velocity on the external surface of PEM-pulse (having radius ${\mathcal R}(t)$), the factor $\bar\gamma$ from (\ref{ga}) is 
\begin{equation}
\bar\gamma = \frac{3}{8 U^3}\left[2U\left(1 + U^2\right)^{3 \over 2} - U\left(1 + U^2\right)^{1 \over 2} - ln\left(U + \sqrt{1 + U^2}\right)\right];
\label{gammamean_1sch}
\end{equation}
this distribution corresponds to a uniform and time decreasing density, like in Friedmann model for the universe;
\item Slab 1. An expansion with thickness of fireball constant ${\mathcal D} = r_{ds} - r_+$ in laboratory frame in which the black hole is at rest, with $U_r (r) = U_r = cost$ and $\bar\gamma = \sqrt{1 + {U_r}^2 }$; this distribution does not require an average;
\item Slab 2. An expansion with thickness of fireball constant in comoving frame of PEM-pulse.
\end{itemize} 
The result has been compared with the one of hydrodynamic equation in general relativity \cite{rswx99} (see Fig. \ref{shells}). Excellent agreement has been found with the scheme in which the thickness of fireball is constant in laboratory frame: what happens is that the thickness in comoving frame increases, but due to the Lorentz contraction, it is kept constant in laboratory frame and equal to ${\mathcal D}=\left(r_{\rm ds}-r_{+}\right)$. In this case $U_r = \sqrt{\bar\gamma^2 - 1}$, where $\bar\gamma$ is computed by conservation equations.\\
A similar situation occurs for the temperature of PEM-pulse. In the comoving frame the temperature decreases as $T'\sim{R^{-1}}$, in accordance with results in literature \cite{p99}. Since $\gamma$ monotonically increases as $\gamma\sim{R}$ \cite{lett1}, in laboratory frame $T=\gamma{T'}\sim{constant}$ \cite{rswx00}; photons are blue-shifted in laboratory frame in such away that, at least in the first phase, the temperature measured by an observer at infinity is constant. The numerical value of the temperature of equilibrium at each instant is found by imposing the equivalence, within a certain precision, of (\ref{entot1}) numerically computed and (\ref{scale2}).

Even if the PEM-pulse is optically thick in the expansion before transparency, photons located at a distance from the external surface less their mean free path can escape and reach the observer at infinity. The mean free path in the comoving frame is given by
\begin{equation}
L_{\gamma} = \frac{1}{\sigma{n_{e^+e^-}}}  \sim  10^{-6}cm 
\label{lambda2}
\end{equation}
while in laboratory frame is given by $\lambda = {L_{\gamma} / \bar\gamma} \sim 10^{-8}$cm.
However the luminosity emitted at this stage is negligible, since the ratio between $\lambda$ and the thickness of the fireball ${\mathcal D}$ in the laboratory frame (with ${\mathcal D}= (r_{ds}- r_+) \sim 10^{9}$cm) is of the order of $\lambda / {\mathcal D} \simeq 10^{-17}$.

\subsection{Era II: interaction of the PEM pulse with remnant}\label{era2}\

The PEM pulse expands initially in a region of very low baryonic contamination created  by the process of gravitational collapse. As it moves outside the baryonic remnant of the progenitor star is swept up.
The existence of such a remnant is necessary in order to guarantee the overall charge neutrality of the system: the collapsing core has the opposite charge of the remnant and the system as a whole is clearly neutral. The number of extra charges in the baryonic remnant negligibly affects the overall charge neutrality of the PEM pulse.

The baryonic matter remnant is assumed to be distributed well outside the dyadosphere in a shell of thickness $\Delta$ between an inner radius $r_{\rm in}$ and an outer radius $r_{\rm out}=r_{\rm in}+\Delta$ at a distance from the black hole not so big that the PEM pulse expanding in vacuum has not yet reached transparency and not so small that the system will reach enoughly high value of Lorentz $\gamma$ in order to not be stopped in the collision (see Fig. \ref{cip2}). For the sake of an example we choose
\begin{equation}
r_{\rm in}=100r_{\rm ds},\hskip 0.5cm \Delta = 10r_{\rm ds}.
\label{bshell_1}
\end{equation}
The total baryonic mass $M_B=N_Bm_p$ is assumed to be a fraction of the dyadosphere initial total energy $(E_{\rm dya})$. The total baryon-number $N_B$ is then expressed as a function of the dimensionless parameter $B$ given by 
\begin{equation}
B=\frac{N_Bm_pc^2}{E_{\rm dya}}\, ,
\label{chimical1}
\end{equation}
where $B$ is a parameter in the range $10^{-8}-10^{-2}$ and $m_p$ is the proton mass. We shall see below the role of $B$ in the determination of the features of the GRBs. We saw in section \ref{fp} the sense in which $B$ and $E_{dya}$ can be considered to be the only two free parameters of the black hole theory for the entire GRB family, the so called ``long bursts''. For the so called ``short bursts'' the black hole theory depends on the two other parameters $\mu$, $\xi$, since in that case $B=0$ since most of the energy, unless the whole energy, in the pulse is emitted at transparency. 
The baryon number density $n^\circ_B$ is assumed to be a constant
\begin{equation}
\bar n^\circ_B={N_B\over V_B},\hskip0.5cm \bar\rho^\circ_B=m_p\bar n^\circ_B c^2.
\label{bnd}
\end{equation}
 
As the PEM pulse reaches the region $r_{\rm in}<r<r_{\rm out}$, it interacts with the baryonic matter which is assumed to be at rest. In our model we make the following assumptions to describe this interaction: 
\begin{itemize}
\item the PEM pulse does not change its geometry during the interaction;
\item the collision between the PEM pulse and the baryonic matter is assumed to be inelastic,
\item the baryonic matter reaches thermal equilibrium with the photons and pairs of the PEM pulse.
\end{itemize}
These assumptions are valid if: (i) the total energy of the PEM pulse is much larger than the total mass-energy of baryonic matter $M_B$, $10^{-8}<B<10^{-2}$, (ii) the ratio of the comoving number density  of pairs and baryons at the moment of collision $n_{e^+e^-}/n^\circ_B$ is very high (e.g., $10^6 <n_{e^+e^-}/ n^\circ_B <10^{12}$) and (iii)  the PEM pulse has a large value of the gamma factor ($100<\bar\gamma $).  
  
In the collision between the PEM pulse and the baryonic matter at $r_{\rm out}>r>r_{\rm in}$ , we impose total conservation of energy and momentum. We consider the collision process between two radii $r_2,r_1$ satisfying 
$r_{\rm out}>r_2>r_1>r_{\rm in}$ and $r_2-r_1\ll \Delta$. The amount of baryonic mass acquired by the PEM pulse is
\begin{equation}
\Delta M = {M_B\over V_B}{4\pi\over3}(r_2^3-r_1^3) ,
\label{mcc_2}
\end{equation}
where $M_B/ V_B$ is the mean-density of baryonic matter at rest.

As for energy density of dyadosphere, here also we choose a simplification for the energy density: in fact during the passage of the shell a deposition of material on the external surface of the fireball creates; however we neglected this effect and assumed that this material after collision diffuses instantaneously in the pulse with a constant density:
\begin{equation}
n'_B={N'_B\over V},
\label{nB}
\end{equation}   
where $N'_B$ is the number of particle of the remnant shell swept up by the pulse and $V$ is the comoving volume of the fireball.

The conservation of total energy leads to the estimate of the corresponding quantities before (with ``$\circ$'') and after such a collision  
\begin{equation}
(\Gamma\bar\epsilon_\circ + \bar\rho^\circ_B)\bar\gamma_\circ^2{\mathcal V}_\circ + \Delta M = (\Gamma\bar\epsilon + \bar\rho_B + {\Delta M\over V} + \Gamma\Delta\bar\epsilon)\bar\gamma^2{\mathcal V},
\label{ecc_2}
\end{equation}
where $\Delta\bar\epsilon$ is the corresponding increase of internal energy due to the collision. Similarly the momentum-conservation gives
\begin{equation}
(\Gamma\bar\epsilon_\circ + \bar\rho^\circ_B)\bar\gamma_\circ U^\circ_r{\mathcal V}_\circ = (\Gamma\bar\epsilon + \bar\rho_B + {\Delta M\over V} + 
\Gamma\Delta\bar\epsilon)\bar\gamma U_r{\mathcal V},
\label{pcc_2}
\end{equation}
where the radial component of the four-velocity of the PEM pulse is $U^\circ_r=\sqrt{\bar\gamma_\circ^2-1}$ and $\Gamma$ is the thermal index. 
We then find 
\begin{eqnarray}
\Delta\bar\epsilon & = & {1\over\Gamma}\left[(\Gamma\bar\epsilon_\circ + \bar\rho^\circ_B) {\bar\gamma_\circ U^\circ_r{\mathcal V}_\circ \over \bar\gamma U_r{\mathcal V}} - (\Gamma\bar\epsilon + \bar\rho_B + {\Delta M\over V})\right],\label{heat_2}\\
\bar\gamma & = & {a\over\sqrt{a^2-1}},\hskip0.5cm a\equiv {\bar\gamma_\circ  \over  
U^\circ_r}+ {\Delta M\over (\Gamma\bar\epsilon_\circ + \bar\rho^\circ_B)\bar\gamma_\circ U^\circ_r{\mathcal V}_\circ}.
\label{dgamma_2}
\end{eqnarray}
These equations determine the gamma factor $\bar\gamma$ and the internal energy density $\bar\epsilon=\bar\epsilon_\circ +\Delta\bar\epsilon$ in the capture process of baryonic matter by the PEM pulse.

The effect of the collision of the PEM pulse with the remnant leads to the following consequences:
\begin{itemize}
	\item a reheating of the plasma in the comoving frame but not in the laboratory frame; an increase of the number of 					$e^+e^-$ pairs and of free electrons originated from the ionization of those atoms remained in the baryonic 						remnant; correspondingly this gives an overall increase of the opacity of the pulse; 
	\item the more the amount of baryonic matter swept up, the more internal energy of the PEMB pulse is converted in 						kinetic energy of baryons.
\end{itemize}

By describing the interaction of PEM pulse with remnant as completely inelastic collision of two particles, one can compute by the energy-momentum conservation equation the decrease of Lorentz $\gamma$ and the increase of internal energy as function of $B$ parameter and also the ultrarelativistic approximation ($\gamma_\circ \rightarrow \infty$):
\begin{enumerate}
	\item an abrupt decrease of the gamma factor given by
				\begin{displaymath}
				\gamma_{coll} = \gamma_\circ \frac{1+B}{\sqrt{ {\gamma_\circ}^2 \left(2B+B^2 \right) +1}} 															\longrightarrow_{\gamma_\circ \rightarrow \infty}  \frac{B+1}{\sqrt{B^2+2B}}\, ,
				\label{gamma_circ}
				\end{displaymath}
				where $\gamma_\circ$ is the gamma factor of the PEM pulse before the collision,
	\item an increase of the internal energy in the comoving frame $E_{coll}$ developed in the collision given by
				\begin{displaymath}
				\frac{E_{coll}}{E_{dya}} =  \frac{\sqrt{ {\gamma_\circ}^2 \left(2B+B^2 \right) +1}}{\gamma_\circ} - 										\left(\frac{1}{\gamma_\circ} + B \right) \longrightarrow_{\gamma_\circ \rightarrow \infty} 	-B+\sqrt{B^2+2B} \, ,
				\label{E_int/E}
				\end{displaymath}
\end{enumerate}
This approximation applies when the final gamma factor at the end of the PEM pulse era is larger than $\gamma_{coll}$, right panel in Fig. \ref{cip2}.

In this phase of expansion, another thermodynamic quantity has not been considered: the chemical potential $\mu$ of the electrons from ionization of baryonic remnant. We remind that the total proper number density of electrons of ionization is given by
\begin{equation}
{n^b_{e^-}}(m, T, \mu) = \frac{aT^3}{k} {7\over8} {1\over A} \int_0^{+\infty} \frac{z^2}{e^{\sqrt{z^2+(mc^2/kT)^2} + \frac{\mu}{kT}}+1} dz\,
\label{num_pote1}
\end{equation}
four equations are imposed to find a formula useful for numerical computation: the first one is the thermodynamical equilibrium of fireball, or
\begin{equation}
\bar n_{e^{\pm}} ({T_{\circ}}) = \bar n_{\gamma} ({T_{\circ}}) ;
\label{equil}
\end{equation} 
the second one is
\begin{equation}
\bar n^b_{e^-} = \bar Z n_B
\label{neb}
\end{equation} 
where $1/2<\bar Z<1$, with $\bar Z=1$ for hydrogen atoms and $\bar Z=1/2$ for baryonic matter in general; the third one derives from the definition of $B$, and states a relation between the two densities $\bar n_B$ and $\bar n_{e^\pm}$: from definition of $B$, we have
\begin{equation}
\frac{N_B}{N_{e^{\pm}}(T_{\circ})} = B \frac{E_{dia}}{m_pc^2} \frac{1}{N_{e^{\pm}}(T_{\circ})} = 10^b 
\label{def_di_b_1}
\end{equation}
where $T_{\circ}$ is the initial temperature of fireball and $b$ is a parameter ($b < 0$) defined by (\ref{def_di_b_1}); so if $V_{\circ}$ is the initial volume of dyadosphere and $w$ the initial volume of the baryonic shell 
\begin{equation}
\bar n^\circ_B = 10^b \bar n_{e^\pm}(T_{\circ}) \frac {V_{\circ}}{w} ;  
\label{def_di_b}
\end{equation}
finally the fourth one is the conservation law of baryonic matter
\begin{equation}
(n^b_{e^-}U^\mu)_{;\mu} = 0.
\label{fourth}
\end{equation}
Therefore the chemical potential $\mu$ is numerically determined at a certain time of expansion if the initial temperature $T_{\circ}$ of fireball and the initial volume of baryonic shell $w$ are known and, at that time, the volume $V$, the temperature $T$ and the Lorentz factor $\bar\gamma$ of the fireball, the volume of the baryonic shell swept up $vb$ and the ratio $\displaystyle {\bar n^b_{e^-}(T) \over \bar n^b_{e^-}}$:
\begin{equation}
2 \zeta (3) {\bar Z  10^b {\bar n^b_{e^-}(T) \over \bar n^b_{e^-}} {{{T_0}^3 w} \over {T^3 V \bar\gamma}} \left({vb \over w}\right)} =
 \int_0^{+\infty} \frac{z^2}{e^{\sqrt{z^2+(mc^2/kT)^2} + \frac{\mu}{kT}}+1} dz
\label{pot_chi_fin}
\end{equation}
where the factor in brackets $\left({vb \over w}\right)$ must be considered only for $r > r_{\rm out}$, while the proportionality factor is the function zeta of Riemann $\zeta (x)$ for computation of $n_{\gamma}$, with $\zeta (3) = 1.202$.\\ 
Therefore the equations for this phase are (\ref{heat_2}), (\ref{dgamma_2}), (\ref{nB}), (\ref{paira}) and (\ref{pot_chi_fin}).

\subsection{Era III: expansion of PEMB pulse}\label{era3}

After the engulfment of the baryonic matter of the remnant the plasma formed of $e^+e^-$-pairs, electromagnetic radiation and baryonic matter expands again as a sharp pulse, namely the PEMB pulse. The calculation is continued as the plasma fluid expands,
cools and the $e^+e^-$ pairs recombine until it becomes optically
thin:
\begin{equation} 
\int_R dr(n_{e^\pm}+\bar
Zn_B)\sigma_T\simeq O(1),
\label{thin_1}
\end{equation}
where $\sigma_T =0.665\cdot 10^{-24}
{\rm cm^2}$ is the Thomson cross-section and the integration is over the radial interval of the PEMB pulse in the
comoving frame. 
In order to study the PEMB pulse expansion the validity of the slab approximation adopted for the PEM pulse phase has to be verified; otherwise the full hydrodynamics relativistic equations should be integrated. The PEMB pulse evolution firstly has been simulated by integrating the general relativistic hydrodynamical equations with the Livermore codes, for a total energy in the dyadosphere of $3.1\times 10^{54}$ erg and a baryonic shell of thickness $\Delta =10 r_{\rm ds}$ at rest at a radius of $100 r_{\rm ds}$ and $B\simeq 1.3\cdot 10^{-4}$. 

In analogy with the special relativistic treatment for the PEM pulse, presented in section~\ref{era1} (see also \cite{rswx99}), for the adiabatic expansion of the PEMB pulse in the constant-slab approximation described by the Rome codes the following hydrodynamical equations with $\bar\rho_B\not=0$ has been found
\begin{eqnarray}
{\bar n_B^\circ\over \bar n_B}&=& { V\over  V_\circ}={ {\mathcal V}\bar\gamma
\over {\mathcal V}_\circ\bar\gamma_\circ},
\label{be'}\\
{\bar\epsilon_\circ\over \bar\epsilon} &=& 
\left({V\over V_\circ}\right)^\Gamma=
\left({ {\mathcal V}\over  {\mathcal V}_\circ}\right)^\Gamma\left({\bar\gamma
\over \bar\gamma_\circ}\right)^\Gamma,
\label{scale1'}\\
\bar\gamma &=&\bar\gamma_\circ\sqrt{{(\Gamma\bar\epsilon_\circ+\bar\rho^\circ_B){\mathcal V}_\circ
\over(\Gamma\bar\epsilon+\bar\rho_B) {\mathcal V}}},
\label{result1'}\\
{\partial \over \partial t}(N_{e^\pm}) &=& -N_{e^\pm}{1\over{\mathcal V}}{\partial {\mathcal V}\over \partial t}+\overline{\sigma v}{1\over\bar\gamma^2}  (N^2_{e^\pm} (T) - N^2_{e^\pm}).
\label{paira'_2}
\end{eqnarray}
In these equations ($r>r_{\rm out}$) the comoving baryonic mass and number densities are $\bar\rho_B=M_B/V$ and $\bar n_B=N_B/V$, where $V$ is the comoving volume of the PEMB pulse.

\begin{figure}
\centering
\includegraphics[width=15cm]{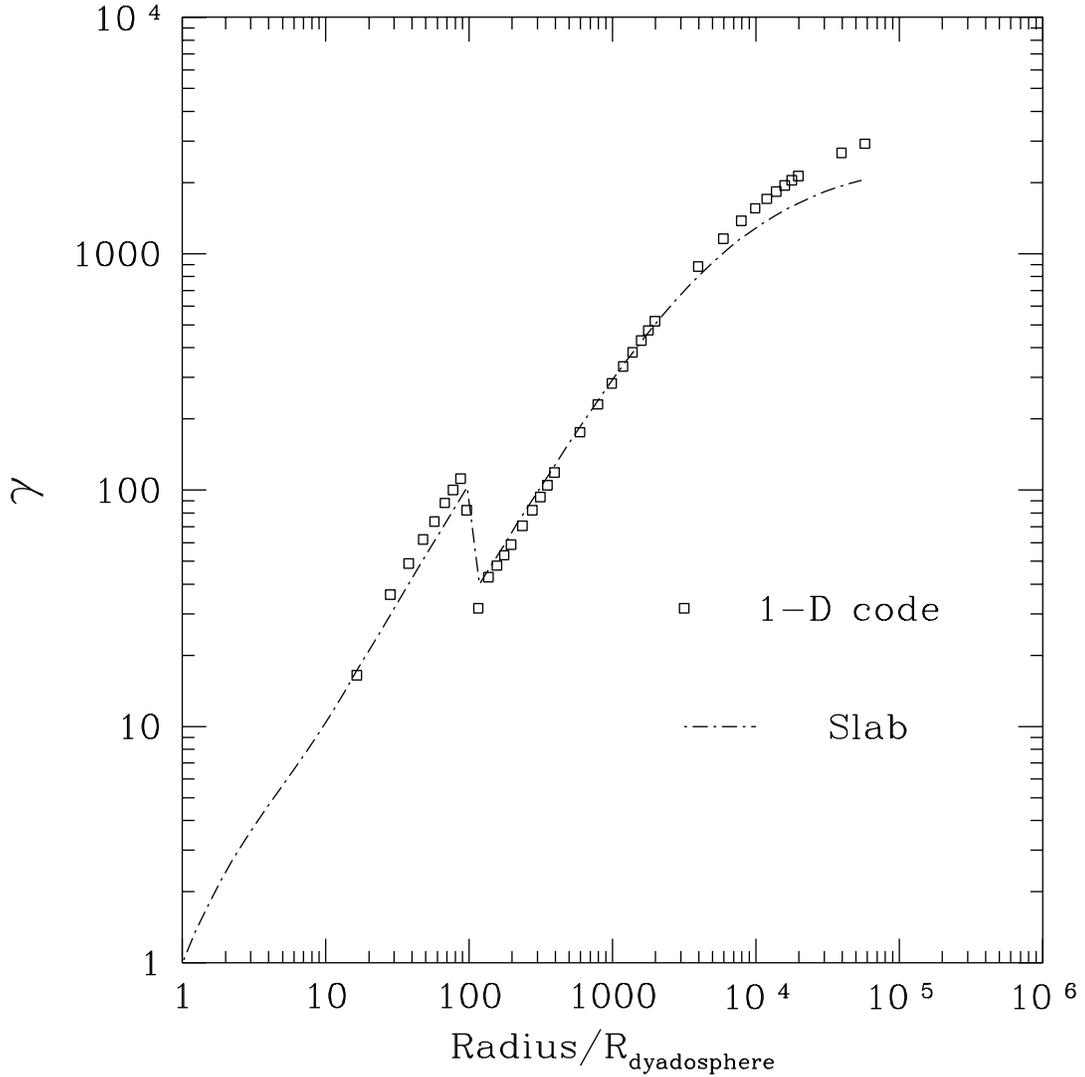}
\caption{{\itshape Lorentz $\bar\gamma$ factor as a function of radial coordinate from the PEMB-pulse simulation is compared with the $\bar\gamma$ factor as solution of hydrodynamics relativistic equations numerically integrated (open squares) for $E_{dya} = 3.1\times 10^{54}$erg  and $B=1.3\times 10^{-4}$, $r_{\rm in}=100r_{\rm ds}$ and $\Delta = 10r_{\rm ds}$. The result is in accordance with the scheme of a fireball with constant thickness in laboratory frame which is valid up to $B= 10^{-2}$}.}
\label{shells_2}
\end{figure}

The result is shown in Fig. \ref{shells_2} \cite{rswx00} where the bulk gamma factor as computed from the Rome and Livermore codes are compared and very good agreement has been found. This validates the constant-thickness approximation in the case of the PEMB pulse as well. On this basis we easily estimate a variety of physical quantities for an entire range of values of $B$.

For the same black hole different cases have been considered \cite{rswx00}. The results of the integration show that for the first parameter range the PEMB pulse propagates as a sharp pulse of constant thickness in the laboratory frame, but already for $B\simeq 1.3\cdot 10^{-2}$ the expansion of the PEMB pulse becomes much more complex, turbulence phenomena can not be neglected any more and the constant-thickness approximation ceases to be valid.

\begin{figure}
\centering
\includegraphics[width=7.3cm]{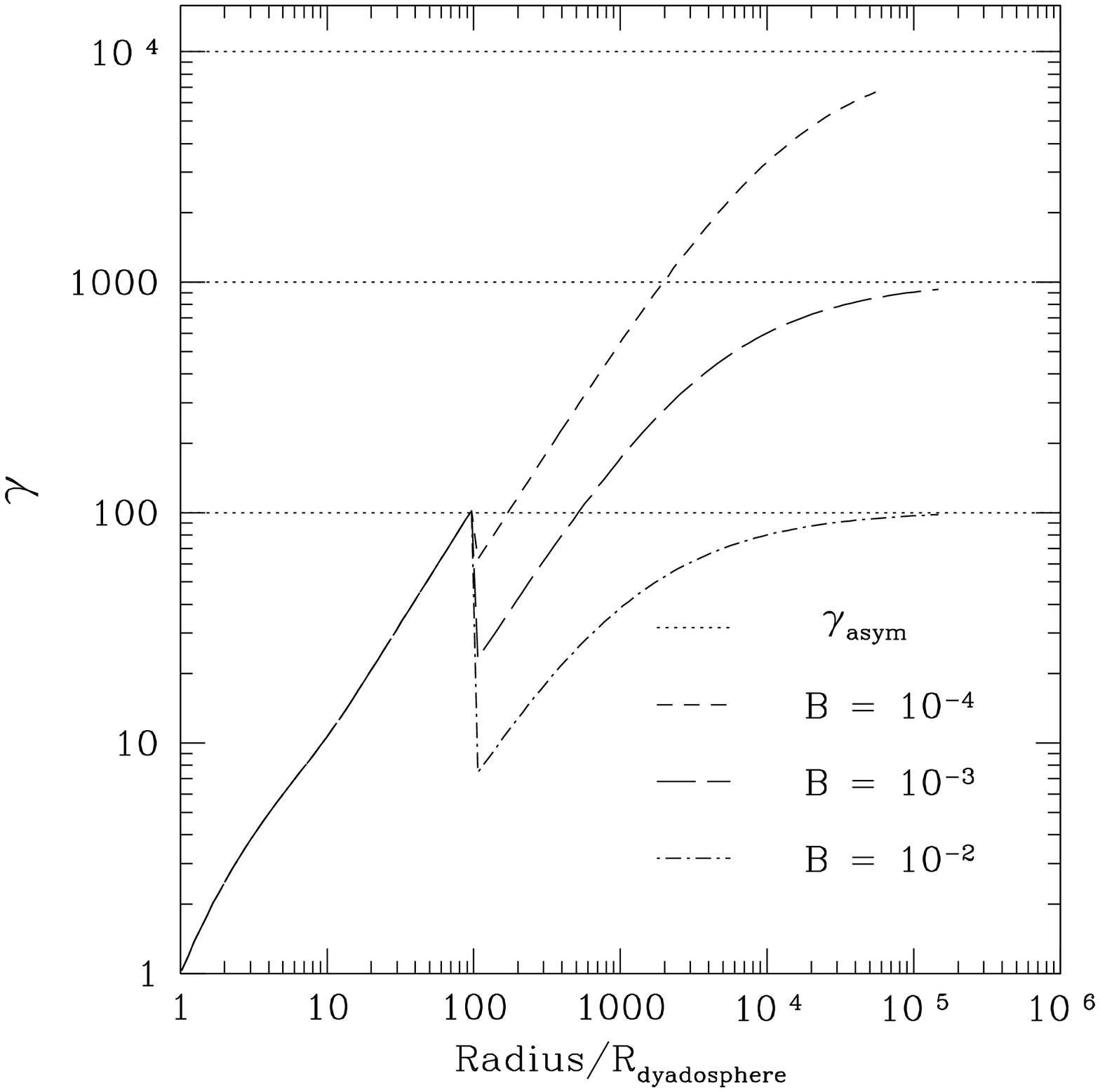}
\includegraphics[width=8.5cm]{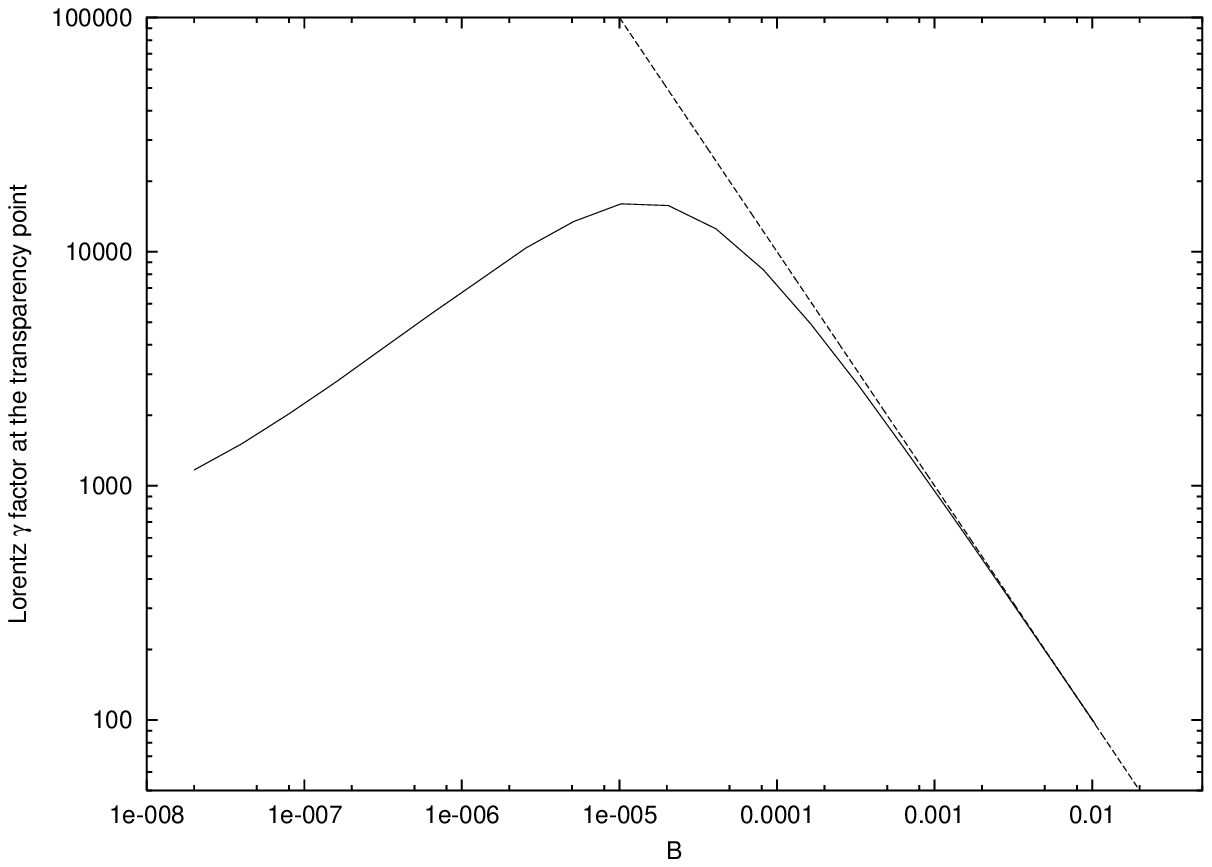}
\caption{{\itshape {\bf Left)} The gamma factors are given as functions of the radius in units of the dyadosphere radius for selected values of $B$ for the typical case $E_{dya}=3.1\times 10^{54}$ erg. The asymptotic values $\gamma_{\rm asym} = E_{\rm dya}/(M_Bc^2)=10^4,10^3,10^2$ are also plotted. The collision of the PEM pulse with the baryonic remnant occurs at $r/r_{ds}=100$ where the jump occurs.{\bf Right)} The $\gamma$ factor (the solid line) at the transparency point is plotted as a function of the $B$ parameter. The asymptotic value (the dashed line) $E_{\rm dya}/ (M_Bc^2)$ is also plotted}.}
\label{3gamma}
\end{figure}

It is also interesting to evaluate the final value of the gamma factor of the PEMB pulse when the transparency condition given by Eq.(\ref{thin_1}) is reached as a function of $B$, see Fig. \ref{3gamma}. For a given black hole, there is a {\em maximum} value of the gamma factor at transparency. By further increasing the value of $B$ the entire $E_{dya}$ is transferred into the kinetic energy of the baryons (see also~\cite{rswx00}).

In Fig. \ref{3gamma}Left we plot the gamma factor of the PEMB pulse as a function of radial distance for different amounts of baryonic matter. The diagram extends to values of the radial coordinate at which the transparency condition given by Eq.(\ref{thin_1}) is reached. The ``asymptotic'' gamma factor
\begin{equation}
\bar\gamma_{\rm asym}\equiv {E_{\rm dya}\over M_B c^2}
\label{asymp}
\end{equation}
is also shown for each curve. The closer the gamma value approaches the ``asymptotic'' value (\ref{asymp}) at transparency, the smaller the intensity of the radiation emitted in the burst and the larger the amount of kinetic energy left in the baryonic matter (see Fig. \ref{3gamma}Right).

\section{On Urca process}\label{gamow}

From G. Gamow \cite{g70}:

``The summer of 1939 I spent with my family vacation on the Copacabana beach in Rio de Janeiro. One evening, visiting the famous Casino da Urca to watch the gamblers, I was introduced to a young theoretical physicist born on an Amazon River plantation, named Mario Schoenberg. We became friends, and I arranged for him a Guggenheim fellowship to spend a year in Washington to work with me in nuclear astrophysics. His visit was very successful, and we hit upon a process which could be responsible for the vast stellar explosions known as supernovae. The trick is done by alternative absorption and reemission of one of the thermal electrons in the very hot (billions of degrees!) stellar interior by various atomic nucleai. Both processes are accompanied by the emission of neutrinos and antineutrinos which, possessing tremendous penetrating power, pass through the body of a star like a swarm of musquitoes through chicken wire and carry with them large amount of energy. Thus, the stellar interior cools rapidly, the pressure drops, and the stellar body collapse with a great explosion of light and heat.

All this is too complicate to explain in non technical words, and I am mentioning it only as background for how we came to give that process is name. We called it the Urca process, partially to commemorate the casino in which we first met, and partially because the Urca process results in a rapid disappearance of thermal energy from the interior of the star, similar to the rapid disappearance of money from the pockets of the gamblers of the Casino d Urca. Sending our article ``On the Urca process'' for publication on the {\em physical review} I was worried that the Editor would ask why we called the process ``Urca''. After much thought I decided to say that this is short for ``UnRecordable Cooling Agent'', but they never asked. Today, there are other known cooling processes involving neutrinos which work even faster than the Urca process. For example, a neutrino pair can be formes instead of two gamma quanta in the annihilation of a positive and negative electrons''.

\section{Casino de Urca today}\label{urca}

\includegraphics[width=\hsize,clip]{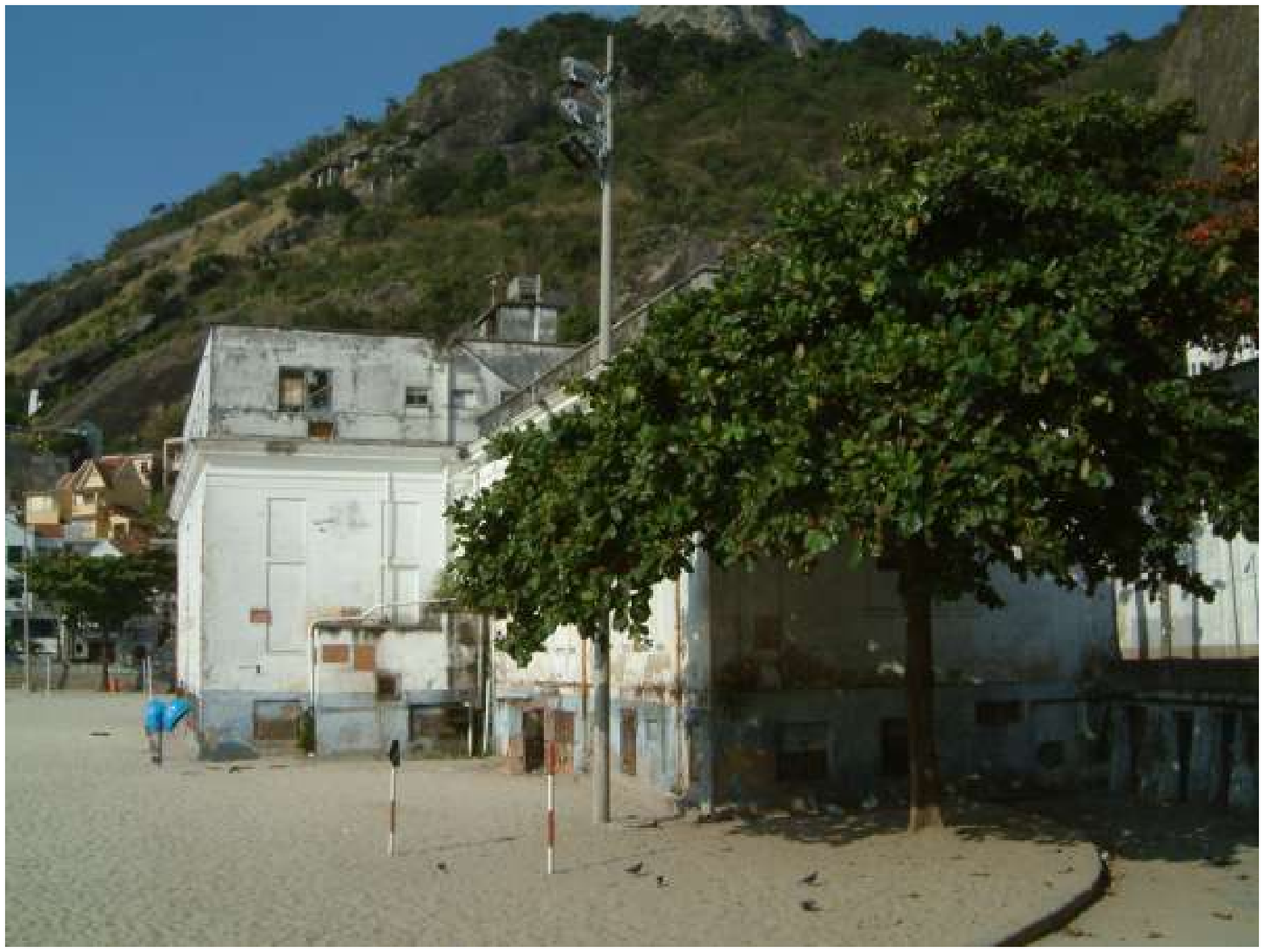}

\end{document}